\pdfoutput=1
\documentclass[apj]{emulateapj}
\slugcomment{Submitted to {\it The Astrophysical Journal}}
\shortauthors{Kirkpatrick et al.}
\shorttitle{AllWISE2 Motion Survey}

\usepackage{color,amsmath,longtable}

\begin{document}

\title{The AllWISE Motion Survey, Part 2}

\author{J.\ Davy Kirkpatrick\altaffilmark{1},
Kendra Kellogg\altaffilmark{1,2},
Adam C.\ Schneider\altaffilmark{3},
Sergio Fajardo-Acosta\altaffilmark{1},
Michael C.\ Cushing\altaffilmark{3},
Jennifer Greco\altaffilmark{3},
Gregory N.\ Mace\altaffilmark{4},
Christopher R.\ Gelino\altaffilmark{1},
Edward L.\ Wright\altaffilmark{5},
Peter R.\ M.\ Eisenhardt\altaffilmark{6},
Daniel Stern\altaffilmark{6},
Jacqueline K.\ Faherty\altaffilmark{7},
Scott S.\ Sheppard\altaffilmark{7},
George B.\ Lansbury\altaffilmark{8},
Sarah E.\ Logsdon\altaffilmark{5},
Emily C.\ Martin\altaffilmark{5},
Ian S.\ McLean\altaffilmark{5},
Steven D.\ Schurr\altaffilmark{1},
Roc M.\ Cutri\altaffilmark{1},
Tim Conrow\altaffilmark{1}
}

\altaffiltext{1}{Infrared Processing and Analysis Center, MS 100-22, California Institute of Technology, Pasadena, CA 91125, USA; davy@ipac.caltech.edu}
\altaffiltext{2}{Department of Physics and Astronomy, Western University, 1151 Richmond Avenue, London, ON N6A 3K7, Canada}
\altaffiltext{3}{Department of Physics and Astronomy, MS 111, University of Toledo, 2801 W. Bancroft St., Toledo, OH 43606-3328, USA}
\altaffiltext{4}{Department of Astronomy, University of Texas at Austin, Austin, TX 78712, USA}
\altaffiltext{5}{Department of Physics and Astronomy, UCLA, 430 Portola Plaza, Box 951547, Los Angeles, CA, 90095-1547, USA}
\altaffiltext{6}{Jet Propulsion Laboratory, California Institute of Technology, 4800 Oak Grove Drive, Pasadena, CA 91109, USA}
\altaffiltext{7}{Department of Terrestrial Magnetism, Carnegie Institution of Washington, Washington, DC 20015, USA}
\altaffiltext{8}{Department of Physics, Durham University, Durham DH1 3LE, UK}

\begin{abstract}

We use the AllWISE Data Release to continue our search for {\it WISE}-detected motions. In this paper, we publish another 27,846 motion objects, bringing the total number to 48,000 when objects found during our original AllWISE motion survey are included. We use this list, along with the lists of confirmed {\it WISE}-based motion objects from the recent papers by Luhman and by Schneider et al.\ and candidate motion objects from the recent paper by Gagn\'e et al.\ to search for widely separated, common-proper-motion systems. We identify 1,039 such candidate systems. All 48,000 objects are further analyzed using color-color and color-mag plots to provide possible characterizations prior to spectroscopic follow-up. We present spectra of 172 of these, supplemented with new spectra of 23 comparison objects from the literature, and provide classifications and physical interpretations of interesting sources. Highlights include: (1) the identification of three G/K dwarfs that can be used as standard candles to study clumpiness and grain size in nearby molecular clouds because these objects are currently moving behind the clouds, (2) the confirmation/discovery of several M, L, and T dwarfs and one white dwarf whose spectrophotometric distance estimates place them 5-20 pc from the Sun, (3) the suggestion that the \ion{Na}{1} `D' line be used as a diagnostic tool for interpreting and classifying metal-poor late-M and L dwarfs, (4) the recognition of a triple system including a carbon dwarf and late-M subdwarf, for which model fits of the late-M subdwarf (giving [Fe/H]${\approx}-$1.0) provide a measured metallicity for the carbon star, and (5) a possible 24-pc-distant K5 dwarf + peculiar red L5 system with an apparent physical separation of 0.1 pc.

\end{abstract}


\section{Introduction}

The utility of motion surveys using data from the NASA {\it Wide-field Infrared Survey Explorer} ({\it WISE}; \citealt{wright2010}).
has been described at length in \cite{kirkpatrick2010} and \cite{schneider2016}. Because of its repeated observations of the entire sky, {\it WISE} is ideally suited to producing a catalog of motion sources detectable in the wavelength regimes covered by its four bands: W1 (3.4 $\mu$m), W2 (4.6 $\mu$m), W3 (12 $\mu$m), and W4 (22 $\mu$m). Most of the repeated sky coverage of {\it WISE} is not in the W3 and W4 bands, which became saturated due to cryogen exhaustion partway through {\it WISE}'s second pass of the sky, but in the W1 and W2 bands, which are largely unaffected by the cryogen loss. The W1 band probes deeply, enabling it to detect stars to great distances. The AllWISE $>$95\% completeness depth of W1 = 17.1 mag at high Galactic latitudes (\citealt{cutri2013}) means that early-L dwarfs can be detected to $\sim$250 pc, early-M dwarfs to $\sim$3 kpc, and earlier-type stars to even greater distances. The W2 band is ideally suited for the detection of cooler objects, since these objects emit their peak energies at this band. In fact, {\it WISE} is capable of detecting Y-type brown dwarfs down to at least $T_{eff}$ = 250K (\citealt{luhman2014b, wright2014}). Hence, the repeated coverage at W1 and W2 makes {\it WISE} an efficient search tool for moving stars and brown dwarfs of all types in the Solar Neighborhood.

For most spots on the sky, {\it WISE} imaging is confined to a span of a few days, with additional coverages possible six months later when that portion of the sky is again visible to the satellite. (At the ecliptic poles, the coverage is nearly continuous because the satellite sees the poles on every orbit.) W1 and W2 data from the original mission have a time baseline of six months for the 80\% of the sky covered twice and a full year for the remaining 20\% that was covered three times. Both the \cite{luhman2014} motion survey and the \cite{kirkpatrick2014} motion survey (the latter hereafter referred to as the ``AllWISE1 Motion Survey'', or just ``AllWISE1'') used this same underlying data set from the original {\it WISE} mission. 

The \cite{luhman2014} survey started with source extractions on single exposure images from the {\it WISE} All-Sky, 3-Band Cryo, and Post-Cryo Releases. The individual astrometric measurements were combined by \cite{luhman2014} into per-epoch measurements to identify objects with significant motion over the time baseline. The entire sky was searched except for small areas near both ecliptic poles totaling only $\sim$20 sq.\ deg. A total of 762 motion objects, not cataloged before, were uncovered.

The AllWISE1 Motion Survey, on the other hand, started with sources detected during AllWISE processing and tabulated in the AllWISE Source Catalog and AllWISE Reject Table (\citealt{cutri2013}). These source detections, from coadds comprised of many individual frames, were then measured on the frame stack itself. At the position of each detection, a point spread function (PSF) fit was performed to measure the source position and flux via a $\chi^2$ minimization procedure. The measurement model for source position included linear motion terms in RA and Dec that could either be set to zero (the ``stationary fit'') or fit fully  (the ``motion fit''). The stationary fit was performed first, and the full set of photometric parameters was computed. Then the motion fit was performed, using the position from the ``stationary fit'' as its initial position estimate. The $\chi^2$ minimization procedure then measured the motion of the object over the frame stack. These tabulated measures were used to search for objects of significant motion, as described in \cite{kirkpatrick2014}. The AllWISE1 Motion Survey covered the full sky and produced 3,525 new motion discoveries along with 16,683 re-discoveries (see section~\ref{uncovered_objects}).

Another {\it WISE}-based survey, combining data from the original {\it WISE} mission with data from the first sky pass of the NEOWISE 2015 Data Release (\citealt{mainzer2014,cutri2015}), uses essentially the same methodology employed by \cite{luhman2014}. This survey, by \cite{schneider2016} (hereafter referred to as the ``NEOWISE Reactivation Motion Survey'', or just ``NEOWISER''), leverages a longer baseline than the previous {\it WISE}-based searches -- namely, $\sim$3.75 years\footnote{Even though the time difference between the earliest {\it WISE} epoch and the first sky pass of the NEOWISE 2015 Data Release is 4.0 years, Schneider et al.\ compared their compiled NEOWISER source catalog to the AllWISE Source Catalog. The astrometry reported for the latter has a mean epoch of $\sim$2010.5.} as opposed to the 0.5-1.0 year baseline available for the AllWISE1 and \cite{luhman2014} surveys. Because the first sky pass contained in the 2015 data release was interrupted by a nineteen-day spacecraft safing operation, the NEOWISER motion survey covers $\sim$90\% of the sky only. The NEOWISER survey uncovered 1,006 new motion objects along with 19,542 re-discoveries.

Because the AllWISE Source Catalog still contains many valid motion sources untapped by the original AllWISE1 Motion Survey, our team performed a second motion search of the AllWISE data, presented here. This survey is hereafter referred to as the ``AllWISE2 Motion Survey'' or just ``AllWISE2''.

\subsection {Why Perform Another Motion Survey?}

The essence of the scientific method is to observe nature, make a testable hypothesis to explain the observations, perform experiments to test the theory, analyze the results, revise the hypothesis if necessary, and repeat. After myriad cycles of this process, scientific methodology helps us develop general theories while allowing us to build an understanding of our surroundings with increasingly greater detail.

Each scientific discipline has its own mechanisms for achieving this. A chemist, for example, might use earlier observations and tested theory to hypothesize what the chemical reaction between two newly created compounds might be. To test this, the chemist need only mix the compounds and observe the result. In this case, the experiment is a tactile process; the chemist is an active participant. 

Astronomy, on the other hand, is generally quite different. Unlike chemists, astronomers have their hands figuratively tied behind their backs. They are unable to force the experiment. There is no picking up of test tubes to bring chemicals into contact. The laboratory itself is even far removed -- light minutes or light years away. The astronomer is left merely to {\it witness}, not participate. Nevertheless, the cosmos is continually performing experiments all around us. The astronomer's challenge is in recognizing these experiments and realizing how they can be used in the scientific method. The satisfaction resulting from this forced ingenuity is, in fact, one of the joys of astronomical research.

Basic observations provide the insight needed to devise new experiments. It was \cite{halley1718} who first showed that three bright stars -- Arcturus, Sirius, and Palilicium (known now as Aldebaran) -- exhibited their own, small motions across the sky, as shown by the fact that their positions had changed dramatically with respect to other stars since the measurements made by Timocharis, Aristyllus, Hipparchus, and Claudius Ptolemy 1600-1800 years previously\footnote{Data from the three earliest observers would have been lost to antiquity had they not been re-recorded by Ptolemy himself in the {\it Almagest} in the second century AD.}. However, it was not until \cite{herschel1783} concluded that the motion of such objects may indicate proximity to the Sun that astronomers realized they had a new tool for distinguishing the nearest stars from the countless points of light in the background. Thereafter, surveys began in earnest to search for proper motion objects whose distances might be measurable through a tool that the earth's yearly orbit about the Sun provides us: trigonometric parallax (e.g., \citealt{bessel1838,henderson1839}). Modern astronomy owes its underpinnings to the successes of these early motion surveys -- from the establishment of the bottom rungs of the ``distance ladder'' (e.g., convergent point analysis of the Hyades cluster; \citealt{boss1908}) to the discovery of degenerate states of matter (via the identification of hot but very low luminosity stars now known as white dwarfs; \citealt{bond1862,adams1914,adams1915,vanmaanen1917}).

By performing the AllWISE2 Motion Survey,
we continue this time-honored tradition. In AllWISE2 we identify nearby objects useful as ``test particles'' for various experiments: 

\begin{itemize}

\item In one experiment, we identify motion stars located within or behind nearby molecular clouds. As these stars move, their flux variations along the line of sight can probe the clumpiness of the cloud material itself since the stars have spectral types appropriate for use as standard candles (section~\ref{early_types}).

\item In a second experiment, we select spectroscopically verified M dwarfs from our follow-up list to hunt for previously missed young objects in the solar vicinity. In this case, we use positional alignments of our motion sources with detections by X-ray and ultraviolet all-sky surveys to identify candidates with high levels of magnetic activity, which can often be tied to youth (section~\ref{young_Mdwarfs}). Young M dwarfs sometimes host young, low-mass brown dwarf companions that can be used as proxies for exoplanet atmospheric studies.

\item  In a third experiment, we search for objects having both unusual colors and higher-than-average motions for their brightness to identify old, low-metallicty late-M and L (sub)dwarfs. The identification of a large collection of such objects spanning the stellar/substellar break will eventually allow us to measure directly the brown dwarf cooling rates at very old ages (section~\ref{subdwarf_analysis}).

\item In a fourth experiment, we search for widely separated common-proper motion binaries having vastly different spectral types in order to constrain physical parameters of the system (section 5).
As one example, we identify a system consisting of a normal M subdwarf and a carbon dwarf. The metallicity of the carbon dwarf, presumably polluted by a close but unseen white dwarf neighbor, can be measured directly using the metallicity of the M subdwarf companion, the first such system where this is possible (section~\ref{carbon_dwarfs} and section~\ref{serendipitous_systems}).

\item  In a fifth experiment, we look for outliers in color-color space to identify unresolved proper motion systems with disparate types. In one example, discussed in Fajardo-Acosta et al.\ (in prep.), an examination of the infrared colors led to the discovery of what we believe is a cold and unique white dwarf member in an unresolved system with a late-M dwarf.

\item In a sixth experiment, discussed in Kellogg et al.\ (in prep.), we comb our list of motion candidates for objects detected by {\it WISE} but not by surveys at shorter wavelengths. The goal is to uncover other very cold brown dwarfs, such as the 250K object WISEA J085510.74$-$071442.5 (\citealt{luhman2014b}), that might have escaped color selection techniques. These cold atmospheres provide much needed empirical data points in the temperature regime just hotter than Jupiter (T$_{eff}\approx$135K; \citealt{aitken1972}).

\end{itemize}

\subsection{Organization of the Paper}

This paper is organized as follows. In section 2 we describe our new motion search using AllWISE and show a number of color-color and color-magnitude diagrams to aid in the characterization of the motion sources. In section 3 we describe our spectroscopic observations and reductions on selected objects. In section 4 we discuss spectral classification and analysis on the resulting spectra, divided into subsections by type: white dwarfs (section 4.1), stars with types $\le$K5 (section 4.2), mid-K through late-M dwarfs (section 4.3), L dwarfs (section 4.4), T dwarfs (section 4.5), and subdwarfs (section 4.6). In section 5 we discuss the search for common-proper-motion systems, with special emphasis on systems having L or T dwarf members (section 5.1), systems with white dwarf members (section 5.2), systems with large magnitude differences (section 5.3), and systems identified serendipitously (section 5.4). Conclusions are given in section 6.

\section{AllWISE2 Motion Search}

\subsection{Criteria for the New Search\label{criteria}}

Section 5 of \cite{kirkpatrick2014} shows that 80\% of the objects found by \cite{luhman2014} but missed by the AllWISE1 Motion Survey were lost because of the criterion $rchi2/rchi2\_pm > 1.03$, where $rchi2$ is the reduced $\chi^2$ value for the stationary fit and $rchi2\_pm$ is the reduced $\chi^2$ value for the motion fit. This criterion was used by AllWISE1 under the assumption that the reduced $\chi^2$ value would be significantly lower for the motion fit relative to the stationary fit if the object were truly moving. For the AllWISE2 Motion Survey, we have used the same criteria used for AllWISE1 -- discussed in section 4.1 of \cite{kirkpatrick2014} -- except that the reduced $\chi^2$ criterion has been changed to  $rchi2/rchi2\_pm \le 1.03$. That is, the combination of AllWISE1 and AllWISE2 drops the $rchi2/rchi2\_pm$ check entirely.

\begin{figure*}
\figurenum{1}
\includegraphics[scale=0.65,angle=0]{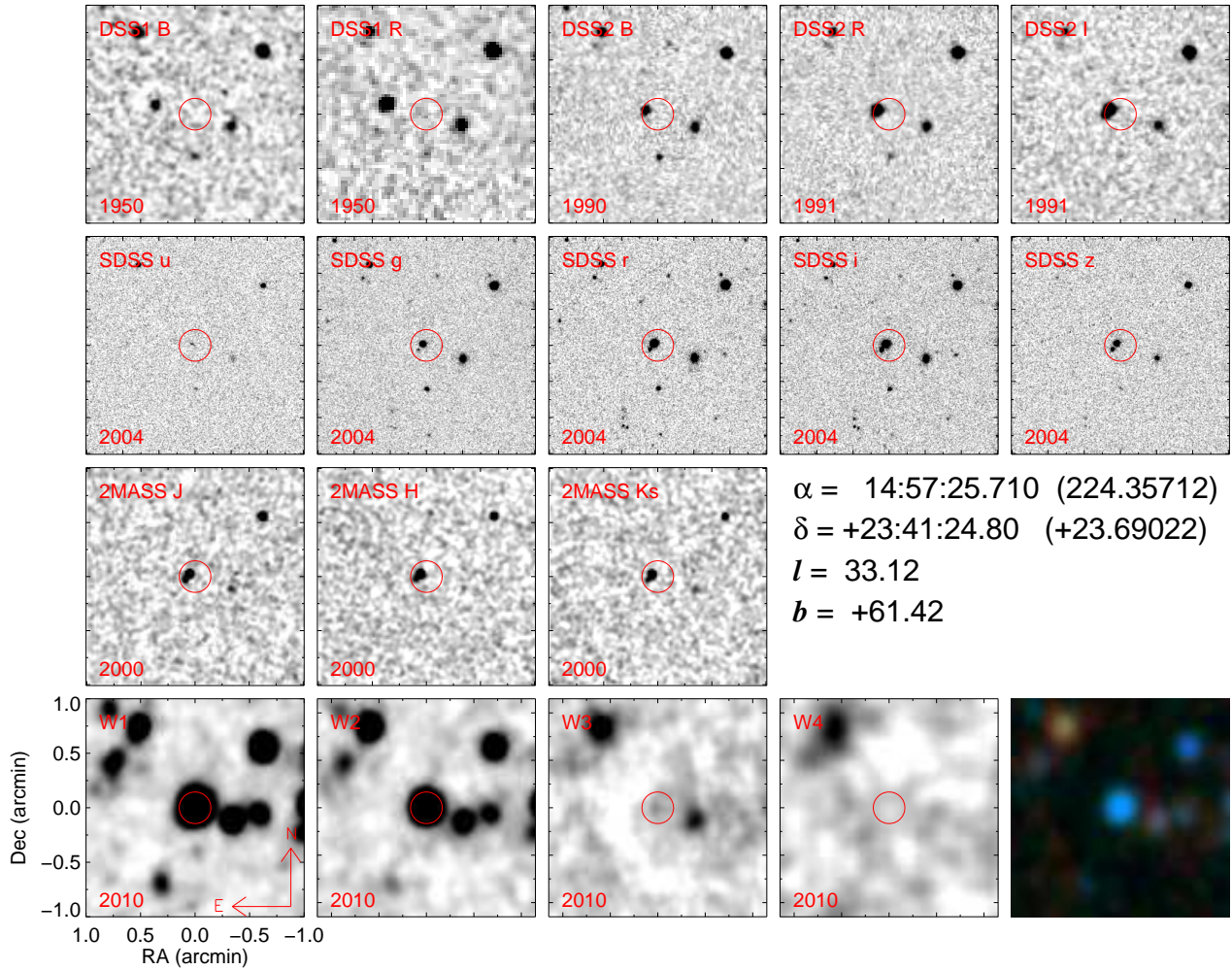}
\caption{Finder chart of a motion candidate identified from AllWISE. The AllWISE position of the candidate is marked by the red circle on all charts, which are two arcminutes on a side with north up and east to the left. Red lettering on each chart indicates the survey, bandpass, and epoch for each image. At lower right is a three-color composite based on the {\it WISE} W1 (blue), W2 (green), and W3 (red) images. Note that the candidate shown is clearly moving. Note also that inspection of the DSS2 $I$-band, SDSS, and 2MASS images reveals that the {\it WISE} source has a fainter, common-proper-motion companion. 
\label{finder_chart}}
\end{figure*}

An unfortunate consequence of this new criterion, and the reason it was not implemented originally, is the large number of new candidates it produces -- another 1,409,845 of them. To scrutinize these, we created finder charts showing a 2$\times$2-arcmin region around each candidate's AllWISE coordinates as imaged by {\it WISE} (in the W1, W2, W3, and W4 bands) and by previous large-area surveys: the Digitized Sky Survey 1 ($B$ and $R$ bands), the Digitized Sky Survey 2 ($B$, $R$, and $I$ bands), the Two Micron All Sky Survey ($J$, $H$, and $K_s$ bands), and, when available, the Sloan Digital Sky Survey ($u$, $g$, $r$, $i$, and $z$ bands). See Figure~\ref{finder_chart} as an example. 

After scrutinzing the first 491,031 charts, sampling different Galactic environments covering 35.8\% of the sky, we found that only 2.2\% of the candidates were valid motion objects. This accrued knowledge allowed us to create additional criteria to further winnow the remaining candidate list. Objects were removed from further consideration if an object from the USNO-B1 catalog (\citealt{monet2003}) was found within 1 arcsec of the candidate's AllWISE position {\it or} if both $w1nm/w1m$ and $w2nm/w2m$ fall below 0.8. The first criterion was imposed because many spurious candidates were found to be either not moving at all between USNO-B1 and {\it WISE} or were barely moving objects whose motions were overestimated in AllWISE. (The $\sim$10-yr time baseline between the USNO-B1 proper moved positions and the AllWISE positions means that this criterion eliminates any legitimate motion objects with $\mu < 0.1$ arcsec/yr.) The second criterion was imposed to eliminate false source detections from AllWISE: $w1nm$ and $w2nm$ give the number of individual exposures on which the source was detected with the profile-fit measurement in band W1 and W2, respectively, whereas $w1m$ and $w2m$ give the total number of individual exposures available in bands W1 and W2. Spurious sources occurring in just a handful of frames are readily tossed out if both $w1nm/w1m$ and $w2nm/w2m$ are significantly less than 1. Running these additional criteria on the original list of 1,409,845 candidates reduces the list to 333,345 objects.

These criteria were then run on the 35.8\% of the sky already scrutinized to see how many valid motion objects would be erroneously eliminated. Only fifty-four of the 10,598 confirmed motion objects (0.5\% of the total) were rejected by the additional criteria. These sources are listed in Table~\ref{valid_motions_tossed}\footnote{The column descriptions are identical to those in Table 2, described in section~\ref{uncovered_objects}.}. Of these fifty-four, fifty-one were rejected because they had a USNO-B1 source lying within 1 arcsec, but this source is a non-moving background object that nearly coincides with the position of the {\it WISE} source. In these cases, the USNO-B1 criterion has unfortunately eliminated a valid motion object. For the remaining three objects (WISE 1222$-$5629, WISE 2046+3358, and WISE 2245+3026)\footnote{Hereafter, we abbreviate object names as WISE hhmm$\pm$ddmm.}, both of the $w1nm/w1m$ and $w2nm/w2m$ values fall below 0.8. All three of these are very bright, heavily saturated sources -- W1 magnitudes of 3.9, $-$1.8, and 4.0 mag, respectively -- and their low $w1nm/w1m$ and $w2nm/w2m$ values are indicative of the fact that very few usable pixels were available for the profile-fit measurement. The mean coverage per pixel was less than 7 for these objects, so measurements were not possible with $>$80\% frequency. 

\begin{center}
\begin{turnpage}
\begin{deluxetable*}{lcccccccccc}
\tabletypesize{\tiny}
\tablewidth{8.0in}
\tablenum{1}
\tablecaption{Valid Motion Objects Not Included in Table 2\label{valid_motions_tossed}}
\tablehead{
\colhead{WISEA Designation} &                          
\colhead{2MASS $J$} &  
\colhead{2MASS $H$} &     
\colhead{2MASS $K_s$} &
\colhead{W1} &
\colhead{W2} &
\colhead{AllWISE} &
\colhead{AllWISE} &
\colhead{Computed} &
\colhead{Computed} &
\colhead{Flag\tablenotemark{b}} \\
\colhead{} &                          
\colhead{(mag)} &  
\colhead{(mag)} &     
\colhead{(mag)} &
\colhead{(mag)} &
\colhead{(mag)} &
\colhead{RA Motion} &
\colhead{Dec Motion} &
\colhead{$\mu_\alpha$\tablenotemark{a}} &
\colhead{$\mu_\delta$\tablenotemark{a}} &
\colhead{} \\
\colhead{} &                          
\colhead{} &  
\colhead{} &     
\colhead{} &
\colhead{} &
\colhead{} &
\colhead{(mas/yr)} &
\colhead{(mas/yr)} &
\colhead{(mas/yr)} &
\colhead{(mas/yr)} &
\colhead{} 
}
\startdata
J000624.35$-$141309.4& 13.395$\pm$0.030& 12.732$\pm$0.029& 12.447$\pm$0.025& 12.275$\pm$0.023& 12.098$\pm$0.024&   200$\pm$56&  -229$\pm$55&     199.8$\pm$7.4&   -79.0$\pm$5.9&  1\\
J001207.46$-$122714.3& 12.686$\pm$0.023& 12.089$\pm$0.024& 11.917$\pm$0.026& 11.796$\pm$0.024& 11.722$\pm$0.022&    27$\pm$52&  -330$\pm$52&      15.6$\pm$7.4&  -141.0$\pm$6.6&  1\\       
J001936.92+492103.2&   12.803$\pm$0.023& 12.202$\pm$0.021& 11.918$\pm$0.021& 11.689$\pm$0.023& 11.487$\pm$0.021&   155$\pm$29&   -57$\pm$29&     140.5$\pm$6.1&   -59.5$\pm$6.0&  0\\  
J002539.83$-$643630.6& 11.418$\pm$0.022& 10.779$\pm$0.021& 10.564$\pm$0.023& 10.363$\pm$0.023& 10.230$\pm$0.020&     5$\pm$35&  -287$\pm$36&     136.9$\pm$7.3&  -184.3$\pm$6.5&  0\\       
J002545.47+730127.9&   13.502$\pm$0.029& 12.881$\pm$0.033& 12.555$\pm$0.025& 12.358$\pm$0.024& 12.185$\pm$0.023&   235$\pm$44&   137$\pm$43&     127.1$\pm$9.3&    -1.5$\pm$8.4&  0\\      
J002952.26+584548.8&   10.108$\pm$0.025&  9.528$\pm$0.026&  9.269$\pm$0.016&  9.123$\pm$0.022&  9.013$\pm$0.020&   -51$\pm$24&  -134$\pm$23&     -34.1$\pm$6.0&  -162.9$\pm$6.0&  1\\
J010352.45$-$155508.3& 13.226$\pm$0.022& 12.636$\pm$0.026& 12.279$\pm$0.024& 12.023$\pm$0.024& 11.833$\pm$0.022&   -30$\pm$54&  -274$\pm$52&     103.7$\pm$8.0&   -21.7$\pm$7.9&  0\\      
J012245.03+530105.2&    9.261$\pm$0.021&  8.636$\pm$0.022&  8.499$\pm$0.020&  8.394$\pm$0.023&  8.439$\pm$0.020&   134$\pm$25&   -33$\pm$24&     172.8$\pm$5.9&   -60.4$\pm$5.9&  1\\      
J012304.83$-$691842.2& 11.979$\pm$0.021& 11.429$\pm$0.025& 11.164$\pm$0.023& 11.030$\pm$0.023& 10.862$\pm$0.020&   191$\pm$36&    31$\pm$38&     237.7$\pm$6.6&    29.5$\pm$5.9&  1\\  
J013218.26$-$120302.7& 11.013$\pm$0.023& 10.469$\pm$0.025& 10.168$\pm$0.026& 10.017$\pm$0.023&  9.827$\pm$0.020&   244$\pm$39&  -128$\pm$38&     426.1$\pm$13.7&  -42.3$\pm$7.0&  1\\    
J013535.39$-$721427.1& 14.026$\pm$0.024& 13.512$\pm$0.021& 13.250$\pm$0.041& 13.070$\pm$0.024& 12.840$\pm$0.023&   203$\pm$48&  -144$\pm$45&      99.8$\pm$9.9&   -21.6$\pm$6.9&  0\\  
J020837.11$-$650516.7& 11.541$\pm$0.024& 10.853$\pm$0.024& 10.663$\pm$0.021& 10.527$\pm$0.022& 10.459$\pm$0.020&   219$\pm$34&    29$\pm$35&     216.4$\pm$6.5&   -10.4$\pm$6.6&  1\\  
J021025.03+622501.8&   11.300$\pm$0.023& 10.660$\pm$0.023& 10.392$\pm$0.016& 10.249$\pm$0.023& 10.222$\pm$0.021&   208$\pm$36&   -76$\pm$35&      80.7$\pm$6.3&   -91.8$\pm$6.2&  0\\      
J030338.48$-$395537.6& 11.840$\pm$0.022& 11.235$\pm$0.021& 11.027$\pm$0.021& 10.932$\pm$0.023& 10.820$\pm$0.020&   184$\pm$26&    58$\pm$26&     219.6$\pm$6.4&    39.8$\pm$6.3&  1\\
J032816.37+575436.0&    9.471$\pm$0.021&  8.797$\pm$0.016&  8.613$\pm$0.019&  8.473$\pm$0.022&  8.495$\pm$0.019&   134$\pm$34&  -140$\pm$32&     147.3$\pm$6.1&   -93.4$\pm$6.0&  1\\ 
J033223.47$-$820014.0& 11.069$\pm$0.021& 10.391$\pm$0.021& 10.196$\pm$0.023& 10.118$\pm$0.023& 10.115$\pm$0.020&   140$\pm$29&    90$\pm$38&     107.8$\pm$8.5&    89.8$\pm$7.7&  0\\  
J033253.99$-$213219.7& 11.114$\pm$0.026& 10.523$\pm$0.025& 10.393$\pm$0.023& 10.333$\pm$0.024& 10.387$\pm$0.020&   137$\pm$26&    19$\pm$25&     150.2$\pm$9.1&     4.1$\pm$8.1&  1\\    
J033952.94$-$374326.8& 13.064$\pm$0.028& 12.515$\pm$0.027& 12.258$\pm$0.024& 12.148$\pm$0.023& 11.982$\pm$0.022&   109$\pm$26&   -85$\pm$26&     145.6$\pm$6.7&   -71.3$\pm$6.6&  0\\
J040131.59$-$771032.8& 14.267$\pm$0.032& 13.727$\pm$0.038& 13.421$\pm$0.039& 13.149$\pm$0.023& 12.919$\pm$0.024&   230$\pm$43&   -94$\pm$48&      86.9$\pm$6.2&   -84.4$\pm$6.2&  0\\ 
J040506.16$-$114423.7& 14.828$\pm$0.038& 14.297$\pm$0.045& 13.952$\pm$0.051& 13.683$\pm$0.025& 13.412$\pm$0.032&   363$\pm$87&  -307$\pm$92&     238.7$\pm$6.9&  -261.0$\pm$6.8&  0\\ 
J041949.59$-$224727.0& 13.850$\pm$0.026& 13.281$\pm$0.029& 13.069$\pm$0.033& 12.874$\pm$0.041& 12.709$\pm$0.044&   158$\pm$91&  -547$\pm$94&     151.4$\pm$11.5& -615.4$\pm$7.7&  1\\      
J045721.97$-$720717.3& 11.842$\pm$0.024& 11.217$\pm$0.025& 10.959$\pm$0.026& 10.845$\pm$0.023& 10.761$\pm$0.021&    99$\pm$29&   133$\pm$31&     105.3$\pm$7.5&   167.9$\pm$6.1&  0\\  
J052552.76$-$624320.6& 14.723$\pm$0.038& 14.028$\pm$0.037& 13.767$\pm$0.054& 13.543$\pm$0.024& 13.342$\pm$0.024&    33$\pm$29&   191$\pm$28&      39.0$\pm$7.1&   193.5$\pm$6.3&  0\\  
J053528.69$-$640321.6& 11.346$\pm$0.024& 10.757$\pm$0.024& 10.477$\pm$0.021& 10.381$\pm$0.023& 10.237$\pm$0.020&    14$\pm$22&   137$\pm$22&      14.3$\pm$6.1&   140.2$\pm$6.0&  0\\
J060433.86$-$371659.7& 13.318$\pm$0.028& 12.738$\pm$0.029& 12.408$\pm$0.025& 12.230$\pm$0.023& 12.023$\pm$0.022&   159$\pm$51&  -246$\pm$52&      78.2$\pm$7.0&  -197.4$\pm$6.9&  0\\     
J070424.11$-$490017.1& 13.704$\pm$0.026& 13.099$\pm$0.031& 12.847$\pm$0.027& 12.628$\pm$0.023& 12.449$\pm$0.022&   172$\pm$39&  -187$\pm$39&      93.5$\pm$6.7&  -190.7$\pm$6.7&  1\\   
J101437.19$-$061934.4& 13.188$\pm$0.027& 12.551$\pm$0.026& 12.329$\pm$0.029& 12.211$\pm$0.022& 12.105$\pm$0.022&   101$\pm$52&  -286$\pm$54&      93.9$\pm$10.1& -218.0$\pm$6.2&  1\\ 
J111050.12$-$732726.0& 11.327$\pm$0.024& 10.735$\pm$0.023& 10.459$\pm$0.021& 10.319$\pm$0.023& 10.168$\pm$0.020&    42$\pm$32&   189$\pm$33&     115.8$\pm$8.5&   104.6$\pm$7.8&  1\\  
J114033.30$-$685844.1& 13.347$\pm$0.027& 12.591$\pm$0.024& 12.363$\pm$0.027& 12.216$\pm$0.024& 12.088$\pm$0.024&  -248$\pm$43&     7$\pm$43&    -183.6$\pm$8.1&    -1.8$\pm$7.3&  0\\      
J115435.82+273806.7&   12.466$\pm$0.022& 11.866$\pm$0.022& 11.593$\pm$0.020& 11.383$\pm$0.023& 11.215$\pm$0.021&   -14$\pm$43&  -275$\pm$43&     -66.0$\pm$6.4&  -172.7$\pm$6.4&  1\\  
J121148.08$-$732828.5& 12.009$\pm$0.023& 11.473$\pm$0.025& 11.203$\pm$0.023& 11.052$\pm$0.023& 10.864$\pm$0.022&  -246$\pm$37&    45$\pm$37&    -158.5$\pm$8.7&    24.7$\pm$7.1&  1\\
J121510.28$-$821830.5& 11.104$\pm$0.023& 10.560$\pm$0.025& 10.336$\pm$0.021& 10.176$\pm$0.023& 10.037$\pm$0.021&  -219$\pm$31&    27$\pm$31&    -214.4$\pm$7.6&   -69.0$\pm$6.9&  1\\
J122258.48$-$562952.0& 11.502$\pm$0.022& 10.975$\pm$0.027& 10.719$\pm$0.021& 10.635$\pm$0.022& 10.479$\pm$0.020&  -154$\pm$26&   -60$\pm$25&    -160.6$\pm$6.4&   -21.8$\pm$6.4&  0\\     
J122833.29$-$562430.4&  4.860$\pm$0.242&  4.167$\pm$0.210&  4.117$\pm$0.264&  3.947$\pm$0.399&  3.637$\pm$0.218&  -183$\pm$27&    50$\pm$27&    -258.7$\pm$16.2& -242.3$\pm$16.3& 1\\ 
J130223.85$-$363400.4& 14.791$\pm$0.045& 14.042$\pm$0.039& 13.627$\pm$0.049& 13.411$\pm$0.025& 13.183$\pm$0.028&  -240$\pm$45&   -35$\pm$48&    -287.8$\pm$6.8&   -80.1$\pm$6.8&  0\\          
J130701.48$-$734518.9& 11.853$\pm$0.023& 11.244$\pm$0.026& 11.020$\pm$0.021& 10.865$\pm$0.024& 10.786$\pm$0.021&  -171$\pm$33&   -64$\pm$34&    -118.7$\pm$19.8& -111.4$\pm$7.8&  0\\         
J131220.16$-$174836.9& 13.455$\pm$0.026& 12.905$\pm$0.024& 12.689$\pm$0.030& 12.529$\pm$0.024& 12.337$\pm$0.025&  -238$\pm$34&   -72$\pm$35&    -216.7$\pm$9.6&   -85.2$\pm$6.6&  1\\     
J132231.09$-$272252.6& 12.840$\pm$0.026& 12.224$\pm$0.023& 12.086$\pm$0.027& 11.950$\pm$0.023& 11.839$\pm$0.022&  -208$\pm$34&  -134$\pm$34&    -202.7$\pm$6.5&  -122.4$\pm$6.4&  1\\
J143840.48+262512.9&   14.615$\pm$0.036& 14.134$\pm$0.042& 13.843$\pm$0.045& 13.767$\pm$0.025& 13.482$\pm$0.029&  -259$\pm$52&  -102$\pm$55&    -259.8$\pm$25.0& -125.6$\pm$8.3&  1\\
J150752.98$-$692400.7& 11.580$\pm$0.022& 10.918$\pm$0.024& 10.720$\pm$0.023& 10.649$\pm$0.023& 10.630$\pm$0.021&  -195$\pm$41&  -153$\pm$40&    -111.0$\pm$7.0&   -79.4$\pm$7.0&  0\\    
J152709.00$-$242601.2& 12.317$\pm$0.026& 11.801$\pm$0.021& 11.563$\pm$0.027& 11.428$\pm$0.024& 11.237$\pm$0.021&  -242$\pm$47&  -137$\pm$46&    -175.1$\pm$6.8&   -95.7$\pm$6.0&  1\\       
J152814.16$-$663149.6& 12.464$\pm$0.025& 11.935$\pm$0.023& 11.636$\pm$0.026& 11.498$\pm$0.023& 11.345$\pm$0.021&  -275$\pm$45&   -30$\pm$47&    -143.0$\pm$16.4& -127.2$\pm$7.1&  0\\
J160056.49$-$714144.2& 11.767$\pm$0.024& 11.236$\pm$0.027& 10.951$\pm$0.021& 10.772$\pm$0.023& 10.609$\pm$0.020&  -258$\pm$35&  -100$\pm$38&    -143.8$\pm$7.9&  -154.0$\pm$6.9&  0\\   
J161321.81$-$412331.3& 12.932$\pm$0.024& 12.450$\pm$0.025& 12.190$\pm$0.025& 11.985$\pm$0.024& 11.834$\pm$0.023&  -302$\pm$54&   -28$\pm$55&    -134.7$\pm$7.3&    -6.7$\pm$7.1&  0\\
J164154.48$-$345205.4& 13.284$\pm$0.027& 12.818$\pm$0.026& 12.533$\pm$0.035& 12.392$\pm$0.023& 12.231$\pm$0.025&  -336$\pm$63&   -19$\pm$70&    -269.2$\pm$6.7&    11.5$\pm$6.7&  0\\
J170016.84$-$510421.7& 12.212$\pm$0.024& 11.571$\pm$0.022& 11.422$\pm$0.023& 11.135$\pm$0.021& 11.110$\pm$0.021&   -63$\pm$46&  -325$\pm$46&     -47.8$\pm$6.5&  -159.0$\pm$6.4&  0\\    
J170551.78$-$515449.2& 12.325$\pm$0.024& 11.808$\pm$0.024& 11.474$\pm$0.023& 11.226$\pm$0.024& 11.060$\pm$0.021&  -201$\pm$47&  -146$\pm$46&     -99.9$\pm$6.6&  -131.5$\pm$6.5&  0\\      
J174549.22$-$361257.9& 11.351$\pm$0.023& 10.766$\pm$0.021& 10.494$\pm$0.025& 10.349$\pm$0.025& 10.258$\pm$0.023&  -153$\pm$47&  -229$\pm$47&      -9.9$\pm$6.3&  -139.3$\pm$6.2&  0\\
J174639.95+225834.7&   11.494$\pm$0.021& 10.823$\pm$0.019& 10.691$\pm$0.023& 10.607$\pm$0.023& 10.606$\pm$0.020&   -15$\pm$40&  -220$\pm$39&      -2.7$\pm$6.5&  -286.4$\pm$6.3&  1\\
J174805.58$-$450851.9& 13.479$\pm$0.024& 13.030$\pm$0.027& 12.765$\pm$0.029& 12.626$\pm$0.025& 12.419$\pm$0.025&  -150$\pm$74&  -381$\pm$75&     -69.2$\pm$6.8&  -185.3$\pm$6.8&  0\\     
J193231.49+403052.2&   11.089$\pm$0.025& 10.441$\pm$0.022& 10.268$\pm$0.020& 10.107$\pm$0.023& 10.086$\pm$0.020&  -119$\pm$32&  -152$\pm$31&     -76.7$\pm$5.8&  -173.2$\pm$5.7&  1\\
J204612.98+335816.2&    0.641$\pm$0.218&  0.104$\pm$0.160& -0.007$\pm$0.204& -1.763$\pm$  NaN& -0.936$\pm$  NaN&  -446$\pm$67&  -149$\pm$82&     357.2$\pm$26.9&  312.9$\pm$27.1& 1\\
J205135.26$-$253238.2& 12.103$\pm$0.027& 11.484$\pm$0.026& 11.384$\pm$0.025& 11.295$\pm$0.023& 11.333$\pm$0.021&  -172$\pm$47&  -251$\pm$48&    -135.0$\pm$6.0&  -258.4$\pm$5.9&  1\\      
J224534.23+302629.5&    5.113$\pm$0.246&  4.427$\pm$0.196&  4.497$\pm$0.320&  4.002$\pm$0.445&  3.526$\pm$0.170&  -839$\pm$41&    61$\pm$46&    -298.4$\pm$25.1& -332.3$\pm$25.2& 1\\
\enddata
\tablenotetext{a}{This is the motion measured between the 2MASS and AllWISE epochs.}
\tablenotetext{b}{If the source is a motion discovery unique to AllWISE, this column is ``0''. For previous discoveries the column is ``1''.}
\end{deluxetable*}
\end{turnpage}
\end{center}

As noted in Table~\ref{valid_motions_tossed}, twenty-seven of these fifty-four objects are new discoveries. Extrapolating to the remaining 64.2\% of the sky means that our additional criteria could potentially eliminate $\sim$100 valid motion objects, roughly 50 of which would be new discoveries. This loss was deemed acceptable since the additional criteria reduce the number of remaining candidates by a factor of $\sim$5. Thus, for the remaining 64.2\% of sky, the additional criteria were employed. In the remainder of this paper, only those objects meeting the full set of criteria are discussed.

\subsection{Motion Objects Uncovered and Comparison to Other {\it WISE} Motion Searches\label{uncovered_objects}}

Table 2 gives the AllWISE coordinates, W1 and W2 magnitudes, and AllWISE-measured motions for all 27,846 verified motion objects from the AllWISE2 survey along with 2MASS $J$, $H$, and $K_s$ magnitudes\footnote{For the L9 pec (v. red) dwarf WISEA J173859.25+614242.1, the $J$, $H$, and $K_s$ magnitudes listed are from \cite{mace2013}.} and our measurement of the 2MASS-to-AllWISE proper motion. This table includes 11,287 new discoveries as well as 16,559 previously identified motion objects\footnote{The AllWISE2 list was checked not only against published motion objects listed in SIMBAD as of early November 2015 but also against three recent papers whose discoveries may not have been fully incorporated into SIMBAD at that time: \cite{luhman2014}, \cite{luhmansheppard2014}, and \cite{schneider2016}.}, the distinction between which can be found in the Flag column. 

Because they were not published in \cite{kirkpatrick2014}, the 16,628 re-discovered motion objects found as part of our previous AllWISE1 survey\footnote{This number differs from the value -- 18,862 -- given in section 4.2 of \cite{kirkpatrick2014} because that list included many saturated sources (W1 $<$ 8.1 mag or W2 $<$ 7.0 mag) with true motions so small that AllWISE should not have been able to detect them. For these objects, the AllWISE motion uncertainty was underestimated by the pipeline, inflating values of the motion significance. Such objects have been dropped from the re-discovery list here.} are listed in Table 3\footnote{For the T9 dwarf WISEA J121756.92+162640.3, the $J$, and $H$, magnitudes listed in the table are not from 2MASS but from \cite{lodieu2013}.}. These two tables, together with the list of 3,525 new discoveries from \cite{kirkpatrick2014} (their Table 3) and WISEA J085510.74$-$071442.5 (their Table 4), represent exactly 48,000 verified motion objects identified by the AllWISE1+AllWISE2 surveys. In addition to these, a small number of possible AllWISE motion objects lacking DSS and 2MASS counterparts are discussed separately in Kellogg et al.\ (in prep.).

\begin{turnpage}
\begin{deluxetable*}{lcccccccccc}
\tabletypesize{\tiny}
\tablewidth{8.0in}
\tablenum{2}
\tablecaption{Motion Objects Identified by the AllWISE2 Motion Survey\label{allwise2_motion_list}}
\tablehead{
\colhead{WISEA Designation} &                          
\colhead{2MASS $J$} &  
\colhead{2MASS $H$} &     
\colhead{2MASS $K_s$} &
\colhead{W1} &
\colhead{W2} &
\colhead{AllWISE} &
\colhead{AllWISE} &
\colhead{Computed} &
\colhead{Computed} &
\colhead{Flag\tablenotemark{b}} \\
\colhead{} &                          
\colhead{(mag)} &  
\colhead{(mag)} &     
\colhead{(mag)} &
\colhead{(mag)} &
\colhead{(mag)} &
\colhead{RA Motion} &
\colhead{Dec Motion} &
\colhead{$\mu_\alpha$\tablenotemark{a}} &
\colhead{$\mu_\delta$\tablenotemark{a}} &
\colhead{} \\
\colhead{} &                          
\colhead{} &  
\colhead{} &     
\colhead{} &
\colhead{} &
\colhead{} &
\colhead{(mas/yr)} &
\colhead{(mas/yr)} &
\colhead{(mas/yr)} &
\colhead{(mas/yr)} &
\colhead{} \\
\colhead{(1)} &                          
\colhead{(2)} &  
\colhead{(3)} &     
\colhead{(4)} &
\colhead{(5)} &
\colhead{(6)} &
\colhead{(7)} &
\colhead{(8)} &
\colhead{(9)} &
\colhead{(10)} &
\colhead{(11)}
}
\startdata
J000003.89+341118.1  &  7.249$\pm$0.017&  6.940$\pm$0.016&  6.885$\pm$0.017&  6.851$\pm$0.062&  6.871$\pm$0.019&  -231$\pm$36&   15$\pm$34&  -224.7$\pm$7.4&    -69.1$\pm$6.6&  1 \\
J000004.53+335248.8  & 12.712$\pm$0.023& 12.096$\pm$0.023& 11.867$\pm$0.019& 11.745$\pm$0.024& 11.619$\pm$0.021&   231$\pm$46&  -89$\pm$46&   170.7$\pm$11.4&     7.1$\pm$8.9&  1 \\
J000005.54+134759.5  & 12.332$\pm$0.026& 11.760$\pm$0.028& 11.549$\pm$0.025& 11.298$\pm$0.023& 11.112$\pm$0.021&   200$\pm$45& -229$\pm$45&   207.9$\pm$7.4&   -109.1$\pm$7.3&  1 \\
J000012.91$-$545452.7& 10.722$\pm$0.020& 10.475$\pm$0.025& 10.449$\pm$0.023& 10.383$\pm$0.023& 10.379$\pm$0.021&   208$\pm$38&  -75$\pm$37&   249.3$\pm$7.3&    -89.8$\pm$6.5&  1 \\
J000015.91$-$481258.5&  9.015$\pm$0.029&  8.593$\pm$0.033&  8.583$\pm$0.023&  8.493$\pm$0.024&  8.551$\pm$0.020&   196$\pm$36& -102$\pm$35&   161.7$\pm$7.9&    -49.1$\pm$7.8&  0 \\
J000017.32+203312.5  & 13.426$\pm$0.027& 12.878$\pm$0.035& 12.578$\pm$0.026& 12.443$\pm$0.023& 12.272$\pm$0.023&   243$\pm$59& -224$\pm$60&   144.5$\pm$9.1&    -46.0$\pm$7.4&  0 \\
J000018.94+495447.2  & 13.263$\pm$0.025& 12.631$\pm$0.024& 12.361$\pm$0.023& 12.220$\pm$0.023& 12.005$\pm$0.021&   172$\pm$32&   88$\pm$32&   138.7$\pm$6.1&     77.0$\pm$6.1&  1 \\
J000021.98+314939.9  & 10.705$\pm$0.020& 10.119$\pm$0.015&  9.863$\pm$0.020&  9.747$\pm$0.023&  9.649$\pm$0.020&   275$\pm$43&  -22$\pm$41&   279.3$\pm$6.6&    -35.7$\pm$6.5&  1 \\
J000022.32$-$050305.3&  9.860$\pm$0.026&  9.478$\pm$0.025&  9.410$\pm$0.025&  9.337$\pm$0.022&  9.404$\pm$0.019&    84$\pm$39& -204$\pm$37&    34.1$\pm$11.4&   -92.4$\pm$7.9&  1 \\
J000022.88+375803.2  & 11.006$\pm$0.022& 10.402$\pm$0.031& 10.139$\pm$0.023&  9.924$\pm$0.022&  9.774$\pm$0.020&   186$\pm$40& -119$\pm$39&   258.7$\pm$7.2&    -66.5$\pm$7.1&  1 \\
J000023.39+534126.8  & 11.357$\pm$0.023& 10.731$\pm$0.023& 10.552$\pm$0.021& 10.465$\pm$0.023& 10.444$\pm$0.020&    38$\pm$26& -161$\pm$25&    65.1$\pm$6.7&   -186.1$\pm$5.9&  1 \\
J000024.05+395157.0  & 12.053$\pm$0.022& 11.504$\pm$0.030& 11.304$\pm$0.022& 11.167$\pm$0.023& 11.004$\pm$0.021&   193$\pm$33&  -12$\pm$32&   229.1$\pm$7.4&      5.0$\pm$6.5&  1 \\
J000028.26$-$360909.7& 13.653$\pm$0.024& 13.105$\pm$0.029& 12.865$\pm$0.029& 12.707$\pm$0.024& 12.494$\pm$0.024&  -207$\pm$56& -258$\pm$56&  -137.3$\pm$7.5&   -103.2$\pm$6.6&  0 \\
J000029.88+334822.6  & 12.195$\pm$0.022& 11.595$\pm$0.021& 11.343$\pm$0.023& 11.198$\pm$0.024& 11.062$\pm$0.021&  -209$\pm$47& -134$\pm$46&  -119.7$\pm$12.1&   -77.4$\pm$9.7&  0 \\
J000032.33$-$565009.7&  8.575$\pm$0.024&  7.968$\pm$0.031&  7.867$\pm$0.024&  7.767$\pm$0.027&  7.853$\pm$0.020&  -181$\pm$32&  -69$\pm$31&   -45.7$\pm$7.3&   -118.1$\pm$7.3&  1 \\
J000032.37$-$244231.3&  7.981$\pm$0.027&  7.484$\pm$0.047&  7.383$\pm$0.024&  7.248$\pm$0.035&  7.387$\pm$0.020&   220$\pm$36&   15$\pm$35&   104.5$\pm$7.9&    -58.0$\pm$7.9&  1 \\
J000033.07$-$532606.9& 12.641$\pm$0.023& 12.051$\pm$0.026& 11.719$\pm$0.023& 11.522$\pm$0.023& 11.335$\pm$0.021&    86$\pm$43& -291$\pm$43&   169.8$\pm$8.1&   -133.1$\pm$7.1&  1 \\
J000035.38$-$011248.8& 13.822$\pm$0.030& 13.226$\pm$0.022& 12.899$\pm$0.027& 12.724$\pm$0.024& 12.506$\pm$0.024&  -256$\pm$64& -251$\pm$66&   -24.9$\pm$9.9&   -246.1$\pm$8.2&  0 \\
J000035.78$-$451506.2& 11.798$\pm$0.026& 11.212$\pm$0.022& 11.112$\pm$0.023& 11.040$\pm$0.023& 11.048$\pm$0.020&   248$\pm$42&   12$\pm$40&   176.5$\pm$6.5&    -11.7$\pm$6.4&  1 \\
J000040.37+162804.4  & 14.061$\pm$0.031& 13.519$\pm$0.041& 13.159$\pm$0.037& 12.985$\pm$0.024& 12.738$\pm$0.026&   383$\pm$72&  -34$\pm$73&   441.2$\pm$7.7&    -27.8$\pm$6.8&  1 \\
J000040.56+031339.3  & 13.711$\pm$0.026& 13.212$\pm$0.031& 12.964$\pm$0.030& 12.849$\pm$0.024& 12.618$\pm$0.026&   -20$\pm$70& -414$\pm$71&   167.5$\pm$12.7&  -307.3$\pm$8.2&  1 \\
J000043.95$-$270016.2& 10.094$\pm$0.023&  9.466$\pm$0.021&  9.292$\pm$0.019&  9.220$\pm$0.023&  9.188$\pm$0.020&    63$\pm$40& -212$\pm$42&    67.6$\pm$7.9&   -161.6$\pm$7.8&  1 \\
J000047.16$-$351007.1&  9.117$\pm$0.029&  8.480$\pm$0.040&  8.282$\pm$0.027&  8.109$\pm$0.022&  8.072$\pm$0.021&   218$\pm$34& -119$\pm$33&   343.2$\pm$7.7&   -111.5$\pm$6.8&  1 \\
J000048.67+295109.0  &  8.987$\pm$0.023&  8.597$\pm$0.027&  8.519$\pm$0.021&  8.464$\pm$0.023&  8.511$\pm$0.020&   -64$\pm$36&  198$\pm$35&    -7.7$\pm$8.1&    115.4$\pm$6.5&  1 \\
J000050.33+624625.8  & 14.513$\pm$0.035& 14.090$\pm$0.051& 13.801$\pm$0.053& 13.567$\pm$0.024& 13.438$\pm$0.027&   245$\pm$44&   44$\pm$44&   162.3$\pm$8.6&     -9.9$\pm$7.0&  1 \\
J000051.18$-$163804.3& 13.104$\pm$0.023& 12.556$\pm$0.024& 12.171$\pm$0.023& 12.012$\pm$0.023& 11.798$\pm$0.023&  -209$\pm$51& -220$\pm$52&   -74.8$\pm$6.9&    -87.5$\pm$6.1&  0 \\
J000056.18+385206.3  & 13.543$\pm$0.024& 12.993$\pm$0.035& 12.842$\pm$0.023& 12.779$\pm$0.024& 12.792$\pm$0.027&   337$\pm$65& -123$\pm$62&   153.3$\pm$7.7&    -63.3$\pm$6.9&  1 \\
J000102.21$-$224639.9&  6.924$\pm$0.018&  6.459$\pm$0.033&  6.357$\pm$0.020&  6.337$\pm$0.081&  6.289$\pm$0.025&   237$\pm$37&   31$\pm$36&    63.0$\pm$6.4&    -87.9$\pm$6.4&  1 \\
J000103.04+610219.1  & 13.904$\pm$0.026& 13.303$\pm$0.036& 13.084$\pm$0.032& 12.860$\pm$0.023& 12.718$\pm$0.025&   170$\pm$36&  -87$\pm$35&   134.9$\pm$7.6&    -38.1$\pm$6.8&  0 \\
J000104.01$-$064310.4& 13.190$\pm$0.031& 12.606$\pm$0.021& 12.296$\pm$0.024& 12.064$\pm$0.025& 11.839$\pm$0.023&  -190$\pm$56& -275$\pm$57&  -104.2$\pm$6.1&   -193.2$\pm$6.1&  1 \\
J000104.97$-$350502.6& 11.515$\pm$0.023& 10.931$\pm$0.025& 10.725$\pm$0.023& 10.586$\pm$0.023& 10.500$\pm$0.020&  -145$\pm$36& -202$\pm$36&   -39.4$\pm$8.8&    -72.1$\pm$6.9&  0 \\
J000107.62$-$811038.6& 13.402$\pm$0.028& 12.879$\pm$0.025& 12.633$\pm$0.031& 12.465$\pm$0.023& 12.244$\pm$0.023&   250$\pm$44& -151$\pm$45&   225.0$\pm$20.1&   -57.6$\pm$6.9&  0 \\
\enddata
\tablecomments{Only a portion of this table is shown here to demonstrate its form and content. A machine-readable version of the full table is available online.}
\tablenotetext{a}{This is the motion measured between the 2MASS and AllWISE epochs.}
\tablenotetext{b}{If the source is a motion discovery unique to AllWISE, this column is ``0''. For previous discoveries the column is ``1''.}
\end{deluxetable*}
\end{turnpage}

\begin{turnpage}
\begin{deluxetable*}{lccccccccc}
\tabletypesize{\tiny}
\tablewidth{8.0in}
\tablenum{3}
\tablecaption{Previously Known Motion Objects Identified by the AllWISE1 Motion Survey\label{allwise1_rediscoveries}}
\tablehead{
\colhead{WISEA Designation} &                          
\colhead{2MASS $J$} &  
\colhead{2MASS $H$} &     
\colhead{2MASS $K_s$} &
\colhead{W1} &
\colhead{W2} &
\colhead{AllWISE} &
\colhead{AllWISE} &
\colhead{Computed} &
\colhead{Computed} \\
\colhead{} &                          
\colhead{(mag)} &  
\colhead{(mag)} &     
\colhead{(mag)} &
\colhead{(mag)} &
\colhead{(mag)} &
\colhead{RA Motion} &
\colhead{Dec Motion} &
\colhead{$\mu_\alpha$\tablenotemark{a}} &
\colhead{$\mu_\delta$\tablenotemark{a}} \\
\colhead{} &                          
\colhead{} &  
\colhead{} &     
\colhead{} &
\colhead{} &
\colhead{} &
\colhead{(mas/yr)} &
\colhead{(mas/yr)} &
\colhead{(mas/yr)} &
\colhead{(mas/yr)} \\
\colhead{(1)} &                          
\colhead{(2)} &  
\colhead{(3)} &     
\colhead{(4)} &
\colhead{(5)} &
\colhead{(6)} &
\colhead{(7)} &
\colhead{(8)} &
\colhead{(9)} &
\colhead{(10)} 
}
\startdata
J000010.31+412141.3  & 12.506$\pm$0.022& 11.990$\pm$0.031& 11.750$\pm$0.021& 11.624$\pm$0.022& 11.446$\pm$0.022&   221$\pm$32&    -6$\pm$32&     239.0$\pm$6.6&   -17.6$\pm$6.5 \\
J000015.37+390251.1  &  9.264$\pm$0.024&  8.872$\pm$0.032&  8.763$\pm$0.023&  8.717$\pm$0.023&  8.768$\pm$0.019&   236$\pm$35&   176$\pm$33&     207.7$\pm$7.2&    65.2$\pm$6.3 \\
J000019.27+431242.4  &  9.924$\pm$0.020&  9.412$\pm$0.019&  9.358$\pm$0.019&  9.282$\pm$0.022&  9.360$\pm$0.020&   189$\pm$26&   -38$\pm$26&     195.3$\pm$6.0&   -73.5$\pm$5.9 \\
J000027.09+575404.9  & 12.497$\pm$0.024& 12.010$\pm$0.031& 11.855$\pm$0.028& 11.734$\pm$0.023& 11.733$\pm$0.023&   371$\pm$29&   239$\pm$28&     381.1$\pm$6.7&   219.1$\pm$6.7 \\
J000028.04$-$412531.3& 13.545$\pm$0.026& 12.974$\pm$0.021& 12.834$\pm$0.032& 12.685$\pm$0.023& 12.548$\pm$0.024&   508$\pm$58&   -78$\pm$57&     506.6$\pm$7.4&   -33.8$\pm$6.6 \\
J000028.55$-$124516.4& 13.200$\pm$0.026& 12.445$\pm$0.023& 11.973$\pm$0.023& 11.707$\pm$0.024& 11.496$\pm$0.021&  -300$\pm$49&  -140$\pm$49&    -156.5$\pm$7.6&   -95.2$\pm$6.0 \\
J000031.00$-$261352.0& 10.400$\pm$0.029&  9.753$\pm$0.031&  9.523$\pm$0.024&  9.328$\pm$0.023&  9.286$\pm$0.020&   320$\pm$38&   230$\pm$37&     297.9$\pm$7.9&   124.7$\pm$7.8 \\
J000031.98+650427.7  & 12.126$\pm$0.022& 11.558$\pm$0.031& 11.393$\pm$0.021& 11.285$\pm$0.023& 11.144$\pm$0.020&   290$\pm$26&   -92$\pm$26&     274.3$\pm$6.5&   -86.0$\pm$6.5 \\
J000034.69$-$365006.8& 11.698$\pm$0.022& 11.095$\pm$0.023& 10.912$\pm$0.023& 10.809$\pm$0.023& 10.718$\pm$0.020&   391$\pm$40&   101$\pm$38&     428.0$\pm$7.3&   105.7$\pm$7.2 \\
J000037.12$-$243830.3& 11.692$\pm$0.023& 11.119$\pm$0.021& 10.862$\pm$0.021& 10.691$\pm$0.022& 10.525$\pm$0.020&  -223$\pm$42&  -321$\pm$41&    -166.9$\pm$8.9&  -188.1$\pm$7.9 \\
J000037.66+420712.8  & 12.581$\pm$0.022& 11.958$\pm$0.024& 11.800$\pm$0.024& 11.682$\pm$0.024& 11.614$\pm$0.021&   340$\pm$32&    47$\pm$31&     300.2$\pm$6.9&    49.1$\pm$6.1 \\
J000044.53$-$502924.7& 11.215$\pm$0.030& 10.726$\pm$0.026& 10.486$\pm$0.024& 10.387$\pm$0.023& 10.230$\pm$0.020&   359$\pm$37&    -6$\pm$36&     394.2$\pm$7.9&     6.3$\pm$7.0 \\
J000048.86+450558.8  & 12.225$\pm$0.021& 11.612$\pm$0.023& 11.348$\pm$0.018& 11.154$\pm$0.024& 11.005$\pm$0.021&    85$\pm$36&  -262$\pm$35&     147.3$\pm$6.1&  -184.2$\pm$6.0 \\
J000052.23+143402.2  & 10.014$\pm$0.019&  9.382$\pm$0.028&  9.155$\pm$0.023&  9.057$\pm$0.024&  9.009$\pm$0.020&   266$\pm$39&   -65$\pm$41&     345.4$\pm$10.7&  -60.1$\pm$7.2 \\
J000101.74$-$214857.3& 11.671$\pm$0.025& 11.148$\pm$0.022& 10.956$\pm$0.022& 10.792$\pm$0.022& 10.592$\pm$0.020&    33$\pm$42&  -301$\pm$41&      71.7$\pm$6.5&  -218.1$\pm$6.3 \\
J000114.58+573310.6  & 12.705$\pm$0.026& 12.280$\pm$0.031& 12.150$\pm$0.024& 11.876$\pm$0.022& 11.770$\pm$0.021&   435$\pm$30&  -409$\pm$29&     397.4$\pm$6.6&  -449.9$\pm$6.5 \\
J000115.50+065934.5  & 11.286$\pm$0.022& 10.741$\pm$0.028& 10.418$\pm$0.021& 10.219$\pm$0.023& 10.042$\pm$0.021&  -623$\pm$43&  -151$\pm$41&    -429.6$\pm$9.8&   -99.6$\pm$8.8 \\
J000121.22+773801.9  &  7.006$\pm$0.027&  6.533$\pm$0.071&  6.417$\pm$0.026&  6.322$\pm$0.055&  6.367$\pm$0.022&   368$\pm$25&   -90$\pm$30&     155.1$\pm$9.8&    12.5$\pm$9.8 \\
J000125.11+523021.2  & 11.171$\pm$0.022& 10.631$\pm$0.021& 10.398$\pm$0.018& 10.262$\pm$0.022& 10.086$\pm$0.020&   261$\pm$25&    90$\pm$25&     301.4$\pm$5.9&    60.2$\pm$5.9 \\
J000133.02+430024.9  & 11.660$\pm$0.022& 11.048$\pm$0.030& 10.810$\pm$0.019& 10.651$\pm$0.023& 10.488$\pm$0.020&   360$\pm$31&  -140$\pm$29&     336.9$\pm$7.4&   -49.4$\pm$6.5 \\
J000144.46$-$352834.1&  9.822$\pm$0.026&  9.189$\pm$0.021&  8.932$\pm$0.023&  8.799$\pm$0.022&  8.742$\pm$0.019&   456$\pm$32&     2$\pm$32&     505.8$\pm$5.9&   -24.6$\pm$5.8 \\
J000204.45+022140.7  & 12.685$\pm$0.026& 12.131$\pm$0.030& 11.891$\pm$0.026& 11.688$\pm$0.023& 11.515$\pm$0.021&  -254$\pm$53&  -170$\pm$52&    -140.3$\pm$10.7& -107.9$\pm$7.0 \\
J000211.31$-$431002.6& 12.597$\pm$0.026& 12.425$\pm$0.023& 12.445$\pm$0.024& 12.471$\pm$0.024& 12.515$\pm$0.024&   116$\pm$57&  -926$\pm$56&     607.1$\pm$8.3&  -668.8$\pm$7.4 \\
J000220.39+423401.4  & 12.357$\pm$0.022& 11.817$\pm$0.029& 11.617$\pm$0.020& 11.455$\pm$0.023& 11.266$\pm$0.021&   234$\pm$30&   148$\pm$29&     310.8$\pm$6.7&   162.4$\pm$6.6 \\
J000233.87+434315.3  & 12.584$\pm$0.022& 12.031$\pm$0.029& 11.877$\pm$0.019& 11.827$\pm$0.023& 11.829$\pm$0.022&   357$\pm$35&    29$\pm$34&     308.3$\pm$7.5&    68.8$\pm$6.6 \\
J000234.55$-$391031.2& 12.711$\pm$0.025& 12.136$\pm$0.025& 11.802$\pm$0.027& 11.589$\pm$0.022& 11.384$\pm$0.021&   189$\pm$44&  -369$\pm$43&     287.5$\pm$6.5&  -159.3$\pm$6.4 \\
J000240.21$-$341347.6& 14.117$\pm$0.024& 14.024$\pm$0.038& 13.919$\pm$0.063& 13.794$\pm$0.025& 13.732$\pm$0.033&   -33$\pm$101& -924$\pm$101&    143.3$\pm$6.8&  -764.8$\pm$6.7 \\
J000252.49+380057.4  & 10.570$\pm$0.022&  9.891$\pm$0.028&  9.795$\pm$0.022&  9.698$\pm$0.022&  9.706$\pm$0.020&   -94$\pm$38&  -265$\pm$36&      -0.6$\pm$7.3&  -266.5$\pm$6.4 \\
J000303.69+564400.3  &  8.153$\pm$0.021&  7.734$\pm$0.049&  7.656$\pm$0.017&  7.575$\pm$0.031&  7.669$\pm$0.020&   199$\pm$26&    96$\pm$26&     194.4$\pm$7.2&    36.8$\pm$6.4 \\
J000307.25+061633.8  & 11.039$\pm$0.021& 10.533$\pm$0.028& 10.295$\pm$0.019& 10.111$\pm$0.022&  9.896$\pm$0.020&     6$\pm$45&  -583$\pm$40&     242.1$\pm$10.7& -506.2$\pm$8.7 \\
J000311.05+035042.3  & 11.292$\pm$0.027& 10.723$\pm$0.023& 10.445$\pm$0.023& 10.254$\pm$0.023& 10.087$\pm$0.021&  -242$\pm$42&  -349$\pm$40&     -79.5$\pm$13.4& -302.2$\pm$8.7 \\
J000313.38$-$171246.0& 13.131$\pm$0.022& 12.509$\pm$0.024& 12.278$\pm$0.024& 12.059$\pm$0.022& 11.899$\pm$0.022&   138$\pm$53&  -238$\pm$53&     158.3$\pm$6.1&  -113.7$\pm$6.1 \\
\enddata
\tablecomments{Only a portion of this table is shown here to demonstrate its form and content. A machine-readable version of the full table is available online.}
\tablenotetext{a}{This is the motion measured between the 2MASS and AllWISE epochs.}
\end{deluxetable*}
\end{turnpage}

\begin{figure}
\figurenum{2}
\includegraphics[scale=0.85,angle=0]{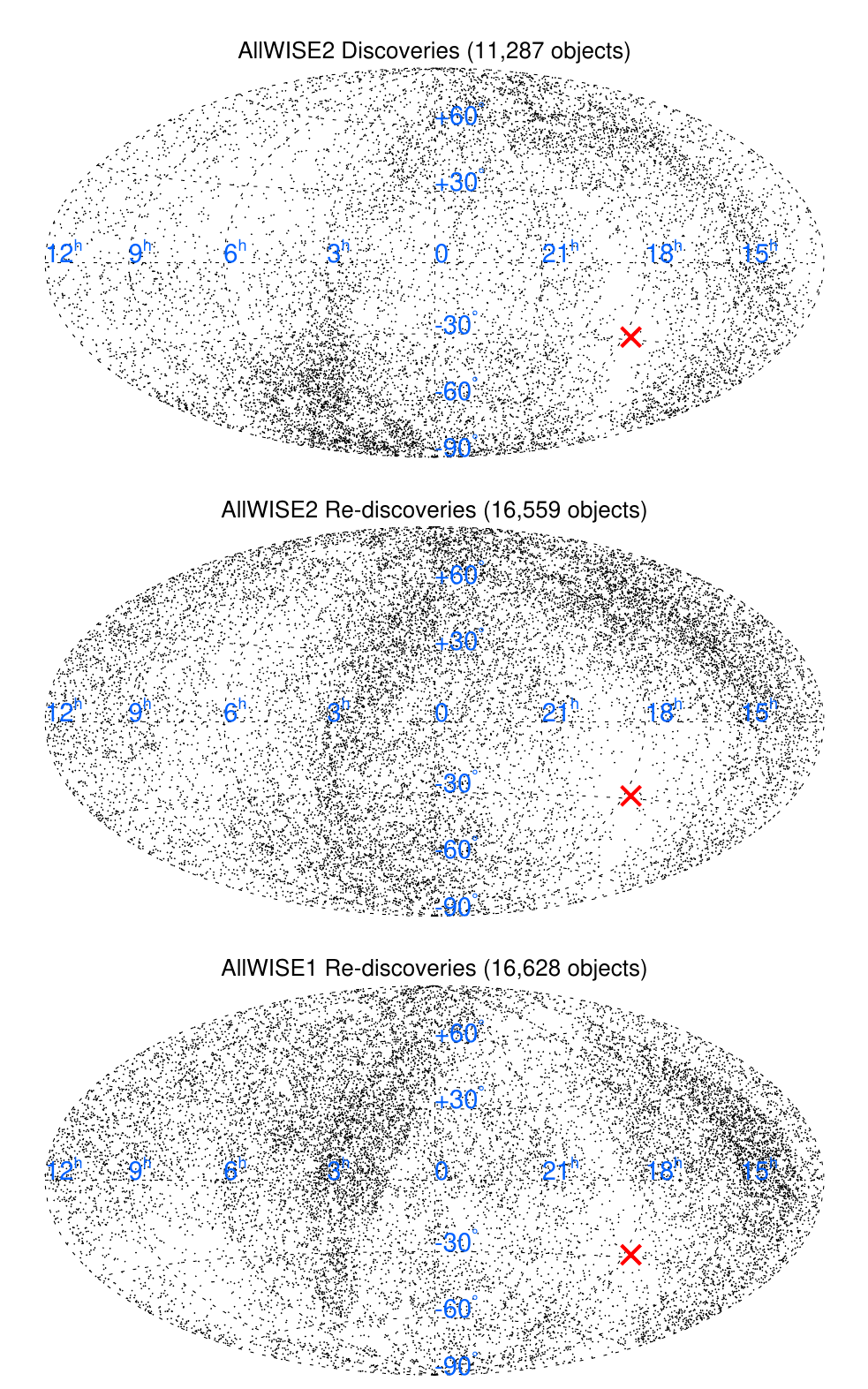}
\caption{Equatorial projections of the sky, with the vernal equinox at center, showing the distribution of motion sources in the AllWISE2 list (discoveries in the top panel and re-discoveries in the middle panel) and the AllWISE1 re-discovery list (bottom panel). The location of the Galactic Center is shown by the red ``X''. 
\label{sky_plots}}
\end{figure}

Figure~\ref{sky_plots} shows the sky distribution of AllWISE2 discoveries along with AllWISE1 and AllWISE2 rediscoveries. These plots can be compared to those shown in Figure 14 of \cite{kirkpatrick2014}. Both sets show an overdensity of sources in the ecliptic longitude zones covered at three epochs by {\it WISE}, as well as an underdensity in the zone of high backgrounds and source confusion along the Galactic Plane, particularly on either side of the Galactic Center. More evident in the AllWISE2 plots is the overdensity of sources toward either ecliptic pole, which is the most repeatedly observed region of the sky for {\it WISE}. The discoveries in the upper plot of Figure~\ref{sky_plots} are concentrated toward the deep southern hemisphere since that area has historically seen fewer motion surveys than the rest of the sky. (For confirmation of this, see the sky distribution of sources from the New Luyten Two Tenths Catalog illustrated in the upper panel of Figure 15 of \citealt{kirkpatrick2014}.)

\begin{figure}
\figurenum{3}
\includegraphics[scale=0.30,angle=270]{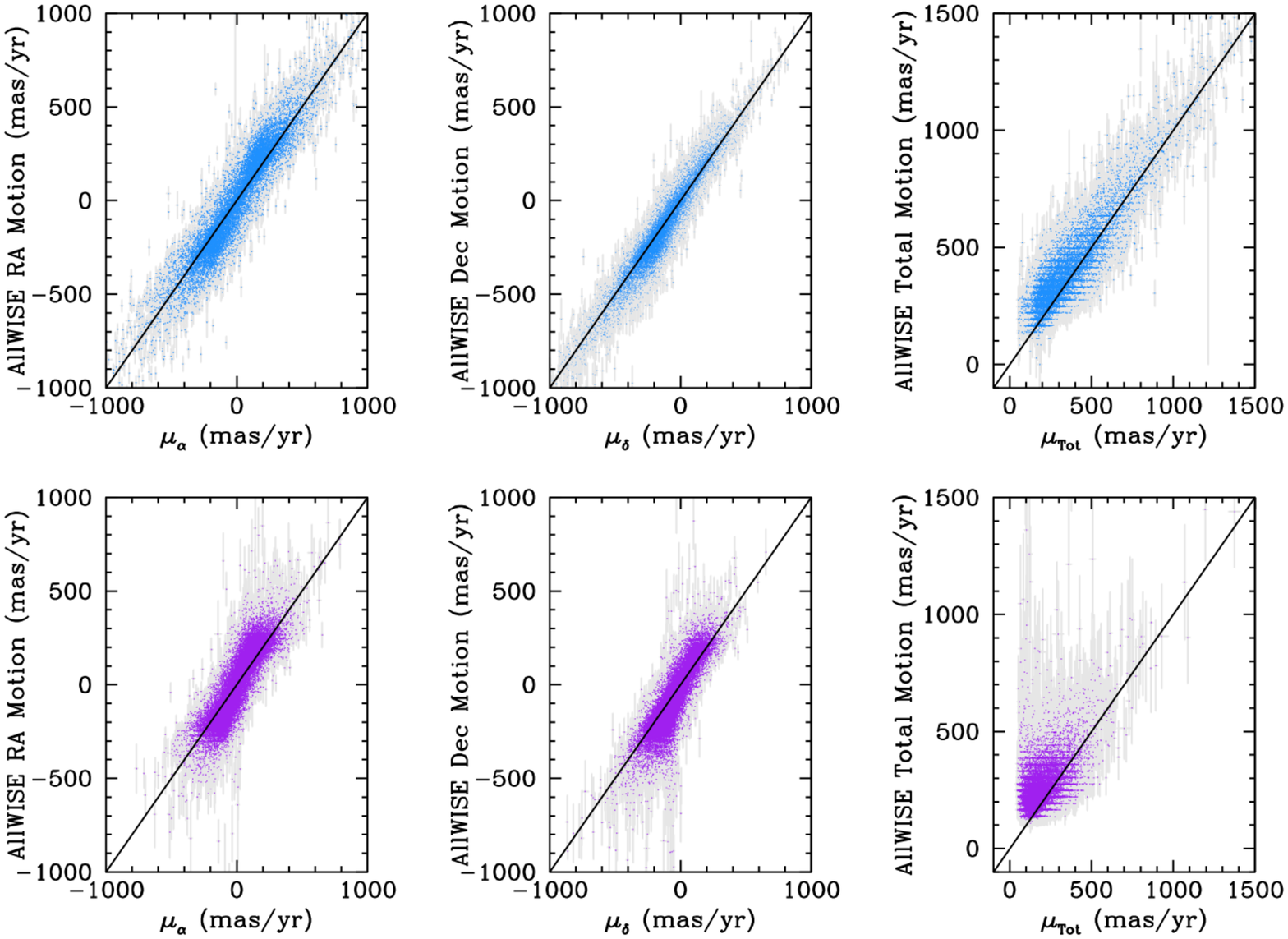}
\caption{Comparison of the motion measured by AllWISE ($y$-axes) to the proper motion derived from the ten-year 2MASS-to-AllWISE time baseline ($x$-axes). Left panels show the RA component of the motion, middle panels show the Dec component, and right panels show the total motion. The upper row shows the results for the AllWISE1 re-discovery list (light blue points with light grey error bars), and the lower row shows all sources from AllWISE2 (purple with light grey error bars). Lines of one-to-one correspondence are shown in black
\label{motion_comparison}}
\end{figure}

A comparison of AllWISE-measured motions to the 2MASS-to-AllWISE motions is shown in Figure~\ref{motion_comparison} for both the AllWISE1 re-discovery list and the list of all motion sources identified by AllWISE2. There is generally good agreement between the short-baseline AllWISE measurements and the longer-baseline 2MASS-to-AllWISE measurements. A small bias is present, as expected, between the two sets, in the sense that the AllWISE measurement is slightly larger on average than the 2MASS-to-AllWISE one. This bias exists because of our selection criteria. At a given, true motion value, we preferentially choose those objects whose measurement values scatter higher than the mean because objects with measurements scattering lower than the mean may not meet our criterion for motion significance. This is the same effect noted in Figure 16 of \cite{kirkpatrick2014}. 

Furthermore, we note that the correlation between the AllWISE-measured motions and the 2MASS-to-AllWISE ones is less tight for AllWISE2 than for AllWISE1. This is also expected since AllWISE2 is preferentially selecting sources with smaller motions, which is evident in subsequent plots. For the remainder of this paper, we use our measurement of the 2MASS-to-AllWISE motion rather than the shorter baseline AllWISE-only one.

\begin{figure}
\figurenum{4}
\includegraphics[scale=0.30,angle=270]{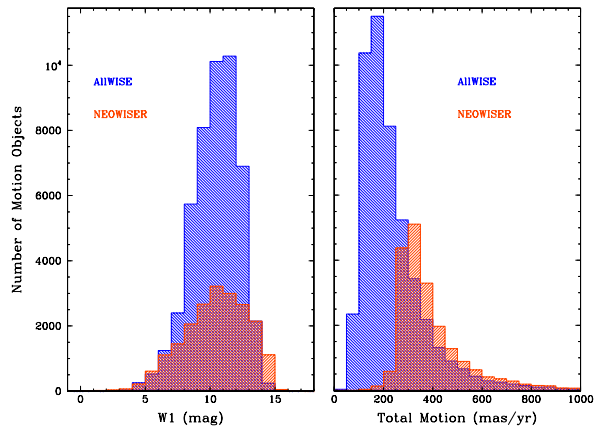}
\caption{Histograms of the total number of motion objects identified by the AllWISE1 + AllWISE2 Motion Surveys (labeled as ``AllWISE'') and the NEOWISER Motion Survey as a function of W1 magnitude and total motion. For clarity, only those discoveries with total motions less than 1000 mas yr$^{-1}$ are shown in the right panel.
\label{histograms_total}}
\end{figure}

A comparison of the parameter space explored by the AllWISE and NEOWISER motion surveys is given in Figure~\ref{histograms_total}, which shows the W1 magnitude and total motion distribution for all identified motion objects\footnote{\cite{luhman2014} did not publish a list of re-discoveries, so we are unable to make the same comparison for his survey.}. By design, the NEOWISER survey was limited to motions exceeding $\sim$250 mas yr$^{-1}$, whereas the AllWISE1 and AllWISE2 surveys identified many thousands of objects with motions below this value. The power of the NEOWISER survey over the AllWISE1 and AllWISE2 surveys is in its extended time baseline -- $\sim$3.75 years for NEOWISER as opposed to $\sim$0.5 years for AllWISE1 and AllWISE2 -- as evidenced by the fact that the NEOWISER survey was much more adept at selecting motion objects at magnitudes fainter than W1 $\approx$ 14 mag. 

\begin{figure}
\figurenum{5}
\includegraphics[scale=0.30,angle=270]{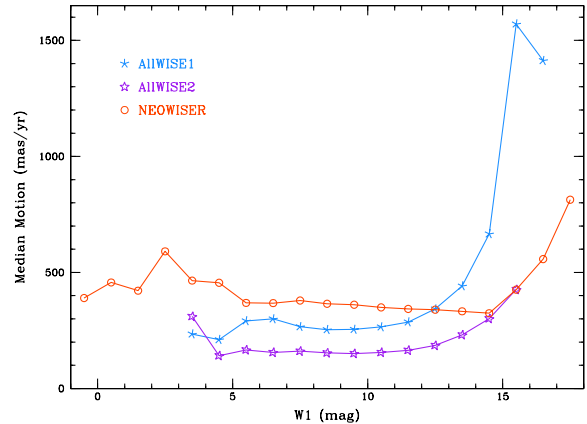}
\caption{The median motion value in each integral W1 magnitude bin for motion objects identified in AllWISE1 (light blue), AllWISE2 (purple), and NEOWISER (orange red).
\label{histograms_total_medPM}}
\end{figure}

This point is further demonstrated in Figure~\ref{histograms_total_medPM}. The only faint (W1 $>$ 15 mag) objects identifiable by AllWISE1 or AllWISE2 are rare objects of {\it very} high motion; those identified have motions typically $>$ 1500 mas yr$^{-1}$. With the longer time baseline, the same motion significance corresponds to a smaller absolute motion for NEOWISER ($\sim$500 mas yr$^{-1}$). Because there is a larger population of sources with motions $>$ 500 mas yr$^{-1}$ compared to the (very sparse) population  having motions $>$ 1500 mas yr$^{-1}$, NEOWISER finds many more objects at these fainter magnitudes. Note also from Figure~\ref{histograms_total_medPM} that AllWISE2 identifies objects with much smaller motions on average than AllWISE1, which is the expected behavior given our $rchi2/rchi2\_pm \le 1.03$ criterion.

\begin{figure}
\figurenum{6}
\includegraphics[scale=0.30,angle=270]{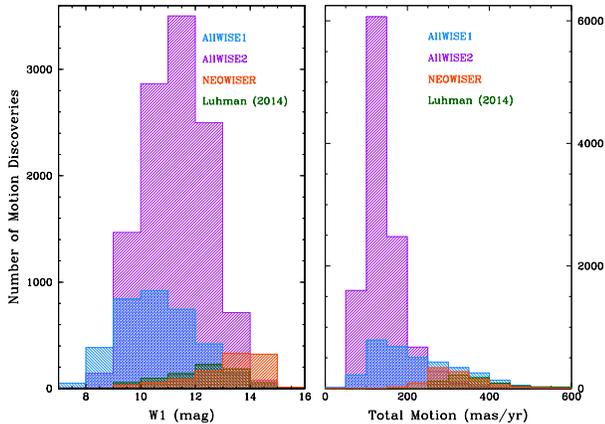}
\caption{Histograms of the total number of motion discoveries identified by the AllWISE1, AllWISE2, NEOWISER, and \cite{luhman2014} motion surveys as a function of W1 magnitude and total motion. For clarity, only those discoveries with total motions less than 600 mas yr$^{-1}$ are shown in the right panel.
\label{histograms_discoveries}}
\end{figure}

Histograms as a function of W1 magnitude and total motion are shown in Figure~\ref{histograms_discoveries} for the discovery lists from AllWISE1, AllWISE2, \cite{luhman2014}, and NEOWISER. The power of the \cite{luhman2014} and NEOWISER surveys is their ability to pick up new motion objects at very faint magnitudes, and this is particularly evident in NEOWISER, where the extended time baseline has made the identification of those faint motion objects easier. The power of the AllWISE1 and AllWISE2 surveys is in picking up new discoveries of smaller motion at brighter magnitudes.

\subsection{Characterizing Motion Sources using Color-color and Color-magnitude Diagrams}

\begin{figure*}
\figurenum{7}
\includegraphics[scale=0.65,angle=270]{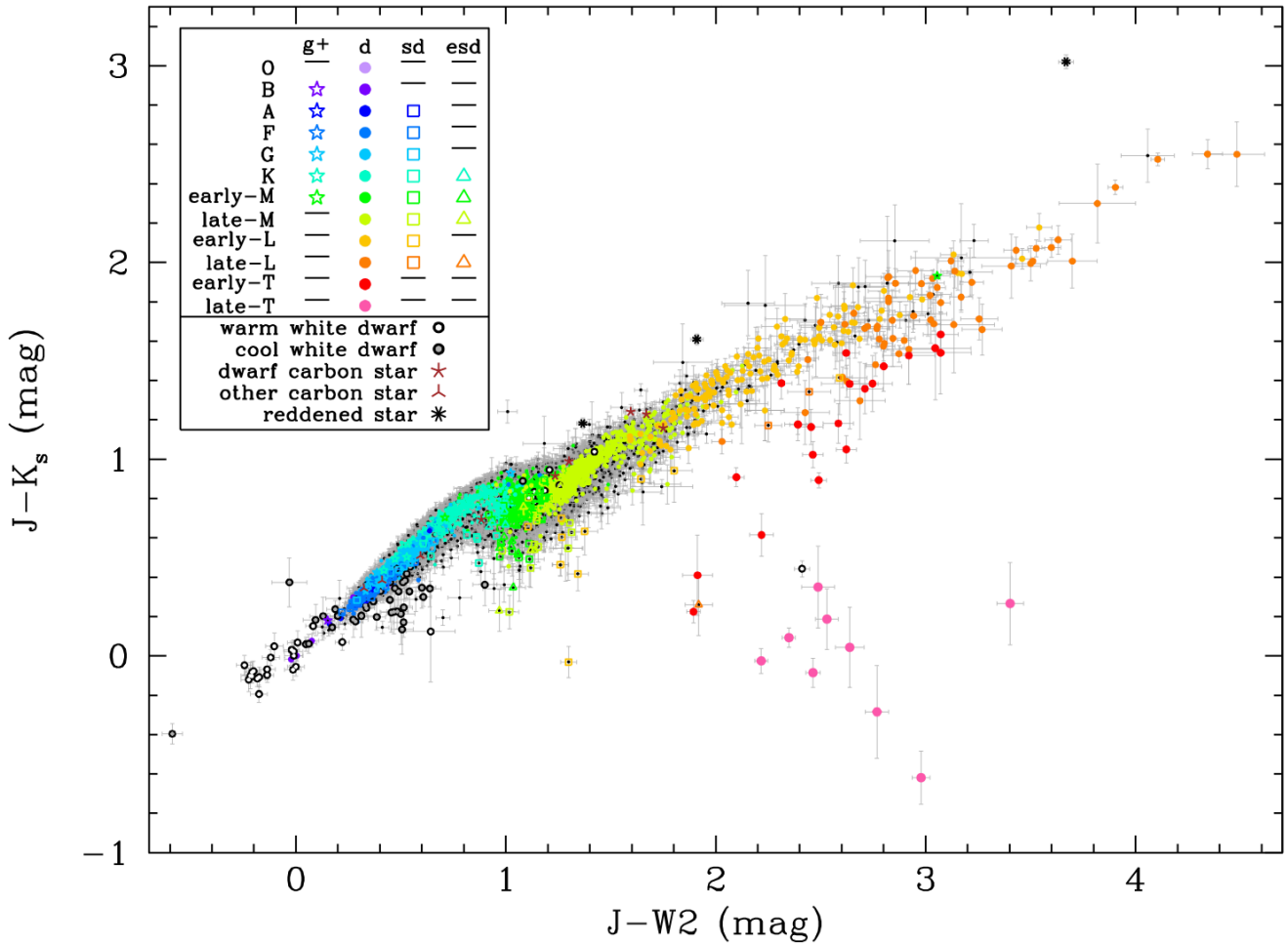}
\caption{The $J-K_s$ vs.\ $J-$W2 diagram for all identified motion objects in the AllWISE1 and AllWISE2 surveys with magnitudes fainter than the nominal {\it WISE} W1 saturation limit of 8.1 mag. Objects with spectroscopic classifications are shown by the colored symbols, as explained in the legend and in the text. Not shown in the figure are motion-selected late-T and Y dwarfs, most of which are too faint for ground-based $K_s$ observations. The most extreme of these, WISEA J085510.74$-$071442.5, falls well outside the bounds of this plot, at $J-$W2 = 10.98$^{+0.53}_{-0.35}$ mag (\citealt{faherty2014}).
\label{color_color_JKs_JW2}}
\end{figure*}

\begin{figure*}
\figurenum{8}
\includegraphics[scale=0.65,angle=270]{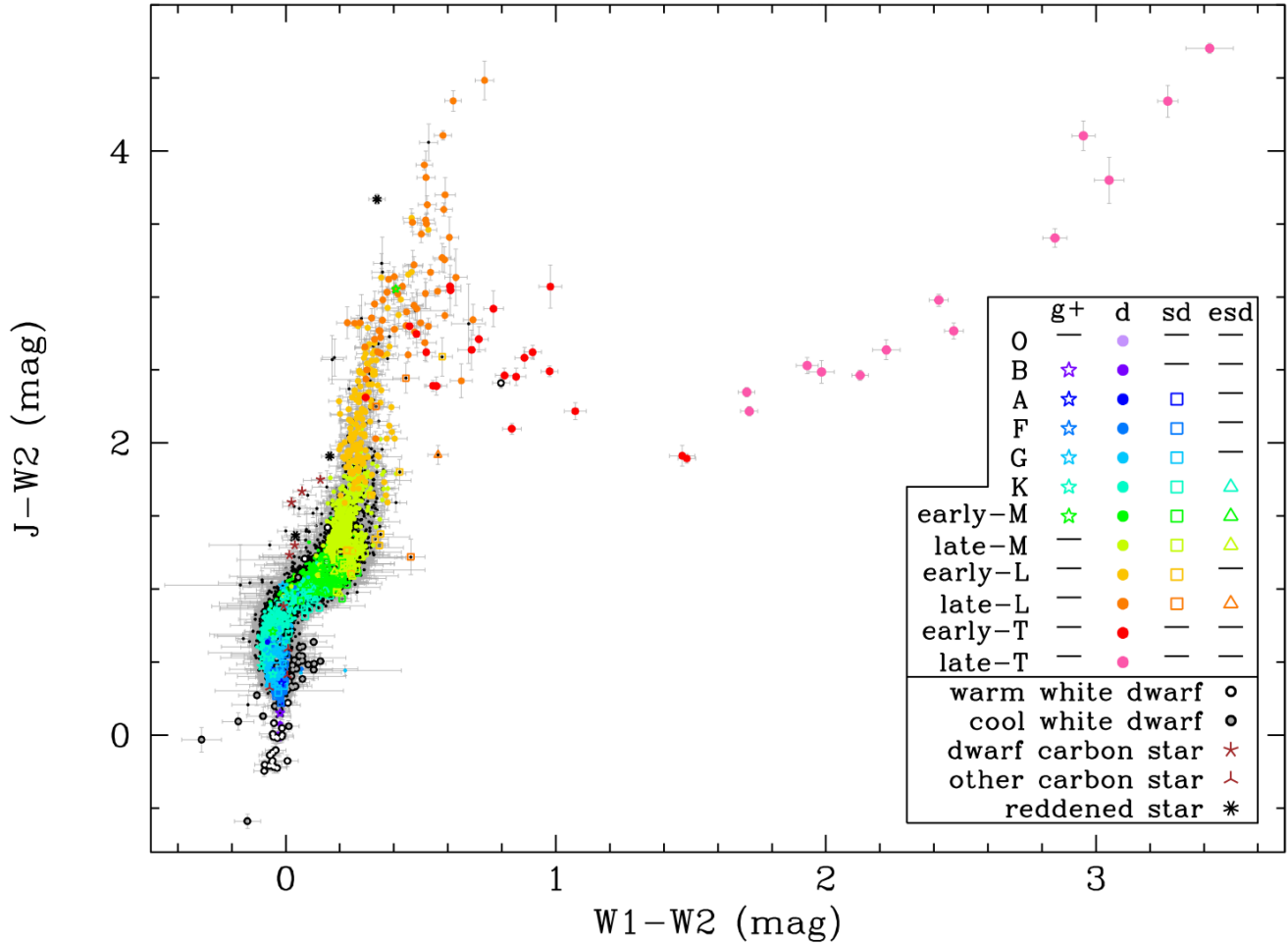}
\caption{The $J-$W2 vs.\ W1$-$W2 diagram for all identified motion objects in the AllWISE1 and AllWISE2 surveys with magnitudes fainter than the nominal {\it WISE} W1 saturation limit of 8.1 mag. Objects with spectroscopic classifications are shown by the colored symbols, as explained in the legend and in the text. Not shown in the figure are motion-selected Y dwarfs, the most extreme of which, WISEA J085510.74$-$071442.5, falls well outside the bounds of this plot, at ($J-$W2, W1$-$W2) = (10.98$^{+0.53}_{-0.35}$, 5.9$\pm$0.8) mag (\citealt{faherty2014}; Wright, priv.\ comm.).
\label{color_color_JW2_W1W2}}
\end{figure*}

\begin{figure*}
\figurenum{9}
\includegraphics[scale=0.80,angle=0]{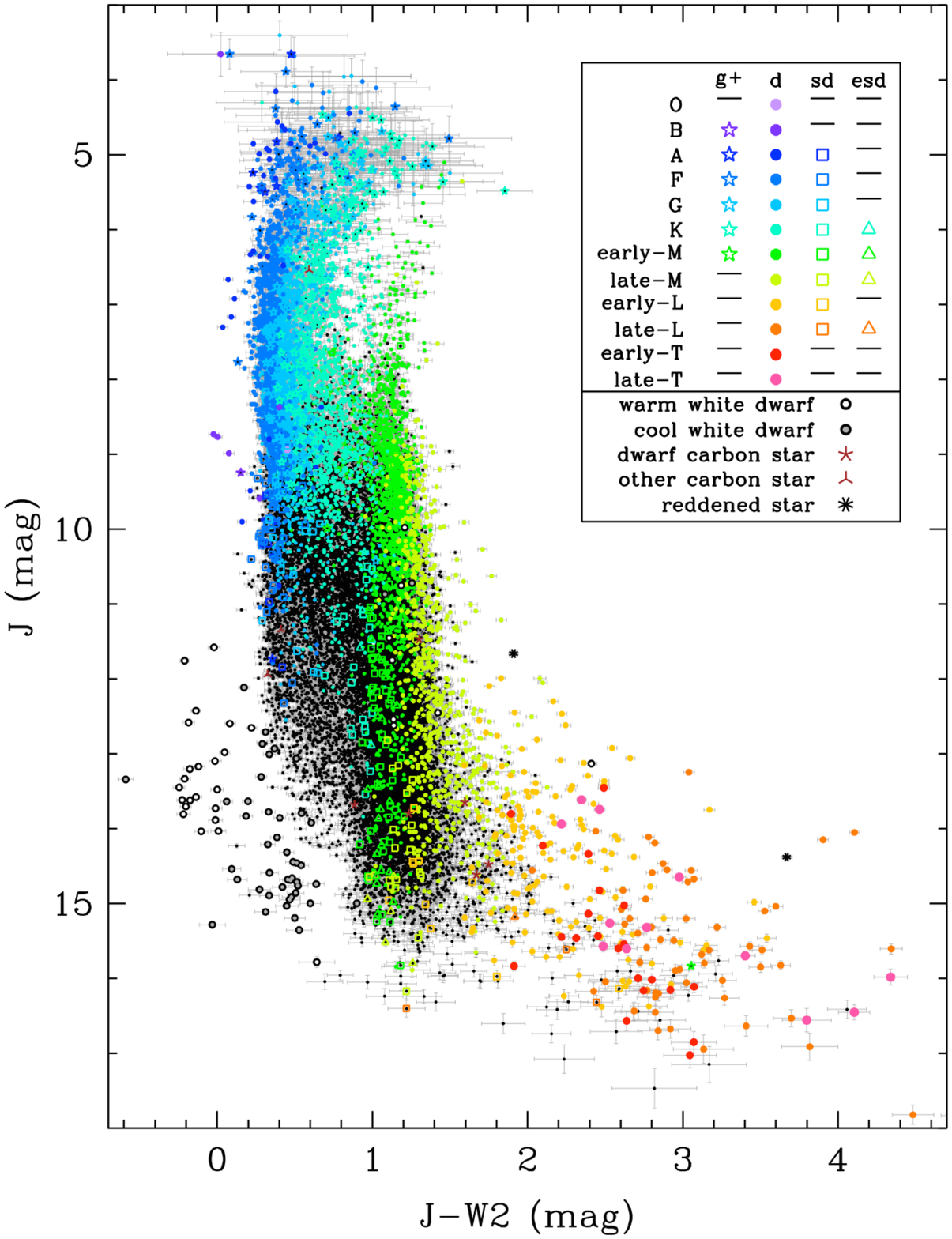}
\caption{The $J$ vs.\ $J-$W2 diagram for all identified motion objects in the AllWISE1 and AllWISE2 surveys. Objects with spectroscopic classifications are shown by the colored symbols, as explained in the legend and in the text. Not shown in the figure are motion-selected Y dwarfs, the most extreme of which, WISEA J085510.74$-$071442.5, falls well outside the bounds of this plot, at $J-$W2 = 10.98$^{+0.53}_{-0.35}$ mag (\citealt{faherty2014}).
\label{color_mag_J_JW2}}
\end{figure*}

\begin{figure*}
\figurenum{10}
\includegraphics[scale=0.80,angle=0]{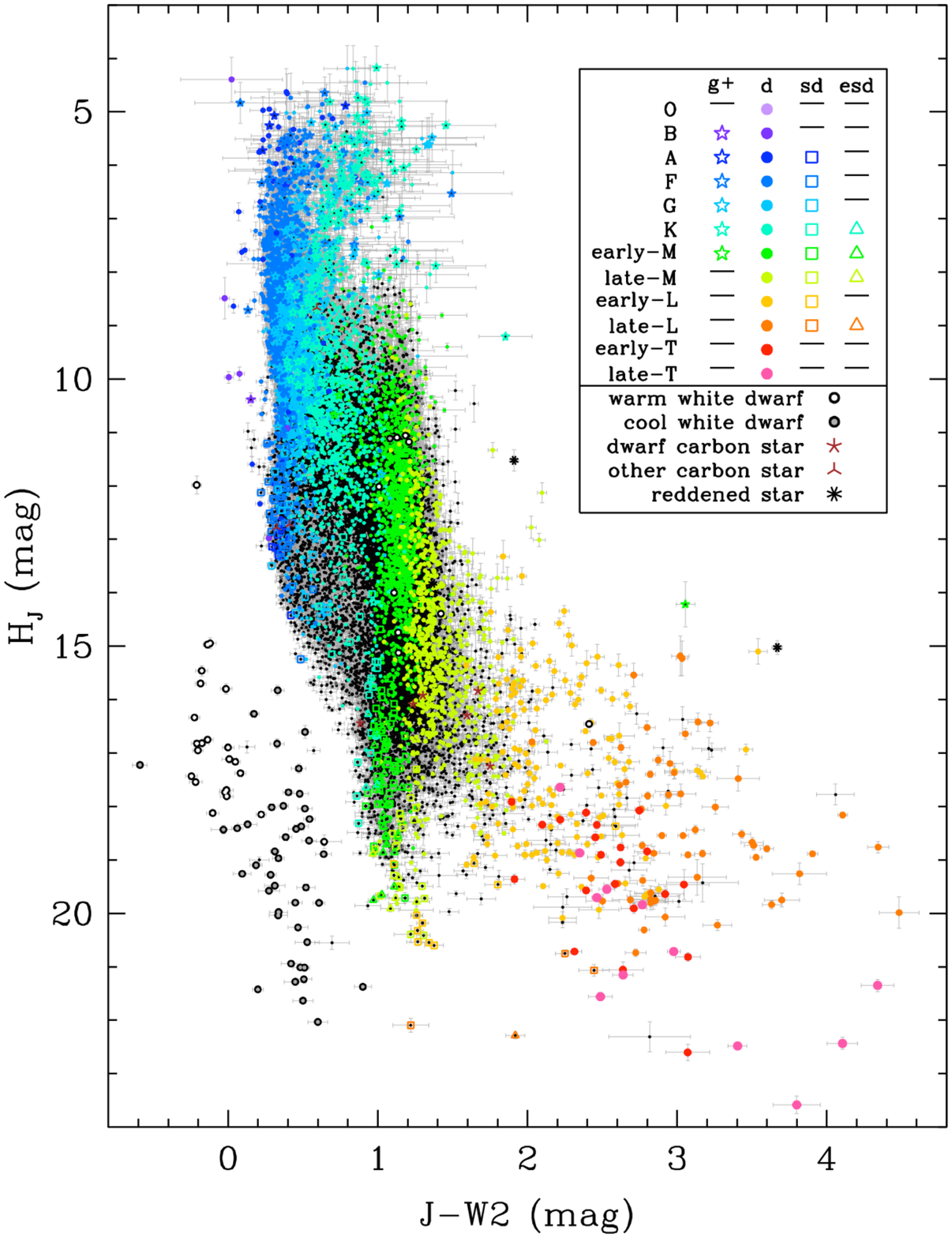}
\caption{The reduced proper motion at $J$-band ($H_J$) vs.\ $J-$W2 diagram for all identified motion objects in the AllWISE1 and AllWISE2 surveys. Objects with spectroscopic classifications are shown by the colored symbols, as explained in the legend and in the text. Not shown in the figure are motion-selected Y dwarfs, the most extreme of which, WISEA J085510.74$-$071442.5, falls well outside the bounds of this plot, at $J-$W2 = 10.98$^{+0.53}_{-0.35}$ mag (\citealt{faherty2014}).
\label{color_mag_HJ_JW2}}
\end{figure*}

Motion objects from the combined AllWISE1 and AllWISE2 surveys are further characterized in Figures~\ref{color_color_JKs_JW2} through~\ref{color_mag_HJ_JW2}. Using SIMBAD, we have searched for published spectral types of previously known objects\footnote{SIMBAD spectral types having only a spectral class but no subclass may have originated from color classes originally assigned by \cite{luyten1979,luyten1980a,luyten1980b} based on his assessment of the color difference between the $B$- and $R$-band photographic plates. Examples of such suspect types include ``K'' and ``M:''. Since they are not based on spectroscopic data, we dropped all such types from our compilation.} and supplemented these with spectral types of new discoveries provided by \cite{kirkpatrick2014}, \cite{luhmansheppard2014}, and this paper. Each spectral type was assigned a different color, although the wide locus spanned by M, L and T dwarfs made it possible to divide each of these into early (e.g., M0-M4.5) and late (e.g., M5-M9.5) divisions. Colors were assigned so that the entire visible palette was covered, with O-type stars at the violet end and late-T dwarfs at the red end. (Because Y dwarfs would have required a dramatic rescaling of the axes, these are not shown.) Objects were further subdivided by luminosity or metallicity types. Objects with luminosity classes of I, II, III, or IV were grouped together as giants or other high-luminosity objects (``g+'') in the figures, and classes of IV-V and V were assigned to dwarfs (``d''). Luminosity class VI was assigned to subdwarfs (``sd'') as were objects with a luminosity or metallicity prefix of sd. Metallicity classes esd and usd were assigned to extreme subdwarfs (``esd''). Objects of type O through T lacking a luminosity or metallicity class were assumed to be dwarfs of solar metallicity (``d''). White dwarfs with temperature types $<$6.7 (\citealt{sion1983}; $T_{eff} > 7500$K) were assigned to the ``warm white dwarf'' class, and those with temperature types $\ge$6.7 ($T_{eff} < 7500$K) were assigned to the ``cool white dwarf'' class. Carbon stars were assigned to either a ``dwarf carbon star'' class for objects like G 77-61 or an ``other carbon star'' class to cover carbon-enhanced metal poor stars (HE 0440$-$1049 and 2MASS J04442200$-$4513542) and the CH-enhanced star HD 26. Three objects discovered to be highly reddened G and K dwarfs (see section~\ref{early_types}) are also distinguished. Legends on each figure show the mapping between our assigned divisions and the colored symbols.

Figure~\ref{color_color_JKs_JW2} shows the location of objects in the $J-K_s$ vs.\ $J-$W2 diagram. The densest locus of objects, which runs redward in both colors with decreasing temperature, is composed of stars with types of F, G, K, or M. The kink near the boundary between K and M stars marks the onset of H$_2$O absorption at the blue side of $K_s$ band\footnote{The 2MASS filter profiles can be found in Figure 2 of \cite{skrutskie2006} and the {\it WISE} ones can be found in Figures 6 and 7 of \cite{wright2010}.} (see Figure 12 of \citealt{cushing2005}) and CO at W2, although the onset of CO likely occurs at a somewhat lower temperature than that of H$_2$O (\citealt{allard1990,yamamura2010,sorahana2012}). From this main locus, the L dwarfs proceed redward in both colors because of decreasing temperature and the production of dust in the photosphere, the latter of which dramatically reddens the $J-K_s$ colors (e.g., \citealt{marley2002,tsuji2004}); the reddest L dwarfs extend to values of ($J-$W2, $J-K_s$) $\approx$ (4.5, 2.6). As the dust clears, the later L dwarfs turn back to the blue in both colors. The appearance of CH$_4$ at $J$ and $K_s$ marks the beginning of the T dwarf sequence (\citealt{burgasser2006}) starting near ($J-$W2, $J-K_s$) $\approx$ (3.0, 1.6) and proceeding blueward in both colors until mid-T types near ($J-$W2, $J-K_s$) $\approx$ (2.0, 0.4). Here, at T$_{eff} < 1500$K (\citealt{kirkpatrick2005}), the T dwarfs change course by turning redward in $J-$W2 color. This effect is caused both by decreasing temperature, which causes the $J$-band flux to plummet since this wavelength is now on the flux-poor Wien tail of the spectral energy distribution, and by increasing absorption by H$_2$O and CH$_4$ throughout most of the near-infrared spectrum, which squeezes most of the flux through the opacity hole on which the {\it WISE} W2 band is centered (\citealt{burrows1997,mainzer2011}).

Falling below the blue end of the main locus, from ($J-$W2, $J-K_s$) = ($-$0.6, $-$0.4) to (1.0, 0.4) are the bulk of the white dwarfs, at least those bright enough to be detectable by {\it WISE}. A small number of white dwarfs, however, appear to be superimposed on the locus of normal M dwarfs; most of these are known white dwarf + M dwarf doubles.\footnote{The exception is the DA6.8 white dwarf WD 2213+317, for which we were unable to find confirmation of a binary nature in the published literature. This object is worthy of additional follow-up.} Subdwarfs of type M and L fall below the central, most densely populated locus, with the later L subdwarfs commingling with normal T dwarfs. These objects are bluer in $J-K_s$ than their solar-metallicity counterparts because of the increased relative importance of pressure-induced H$_2$ opacity (\citealt{borysow1997}).

Two objects from the literature stand out as having unusual locations compared to objects of similar published type. The first of these, shown by the open green star at ($J-$W2,$J-K_s$) = (3.06,1.93), is 2MASS J08273118$-$1100029 classified as M0 III by \cite{cruz2003}. This object has a motion significance of only 5.2$\sigma$ and thus barely passes our test for inclusion in the motion list; because it is too faint to be seen in the DSS1 or DSS2 images, its measured motion of 47$\pm$9 mas yr$^{-1}$ cannot be independently verified and may be fictitious. Even if it not actually moving, its colors do not match those of the other early-M giants on the figure. Interstellar reddening may explain its colors, or the \cite{cruz2003} spectrum may be in error. This object is worthy of further follow-up. The second unusual object, at ($J-$W2,$J-K_s$) = (2.41,0.44), is the dust-enshrouded white dwarf G 29-38 (\citealt{zuckerman1987,wickramasinghe1988,graham1990}), a well known oddball in color space.

Figure~\ref{color_color_JW2_W1W2} shows the $J-$W2 vs.\ W1$-$W2 diagram. As with the previous figure, the densest locus of objects is composed of F, G, K, and M stars. The redward jog in W1$-$W2 color near the K/M boundary is probably caused by the appearance of H$_2$O absorption at W1 (\citealt{yamamura2010,sorahana2012}), which appears somewhat earlier in the temperature sequence than the CO absorption at W2. Running redward of these types in both colors are the L dwarfs, which reach values as red as $J-$W2 $\approx$ 4.5 at mid-L types. After this point, the later L dwarfs turn back to the blue in $J-$W2; here, the $J$-band flux increases relative to W2. This $J$-band brightening has been the subject of much speculation (see \citealt{kirkpatrick2005} for an overview), but is likely related to the disappearnce of photospheric dust that blankets the $J$-band region\footnote{The disappearance of Na and K into chloride and/or sulfide compounds robs at least the far optical portion of the spectrum of absorption by the pressure-broadened absorption wings of \ion{K}{1} and \ion{Na}{1}. However, \cite{burrows2003} find that these wings influence the continuum shape only out to $\sim$1.0 $\mu$m, so probably do not impact $J$-band.}. At this same point in the temperature sequence, the fundamental band of CH$_4$ at 3.3 $\mu$m begins to depress the W1 flux (\citealt{oppenheimer1998,noll2000,cushing2005}), first via the $Q$-branch at mid-L types and then with wider, deeper absorption in the $P$- and $R$-branches by mid-T. Also, as mentioned previously, blanketing by ever strengthening H$_2$O and CH$_4$ absorption creates a flux excess through the opacity hole in the W2 band. This combination of effects causes the W1$-$W2 colors to shift dramatically to the red. The reversal of the $J-$W2 color to the red at mid-T types is a consequence both of the excess flux escaping through the W2 opacity hole as well as the $J$-band flux plummeting as $J$-band falls further down the Wien tail of the flux distribution at cooler temperatures.

Falling below the blue end of the main locus, from (W1$-$W2, $J-$W2) = ($-$0.3, $-$0.6) to (0.2, 0.6) are the bulk of the white dwarfs. Subdwarfs of type M and L typically fall blueward in $J-$W2 and redward in W1$-$W2 of their solar-metallicity counterparts. This shift in colors is almost entirely modulated by the increased relative importance of collision-induced H$_2$ absorption, which is present at $J$, stronger at W2, and strongest at W1 (see Figures 3 and 6 of \citealt{borysow1997}). Oddly, the carbon dwarfs, which are also low-metallicity, fall to the {\it left} of the main locus and differentiate themselves far better from the bulk of other objects on this diagram than in Figure~\ref{color_color_JKs_JW2}. Absorption bands of C$_2$ and CN at $J$-band (\citealt{joyce1998}) may help to modulate what would otherwise be a blueward trend in $J-$W2 color by H$_2$ absorption alone; likewise, the expected redward trend in W1$-$W2 color may be pushed blueward of the main locus by an unknown carbon-bearing species dominating at W2 band.

The next plot, Figure~\ref{color_mag_J_JW2}, shows the $J$ vs.\ $J-$W2 diagram. Here, each spectral subtype separates out well in $J-$W2 color, with the exception of the L and T dwarfs which are known not to follow a monotonic relation (Figure 7 of \citealt{kirkpatrick2011}). Most of the white dwarfs fall to the left and below the main sequence, at values of $J-$W2 $<$ 0.6 mag. White dwarfs have colors similar to or bluer than the O, B, A, and F stars but generally fall at much fainter magnitudes, revealing their lower luminosities.

The final plot, Figure~\ref{color_mag_HJ_JW2}, shows the $H_J$ vs.\ $J-$W2 diagram, where $H_J$ is the reduced proper motion\footnote{This can be thought of as a poor astronomer's absolute magnitude, computed when parallax info may be lacking. In this case, the motion, $\mu$, is substituted for the parallax, $\pi$, in the distance modulus equation, the assumption being that objects with closer distances should also exhibit higher proper motions.} at $J$-band, defined by $H_J = J + 5log(\mu) + 5$, where $J$ is the apparent $J$-band magnitude and $\mu$ is the total proper motion in arcsec yr$^{-1}$. Here, the instrinsically fainter white dwarfs separate much more cleanly from the bulk of early-type stars than they do in Figure~\ref{color_mag_J_JW2}. The main locus of F, G, K, and M stars, along with the sequence of L and T dwarfs, shows a broad distribution in $H_J$ values for a given $J-$W2 color, indicative of the randomness of the $V_{tan}$ component of each object's space velocity. However, those random $V_{tan}$ components are generally higher for old objects with high kinematics compared to solar-age field stars. Therefore, metal-poor objects of high kinematics are easy to identify on this diagram since they generally fall much lower on the diagram (larger $H_J$ values). This plot, in concert with the color-color plots shown in Figure~\ref{color_color_JKs_JW2} and Figure~\ref{color_color_JW2_W1W2}, is extremely useful in selecting subdwarf candidates for follow-up.

\section{Spectroscopic Follow-up\label{spec_followup}}

As the AllWISE2 survey was progressing, we made early versions of the four color-color and color-magnitude plots discussed above in order to select potentially interesting sources for spectroscopic follow-up. We also made the same diagrams for the \cite{luhman2014} and NEOWISER lists to look for other interesting sources still lacking spectroscopic classifications. These observations are listed in Table 4, along with supporting observations of other literature sources that were used primarily as comparison objects, as listed in Table 5.
The acquisition and reduction of these spectra are discussed below.

\subsection{Optical}

\subsubsection{Palomar/DSpec\label{dspec}}


The Double Spectrograph (DSpec; \citealt{oke1982}) on the Hale 5m telescope on Palomar Mountain, California, served as our primary optical spectrograph in the northern hemisphere. It was used during the UT nights of 2014 Jan 26, Feb 23/24, Apr 22, Jun 25/26, Jul 21, Sep 27, Oct 24, and Nov 15 as well as 2015 Jun 08, Sep 07, and Dec 10. The blue-side detector is a CCD with $\sim$2800 pixels along the dispersion axis, and the red-side detector is an LBNL deep-depletion 4k$\times$2k CCD. For most of these nights, the 600 line mm$^{-1}$ grating blazed at 4000 \AA\ was used in the blue arm and the 600 line mm$^{-1}$ grating blazed at 10000 \AA\ was used in the red arm, with the D68 dichroic inserted to split the light near 6800 \AA\ between the two arms. Grating angles were selected so that coverage was obtained from $\sim$4000 to $\sim$10000 \AA\ with an overlap of $\sim$450 \AA\ between the two sides to allow seamless spectra to be produced across the dichroic break. (For the 2014 Feb nights, the grating angles were mis-set, resulting in a small gap of $\sim$150 \AA\ between the arms.) Use of the 1.0 slit gives a resolution typically around 3.6 \AA\ on the blue side and 2.4 \AA\ on the red.

For the nights of 2014 Jan 26, Apr 22, Jul 21, and Nov 15, the D55 dichroic was used instead to split the light near 5500 \AA, with the 600 line mm$^{-1}$ grating blazed at 4000 \AA\ in the blue arm and the 316 line mm$^{-1}$ grating blazed at 7500 \AA\ in the red arm. Grating angles were selected so that there was an overlap of 200 \AA\ between the two spectral pieces, resulting in a total coverage from 3400 to 10250 \AA. Use of the 1.5 arcsec slit gives a resolution of $\sim$3.3 \AA\ on the blue side and 5.3 \AA\ on the red. 

Reduction and calibration procedures for CCD data were employed using standard IRAF routines. A series of bias frames were median combined, and this median image was then subtracted from all flats and target frames. A series of dome flat frames were then median combined, and the resulting median image was normalized at its mean value. This normalized image was used as the flat field correction for all target images. (In some cases, flat field structure introduced by the dome lamps themselves was mitigated prior to the normalization step using the methodology described in section 4.1.4 of \citealt{kirkpatrick2010}.) Spectra were optimally extracted using variance weighting. Lamps of Fe and Ar were used for wavelength calibration for the blue-side data, and He, Ne, and Ar lamps were used for the red-side data. Flux calibration was accomplished using standard stars from \cite{hamuy1994}. The resulting blue-side and red-side spectra were then renormalized in their region of overlap near 6800 \AA\ if necesssary (normalizations usually consisted only of small tweaks around one) and the spectra stitched together at the normalization point to create a single spectrum.

\subsubsection{duPont/BCSpec}


The Boller \& Chivens Spectrograph (BCSpec\footnote{See {\url http://www.lco.cl/telescopes-information/irenee-du-pont/instruments/website/boller-chivens-spectrograph-manuals/user-manual/the-boller-and-chivens-spectrograph}.}) on the 2.5m Ir{\'e}n{\'e}e duPont telescope at Las Campanas Observatory, Chile, served as our primary optical spectrograph in the southern hemisphere and was used on the UT nights of 2014 Apr 30, May 01-04, and Nov 16-20. The camera uses a thinned, backside-illuminated Marconi 2048$\times$515 CCD, and our observations employed a 1.5-arcsec slit and the 300 line mm$^{-1}$ grism blazed at 5000 \AA\ to produce spectra from 3900 to 10100 \AA\ with a resolution of 6.9 \AA. Standard reduction procedures, like those described in section~\ref{dspec}, were employed, with one notable exception. The red fringing on this CCD is pronounced, so during our April and May runs we took dome flats after many of the targets while preserving the telescope pointing of the target, since this is the procedure recommended in the observers' manual for mitigating this effect. However, we found that this procedure does not actually work in practice. Dome flats taken just after a target and with a preserved telecope pointing rarely remove the fringing because the spacing between fringes and the fringe pattern itself have changed. Because we had a number of dome flats taken at different times and telescope positions during the nights, we looked for correlations with time or with telescope elevation, and found no obvious pattern. To remove the fringing, therefore, we employed an empirical approach in which we built a suite of flats, each having a noticeably different fringe spacing, and created different flat fielded versions of our targets using each of the flats in the suite. We then chose by eye the version for which the fringing was best removed. (Because of the significant overhead involved in both acquiring these nighttime dome flats and reducing the data in this fashion, the fringing in our November data was left uncorrected.)

Another issue also needed special attention. Because our setup results in a spectrum across 3900 to 10100 \AA, objects with signifcant flux below 5050 \AA\ will have the red ends of their spectra contaminated by second-order light. Most of our objects are M dwarfs for which this effect is minimal, but such is not the case for the \cite{hamuy1994} flux standards we employed. Thus, the flux calibration derived from the blue stars will incorrectly calibrate the red end of the red stars. To correct for this, we employed another empirical approach, where we found late-M dwarfs with identical spectral types between the duPont runs and the Palomar runs. Because the Palomar data do not suffer from second-order light (the dichroic between the two arms serves as an order-blocking filter), we can use them to derive the red-end tweak to the flux calibration of our targets. Specifically, we compared the fluxed spectrum of the Palomar-observed M7.5 dwarf WISE 0852+5319 to the duPont-observed M7.5 dwarf WISE 1218+1140 and the duPont-observed M8 dwarfs WISE 1546$-$5534 and WISE 2002$-$4433 to derive a correction to be multiplied back into all the flux-calibrated duPont data. This correction is identically one below 9600 \AA\ and rises linearly to 1.81 at 10100 \AA. This correction should be sufficient for all of the M stars observed, although data longward of 8800 \AA\ in much bluer targets (WISE 0345$-$0348AB, WISE 1028$-$6327, and WISE 1747$-$5436) should be used with caution.

\subsubsection{NTT/EFOSC2}


Spectra of ten objects were obtained on the UT nights of 2014 Jul 03-04 and 2015 Dec 07-10 at the European Southern Observatory (ESO) 3.58m New Technology Telescope (NTT) at La Silla, Chile, with the ESO Faint Object Spectrograph and Camera, version 2 (EFOSC2)\footnote{See {\url http://www.ls.eso.org/lasilla/Telescopes/ntt/efosc/ manual/EFOSC2manual\_v3.1.pdf}.}. EFOSC2 uses a Loral 2048$\times$2048 CCD, and observations employed an on-chip 2$\times$2 binning. The spectograph was used in long-slit mode with a 1-arcsec slit and grism \#13 (236 grooves mm$^{-1}$ blazed at 4400 \AA), which resulted in spectra running from 4000 to 9300 \AA\ at a resolution of 36 \AA. Standard reduction procedures, like those described in section~\ref{dspec}, were employed. No mitigation for fringing in the far red, most noticeable in the high signal-to-noise spectrum of WISE 2101$-$4907, was attempted due to lack of adequate calibrations for this.

\begin{center}
\begin{longtable*}{llccccc}
\tablenum{4}
\tablecaption{Spectroscopic Follow-up of {\it WISE} Motion Sources\label{spectra_allwise}}
\tablehead{
\colhead{WISEA Designation} &  
\colhead{Other Name}&                        
\colhead{Opt} &  
\colhead{NIR} &  
\colhead{Telescope/} &     
\colhead{Obs.\ Date} &
\colhead{Exp.\ Time\tablenotemark{a}} \\
\colhead{} &                          
\colhead{} &                          
\colhead{Sp.\ Type} &  
\colhead{Sp.\ Type} &  
\colhead{Instrument} &     
\colhead{(UT)} &
\colhead{(s)} \\
\colhead{(1)} &                          
\colhead{(2)} &  
\colhead{(3)} &  
\colhead{(4)} &     
\colhead{(5)} &
\colhead{(6)} &
\colhead{(7)} 
}
\hline

\multicolumn{7}{l}{$^a$For double-armed spectrographs, the first value is the exposure time of the blue arm and the second is the red arm.} \\
\multicolumn{5}{l}{$^b$These two objects are unresolved by AllWISE.} \\
\multicolumn{5}{l}{$^c$Serendipitous flare observation.} \\
\multicolumn{7}{l}{$^d$Object is best fit by the esdM3 standard, except that its Na ``D'' doublet is exceptionally broad and deep and the \ion{Na}{1} doublet at} \\
\multicolumn{7}{l}{8183 and 8195 \AA\ is much stronger than the corresponding lines observed in the standard. See text for details.} \\
\multicolumn{5}{l}{$^e$Spectrum is reddened.}

\endlastfoot

J000149.00$-$221347.8 & LEHPM 50                  & esdK7   & --- & duPont/BCSpec & 2014 Nov 19  &   3600    \\  
J000538.83+020951.2   & NLTT 177                  & M6      & --- & duPont/BCSpec & 2014 Nov 19  &   1800    \\  
J000622.67$-$131955.6 & ---                       & L7:     & --- & Keck/LRIS     & 2014 Jun 26  &  ---/400  \\  
J003449.93+551352.8   & ---                       & M4.5    & --- & Palomar/DSpec & 2015 Dec 10  & 1800/1860 \\  
J005122.35$-$225133.5 & LEHPM 1029                & M8      & --- & duPont/BCSpec & 2014 Nov 18  &   3600    \\  
J011157.18+121132.5   & ---                       & M8      & --- & Palomar/DSpec & 2014 Nov 15  &  ---/1860 \\  
J013012.66$-$104732.4 & ---                       & M9      & --- & Palomar/DSpec & 2015 Dec 10  &  ---/3600 \\  
J013028.28$-$383639.2 & ---                       & M5      & --- & duPont/BCSpec & 2014 Nov 16  &   2400    \\  
J013042.06$-$064705.1 & ---                       & sdM5:   & --- & Palomar/DSpec & 2015 Dec 10  & 2400/2460 \\  
J013407.74+052503.8   & NLTT 5216                 & M4.5    & --- & duPont/BCSpec & 2014 Nov 19  &   2400    \\  
J015743.59$-$094810.5 & ---                       & M4      & --- & Palomar/DSpec & 2014 Nov 15  & 1360/910  \\  
J015812.03+323157.9   & ---                       & ---     & L4.5& IRTF/SpeX     & 2014 Nov 11  &   1800    \\  
J020645.37$-$242823.9 & LEHPM 2189                & sdM5.5  & --- & Palomar/DSpec & 2015 Sep 07  & 2400/2460 \\  
J022855.32$-$043950.4 & PM J02289$-$0439          & sdM2    & --- & duPont/BCSpec & 2014 Nov 17  &   2400    \\  
J023119.81+281144.4   & ---                       & M9      & --- & Palomar/DSpec & 2014 Sep 27  &  ---/2400 \\  
J023624.90$-$204105.9 & LP 830-18                 & dCarbon & --- & Palomar/DSpec & 2014 Sep 27  & 1260/1260 \\  
...                   & ...                       & dCarbon & --- & Palomar/DSpec & 2014 Oct 24  & 1260/1260 \\  
J023852.70+361733.6   & LP 245-62& esdM3 pec\tablenotemark{d}& ---& Palomar/DSpec & 2014 Nov 15  & 1860/1860 \\  
J030706.80$-$591444.6 & ---                       & sdM1    & --- & duPont/BCSpec & 2014 Nov 16  &   2400    \\  
J030933.60$-$135431.4 & ---                       & M6:     & --- & duPont/BCSpec & 2014 Nov 16  &   3600    \\  
J033039.63$-$234845.6 & ---                       & esdM8:  & --- & Palomar/DSpec & 2014 Sep 27  &  ---/1600 \\  
J033613.83+001011.0   & ---                       & sdM7.5  & --- & Palomar/DSpec & 2014 Oct 24  & 1800/1860 \\  
J033806.20$-$491654.3 & ---                       & sdM3    & --- & duPont/BCSpec & 2014 Nov 17  &   3600    \\  
J034122.19$-$495130.0 & LEHPM 3466                & sdM1    & --- & duPont/BCSpec & 2014 Nov 18  &   3600    \\  
J034501.64$-$034848.6A& ---                       & DA?     & --- & duPont/BCSpec & 2014 Nov 16  &   3600    \\  
J034501.64$-$034848.6B& ---                       & DA?     & --- & duPont/BCSpec & 2014 Nov 16  &   3600    \\  
J040432.39$-$625918.7 & L 129-35                  & M4      & --- & duPont/BCSpec & 2014 Nov 20  &   1800    \\  
J040546.68+371941.9   & ---                       & M5      & --- & Palomar/DSpec & 2014 Feb 23  &  180/180  \\  
J041847.95+252001.8   & ---     & ---     & mid-K\tablenotemark{e}  & IRTF/SpeX   & 2015 Jan 27  &   1200    \\  
J042225.13+033708.3   & 1RXS J042224.0+033710     & M4.5    & --- & Palomar/DSpec & 2014 Feb 23  &  120/120  \\  
J044755.76+253446.5   & ---                       & M4      & --- & Palomar/DSpec & 2014 Feb 23  &  120/120  \\  
J045806.65$-$280604.9 & LP 891-41                 & usdK8:  & --- & duPont/BCSpec & 2014 Nov 19  &   2400    \\  
J050011.70+191625.9   & ---                       & M1      & --- & Palomar/DSpec & 2014 Feb 23  &  120/120  \\  
J050048.17+044214.2   & ---                       & L0.5 pec& --- & Palomar/DSpec & 2014 Oct 24  &  900/960  \\  
J050603.86$-$381647.6 & ---                       & esdM3.5 & --- & duPont/BCSpec & 2014 Nov 16  &   3600    \\  
J050750.72$-$034245.8 & ---                       & M9 pec? & --- & Palomar/DSpec & 2015 Dec 10  &  ---/4800 \\  
J050854.88+331920.6   & ---                       & L2      & --- & Palomar/DSpec & 2014 Feb 23  & 1260/1260 \\  
J052452.57+463202.9   & ---                       & sdM3    & --- & Palomar/DSpec & 2015 Dec 10  & 2400/2460 \\  
J053641.86$-$000600.2 & 1RXS J053641.2$-$000555   & M4.5    & --- & Palomar/DSpec & 2014 Feb 24  &  120/120  \\  
J054608.21$-$044011.4 & ---                       & M4.5    & --- & Palomar/DSpec & 2014 Feb 24  &  120/120  \\  
J055929.26+584414.7   & ---                       & M9      & --- & Palomar/DSpec & 2014 Feb 23  & 1860/1860 \\  
J063228.30+264347.3   & ---                       & M7      & --- & Palomar/DSpec & 2015 Dec 10  &  ---/4800 \\  
J063226.88+264404.5   & LP 363-4                  & K5      & --- & Palomar/DSpec & 2015 Dec 10  &  120/120  \\  
J065451.27+161106.9   & ---                       & K0      & --- & Palomar/DSpec & 2014 Feb 24  &  120/120  \\  
J070128.35$-$013713.1 & ---                       & M4      & --- & Palomar/DSpec & 2014 Feb 24  &  120/120  \\  
J070512.04$-$100751.9 & ---                       & M5      & --- & Palomar/DSpec & 2014 Feb 24  &  120/120  \\  
J073053.34$-$633522.0 & SCR J0730$-$6335          & M4      & --- & duPont/BCSpec & 2014 Nov 20  &   1800    \\  
J075353.01$-$645920.4 & ---                       & usdM1:  & --- & duPont/BCSpec & 2014 Nov 19  &   3000    \\  
J082640.45$-$164031.8 & ---                       & L8:     & --- & Palomar/DSpec & 2014 Feb 23  &  ---/1200 \\  
J085224.36+513925.5   & ---                       & M7.5    & --- & Palomar/DSpec & 2014 Feb 24  & 1260/1260 \\  
J085512.39$-$023356.8 & ---                       & L7      & --- & Keck/LRIS     & 2015 Dec 05  &  ---/600  \\  
J091250.93+220511.1   & PYC J09128+2205           & M3      & --- & Palomar/DSpec & 2014 Feb 24  &  120/120  \\  
J092026.98$-$755724.5 & SCR J0920$-$7557          & M4.5    & --- & duPont/BCSpec & 2014 Apr 30  &   1500    \\  
J093513.92$-$030111.6 & ---                       & M3      & --- & Palomar/DSpec & 2014 Feb 24  &  120/120  \\  
J094812.96$-$190905.7 & LP 788-38 (LHS 2192)      & usdK7   & --- & duPont/BCSpec & 2014 May 02  &   1200    \\  
J094904.92+023251.4   & ---                       & sdM5    & --- & Palomar/DSpec & 2015 Dec 10  & 2400/2400 \\  
J094929.34$-$010309.5 & ---                       & M0      & --- & Palomar/DSpec & 2014 Feb 24  &  120/120  \\  
J101329.72$-$724619.2 & ---                       & ---   & sdL2? & Magellan/FIRE & 2015 Feb 08  &    960    \\  
J101923.16+392259.9   & ---                       & M4.5    & --- & Palomar/DSpec & 2014 Feb 24  &  120/120  \\  
J102807.87$-$632708.3 & ---                       & DAZ?    & --- & duPont/BCSpec & 2014 May 01  &   3600    \\  
J102940.10+571543.7   & SDSS J102939.69+571544.3& ---& L6 pec (blue) & IRTF/SpeX  & 2015 May 08  &   1200    \\  
J102944.49+254536.5   & ---                       & M4.5    & --- & Palomar/DSpec & 2014 Feb 24  &  120/120  \\  
J105515.71$-$735611.3 & ---                       & M7      & --- & NTT/EFOSC2    & 2015 Dec 07  &   1200    \\  
J105536.09$-$575042.1 & ---                       & M4      & --- & duPont/BCSpec & 2014 Apr 30  &   1200    \\  
J105607.88$-$575041.1 & UPM J1056$-$5750          & M4      & --- & duPont/BCSpec & 2014 Apr 30  &   1200    \\  
J105938.05+150906.0   & ---                       & M3.5    & --- & Palomar/DSpec & 2014 Feb 24  &  120/120  \\  
J111431.54+570315.9   & ---                       & M0      & --- & Palomar/DSpec & 2014 Feb 24  &  120/120  \\  
J112152.91$-$264937.3 & ---                       & sdM2    & --- & NTT/EFOSC2    & 2015 Dec 10  &   1800    \\  
J113333.55$-$413954.4 & CD$-$40 6796              & K0      & --- & duPont/BCSpec & 2014 May 02  &   120     \\  
J113333.67$-$414016.7 & ---                       & M0.5    & --- & duPont/BCSpec & 2014 May 02  &   600     \\  
J114025.30$-$062412.1 & LP 673-26                 & M5      & --- & Palomar/DSpec & 2014 Feb 24  &  120/120  \\  
J120146.62$-$132404.8 & ---                       & M9.5    & --- & Palomar/DSpec & 2014 Apr 22  &  ---/900  \\  
J120224.36$-$011145.7 & ---                       & M4      & --- & Palomar/DSpec & 2014 Feb 24  &  120/120  \\  
J120352.53+181057.9   & ---                       & M0      & --- & Palomar/DSpec & 2014 Feb 24  &  120/120  \\  
J120624.97+001601.0   & ---                       & M5      & --- & Palomar/DSpec & 2014 Feb 24  &  120/120  \\  
J120641.18+184137.9   & TYC 1442-929-1            & K2      & --- & Palomar/DSpec & 2014 Jun 25  &  120/120  \\  
J120820.12$-$052856.7 & ---                       & K7      & --- & Palomar/DSpec & 2014 Feb 24  &  120/120  \\  
J121058.26$-$461204.5 & RAVE J121058.2$-$461207   & M1      & --- & duPont/BCSpec & 2014 May 04  &   120     \\  
J121058.01$-$461917.3 & LTT 4560                  & G0      & --- & duPont/BCSpec & 2014 May 04  &    20     \\  
J121254.00$-$415932.3 & WT 320                    & esdK7   & --- & duPont/BCSpec & 2014 May 02  &   2400    \\  
J121556.24$-$501419.2 & PM J12159$-$5014          & sd/esdM2& --- & duPont/BCSpec & 2014 Apr 30  &   3600    \\  
J121830.24+114012.9   & ULAS2MASS J1218+1140      & M7.5    & --- & duPont/BCSpec & 2014 May 01  &   5400    \\  
J122208.49$-$844905.6 & ---                       & M3      & --- & duPont/BCSpec & 2014 May 02  &   120     \\  
J122248.13$-$211638.6 & ---                       & L7:     & --- & Keck/LRIS     & 2015 Dec 05  &  ---/600  \\  
J122355.12+551050.3   & ---                       & M8 pec  & --- & Palomar/DSpec & 2015 Dec 10  &  ---/2460 \\  
J122745.55$-$454114.1 & SCR J1227$-$4541          & usdK?   & --- & duPont/BCSpec & 2014 May 02  &   1200    \\  
J123517.17+445043.6   & ---                       & M4.5    & --- & Palomar/DSpec & 2014 Feb 24  &  120/120  \\  
J124007.18+204828.9   & G 59-32                   & K2      & --- & duPont/BCSpec & 2014 May 04  &   20      \\  
J124014.80+204752.7   & ---                       & M7      & --- & duPont/BCSpec & 2014 May 04  &   1800    \\  
...                   & ...                       & M7      & --- & duPont/BCSpec & 2014 May 04  &   1800\tablenotemark{c}    \\  
J124135.42$-$245748.9 & ---            & ---     & L2.5p (sl.bl.) & IRTF/SpeX     & 2015 May 08  &   1200    \\  
J124352.14$-$405829.9 & WT 338                    & esdM2.5 & --- & duPont/BCSpec & 2014 May 02  &   1200    \\  
J124725.86$-$434353.2 & LTT 4892                  & G0      & --- & duPont/BCSpec & 2014 May 04  &   20      \\  
J124726.75$-$434441.8 & SCR J1247$-$4344B         & M5      & --- & duPont/BCSpec & 2014 May 04  &   1200    \\  
J124715.16$-$444149.1 & PM J12472$-$4441          & sdM5    & --- & duPont/BCSpec & 2014 May 02  &   3600    \\  
...                   & ...                       & sdM5    & --- & NTT/EFOSC2    & 2014 Jul 04  &   1000    \\  
J130556.06$-$101928.9 & LP 736-24                 & M5      & --- & Palomar/DSpec & 2014 Feb 23  & 1260/1260 \\  
J130813.94$-$125028.0 & LP 737-1                  & ...     & N/A & Keck/NIRSPEC  & 2014 Apr 12  &    720    \\  
...                   & ...                       & d/sdM2  & --- & duPont/BCSpec & 2014 May 02  &   1200    \\  
J133316.07+374422.4   & SDSS J133316.06+374421.7  & ---     & L5  & IRTF/SpeX     & 2015 May 09  &   1200    \\  
J134310.42$-$121628.9 & WISE J134310.44$-$121628.8& L4      & --- & Palomar/DSpec & 2014 Jan 26  &  ---/600  \\  
J134824.42$-$422744.9 & ---                       & ---     & L2  & IRTF/SpeX     & 2015 May 09  &   1200    \\  
J135501.90$-$825838.9 & ---                       & ---     &sdL5?& Magellan/FIRE & 2015 May 31  &   3600    \\  
J140350.20$-$592348.0 & ---                       & M3      & --- & duPont/BCSpec & 2014 May 01  &   1200    \\  
J140400.30$-$592400.5 & L 197-165                 & M3      & --- & duPont/BCSpec & 2014 May 01  &   1200    \\  
J140458.33$-$472632.2 & ---                       & M6      & --- & duPont/BCSpec & 2014 May 01  &   3600    \\  
J141143.25$-$452418.4 & ---                       & sdM9    & --- & duPont/BCSpec & 2014 May 04  &   5400    \\  
...                   & ...                       & sdM9    & --- & NTT/EFOSC2    & 2014 Jul 04  &   1000    \\  
...                   & ...                       & ---     & sdM9& IRTF/SpeX     & 2015 Jul 03  &   1500    \\  
J141144.12$-$140300.6 & ---                       & M9      & --- & Palomar/DSpec & 2015 Jun 08  & ---/4860  \\  
J142350.08$-$164612.5 & SCR J1423$-$1646B         & M0.5    & --- & Palomar/DSpec & 2015 Jun 08  &  190/190  \\  
J142350.33$-$164603.8 & PPM 228725                & K5 pec  & --- & Palomar/DSpec & 2015 Jun 08  &  190/190  \\  
J145052.05$-$212547.1 & LP 801-17                 & sdM2.5  & --- & Palomar/DSpec & 2015 Jun 08  &  660/660  \\  
J145131.36+335222.0   & ---                       & ---     &sdM4?& IRTF/SpeX     & 2015 May 09  &   1200    \\  
J145255.48+272322.7   & SDSS J145255.58+272324.4  & L0      & --- & Palomar/DSpec & 2015 Jun 08  & ---/3660  \\  
J145408.03+005325.7   & TVLM 868-20073            & M9      & --- & Palomar/DSpec & 2015 Jun 08  & ---/3660  \\  
J145409.18+005338.8   & Wolf 559                  & M3      & --- & Palomar/DSpec & 2015 Jun 08  &  120/120  \\  
J145725.71+234124.8\tablenotemark{b} & LSPM J1457+2341N & dCarbon & --- & Palomar/DSpec & 2014 Jun 25-26  & 3060/3060\\  
...                   & ...                       & ---     & dCarbon & IRTF/SpeX & 2015 May 08  &   1200    \\  
J145725.71+234124.8\tablenotemark{b} & LSPM J1457+2341S & sdM8    & --- & Palomar/DSpec & 2014 Jun 25-26  & 6000/4860\\  
...                   & ...                       & ---     & sdM8 & IRTF/SpeX    & 2015 May 08  &   1200    \\  
J150712.17+603038.5   & ---                       & L2:     & --- & Palomar/DSpec & 2014 Jun 25-26 & ---/6120 \\  
J151652.40$-$283219.6 & ---                       & M5      & --- & Palomar/DSpec & 2014 Feb 23  &  120/120  \\  
J153914.96$-$535241.5 & WISE J153914.96-535241.5  & M4      & --- & duPont/BCSpec & 2014 May 02  &    600    \\  
J154045.67$-$510139.3 & ---                       & M6.5    & --- & duPont/BCSpec & 2014 May 01  &    600    \\  
J154225.49$-$100708.5 & SIPS J1542$-$1007         & M7.5    & --- & Palomar/DSpec & 2014 Feb 23-24 &1200/2460 \\  
J154644.64$-$525438.5 & WISE J154644.64$-$525438.5& M5      & --- & duPont/BCSpec & 2014 May 01  &    900    \\  
...                   & ...                       & M5      & --- & duPont/BCSpec & 2014 May 02  &   1800    \\  
J161417.08$-$815111.9 & ---                       & usdM9   & --- & duPont/BCSpec & 2014 May 02  &   3600    \\  
...                   & ...                       & late-usdM&--- & NTT/EFOSC2    & 2014 Jul 04  &    800    \\  
...                   & ...                       & ---     &usdM9& Magellan/FIRE & 2014 Aug 07  &    720    \\  
J161519.28+033601.8   & ---                       & M7      & --- & Palomar/DSpec & 2014 Feb 24  & 1200/1200 \\  
J163035.70$-$201751.4 & ---            & K5\tablenotemark{e}& --- & duPont/BCSpec & 2014 Apr 30  &   1800    \\  
J163419.49+482758.6   & G 202-59                  & M4      & --- & Palomar/DSpec & 2015 Jun 08  &  120/120  \\  
J170221.19+715841.7A  & LP 43-310                 & M4.5    & --- & Palomar/DSpec & 2015 Jun 08  & 3660/3660 \\  
J170221.19+715841.7B  & LSPM J1702+7158N          & cold wd & --- & Palomar/DSpec & 2015 Jun 08  & 3660/3660 \\  
J171257.92+064540.2   & ---                    & ---     & T2 pec & Magellan/FIRE & 2015 May 31  &   3000    \\  
J171826.98$-$224543.5 & UPM J1718$-$2245A         & M3.5    & --- & duPont/BCSpec & 2014 Apr 30  &   1200    \\  
J171828.99$-$224630.2 & UPM J1718$-$2245B         & M4.5    & --- & duPont/BCSpec & 2014 Apr 30  &   1200    \\  
J172230.07$-$695119.2 & 1RXS J172231.8$-$695121   & M3      & --- & duPont/BCSpec & 2014 May 01  &   1200    \\  
J172237.14$-$695112.2 & ---                       & M4      & --- & duPont/BCSpec & 2014 May 01  &   2400    \\  
J174019.07$-$550726.9 & ---                       & M2.5    & --- & duPont/BCSpec & 2014 May 04  &   3600    \\  
J174102.52$-$423454.8 & ---                       & M4      & --- & duPont/BCSpec & 2014 May 01  &   1200    \\  
J174344.00+631322.7   & ---                       & M9      & --- & Palomar/DSpec & 2014 Jun 25  & 1800/1860 \\  
J174634.81+510011.0   & ---                       & L0:     & --- & Palomar/DSpec & 2014 Apr 22  &  ---/1200 \\  
J174724.76+400851.4   & LSPM J1747+4008           & M4      & --- & Palomar/DSpec & 2014 Jun 26  &  660/660  \\  
J174736.29$-$543634.4 & PM J17476$-$5436          & DC      & --- & duPont/BCSpec & 2014 Apr 30  &   1800    \\  
...                   & ...                       & DC      & --- & NTT/EFOSC2    & 2014 Jul 04  &    400    \\  
J175839.18$-$583931.9 & ---                       & M6      & --- & duPont/BCSpec & 2014 Apr 30  &   1800    \\  
J180001.15$-$155927.2 & 2MASS J18000116$-$1559235 & L4.5    & --- & Palomar/DSpec & 2014 Jun 26  &  ---/1260 \\  
J180405.05+562134.2   & NLTT 45912                & dCarbon & --- & Palomar/DSpec & 2015 Jun 08  &  180/180  \\  
...                   & ...                       & dCarbon & --- & Palomar/DSpec & 2015 Sep 07  & 1260/1260 \\  
J183555.14$-$222614.5 & ---           & G8\tablenotemark{e} & --- & Palomar/DSpec & 2014 Sep 27  &  900/900  \\  
J183544.27$-$791212.6 & ---                       & M5      & --- & duPont/BCSpec & 2014 May 04  &    600    \\  
J183942.17+124910.3   & ---         & mid-G\tablenotemark{e}& --- & Palomar/DSpec & 2015 Jun 08  &  120/120  \\  
J184202.04+210428.7   & ---                       & L1.5    & --- & Palomar/DSpec & 2014 Jun 25-26&6000/6120 \\  
J184333.42$-$635550.0 & WT 583                    & M4.5    & --- & NTT/EFOSC2    & 2014 Jul 04  &    400    \\  
J190520.39$-$543445.0 & WT 625                    & M4      & --- & duPont/BCSpec & 2014 May 02  &    240    \\  
J191636.63$-$154008.6 & ---                       & usdM0   & --- & duPont/BCSpec & 2014 May 04  &   2400    \\  
J194023.28+634602.5   & ---                       & M9.5    & --- & Palomar/DSpec & 2014 Jun 25  & 1800/1860 \\  
J194152.72$-$020856.5 & PM J19418$-$0208          & sdM8    & --- & Palomar/DSpec & 2014 Jun 26  & 1200/1260 \\  
...                   & ...                       & ---     & sdM8& IRTF/SpeX     & 2015 Jun 27  &   1200    \\  
...                   & ...                       & ---     & N/A & Keck/NIRSPEC  & 2015 Jul 03  &   1200    \\  
J200201.12+794206.0   & NLTT 48736                & sdM2    & --- & Palomar/DSpec & 2014 Oct 24  & 2460/2460 \\  
J200252.08$-$443313.0 & ---                       & M8      & --- & duPont/BCSpec & 2014 May 04  &   2400    \\  
J200403.17$-$263751.9 & ---                       & M7.5:   & --- & Palomar/DSpec & 2014 Sep 27  &  ---/2460 \\  
J200756.41+700100.5   & NLTT 48835                & M6      & --- & Palomar/DSpec & 2014 Jun 25  & 1200/1260 \\  
J200825.00+703058.5   & ---                       & M7:     & --- & Palomar/DSpec & 2014 Sep 27  &  ---/2400 \\  
J204543.15$-$141140.3 & PM J20457$-$1411          & usdM5   & --- & Palomar/DSpec & 2014 Jun 25-26&3000/3120 \\  
...                   & ...                       & ---     &usdM5& IRTF/SpeX     & 2015 Jun 27  &   1200    \\  
J210104.88$-$490626.5 & WT 765                    & DA?     & --- & NTT/EFOSC2    & 2014 Jul 03  &    400    \\  
J210107.08$-$490727.1 & WT 766                    & M4.5    & --- & NTT/EFOSC2    & 2014 Jul 03  &    400    \\  
J212021.16+265218.7   & ---                       & M7      & --- & Palomar/DSpec & 2014 Oct 24  & 3600/3720 \\  
J212100.87$-$623921.6 & ---                       & ---     & T2  & Magellan/FIRE & 2014 Aug 07  &    720    \\  
J213322.53+731943.0   & ---                       & M4.5    & --- & Palomar/DSpec & 2014 Oct 24  & 1260/1260 \\  
J213409.15+713236.1   & ---                       & sdM9    & --- & Palomar/DSpec & 2014 Oct 24  & 4800/5040 \\  
...                   & ...                       & ---     & N/A & Keck/NIRSPEC  & 2014 Dec 03  &   2400    \\  
...                   & ...                       & sdM9    & --- & Keck/LRIS     & 2015 Aug 13  &  ---/600  \\  
J213512.67+731236.9   & ---                       & L2:     & --- & Palomar/DSpec & 2014 Jun 26  &  ---/1260 \\  
J220041.80$-$463613.2 & ---                       & M4.5    & --- & NTT/EFOSC2    & 2015 Dec 07  &   3600    \\  
J220139.54$-$411208.6 & ---                       & M4:     & --- & duPont/BCSpec & 2014 Nov 19  &   3600    \\  
J222355.08$-$222851.4 & LEHPM 4665                & sdK7::  & --- & duPont/BCSpec & 2014 Nov 16  &   1800    \\  
J223006.48$-$272007.9 & ---                       & L0:     & --- & Palomar/DSpec & 2014 Jun 25  &  ---/1860 \\  
J223041.19$-$095048.4 & ---                       & esdK7:: & --- & duPont/BCSpec & 2014 Nov 17  &   3600    \\  
J224909.81+320547.0   & 2MASSI J2249091+320549    & L4:     & --- & Keck/LRIS     & 2014 Jun 26  &  ---/400  \\  
J225830.54$-$260110.9 & ---                       & sdM6.5  & --- & Palomar/DSpec & 2014 Jun 25  &  600/600  \\  
J230434.29+211142.5   & NLTT 55748                & M0.5    & --- & Palomar/DSpec & 2015 Jun 08  &  360/360  \\  
J230435.03+211132.4   & NLTT 55750                & M4.5    & --- & Palomar/DSpec & 2015 Jun 08  &  360/360  \\  
J232404.34+161721.9   & ---                       & M8      & --- & Palomar/DSpec & 2014 Jul 21  &  ---/300  \\  
J234441.89+131243.3   & ---                       & M7:     & --- & duPont/BCSpec & 2014 Nov 18  &   3600    \\  
\end{longtable*}
\end{center}

\subsubsection{Keck/LRIS}


Spectra of seven objects were obtained on the UT nights of 2014 Jun 26, 2015 Aug 13, and 2015 Dec 05 with the Low Resolution Imaging Spectrometer (LRIS, \citealt{oke1995}) at the 10m W.\ M.\ Keck Observatory on Mauna Kea, Hawai'i. The blue camera has a two 2k$\times$4k e2v (Marconi) CCDs, and the red camera has two 2k$\times$4k LBNL thick, high-resistivity CCDs. Observations employed a 600 line mm$^{-1}$ grating blazed at 4000 \AA\ in the blue channel and a 400 line mm$^{-1}$ grating blazed at 8500 \AA\ in the red. The 560 dichroic was used to split light between channels near 5600 \AA. The blue channel produced a spectrum covering the wavelength range from 3200 to 5600 \AA, but none of the targets had measurable flux here. The red channel produced a spectrum covering the range from 5440 to 10270 \AA. Use of the 1.0-arcsec slit produced a red-channel resolution of $\sim$7.9 \AA\ for the 2014 Jun 26 data; slightly more degraded resolution was seen for the 2015 data because a 1.5-arscec slit was used. Standard reduction procedures, like those described in section~\ref{dspec}, were employed.

\subsection{Near-infrared}

\subsubsection{IRTF/SpeX}


SpeX (\citealt{rayner2003}) on the NASA 3m Infrared Telescope Facility (IRTF) on Mauna Kea, Hawai'i, served as our primary near-infrared spectrograph in the northern hemisphere. The UT dates of observation were 2014 Nov 11 and 2015 Jan 27, May 08-09, Jun 27, Jul 03-05, and Jul 20. SpeX was used in prism mode with a 0$\farcs$5 or 0$\farcs$8 wide slit to achieve a resolving power of R$\equiv\lambda / \Delta \lambda \approx 100-150$ over the range 0.8-2.5 $\mu$m. All data were reduced using Spextool (\citealt{cushing2004}). A0 stars were used for the telluric correction and flux calibration steps following the technique described in \cite{vacca2003}. 

\subsubsection{Magellan/FIRE}


The Folded-port Infrared Echellette (FIRE; \citealt{simcoe2008}, \citealt{simcoe2010}) at the 6.5m Walter Baade Telescope at Las Campanas Observatory, Chile, served as our primary near-infrared spectrograph in the southern hemisphere. The UT dates of observation were 2014 Aug 07-09, 2015 Feb 08, and 2015 May 31. In prism mode, which was used for WISE 1013$-$7246, WISE 1614$-$8151, and WISE 2121$-$6239, FIRE covers a wavelength range from 0.8 to 2.5 $\mu$m at a resolution ranging from R=500 at $J$-band to R=300 at $K$-band for a slit width of 0$\farcs$6. In echelle mode, which was used for WISE 1355$-$8258 and WISE 1712+0645, FIRE covers a wavelength range from $Y$ through $K$ bands at a resolution of R=6000 to 8000 for a slit width of 0$\farcs$6. Data were reduced using the FIREHOSE pipeline, which is based on the MASE (\citealt{bochanski2009}) and Spextool packages. In the case of the prism data, the Ar line list was supplemented with additional line identifications from \cite{norlen1973} to allow for a more robust wavelength solution in the $K$-band.

\setcounter{LTchunksize}{50}
\begin{center}
\begin{longtable*}{llccccc}
\tabletypesize{\tiny}
\tablenum{5}
\tablecaption{Spectroscopic Follow-up of Literature Sources of Interest\label{spectra_literature}}
\tablehead{
\colhead{WISEA Designation} &  
\colhead{Other Name}&                        
\colhead{Opt} &  
\colhead{NIR} &  
\colhead{Telescope/} &     
\colhead{Obs.\ Date} &
\colhead{Exp.\ Time\tablenotemark{a}} \\
\colhead{} &                          
\colhead{} &                          
\colhead{Sp.\ Type} &  
\colhead{Sp.\ Type} &  
\colhead{Instrument} &     
\colhead{(UT)} &
\colhead{(s)} \\
\colhead{(1)} &                          
\colhead{(2)} &  
\colhead{(3)} &  
\colhead{(4)} &     
\colhead{(5)} &
\colhead{(6)} &
\colhead{(7)} 
}
J001450.17$-$083823.4 & ---                       & ---      & N/A & Keck/NIRSPEC  & 2014 Dec 03  &   1200    \\  
...                   & ...                       & ---      & sdL0& IRTF/SpeX     & 2015 Jul 20  &   1200    \\  
J004326.26+222124.0   & ---                       & M8 pec?  & --- & Palomar/DSpec & 2014 Oct 24  & 2400/2460 \\  
...                   & ...                       & M8       & --- & Keck/LRIS     & 2015 Aug 18  &  ---/900  \\  
J043535.82+211508.9   & ---                       & sdL0     & --- & Palomar/DSpec & 2014 Feb 24  & 1800/1860 \\  
J045210.00$-$224517.0 & LEHPM 2-59                & esdM8.5  & --- & Palomar/DSpec & 2014 Feb 24  & 1260/1260 \\  
J045921.21+154059.2   & ---                       & sdM6     & --- & Palomar/DSpec & 2014 Feb 24  & 1320/1320 \\  
...                   & ...                       & ---      & N/A & Keck/NIRSPEC  & 2014 Dec 03  &   1200    \\  
J055859.23$-$290326.0 & APMPM J0559$-$2903        & esdM7.5  & --- & Palomar/DSpec & 2014 Feb 24  & 1260/1260 \\  
...                   & ...                       & ---      & N/A & Keck/NIRSPEC  & 2014 Dec 03  &   1200    \\  
J072003.20$-$084651.3 & ---                       & M9       & --- & Palomar/DSpec & 2014 Feb 23  &  780/780  \\  
...                   & ...                       & M9       & --- & Palomar/DSpec & 2014 Feb 24  &  660/660  \\  
...                   & ...                       & M9       & --- & Palomar/DSpec & 2015 Dec 10  & 1800/1800 \\  
J072342.97+031617.7   & LSPM J0723+0316           & sdM7.5   & --- & Palomar/DSpec & 2014 Feb 23  & 1260/1260 \\  
J082234.00+170013.6   & LSR J0822+1700            & usdM7.5  & --- & Palomar/DSpec & 2014 Feb 23  & 1260/1260 \\  
J101307.39$-$135632.2 & SSSPM J1013$-$1356        & sdM9     & --- & Palomar/DSpec & 2014 Feb 23  & 1260/1260 \\  
J115826.62+044745.0   & ULAS J115826.62+044746.8  & sdM7.5   & --- & Palomar/DSpec & 2014 Feb 23-24& 3600/3660\\  
J122704.69$-$044718.0 & 2MASS J12270506$-$0447207 & usdM8.5  & --- & Palomar/DSpec & 2014 Feb 23  & 1260/1260 \\  
...                   & ...                       & usdM8.5  & --- & Palomar/DSpec & 2014 Jun 26  & 1800/1920 \\  
J124425.63+102437.3   & ULAS J124425.90+102441.9  & sdM9     & --- & Palomar/DSpec & 2014 Feb 23  &  ---/2760 \\  
J125613.46$-$140851.0 & SSSPM J1256$-$1408        & sdM9     & --- & Palomar/DSpec & 2014 Feb 23  & 1260/1260 \\  
...                   & ...                       & sdM9     & --- & Palomar/DSpec & 2014 Jun 25  & 1200/1260 \\  
J125636.75$-$022455.1 & SDSS J125637.13$-$022452.4& sdL3.5   & --- & Palomar/DSpec & 2014 Feb 23-24&3660/3720 \\  
J133148.63$-$011702.0 & SDSS J133148.90$-$011651.4& L4: pec? & --- & Palomar/DSpec & 2014 Feb 23-24& ---/5520 \\  
J133348.24+273508.8   & SDSS J133348.24+273508.8  & sdL0     & --- & Keck/LRIS     & 2014 Jun 26  &  ---/400  \\  
...                   & ...                       & ---      & sdL0& IRTF/SpeX     & 2015 Jul 05  &   1440    \\  
J141624.14+134827.6   & SDSS J141624.08+134826.7  & sdL7     & --- & Palomar/DSpec & 2014 Feb 23-24&3600/3720 \\  
...                   & ...                       & sdL7     & --- & Palomar/DSpec & 2014 Jun 26  & 1200/1260 \\  
J142503.72+710207.7   & LSR J1425+7102            & sdM8     & --- & Palomar/DSpec & 2014 Feb 24  & 1260/1260 \\  
...                   & ...                       & sdM8     & --- & Palomar/DSpec & 2014 Jun 26  & 1200/1260 \\  
J143435.59+220248.1   & 2MASS J14343616+2202463   & L0.5     & --- & Palomar/DSpec & 2014 Feb 24  & 1800/1860 \\  
J144418.19$-$201945.6 & SSSPM J1444$-$2019        & sdL0     & --- & Palomar/DSpec & 2014 Feb 23  & 1260/1260 \\  
...                   & ...                       & ---      & sdL0& IRTF/SpeX     & 2015 Jul 03  &    720    \\  
J154640.78$-$553446.0 & SCR J1546$-$5534          & M8       & --- & duPont/BCSpec & 2014 May 01  &   1200    \\  
J204027.30+695924.1   & ---                       & sdL0     & --- & Palomar/DSpec & 2014 Jun 25  & 1200/1260 \\  
...                   & ...                       & ---      & N/A & Keck/NIRSPEC  & 2015 Jul 11  &    360    \\  
\end{longtable*}
\end{center}

\subsubsection{Keck/NIRSPEC}


Several sources were also observed with the Near-Infrared Spectrometer (NIRSPEC, \citealt{mclean1998,mclean2000}) at the 10m W.\ M.\ Keck Observatory on Mauna Kea, Hawai'i. The observation dates were UT 2014 Apr 12 and Dec 03, and 2015 Jul 03 and Jul 11. In its low-resolution mode employed here, use of the 42\arcsec$\times$0${\farcs}$38 slit results in a resolving power of R~$\equiv~{\lambda}/{\Delta}{\lambda}~{\approx}~2500$. Our brown dwarf candidates were observed in the N3 configuration (see \citealt{mclean2003}) that covers part of the $J$-band window from 1.15 to 1.35 $\mu$m. Data were reduced using the REDSPEC package, as described in \cite{mclean2003}. 

\section{Spectral Classification and Analysis\label{spec_classification}}

After reduction, we separated the spectra into groups based on by-eye classification -- white dwarf, early type star, late type star (or brown dwarf), subdwarf, etc.\ -- then performed more careful classification and analysis as described in the following sections.

\subsection{White Dwarfs}

\begin{figure}
\figurenum{11}
\includegraphics[scale=0.375,angle=0]{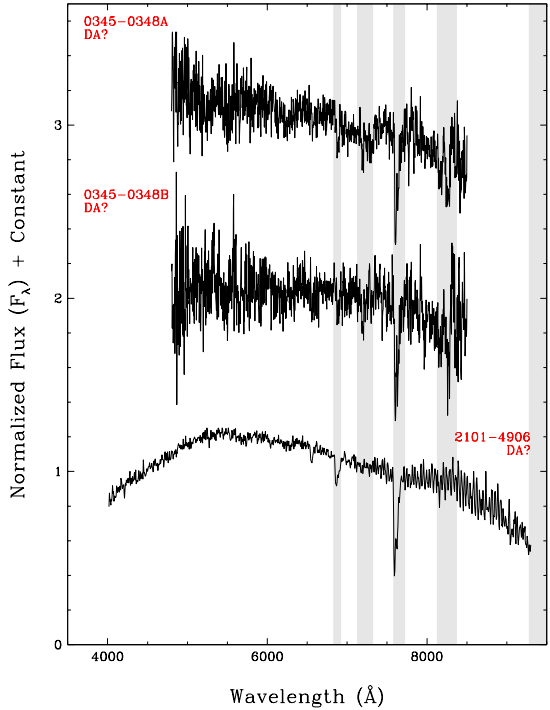}
\caption{Spectra of white dwarfs, part one. Spectra have been normalized at 7500 \AA\ and offsets added to separate the spectra vertically. The light grey bands indicate wavelength zones with uncorrected telluric absorption.
\label{seq_wds1}}
\end{figure}

\begin{figure}
\figurenum{12}
\includegraphics[scale=0.375,angle=0]{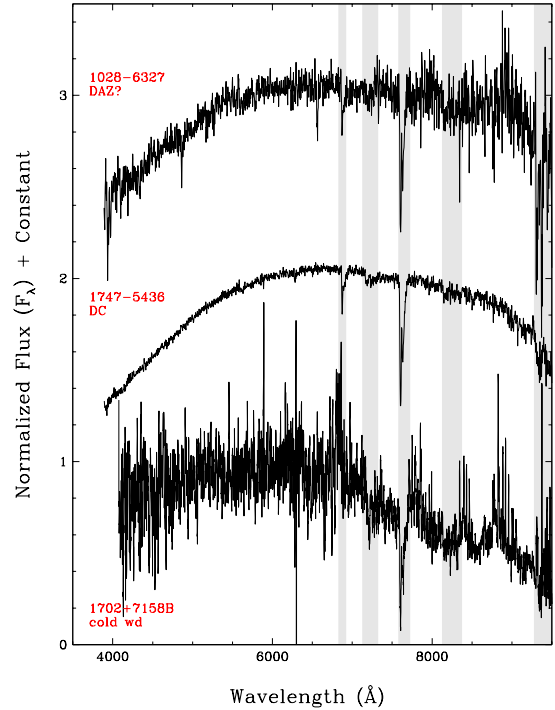}
\caption{Spectra of white dwarfs, part 2. Spectra have been normalized at 7500 \AA\ except for WISE 1702+7158B, which has been normalized at its peak flux. (See the caption of Figure~\ref{seq_wds1} for other details.) The spectrum shown for WISE 1747$-$5436 is the one from duPont/BCSpec.
\label{seq_wds2}}
\end{figure}

Spectra of white dwarf suspects (Figures~\ref{seq_wds1} and~\ref{seq_wds2}) were classified using the scheme described in \cite{sion1983}. We provide only the core type -- DA, DC, etc.\ -- here and not the temperature suffix. Most of our white dwarf spectra have either poor signal-to-noise, which make them difficult to compare to the models used in the temperature classification, and/or they lack data blueward of H$\beta$, which is generally needed to establish accurate temperatures. 

Spectra of the common-proper-motion double WISE 0345$-$0348A and WISE 0345$-$0348B (Figure~\ref{seq_wds1}) are noisy but appear to show H$\alpha$ absorption, so these have tentatively been classified as DA white dwarfs. The signal-to-noise for WISE 2101$-$4906 (Figure~\ref{seq_wds1}) is much better, and while there appears to be a detection of H$\alpha$, other hydrogen lines in the Balmer series are not clearly seen. Hence, this object is only tentatively classified as DA. WISE 1028$-$6327 (Figure~\ref{seq_wds2}) shows H$\alpha$, H$\beta$, H$\gamma$, and H$\delta$ absorption along with the \ion{Ca}{2} H\&K doublet, so this object is tenatively classified as a DAZ pending a spectrum with higher signal-to-noise. Despite the excellent signal-to-noise in the spectrum of WISE 1747$-$5436 (Figure~\ref{seq_wds2}), no clear features are seen, leading to a classification of DC. 

The remaining spectrum, that of WISE 1702+7158B (Figure~\ref{seq_wds2}), is extremely noisy due to the faintness of the object at optical wavelengths. The object itself was not detected by AllWISE, but its common-proper-motion primary, the M4.5 dwarf WISE 1702+7158A, was identified by AllWISE as a high motion source. Analysis of this pair can be found in section 5. 
Curiously, our DA suspect WISE 2101$-$4906 (above) shares proper motion with another M4.5 dwarf, WISE 2101$-$4907, as also discussed in section 5.

\subsection{Early-type ($\le$K5) Stars\label{early_types}}

\begin{figure}
\figurenum{13}
\includegraphics[scale=0.375,angle=0]{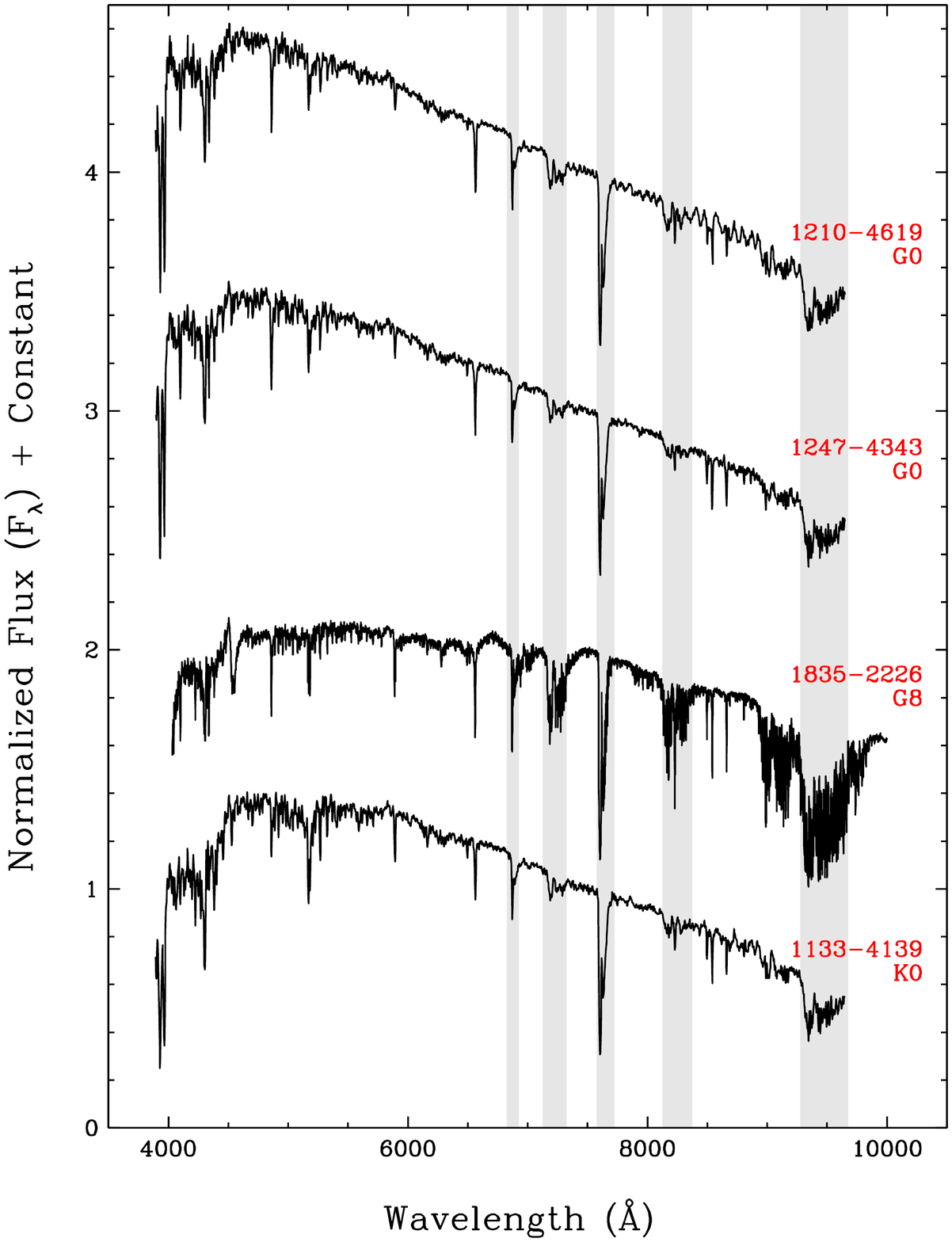}
\caption{Spectra of main sequence objects with types of G0 through K0. The spectrum of WISE 1835$-$2226 exhibits reddening. (See the caption of Figure~\ref{seq_wds1} for other details.)
\label{seq_early_types1}}
\end{figure}

\begin{figure}
\figurenum{14}
\includegraphics[scale=0.375,angle=0]{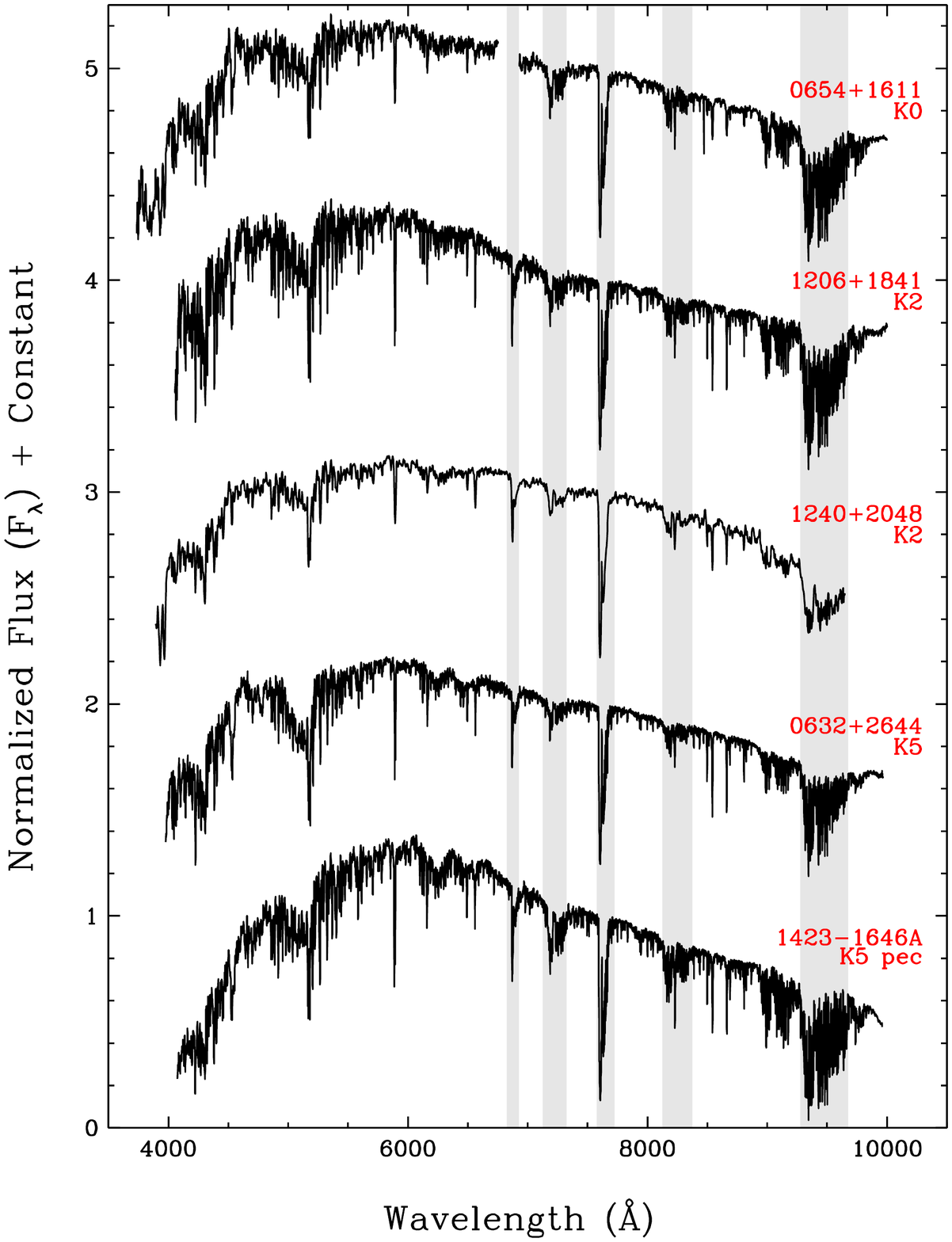}
\caption{Spectra of main sequence objects with types of K0 through K5. (See the caption of Figure~\ref{seq_wds1} for other details.)
\label{seq_early_types2}}
\end{figure}

Spectra of objects with types of K5 or earlier were classified using spectral standards taken as part of the NStars project (see Appendix C of \citealt{gray2009}). These standard spectra cover only the blue end of our spectral region (typically 3800 to 5600 \AA) but cover a number of prominent diagnostic lines -- \ion{Ca}{2} H\&K at 3934 and 3969 \AA; the $\delta$, $\gamma$, and $\beta$ Balmer lines of \ion{H}{1} at 4102, 4340, and 4861 \AA; \ion{Ca}{1} at 4227 \AA; \ion{Fe}{1} at 4308 and 5270 \AA; and the \ion{Mg}{1} b triplet at 5167, 5173, and 5184 \AA. Spectra of nine early-type objects are shown in Figures~\ref{seq_early_types1} and~\ref{seq_early_types2}. 

One object, WISE 1423$-$1646A (Figure~\ref{seq_early_types2}), has a peculiar spectrum best fit by the K5 standard. That is, the line strengths best match the K5 standard even though the overall continuum over our spectral range is somewhat bluer than the standard. This object has common proper motion (section~\ref{serendipitous_systems}) with the normal M0.5 dwarf WISE 1423$-$1646B shown later. We do not know the physical cause for the peculiarity in WISE 1423$-$1646A, but it is not shared by the companion object.

One of the objects in Figure~\ref{seq_early_types1}, WISE 1835$-$2226 is reddened relative to other objects in the G to early-K sequence. Using the IDL program FM\_UNRED that dereddens an observed spectrum using the \cite{fitzpatrick1999} parameterization of interstellar extinction as a function of wavelength\footnote{See \url{http://idlastro.gsfc.nasa.gov/}.}, we find that correcting by a color excess of $E_{B-V} \approx 0.15$ mag gives an overall spectral shape that most closely matches that of an unreddened G8 dwarf, which matches the classification based on the depth of the spectral features alone. This color excess corresponds to a total $V$-band extinction of $A(V) \approx 0.47$ mag if $R_V = A_V/E(B_V) = 3.1$ is assumed (\citealt{schultz1975}).

\begin{figure}
\figurenum{15}
\includegraphics[scale=0.375,angle=0]{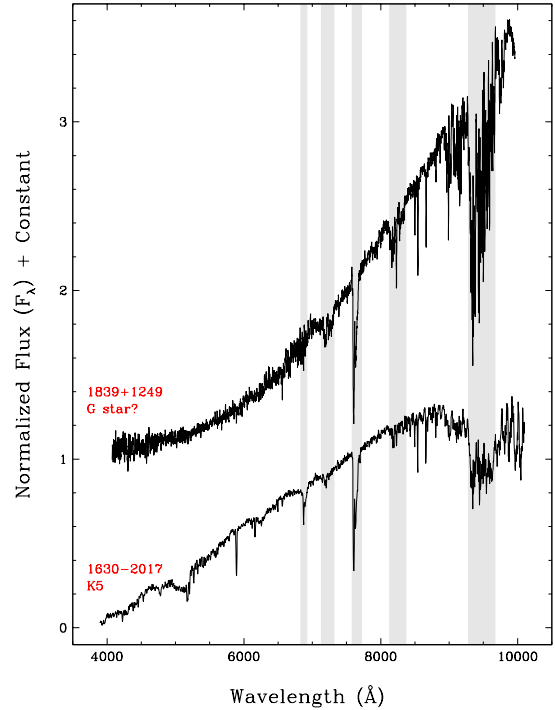}
\caption{Observed spectra of two high motion stars that are heavily reddened. (See the caption of Figure~\ref{seq_wds1} for other details.)
\label{seq_reddened}}
\end{figure}

\begin{figure}
\figurenum{16}
\includegraphics[scale=0.375,angle=0]{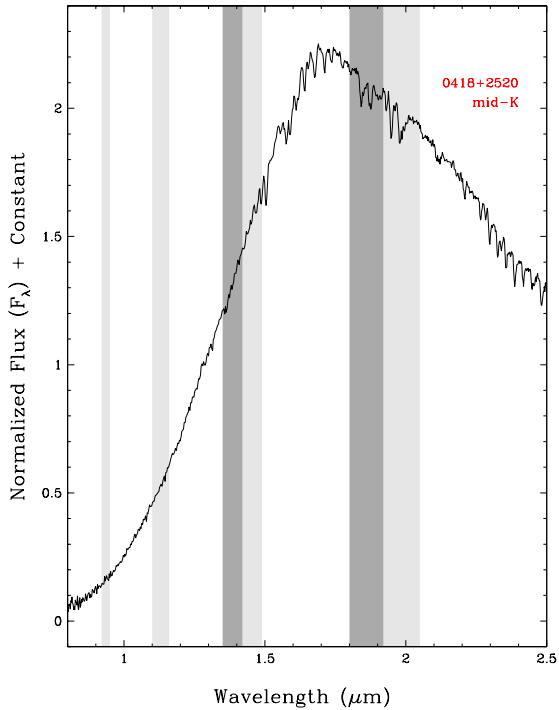}
\caption{Observed spectrum of another high motion object that is heavily reddened. Regions of telluric absorption are shown by the dark grey bands if the atmospheric transmission is less than 20\% and by light grey bands if the atmospheric transmission is between 20\% and 80\% (\citealt{rayner2009}). The spectrum has been normalized at 1.28 $\mu$m.
\label{seq_reddened_IR}}
\end{figure}

Three other early-type stars are even more heavily reddened, as shown in Figures~\ref{seq_reddened} and~\ref{seq_reddened_IR}. WISE 1839+1249 has a continuum that is best fit to a mid-G dwarf if a color excess of $E_{B-V} \approx 1.9$ mag ($A(V) \approx 5.9$ mag) is assumed; the spectral type is uncertain because of the low signal-to-noise in the short-wavelength region, where only questionable detections of \ion{Na}{1} at 5890 \AA\ and H$\alpha$ are seen. In WISE 1630$-$2017, the strengths of the MgH and CaH bands as well as the \ion{Mg}{1}, \ion{Na}{1}, \ion{Ca}{1} lines are best fit by a K5 dwarf, as is the continuum if a color excess of $E_{B-V} \approx 1.0$ mag ($A(V) \approx 3.1$ mag) is assumed. For WISE 0418+2520, the near-infrared continuum shows metal lines throughout the $J$ and $H$ bands, along with CO bandheads at $K$, indicating a mid-K dwarf. (The CO bandheads are not strong enough for a mid-K giant; see the suite of K dwarf and giant spectra in the IRTF SpeX Spectral Library of \citealt{rayner2009}.) The continuum is best fit if a color excess of $E_{B-V} \approx 4.7$ mag ($A(V) \approx 14.6$ mag) is assumed.

With measured extinctions and spectral types in hand, we can calculate the distances to the four reddened stars in Figures~\ref{seq_early_types1},~\ref{seq_reddened} and~\ref{seq_reddened_IR}. Using their apparent $K_s$-band magnitudes from the 2MASS Point Source Catalog\footnote{See \url{http://www.ipac.caltech.edu/2mass/releases/allsky/}.}, assuming $A_K = 0.112A_V$ (\citealt{rieke1985}), and using the absolute $K_s$ magnitudes\footnote{See also \url{http://www.pas.rochester.edu/$\sim$emamajek/ EEM\_dwarf\_UBVIJHK\_colors\_Teff.txt}.} corresponding to their spectral types as derived above (\citealt{pecaut2013}), we find the approximate distances listed in Table~\ref{reddened_distances}. The table also lists the tangential velocity for that distance, using our computed 2MASS-to-AllWISE proper motion values. The average tangential velocity for a disk star is 37 km s$^{-1}$ (\citealt{reid1997}) and for a halo star is 175-215 km s$^{-1}$ (\citealt{reid2000}), suggesting the possibility that WISE 1630$-$2017 is a halo member. Nonetheless, these velocities are typical of stars found during motion surveys, so the derived distances appear reasonable.

\begin{deluxetable}{llllll}
\tablenum{6}
\tablecaption{Derived Parameters for Reddened Motion Stars\label{reddened_distances}}
\tablehead{
\colhead{WISE} & 
\colhead{Approx.} &                         
\colhead{$E(B-V)$} &
\colhead{Dist.} &
\colhead{$v_{tan}$} &
\colhead{Note} \\
\colhead{Object} &   
\colhead{Type} &                       
\colhead{(mag)} &       
\colhead{(pc)} &       
\colhead{(km/s)} &       
\colhead{}  
}
\startdata
0418+2520   & K5 V& 4.7 & 114 &  73 & Taurus \\
1630$-$2017 & K5 V& 1.0 & 227 & 246 & Ophiuchus \\
1835$-$2226 & G8 V& 0.15& 282 &  96 & --- \\
1839+1249   & G5 V& 1.9 & 295 & 131 & Hercules \\
\enddata
\end{deluxetable}

\begin{figure}
\figurenum{17}
\includegraphics[scale=0.20,angle=0]{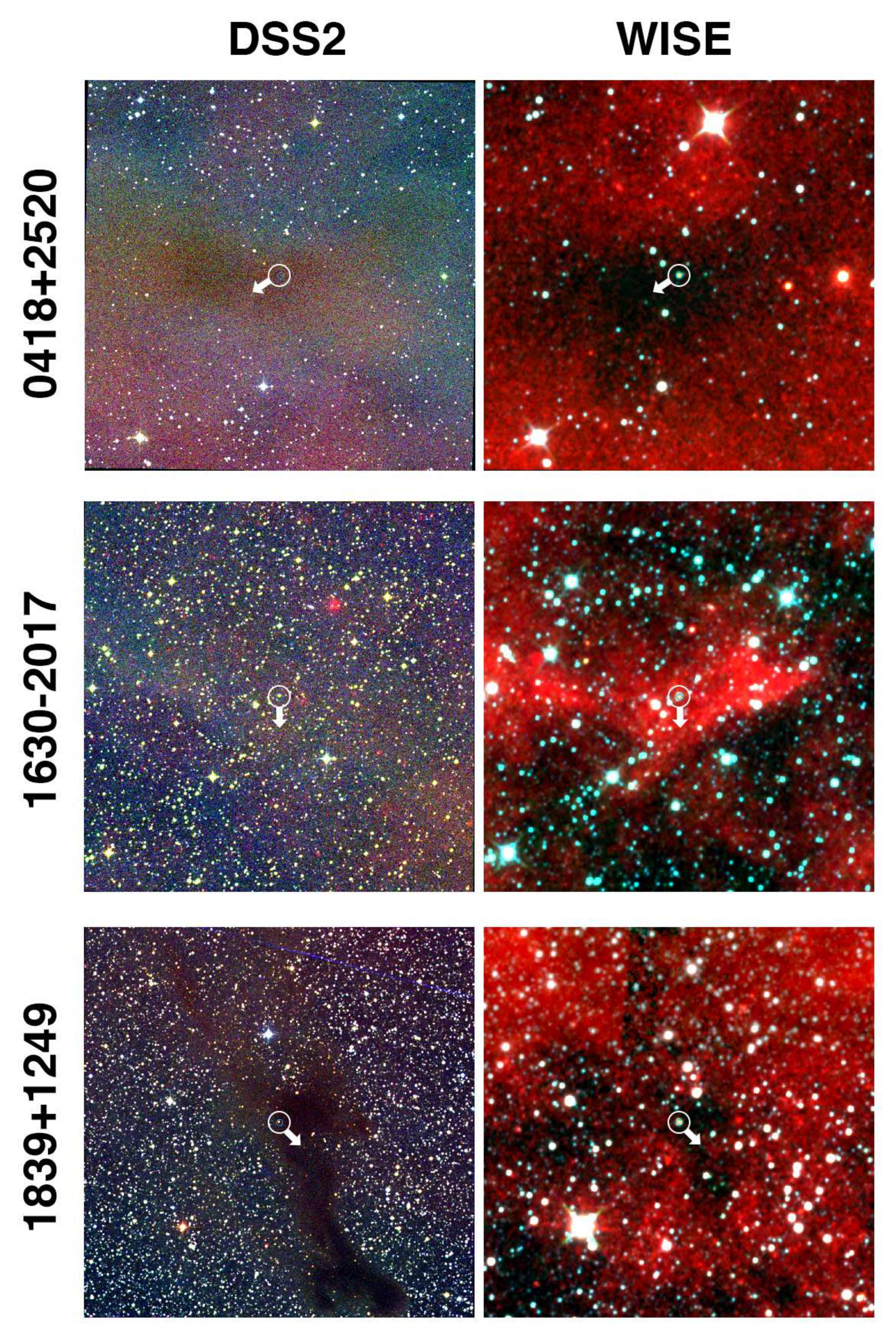}
\caption{20$\times$20 arcmin views, with north up and east to the left, centered at the location of our reddened motion objects. The images are three-color composites of the DSS2 $B$ (coded as blue), $R$ (green), and $I$ (red) bands on the left and the {\it WISE} All-Sky W1 (blue), W2 (green), and W3 (red) bands on the right. Images for WISE 0418+2520 are shown in the top row, WISE 1630$-$2017 in the middle row, and WISE 1839+1249 in the bottom row. White circles mark the {\it WISE}-epoch location of the motion star in all images, and arrows indicate the direction of the motion vector. From the top panel to the bottom, these stars will take approximately 1200, 700, and 2800 years to cross behind the densest portion of the intervening cloud.
\label{finder_reddened_objects}}
\end{figure}

For three of the four objects, however, the total line-of-sight reddening measured in a beam of 5-arcmin radius\footnote{See tool available at \url{http://irsa.ipac.caltech.edu/applications/ DUST/}.} is lower than the value we measure for the motion star -- only for WISE 1835$-$2226 is our value of $E(B-V) = 0.15$ lower than the total column extinction measured by either \cite{schlafly2011} ($0.27{\pm}0.01$) or \cite{schlegel1998} ($0.31{\pm}0.01$). For these three cases, a check of the 100 $\mu$m dust maps produced by \cite{schlegel1998} shows that the star falls squarely along a patch of very high extinction relative to nearby areas, explaining why our values of the extinction are higher than the region average. For all three of these objects, much of the intervening dust is known to be near the Sun. At 114 pc, WISE 0418+2520 is likely moving through the Taurus Molecular Cloud, whose mean distance\footnote{\cite{schlafly2014} (their Table 1) estimate a distance of $\sim$105 pc for the portion of the cloud where WISE 0418+2520 is located.} is $\sim$140 pc (\citealt{kenyon1994}). At 227 pc, WISE 1630$-$2017 is moving behind the Ophiuchus Cloud, whose mean distance is $\sim$120 pc (\citealt{lombardi2008}, \citealt{schlafly2014}). At 295 pc, WISE 1839+1249 is moving behind the Hercules Cloud, whose mean distance is $\sim$200 pc (\citealt{schlafly2014}; see also \citealt{dame2001}). The locations of these three objects with respect to the intervening clouds are shown in Figure~\ref{finder_reddened_objects}.  

These three stars have relatively high velocities, so are presumably middle-aged objects unrelated to the clouds themselves. The mean motions and dispersions for objects belonging to Taurus (\citealt{jones1979}) and Ophiuchus (\citealt{makarov2007}, \citealt{mamajek2008}) preclude membership for both WISE 0418+2520 ($\mu_\alpha = 111.8{\pm}5.7$, $\mu_\delta = -75.1{\pm}5.6$ mas yr$^{-1}$) and WISE 1630$-$2017 ($\mu_\alpha = -2.3{\pm}9.7$, $\mu_\delta  = -228.9{\pm}9.1$ mas yr$^{-1}$). Except for occasional chromospheric activity that primarily affects the ultraviolet and blue optical bandpasses, the vast majority of middle-aged mid-G to mid-K dwarfs are free of intrinsic variability at the millimag level (\citealt{ciardi2011}) and are therefore useful as standard candles. For the three stars in Figure~\ref{finder_reddened_objects}, any observed variability can be attributed to line-of-sight variations in the dust column as the stars trek behind (or within) the cloud.

\begin{figure}
\figurenum{18}
\includegraphics[scale=0.325,angle=270]{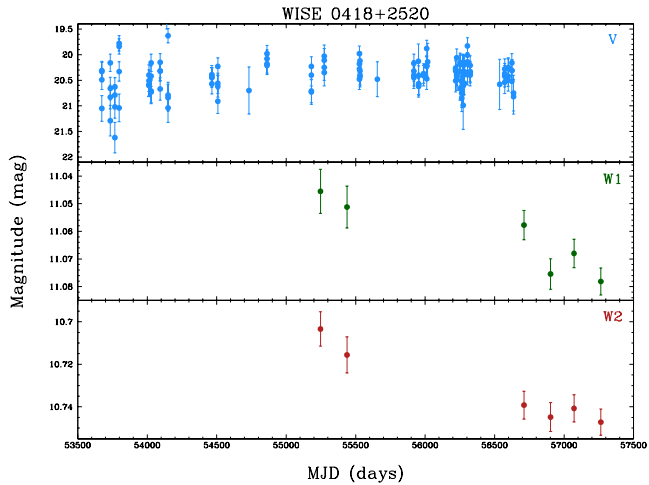}
\caption{Photometric monitoring of the heavily reddened motion star WISE 0418+2520 at $V$, W1, and W2 bands. Although variability is hard to discern at the faint flux levels at $V$ band (from the Catalina Sky Survey), the W1 and W2 bands show a slight diminution in flux over the available 4.5-year {\it WISE} baseline.
\label{reddened_var_0418}}
\end{figure}

\begin{figure}
\figurenum{19}
\includegraphics[scale=0.325,angle=270]{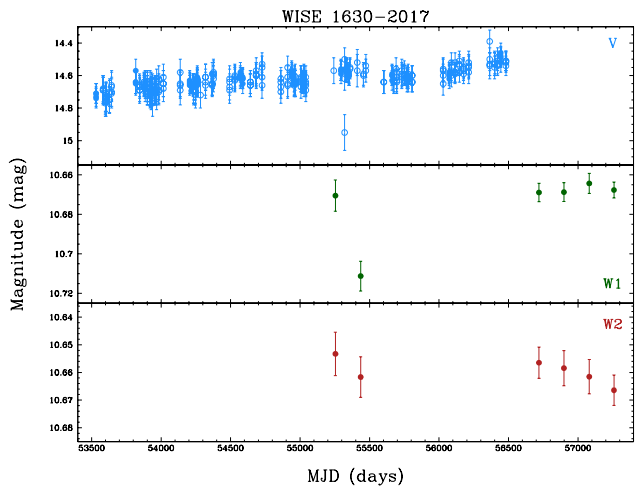}
\caption{Photometric monitoring of the heavily reddened motion star WISE 1630$-$2017 at $V$, W1, and W2 bands. The $V$ band data, from the Catalina Sky Survey (solid blue points) and Siding Spring Survey (open blue points; see text for details), show a small, $\sim$0.2-mag brightening over the eight-year time baseline. With the exception of the discrepant second-epoch W1 point, the {\it WISE} photometry shows no obvious variability, although the {\it WISE} bands are less sensitive than $V$ to intervening dust.
\label{reddened_var_1630}}
\end{figure}

\begin{figure}
\figurenum{20}
\includegraphics[scale=0.325,angle=270]{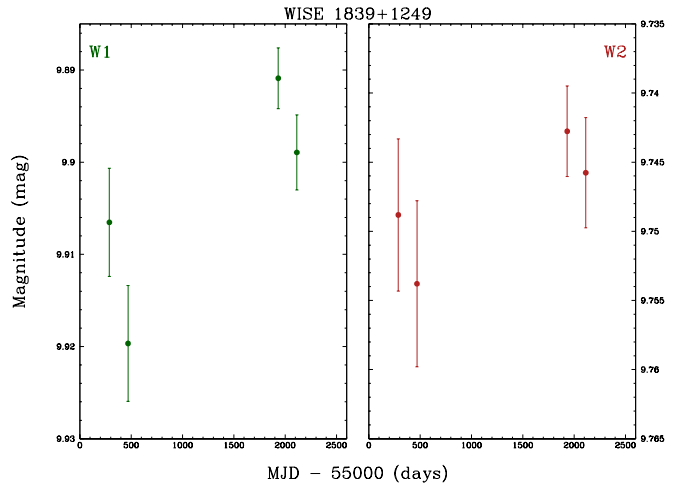}
\caption{Photometric monitoring of the heavily reddened motion star WISE 1839+1249 at W1 (left) and W2 (right) bands.
\label{reddened_var_1839}}
\end{figure}

Figures~\ref{reddened_var_0418} through~\ref{reddened_var_1839} show examples of existing light curves for these objects. Figure~\ref{reddened_var_0418} shows $V$, W1, and W2 data for WISE 0418+2520, where the $V$-band data come from the Catalina Sky Survey\footnote{See {\url http://nesssi.cacr.caltech.edu/DataRelease/}.} (\citealt{drake2009}) and the {\it WISE} data are the per-epoch weighted means computed using the measured magnitudes in each individual W1 and W2 frame. (A few frames with quality score of zero, which generally indicates image smearing, were eliminated prior to the computation.) WISE 0418+2520 is much fainter, by almost ten magnitudes, in $V$ than in W1 and W2, so the photometric errors make it difficult to discern any brightness changes at $V$. At W1 and W2, on the other hand, there is a trend of steady dimming, although the magnitude change over the 4.5-yr baseline is only ~0.03-0.04 mag in both bands. Monitoring this object at intermediate bands, such as $R$, $J$, $H$, and $K$, would be fruitful.

The corresponding light curves for WISE 1630$-$2017 are shown in Figure~\ref{reddened_var_1630}. Data at $V$-band are available from both the Catalina Sky Survey as well as the Siding Spring Survey\footnote{See {\url https://www.mso.anu.edu.au/$\sim$rmn/}.}. Because there appears to be a zero-point offset between the Catalina and Siding Spring calibrations, we have subtracted 0.1 mag from the Catalina values prior to plotting. The $V$-band light curve shows a secular brightening of about $\sim$0.2 mag over the course of eight years. The W1 and W2 photometry show no significant changes in magnitude over the {\it WISE} time baseline, although the second-epoch W1 point is anomalous. Having simultaneous photometry in many bands would help elucidate if anomalies such as this are just random measurement fluctuations or indications of small-scale clumpiness within the cloud.

For WISE 1839+1249 (Figure~\ref{reddened_var_1839}), we have only {\it WISE} photometric monitoring available, which may show a barely discernable brightening over the {\it WISE} time baseline. On-going and future surveys will be able to provide nightly monitoring of such objects at a variety of wavelengths. Notable examples are the Panoramic Survey Telescope and Rapid Response System (Pan-STARRS; \citealt{hodapp2004}) at $g$, $r$, $i$, $z$, and $y$; the Large Synoptic Survey Telescope (LSST; \citealt{ivezic2014}) at those same wavelengths, plus the $u$ band; and the Zwicky Transient Facility (ZTF; \citealt{bellm2014}) at $g^{\prime}$ and $R$. Nightly monitoring is important because the light beam of each star shines through a column of intervening material that is offset between 30 AU (for Taurus) and 100 AU (for Ophiuchus) from the night before, providing exquisite detail on the line-of-sight clumpiness of these clouds. Combined with simultaneous data from multiple wavelengths, we can identify  variations that run contrary to the ``universal'' reddening law, which will give us insight into the average grain sizes along the beam. 

\subsection{Mid-K through Late-M Dwarfs}

\begin{figure*}
\figurenum{21}
\includegraphics[scale=0.65,angle=270]{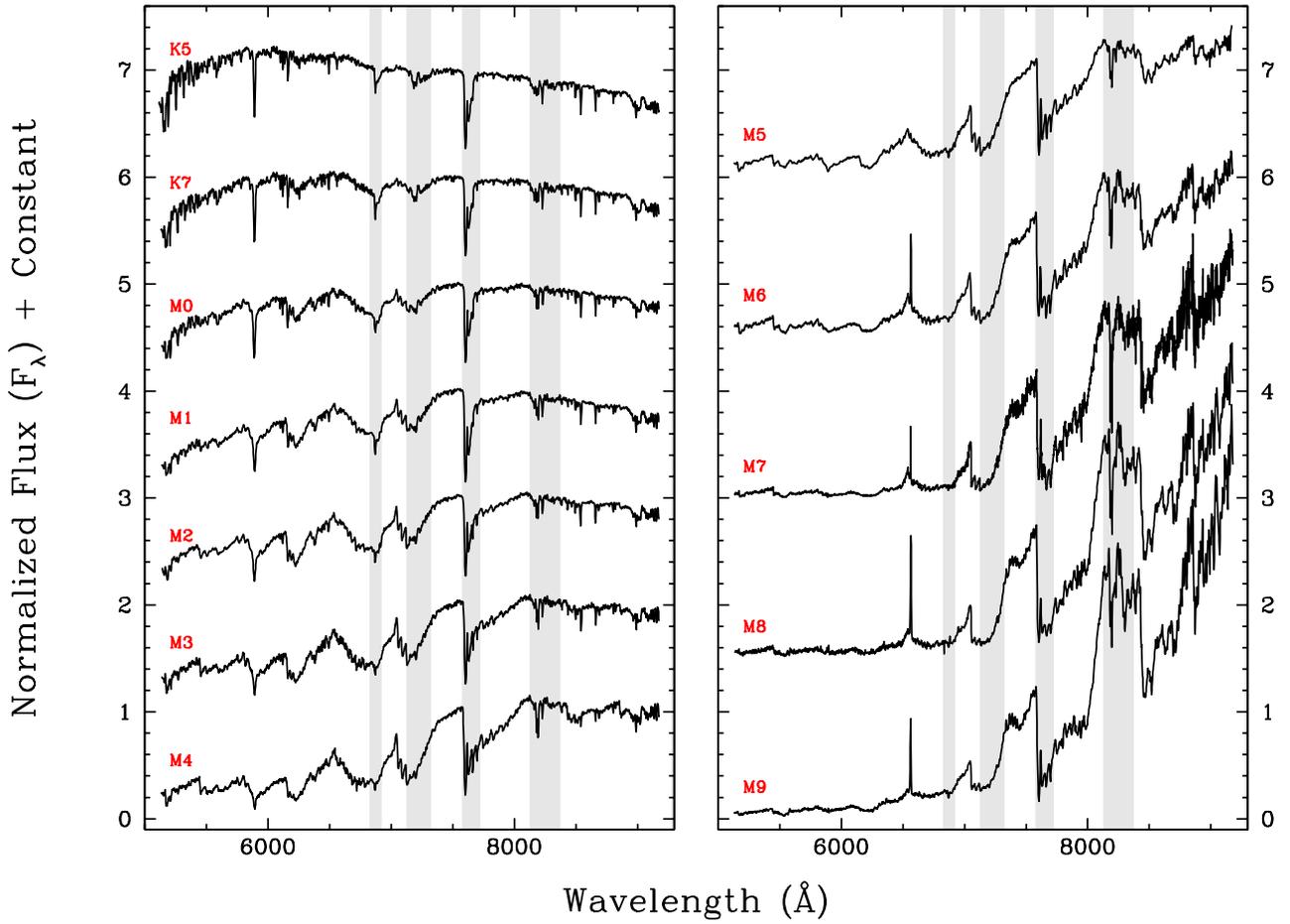}
\caption{Spectra of optical spectral standards from K5 through M9 (Table 7). See the caption of Figure~\ref{seq_wds1} for other details.
\label{opt_stds_spectra}}
\end{figure*}

Mid-K through late-M dwarfs were typed in the optical using the classification system of \cite{kirkpatrick1991}. Table 7
lists the standards employed for the classifications produced in this section. The spectrum of the M7 standard comes from the Deep Imaging 
Multi-object Spectrograph (DEIMOS; \citealt{faber2003})  on the 10m W.\ M.\ Keck Observatory (see Table 3 of \citealt{kirkpatrick2010}). All others come from the Double Spectrograph on the Palomar 5m telescope and originated in a project described in \cite{kirkpatrick1997}, although only those spectra of type M6 and later were discussed in that paper. All of the objects in Table 7
were listed as primary (or secondary, in the case of the M9) spectral standards in \cite{kirkpatrick1991} except for the M5 and the M8. Unlike the two M5 dwarfs originally proposed, this M5 has the advantage of being a single star lying near the celestial equator for ease of observation in both hemispheres; this M8 falls further away from the Galactic Plane than does the only M8 standard listed in \cite{kirkpatrick1991} and is less subject to contamination by background stars. For reference, a plot of these standard spectra is shown in Figure~\ref{opt_stds_spectra}. 

\begin{deluxetable}{lll}
\tablenum{7}
\tablecaption{Mid-K through Late-M Dwarf Spectral Standards in the Optical\label{opt_stds_list}}
\tablehead{
\colhead{Spectral} &                          
\colhead{Object} &
\colhead{Other} \\
\colhead{Type} &                          
\colhead{Name} &       
\colhead{Name}  
}
\startdata
K5 & Gl 820A  & 61 Cyg A \\
K7 & Gl 820B  & 61 Cyg B \\
M0 & Gl 270   & BD+33$^\circ$ 1505 \\
M1 & Gl 229A  & HD 42581 \\
M2 & Gl 411   & HD 95735 \\
M3 & Gl 436   & Ross 905 \\
M4 & Gl 402   & Wolf 358 \\
M5 & GJ 1057  & LHS 168  \\
M6 & Gl 406   & Wolf 359 \\
M7 & Gl 644C  & vB 8     \\
M8 & LP 412-31& ---      \\
M9 & LHS 2065 & ---      \\
\enddata
\end{deluxetable}

\begin{figure}
\figurenum{22}
\includegraphics[scale=0.375,angle=0]{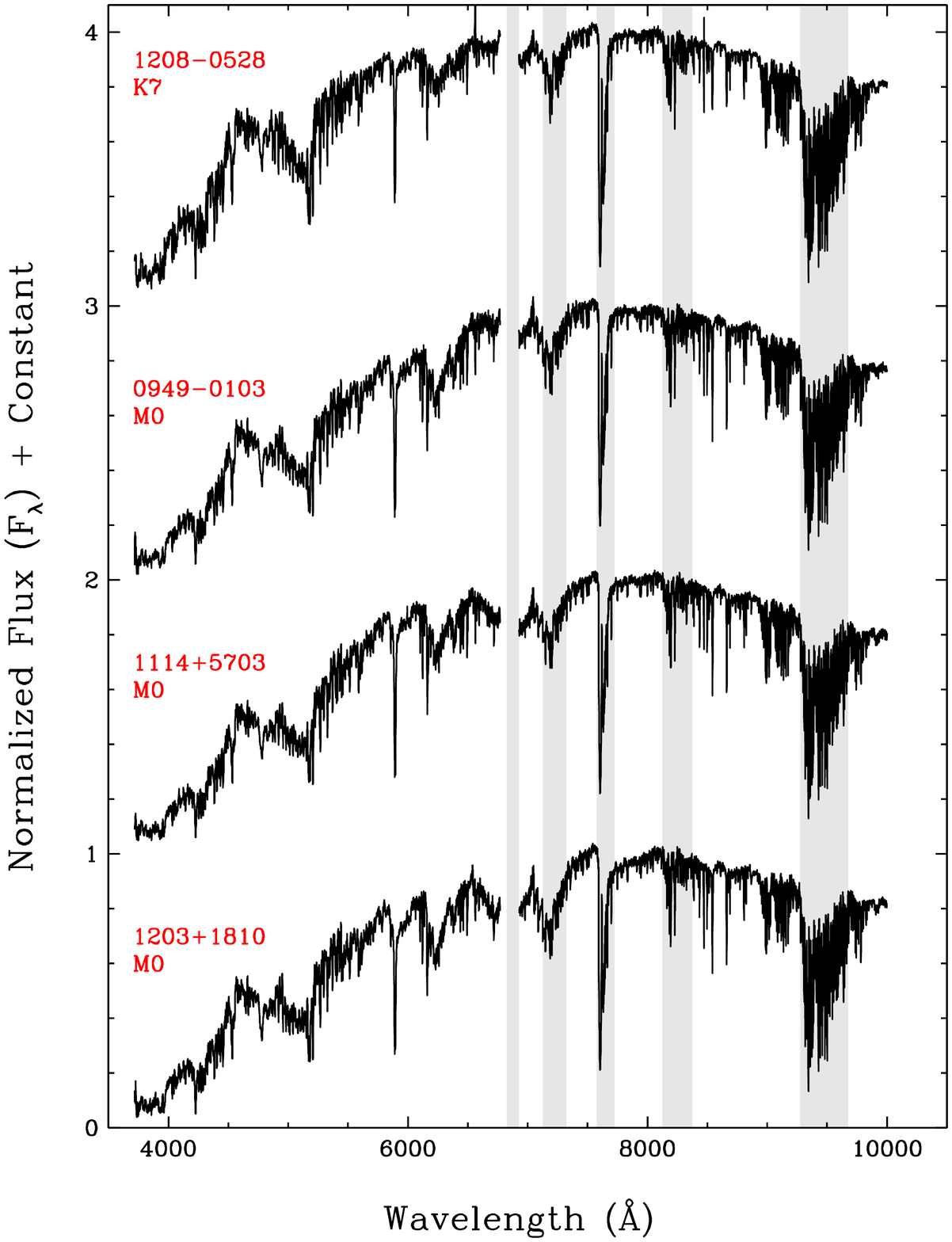}
\caption{Spectra of main sequence objects with types of K7 through M0. (See the caption of Figure~\ref{seq_wds1} for other details.)
\label{seq_K7_M05.1}}
\end{figure}

\begin{figure}
\figurenum{23}
\includegraphics[scale=0.375,angle=0]{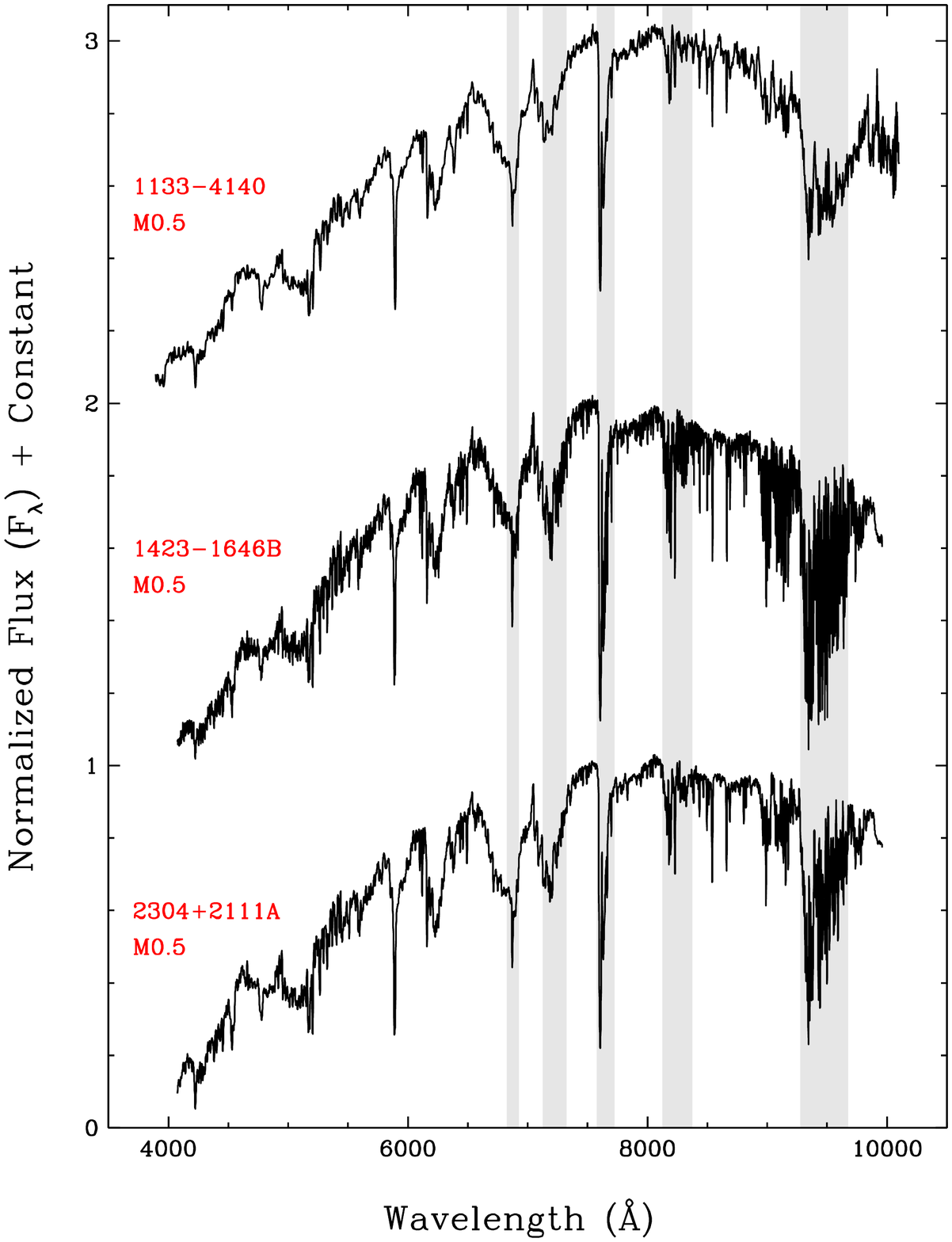}
\caption{Spectra of main sequence objects with types of M0.5. (See the caption of Figure~\ref{seq_wds1} for other details.)
\label{seq_K7_M05.2}}
\end{figure}

\begin{figure}
\figurenum{24}
\includegraphics[scale=0.375,angle=0]{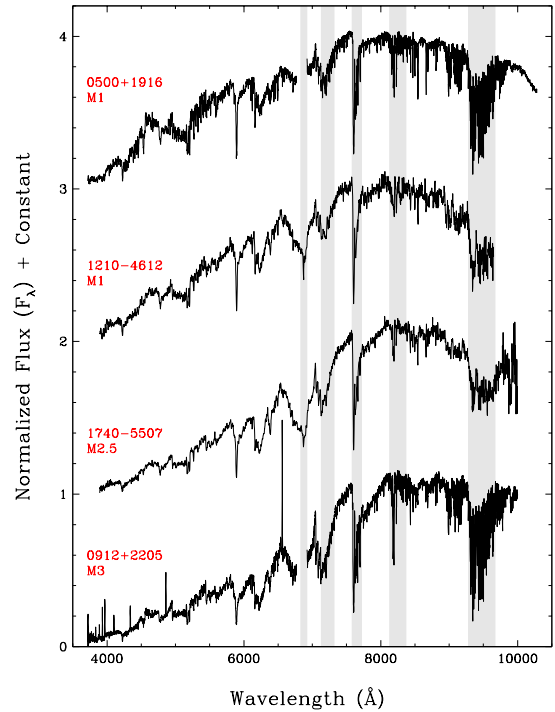}
\caption{Spectra of main sequence objects with types of M1 through M3. (See the caption of Figure~\ref{seq_wds1} for other details.)
\label{seq_M1_M3.1}}
\end{figure}

\begin{figure}
\figurenum{25}
\includegraphics[scale=0.375,angle=0]{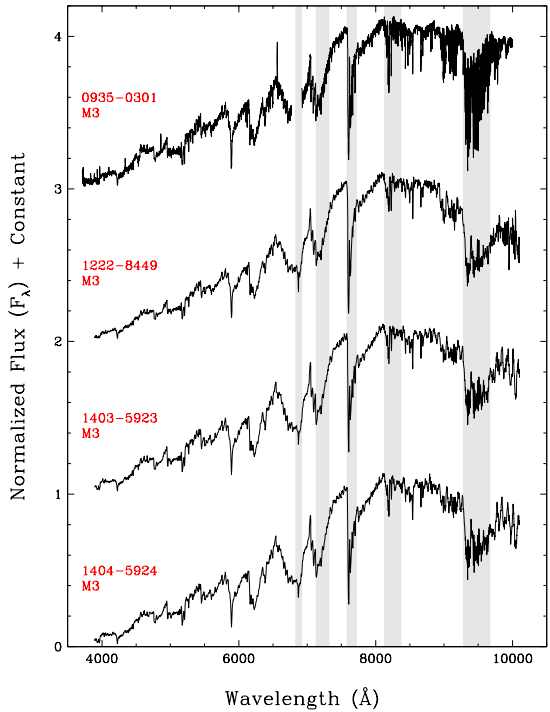}
\caption{Spectra of more main sequence objects with types of M3. (See the caption of Figure~\ref{seq_wds1} for other details.)
\label{seq_M1_M3.2}}
\end{figure}

\begin{figure}
\figurenum{26}
\includegraphics[scale=0.375,angle=0]{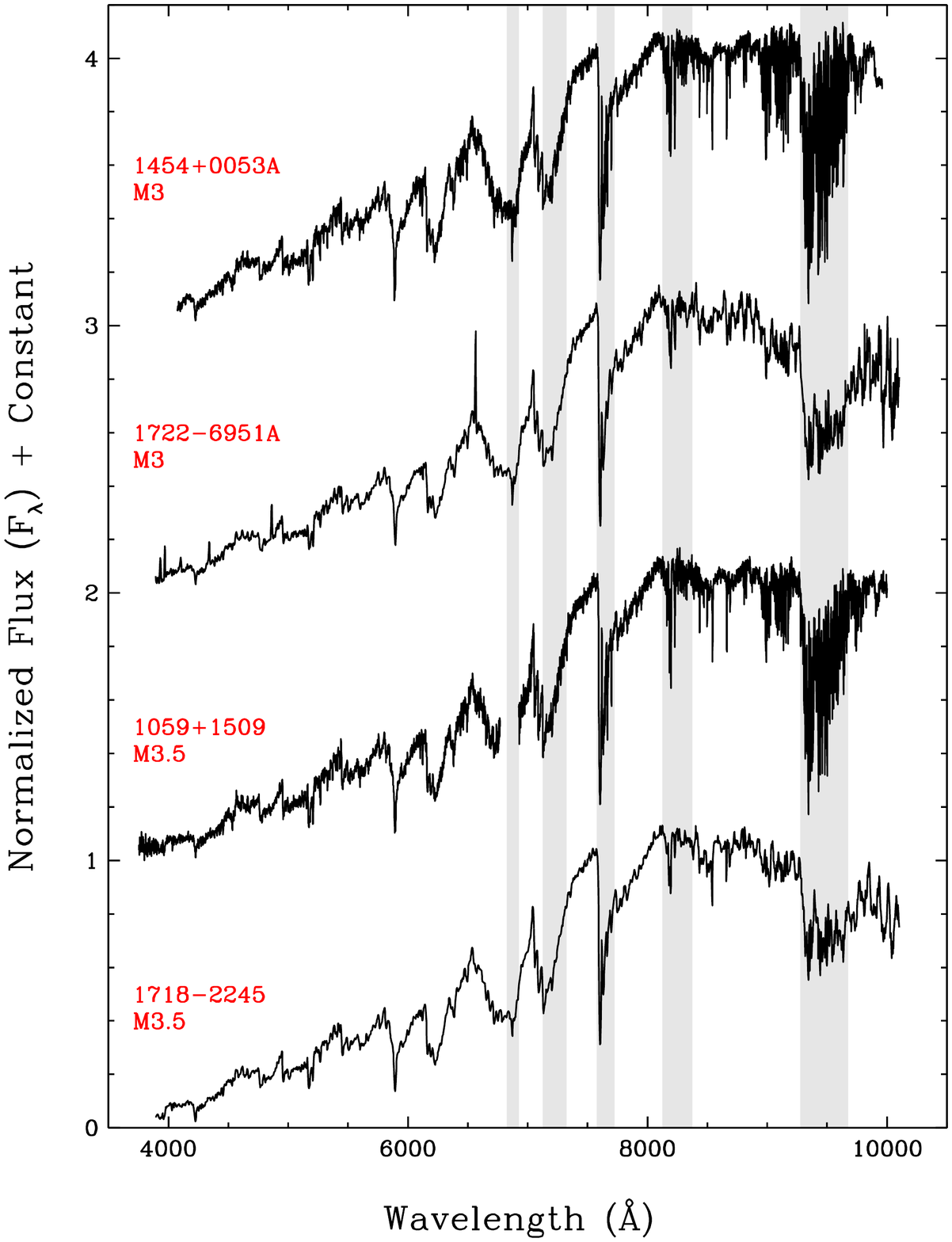}
\caption{Spectra of main sequence objects with types of M3 through M3.5. (See the caption of Figure~\ref{seq_wds1} for other details.)
\label{seq_M3_M4.1}}
\end{figure}

\begin{figure}
\figurenum{27}
\includegraphics[scale=0.375,angle=0]{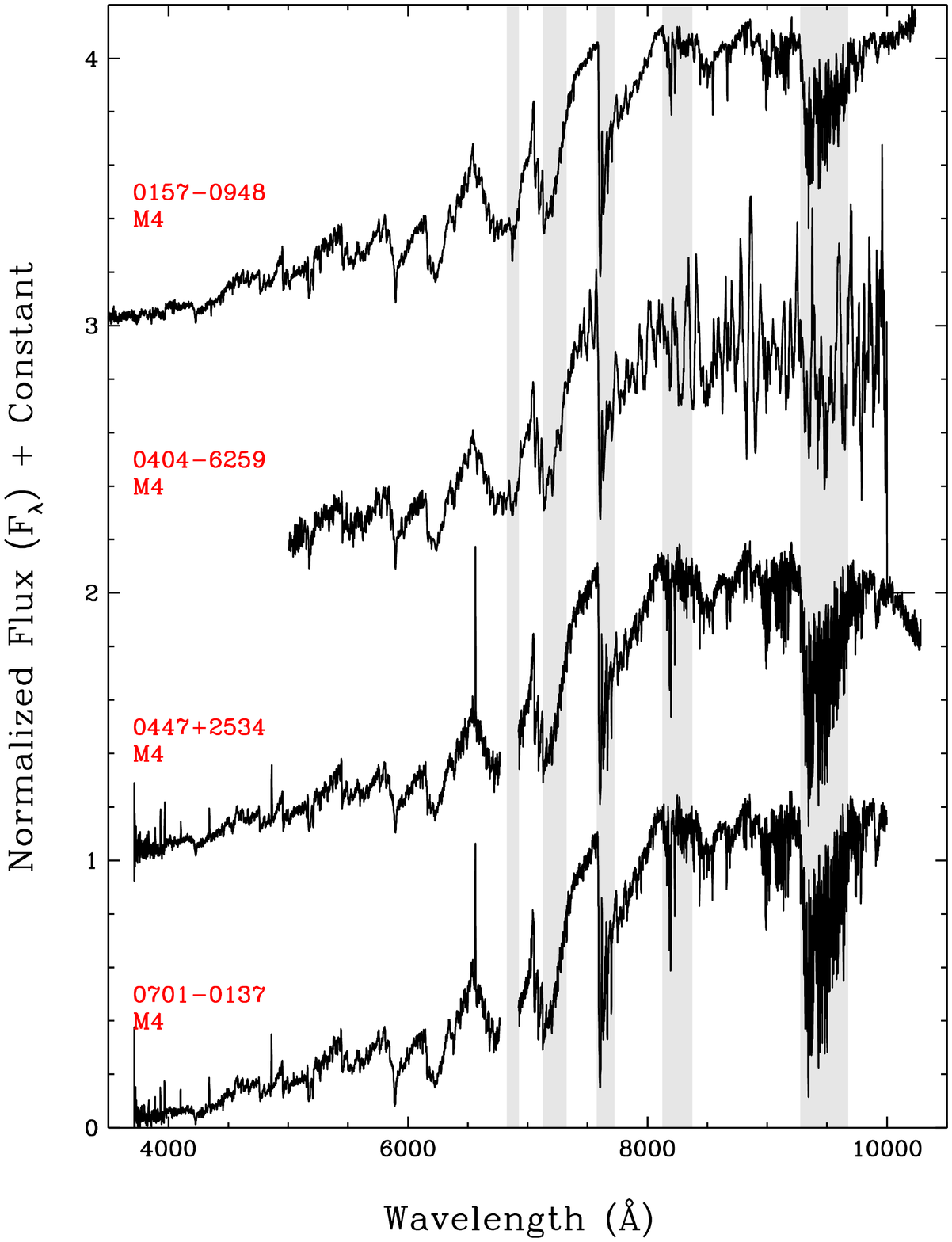}
\caption{Spectra of main sequence objects with types of M4. (See the caption of Figure~\ref{seq_wds1} for other details.)
\label{seq_M3_M4.2}}
\end{figure}

\begin{figure}
\figurenum{28}
\includegraphics[scale=0.375,angle=0]{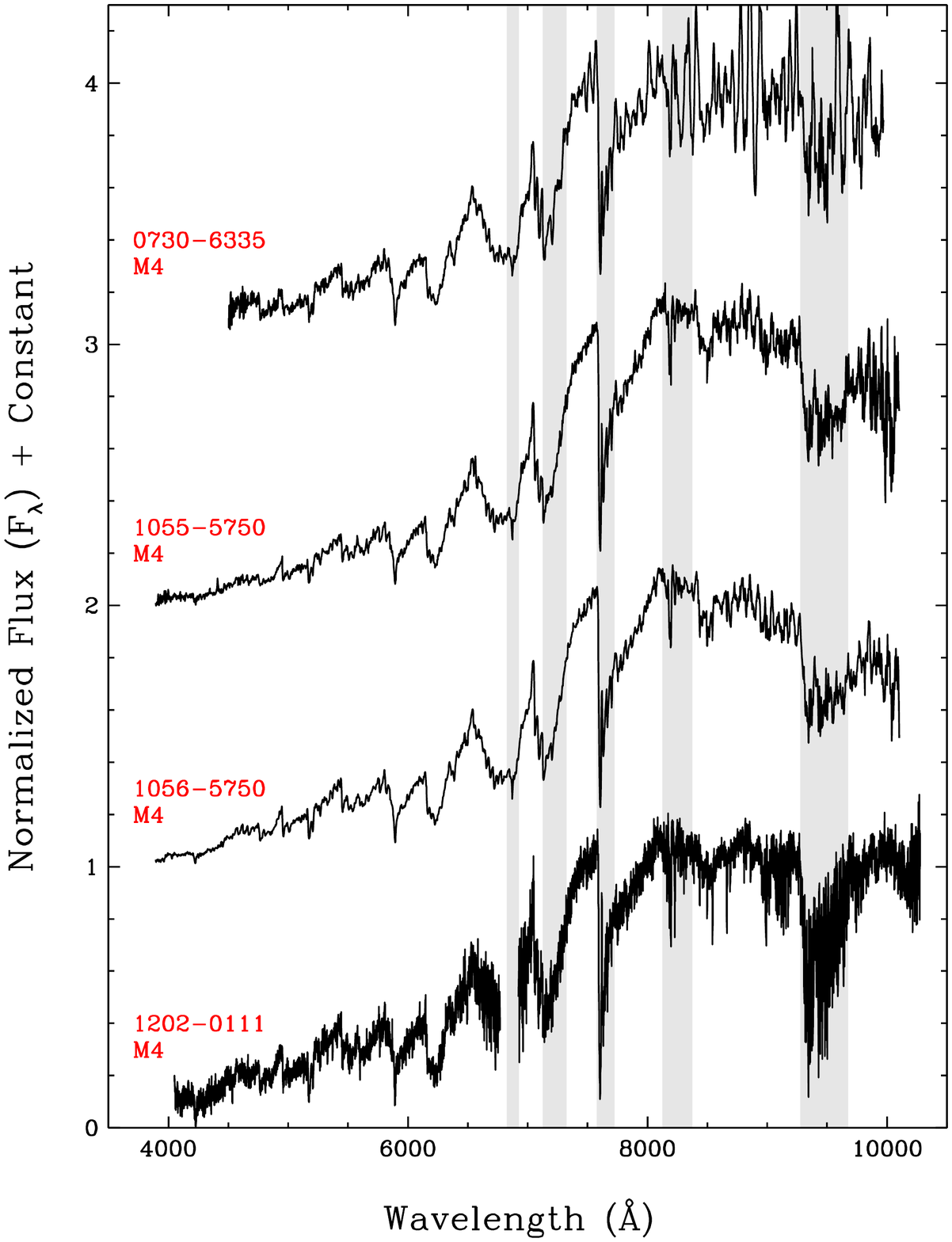}
\caption{Spectra of more main sequence objects with types of M4. (See the caption of Figure~\ref{seq_wds1} for other details.)
\label{seq_M4_M4.1}}
\end{figure}

\begin{figure}
\figurenum{29}
\includegraphics[scale=0.375,angle=0]{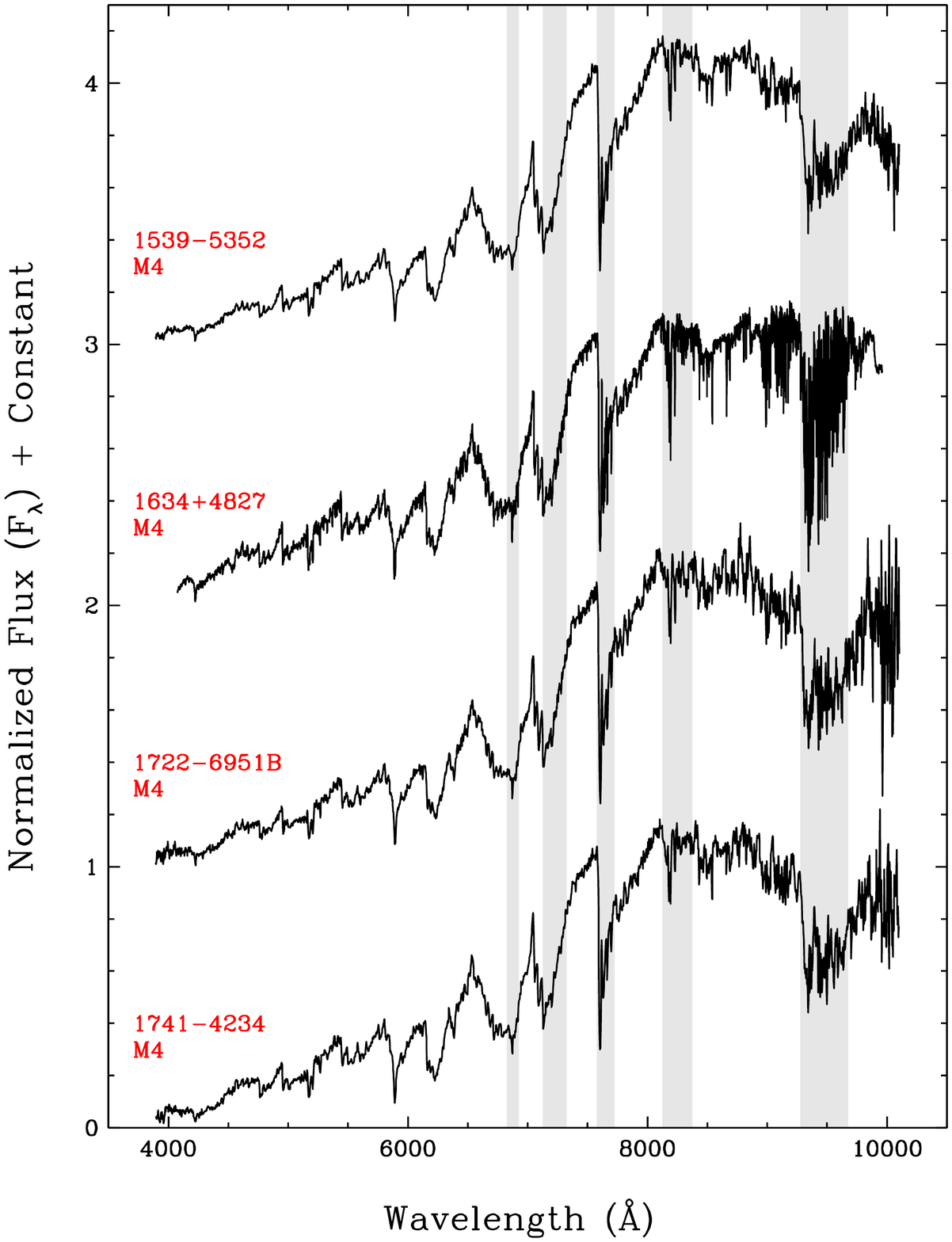}
\caption{Spectra of more main sequence objects with types of M4. (See the caption of Figure~\ref{seq_wds1} for other details.)
\label{seq_M4_M4.2}}
\end{figure}

\begin{figure}
\figurenum{30}
\includegraphics[scale=0.375,angle=0]{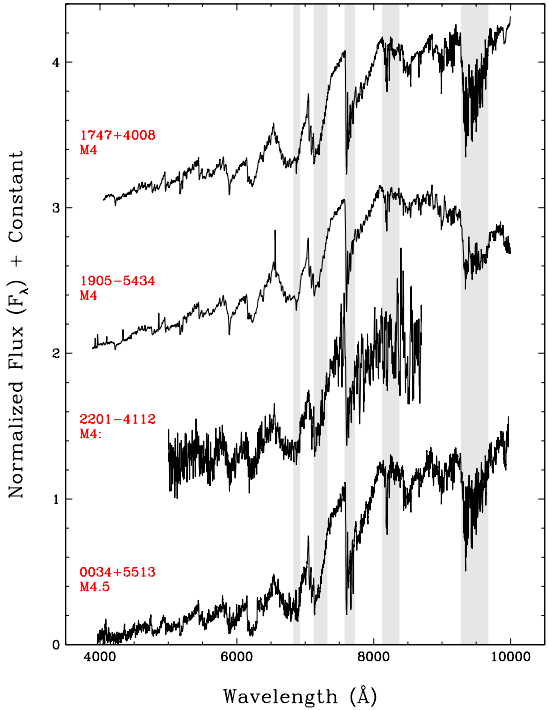}
\caption{Spectra of more main sequence objects with types of M4 along with some objects of type M4.5. (See the caption of Figure~\ref{seq_wds1} for other details.)
\label{seq_M4_M45.1}}
\end{figure}

\begin{figure}
\figurenum{31}
\includegraphics[scale=0.375,angle=0]{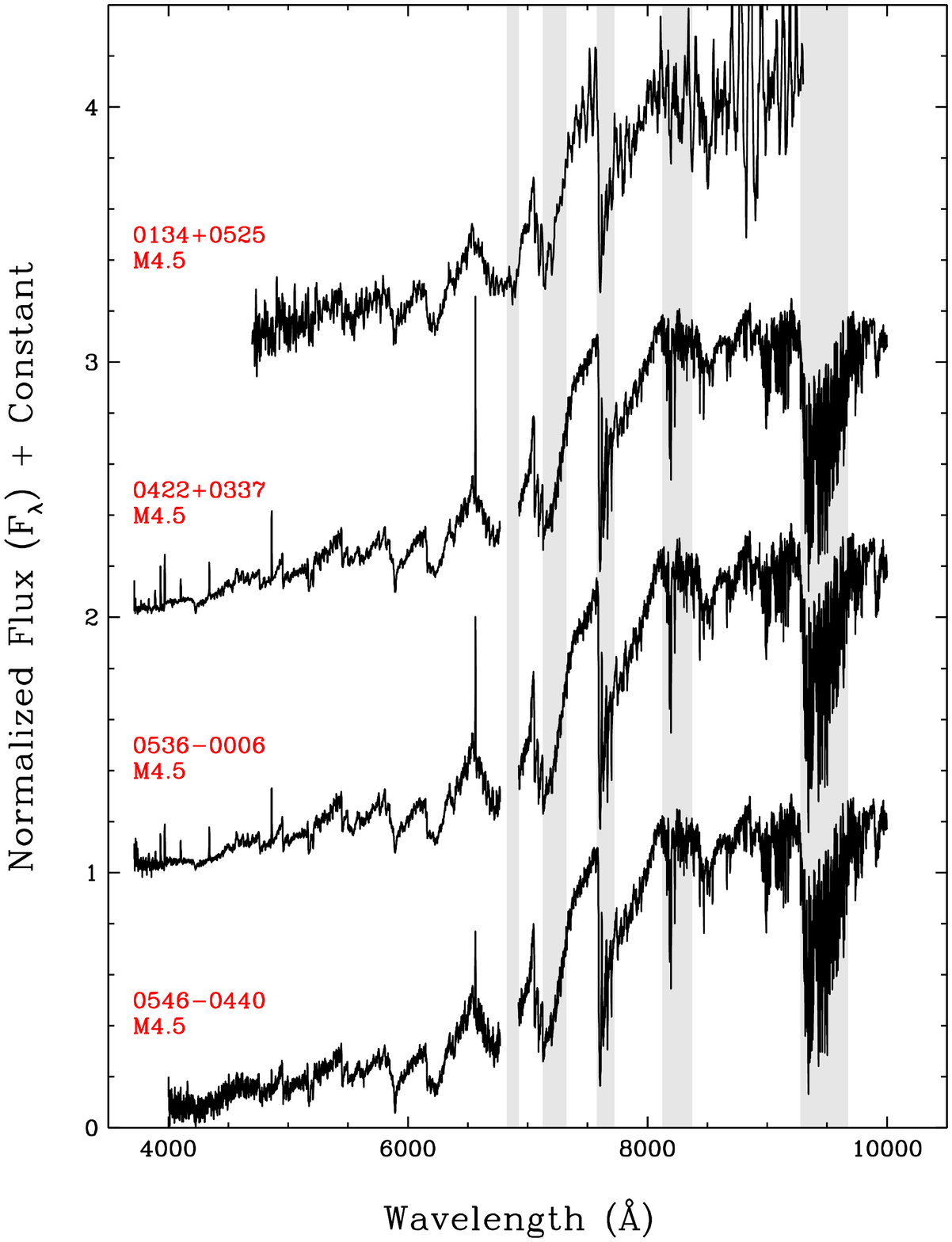}
\caption{Spectra of more main sequence objects with types of M4.5. (See the caption of Figure~\ref{seq_wds1} for other details.)
\label{seq_M4_M45.2}}
\end{figure}

\begin{figure}
\figurenum{32}
\includegraphics[scale=0.375,angle=0]{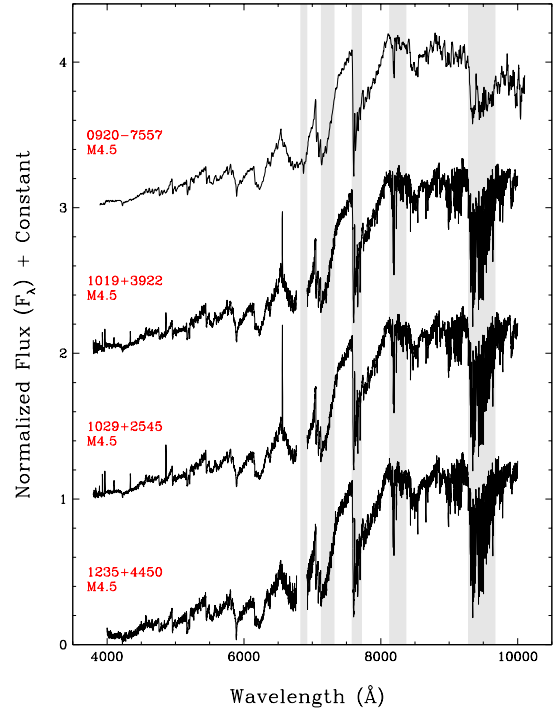}
\caption{Spectra of even more main sequence objects with types of M4.5. (See the caption of Figure~\ref{seq_wds1} for other details.)
\label{seq_M45_M45.1}}
\end{figure}

\begin{figure}
\figurenum{33}
\includegraphics[scale=0.375,angle=0]{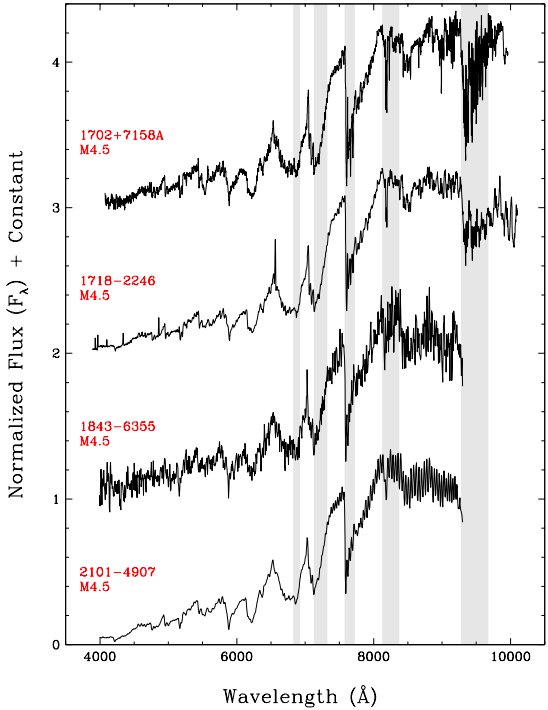}
\caption{Spectra of more main sequence objects with types of M4.5. (See the caption of Figure~\ref{seq_wds1} for other details.)
\label{seq_M45_M45.2}}
\end{figure}

\begin{figure}
\figurenum{34}
\includegraphics[scale=0.375,angle=0]{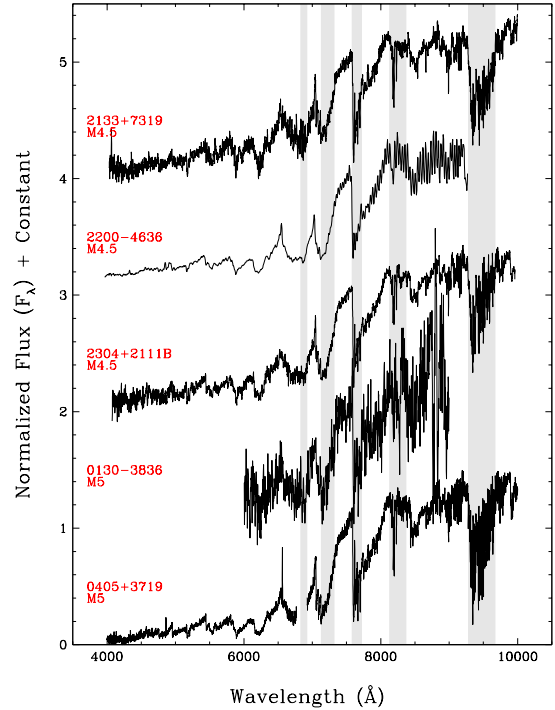}
\caption{Spectra of the remainder of main sequence objects with types of M4.5, along with some of type M5. (See the caption of Figure~\ref{seq_wds1} for other details.)
\label{seq_M45_M5.1}}
\end{figure}

\begin{figure}
\figurenum{35}
\includegraphics[scale=0.375,angle=0]{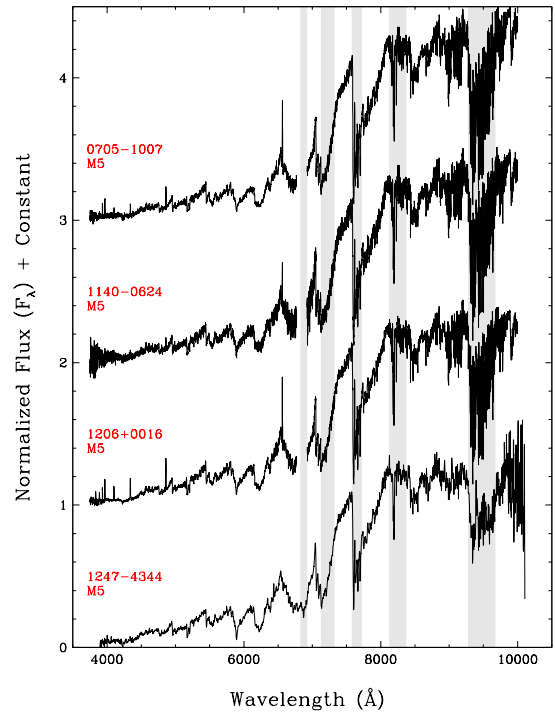}
\caption{Spectra of more main sequence objects of type M5. (See the caption of Figure~\ref{seq_wds1} for other details.)
\label{seq_M45_M5.2}}
\end{figure}

\begin{figure}
\figurenum{36}
\includegraphics[scale=0.375,angle=0]{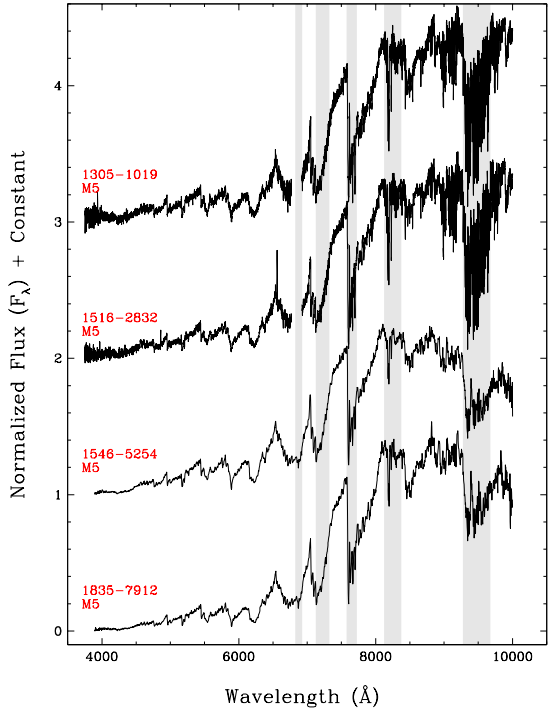}
\caption{Spectra of the remaining main sequence objects with types of M5. (See the caption of Figure~\ref{seq_wds1} for other details.) For WISE 1546$-$5254, the spectrum shown is the one from UT 2014 May 02.
\label{seq_M5_M6.1}}
\end{figure}

\begin{figure}
\figurenum{37}
\includegraphics[scale=0.375,angle=0]{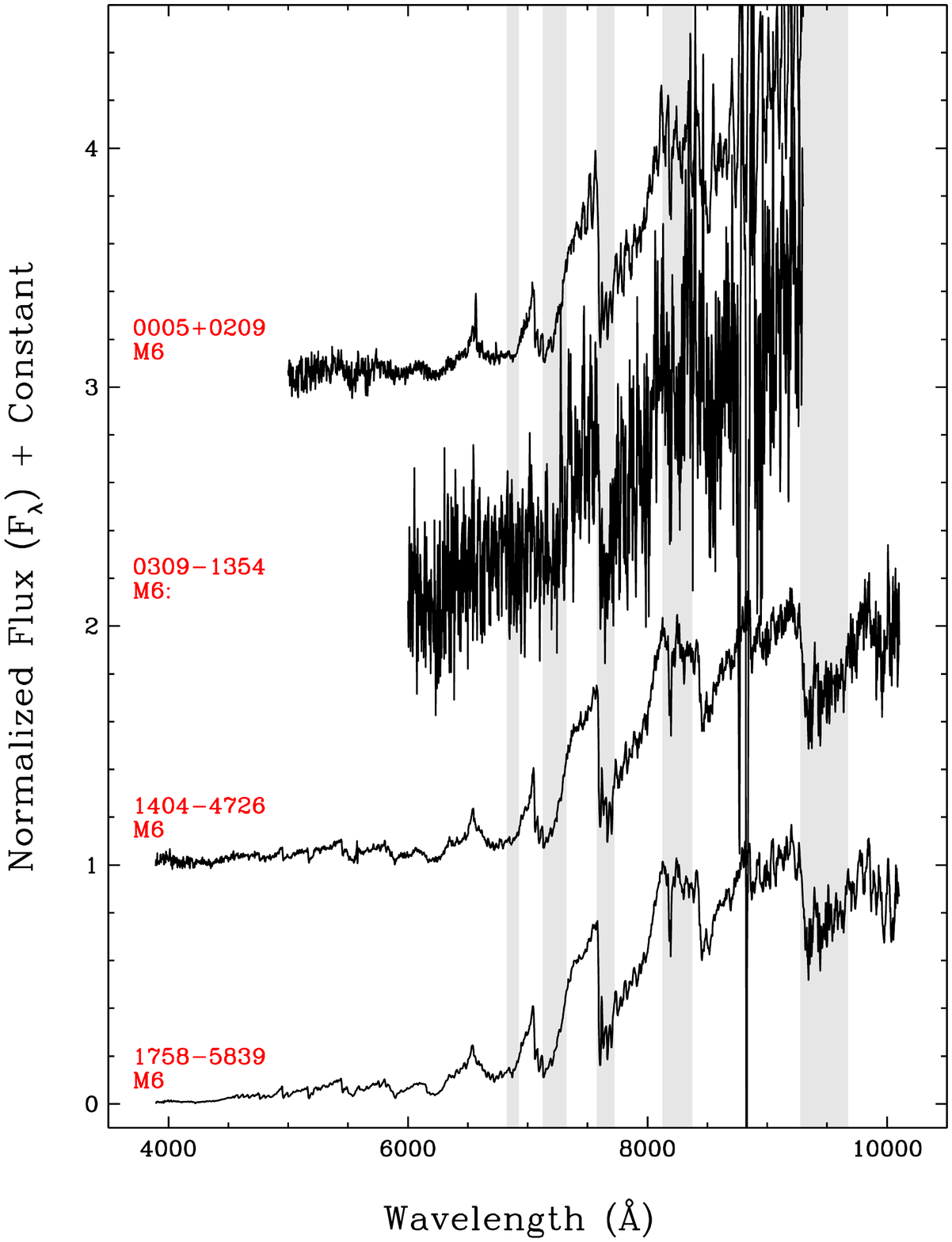}
\caption{Spectra of main sequence objects with types of M6. (See the caption of Figure~\ref{seq_wds1} for other details.)
\label{seq_M5_M6.2}}
\end{figure}

\begin{figure}
\figurenum{38}
\includegraphics[scale=0.375,angle=0]{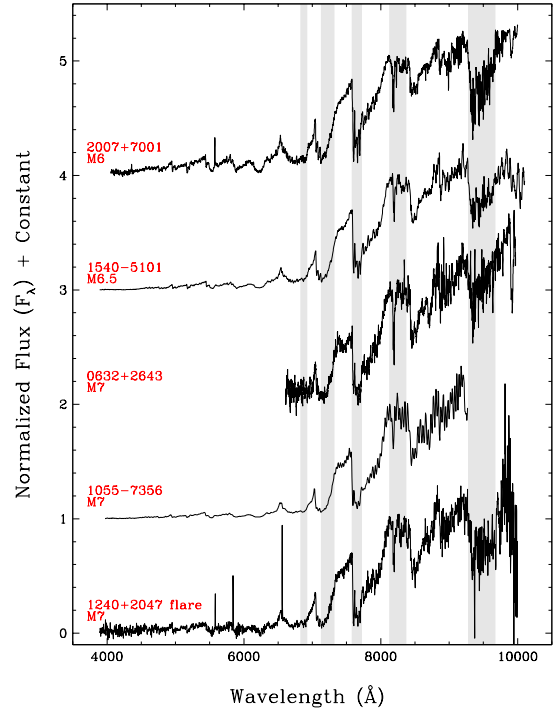}
\caption{Spectra of dwarfs with types of M6 through M7. (See the caption of Figure~\ref{seq_wds1} for other details.) The spectrum shown for WISE 1240+2047 is the second integration from UT 2014 May 04.
\label{seq_M6_M7.1}}
\end{figure}

\begin{figure}
\figurenum{39}
\includegraphics[scale=0.375,angle=0]{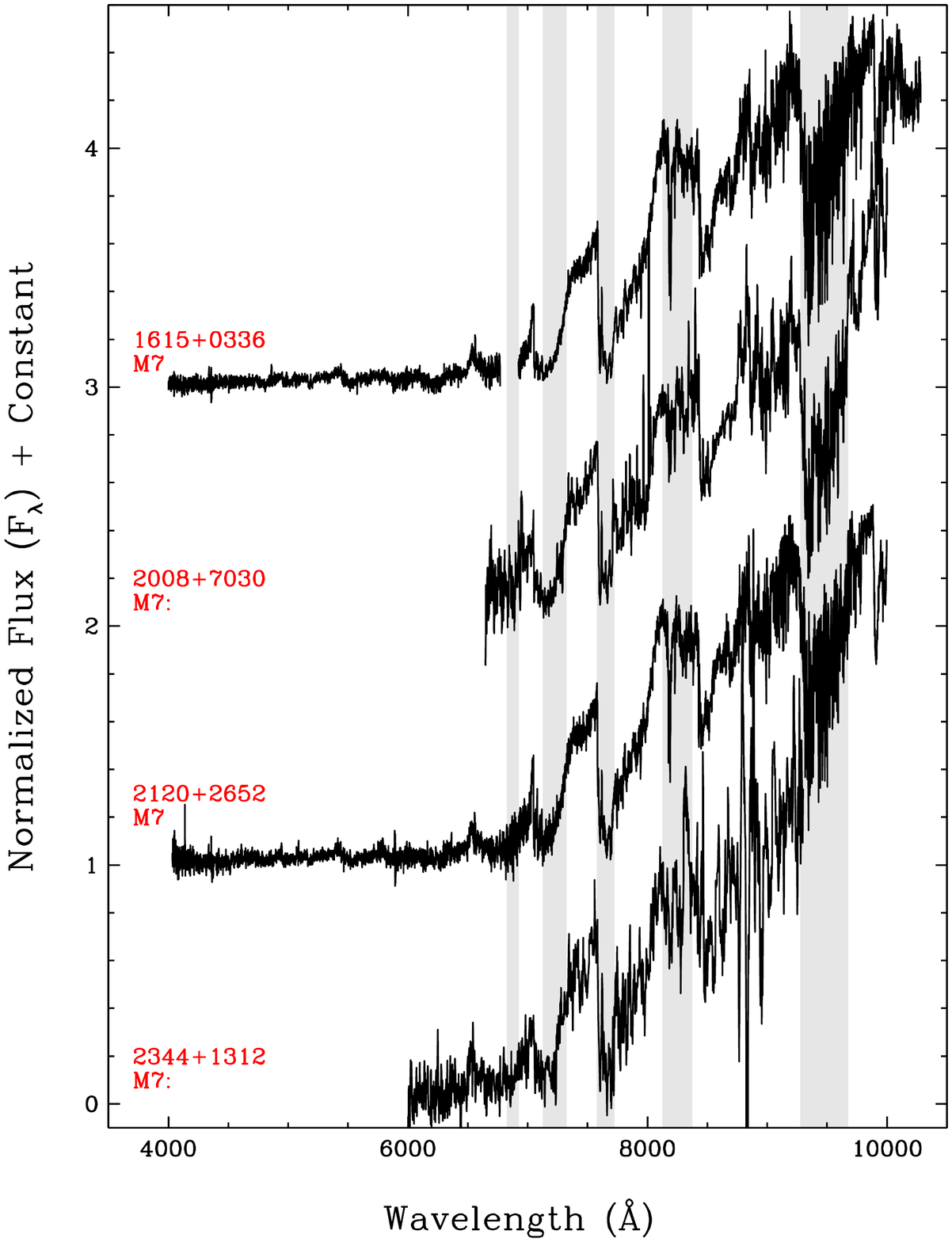}
\caption{Spectra of dwarfs with types of M7. (See the caption of Figure~\ref{seq_wds1} for other details.)
\label{seq_M6_M7.2}}
\end{figure}

\begin{figure}
\figurenum{40}
\includegraphics[scale=0.375,angle=0]{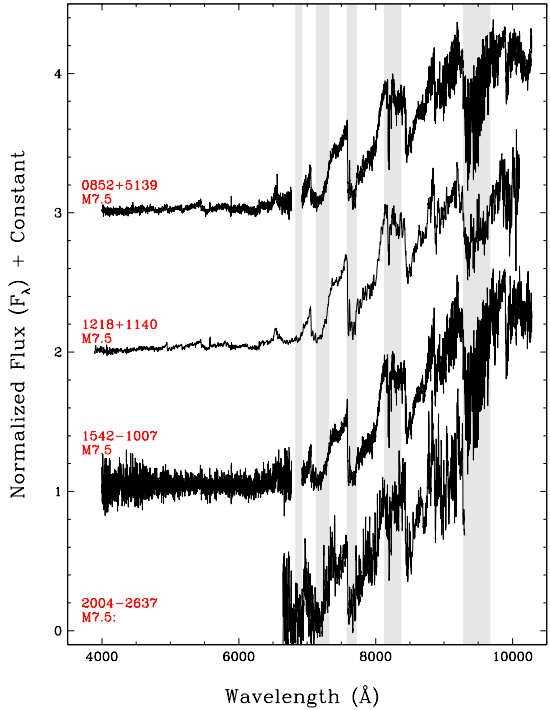}
\caption{Spectra of dwarfs with types of M7.5. (See the caption of Figure~\ref{seq_wds1} for other details.)
\label{seq_M75_M8.1}}
\end{figure}

\begin{figure}
\figurenum{41}
\includegraphics[scale=0.375,angle=0]{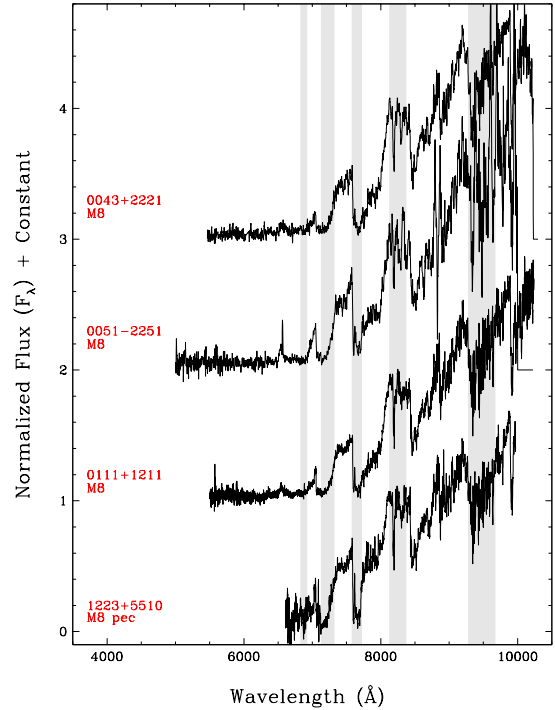}
\caption{Spectra of dwarfs with types of M8. (See the caption of Figure~\ref{seq_wds1} for other details.) The spectrum shown for WISE 0043+2221 is the one from Keck/LRIS.
\label{seq_M75_M8.2}}
\end{figure}

\begin{figure}
\figurenum{42}
\includegraphics[scale=0.375,angle=0]{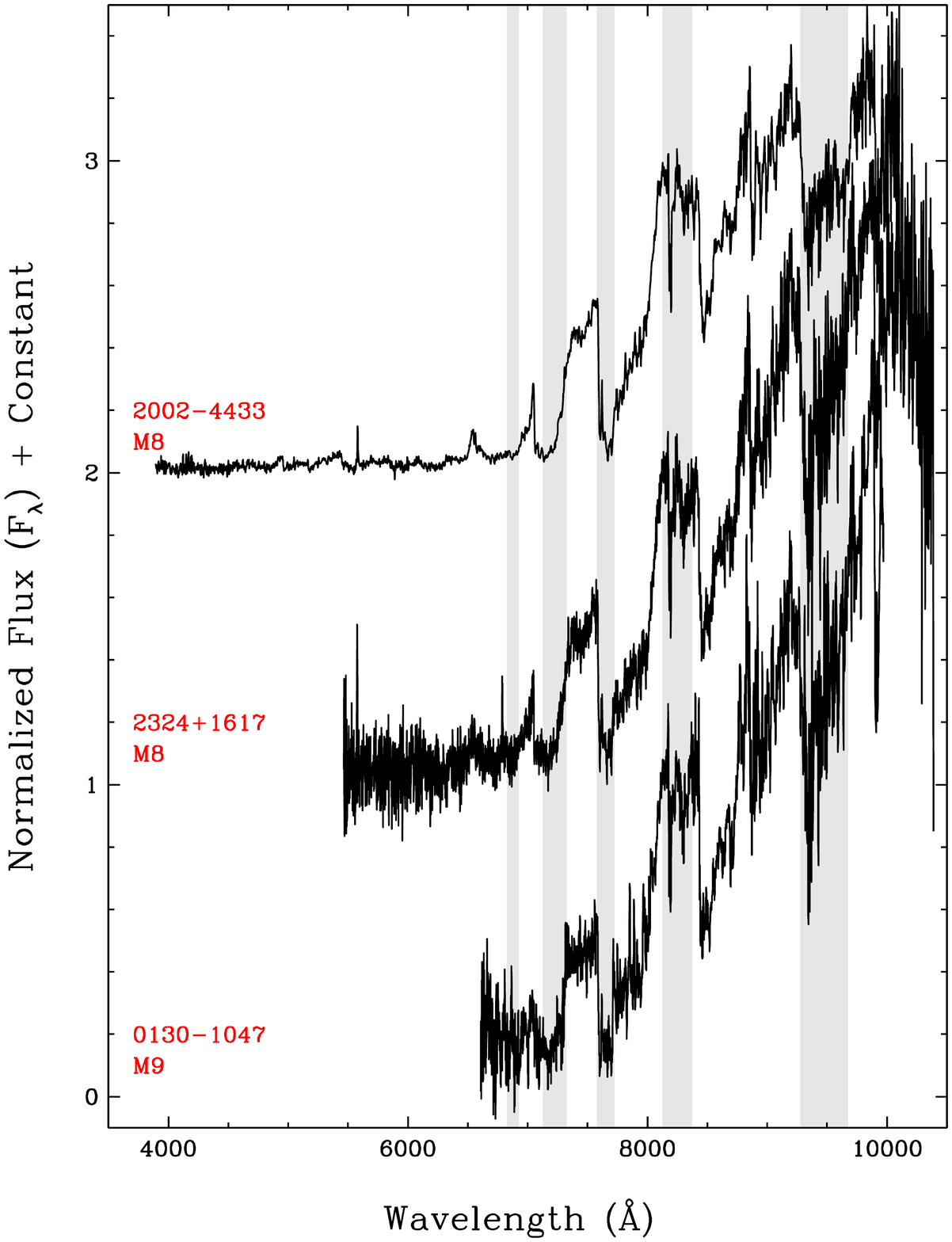}
\caption{Spectra of dwarfs with types of M8 through M9. (See the caption of Figure~\ref{seq_wds1} for other details.)
\label{seq_M75_M8.25}}
\end{figure}

\begin{figure}
\figurenum{43}
\includegraphics[scale=0.375,angle=0]{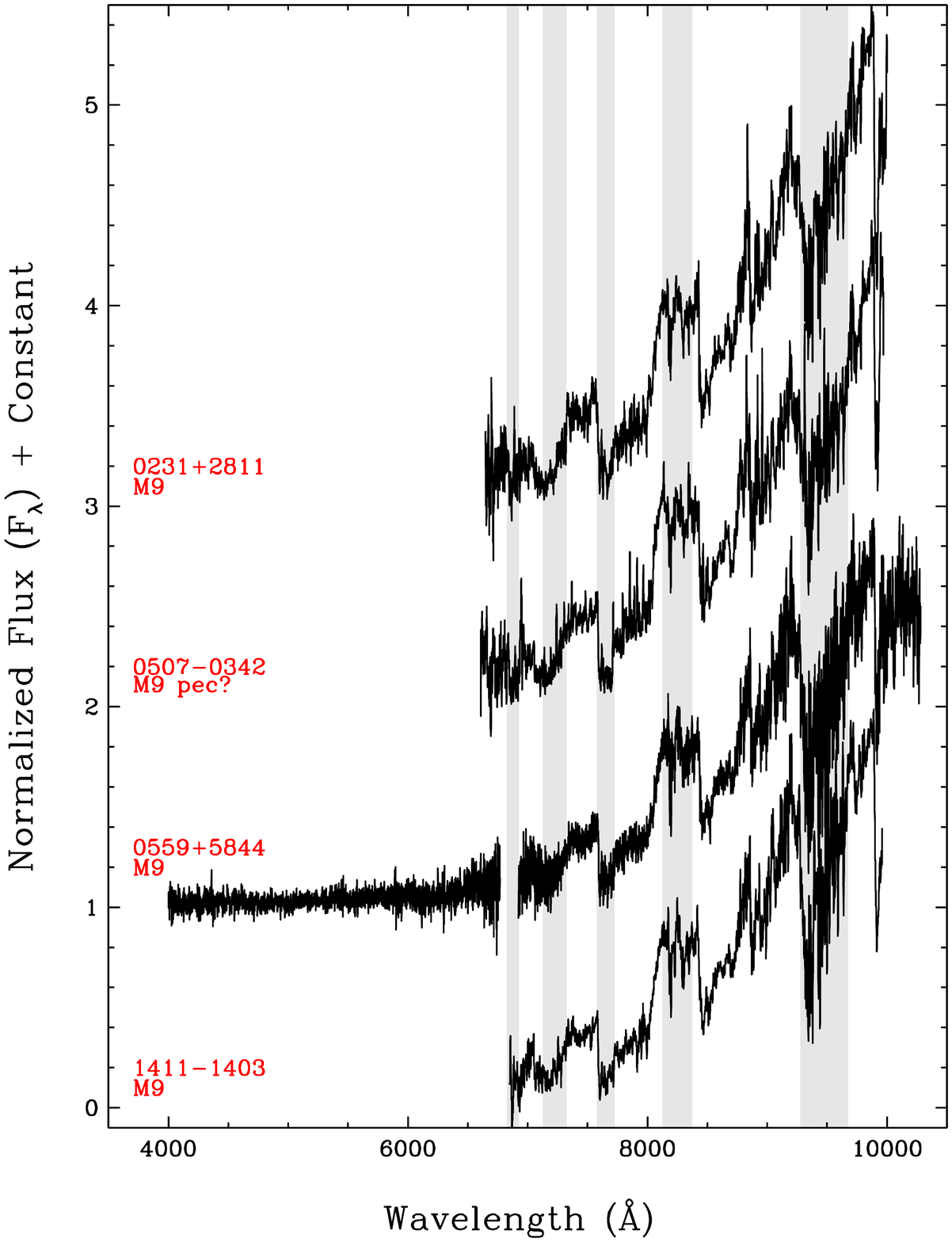}
\caption{Spectra of dwarfs with types of M9. (See the caption of Figure~\ref{seq_wds1} for other details.)
\label{seq_M8_M95.1}}
\end{figure}

\begin{figure}
\figurenum{44}
\includegraphics[scale=0.375,angle=0]{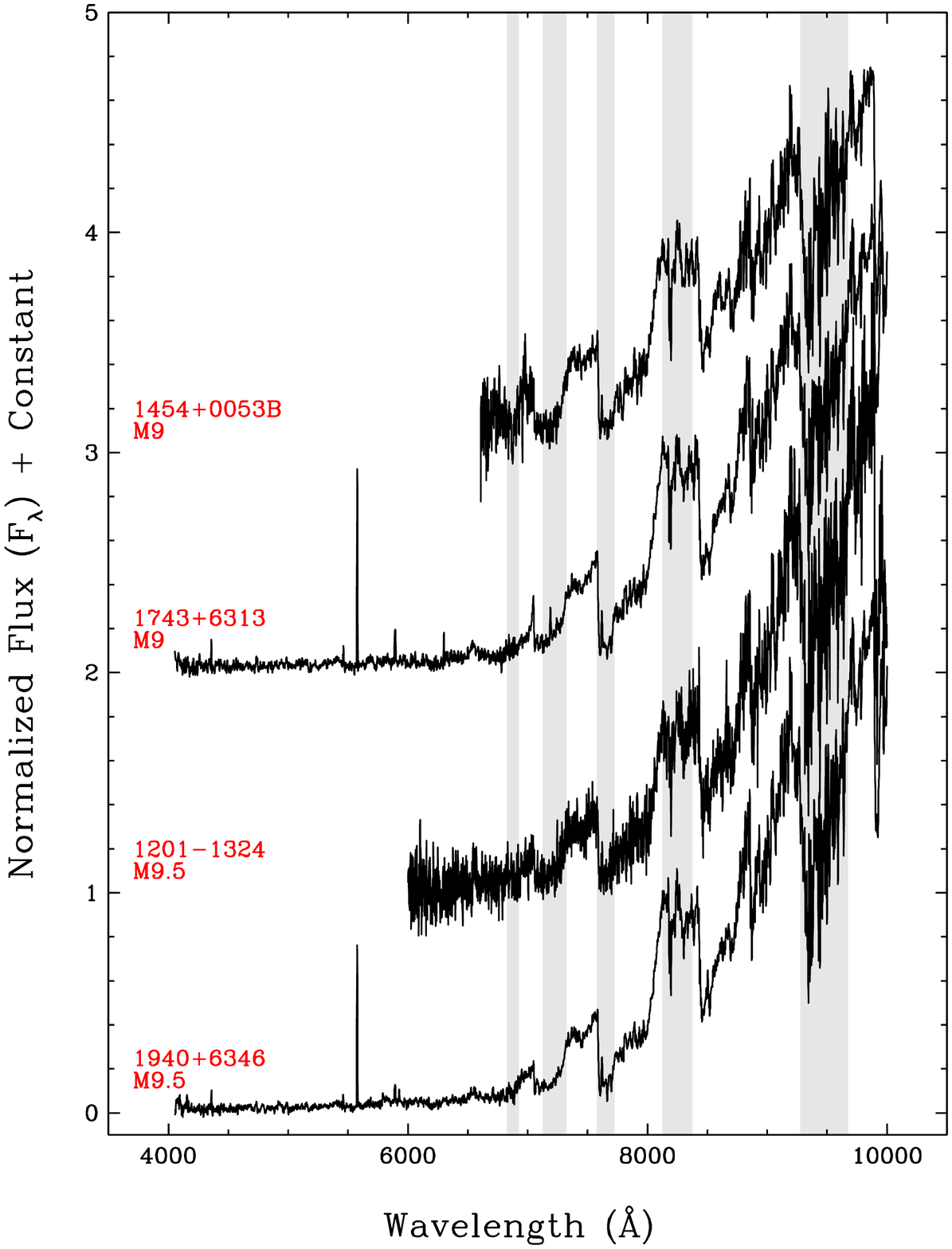}
\caption{Spectra of dwarfs with types of M9 through M9.5. (See the caption of Figure~\ref{seq_wds1} for other details.)
\label{seq_M8_M95.2}}
\end{figure}

Plots of our mid-K through M dwarfs are shown in Figure~\ref{seq_K7_M05.1} through Figure~\ref{seq_M8_M95.2}. We used these observations to search for newly identified active M dwarfs that could be part of nearby young associations, and to identify and/or better characterize newly discovered M dwarfs within $\sim$20 pc of the Sun.

\subsubsection{Searching for Young, Active M Dwarfs\label{young_Mdwarfs}}

For F, G, and K dwarfs, there is a well established correlation between youth and magnetic activity (\citealt{hartmann1987}). As these stars spin down with time, the activity driven by the magnetic dynamo also decreases. Hence, activity levels in the coronae and chromospheres can be used as proxies for age. These indicators, however, are not perfect bellwethers. If an old star is observed only once and during a flare, its high level of activity may be misconstrued as a sign of youth. Similarly, old stars in close binary systems can retain high levels of magnetic activity because tidal locking keeps the rotation speeds high. Nonetheless, such activity indicators can be used to identify young candidates. 

In these stars, the dynamo depends on rotational sheer and an anchor point at the interface between the radiative and convective zones within the star. This same dynamo would not, however, be able to anchor itself in the fully convective atmospheres possessed by mid- to late-M dwarfs (\citealt{reid2000}).  Although, it is unclear what the underlying physical mechanism is, the same activity/age correlations are seen in M dwarfs as in higher mass stars (\citealt{browning2010}), so we can use these same youth indicators despite not fully understanding why they work. 

We have considered the spectroscopically verified M dwarfs from Table 4 and Table~\ref{spectra_literature} to search for previously missed, young M dwarfs in the solar neighborhood. Table 8 lists those M dwarfs with detections in X-ray, ultraviolet, and/or H$\alpha$ emission. Detections in the X-ray or ultraviolet were assumed if our AllWISE object falls within 30 arcsec of a 3$\sigma$ detection from the R\"ontgensatelit\footnote{We included objects from the 1RXS catalog ({\url http://heasarc.gsfc.nasa.gov/db-perl/W3Browse/w3browse.pl}) with 3$\sigma$ detections in the raw count rates, prior to conversion to fluxes.} (ROSAT) or 20 arcsec of a 3$\sigma$ detection in either the far- (FUV) or near-ultraviolet (NUV) from the Galaxy Evolution Explorer\footnote{The {\it GALEX} source catalog is available at {\url http://galex.stsci.edu/GR6/?page=mastform}.} ({\it GALEX}), these radii taking into account the resolution of each survey and the time span between them and AllWISE. The ROSAT count rates have been converted to fluxes using the formula $F_X = ({\rm count\, rate}){\times}(8.31 + 5.30{\times}HR1){\times}10^{-12}$ erg cm$^{-2}$ s$^{-1}$ from \cite{schmitt1995}, where $HR1=(B-A)/(B+A)$, with $A$ being the counts in the soft channel (0.1 - 0.4 keV) and $B$ the counts in the hard channel (0.5 - 2.0 keV). (See also \citealt{voges1999}.) 

\begin{turnpage}
\begin{deluxetable*}{cccccccccccccc}
\tabletypesize{\tiny}
\tablewidth{9.5in}
\tablenum{8}
\tablecaption{Measured Properties of Active M Dwarfs from Table 4 and Table~\ref{spectra_literature}\label{M_dwarf_activity}}
\tablehead{
\colhead{WISE} & 
\colhead{Spec.} &                         
\colhead{USNO $R$} &  
\colhead{USNO $I$} &     
\colhead{2MASS $J$} &
\colhead{$f_J$ } &
\colhead{$F_J$} &
\colhead{$F_X$ } &
\colhead{$f_{\rm FUV}$} &
\colhead{$f_{\rm NUV}$} &
\colhead{H$\alpha$ EW} &
\colhead{Group} &
\colhead{Memb.} &  
\colhead{d$_{\rm pred}$} \\    
\colhead{Object} & 
\colhead{Type} &                         
\colhead{(mag)} &  
\colhead{(mag)} &     
\colhead{(mag)} &
\colhead{(mJy)} &
\colhead{(erg/cm$^{2}$s)} &
\colhead{(erg/cm$^{2}$s)} &
\colhead{($\mu$Jy)} &
\colhead{($\mu$Jy)} &
\colhead{(\AA)} &
\colhead{} &
\colhead{Prob.(\%)} &  
\colhead{(pc)} \\     
\colhead{(1)} &                          
\colhead{(2)} &  
\colhead{(3)} &     
\colhead{(4)} &
\colhead{(5)} &
\colhead{(6)} &
\colhead{(7)} &
\colhead{(8)} &
\colhead{(9)} &
\colhead{(10)} &
\colhead{(11)} &
\colhead{(12)} &
\colhead{(13)} &     
\colhead{(14)} 
}
\startdata
0005+0209   & M6  & 16.56&  13.78& 11.992$\pm$0.029&  25.5&  8.10e-12& ---             &      ---      &   ---          &  $-$10.9$\pm$1.5& AB Dor     & 91.96&   18.1$\pm$1.0 \\ 
0051$-$2251 & M8  & 19.36&  15.37& 13.019$\pm$0.027&  9.88&  3.15e-12& ---             &      ---      &   ---          &  $-$20.6$\pm$2.5& old field  & 98.76&   27.3$\pm$12.8\\ 
0157$-$0948 & M4  & 16.03&  ---  & 13.064$\pm$0.038&  9.48&  3.02e-12& ---             &      ---      &  47.70$\pm$1.07&      ---        & old field  & 96.37&   45.0$\pm$16.8\\ 
0405+3719   & M5  & 15.01&  13.39& 10.915$\pm$0.021&  68.6&  21.9e-12& ---             &      ---      &  14.70$\pm$3.50&   $-$6.2$\pm$0.3& old field  & 70.88&   83.4$\pm$32.8\\ 
0422+0337   & M4.5& 14.58&  10.90&  9.857$\pm$0.023&  182 &  57.9e-12& 18.5$\pm$2.7e-13&      ---      &  42.29$\pm$1.30&   $-$7.0$\pm$0.1& old field  & 79.98&   51.8$\pm$22.4\\ 
0447+2534   & M4  & 14.10&  11.64& 10.631$\pm$0.017&  89.1&  28.4e-12& ---             &      ---      &   ---          &   $-$5.1$\pm$0.4& old field  & 70.55&   83.0$\pm$33.9\\ 
0500+1916   & M1  & 11.21&  10.17&  9.125$\pm$0.021&  357.&  113.e-12& ---             &      ---      &  21.36$\pm$4.31&      ---        & old field  & 85.37&   55.8$\pm$22.0\\ 
0536$-$0006 & M4.5& 12.62&  11.35& 10.630$\pm$0.019&  89.2&  28.4e-12&  3.0$\pm$1.0e-13&      ---      &   ---          &   $-$6.0$\pm$0.4& old field  & 63.73&   34.6$\pm$14.8\\  
0546$-$0440 & M4.5& 13.89&  11.58& 10.366$\pm$0.023&  114.&  36.2e-12& ---             &      ---      &   ---          &   $-$3.9$\pm$0.5& young field& 50.46&   49.8$\pm$24.6\\ 
0701$-$0137 & M4  & 14.24&  12.52& 10.613$\pm$0.020&  90.6&  28.9e-12& ---             &      ---      &   ---          &   $-$4.8$\pm$0.3& old field  & 61.19&   85.8$\pm$35.7\\ 
0705$-$1007 & M5  & 14.11&  11.50& 10.196$\pm$0.021&  133.&  42.4e-12& ---             &      ---      &   ---          &   $-$6.2$\pm$0.3& old field  & 97.08&   33.3$\pm$15.6\\ 
0720$-$0846 & M9  & 16.87&  13.95& 10.628$\pm$0.023&  89.4&  28.5e-12& ---             &      ---      &   ---          &  $-$17.7$\pm$0.3& old field  & 49.16&   52.6$\pm$21.6\\  
0852+5139   & M7.5& 19.23&  16.10& 13.984$\pm$0.024&  4.06&  1.29e-12& ---             &      ---      &   ---          &   $-$5.5$\pm$1.7& old field  & 98.24&   24.5$\pm$9.0 \\ 
0912+2205   & M3  & 13.32&  11.59& 10.733$\pm$0.017&  81.1&  25.9e-12& ---             &      ---      &  18.50$\pm$0.99&   $-$5.9$\pm$0.3& old field  & 77.90&   81.4$\pm$31.6\\ 
0935$-$0301 & M3  & 13.12&  11.38& 10.555$\pm$0.025&  95.6&  30.4e-12& ---             &      ---      &  10.75$\pm$2.25&   $-$1.9$\pm$0.2& old field  & 94.83&   67.4$\pm$27.4\\  
1019+3922   & M4.5& 14.69&  12.88& 11.143$\pm$0.014&  55.6&  17.7e-12& ---             &  1.55$\pm$0.12&   6.16$\pm$0.21&   $-$4.3$\pm$0.3& old field  & 98.44&   42.2$\pm$20.4\\ 
1029+2545   & M4.5& 13.48&  11.31& 10.358$\pm$0.018&  115.&  36.5e-12&  3.7$\pm$1.2e-13& 10.33$\pm$2.68&  20.63$\pm$2.79&   $-$7.5$\pm$0.2& old field  & 96.27&   69.0$\pm$30.6\\ 
1055$-$5750 & M4  & ---  &  12.60& 11.424$\pm$0.021&  42.9&  13.7e-12& ---             &      ---      &   ---          &   $-$1.6$\pm$0.4& old field  & 93.39&   52.2$\pm$21.6\\  
1059+1509   & M3.5& 14.18&  12.06& 11.154$\pm$0.018&  55.1&  17.5e-12& ---             &      ---      &   4.41$\pm$1.38&      ---        & old field  & 96.81&   69.4$\pm$28.8\\  
1114+5703   & M0  & 12.08&  10.79& 10.258$\pm$0.015&  126.&  40.0e-12& ---             &      ---      &  11.63$\pm$1.27&      ---        & old field  & 93.06&   56.6$\pm$23.8\\ 
1140$-$0624 & M5  & 15.61&  13.27& 11.957$\pm$0.021&  26.3&  8.37e-12& ---             &      ---      &   ---          &   $-$2.5$\pm$0.5& old field  & 96.69&   40.2$\pm$15.2\\ 
1202$-$0111 & M4  & 16.27&  14.72& 13.256$\pm$0.021&  7.94&  2.53e-12& ---             &      ---      &   1.27$\pm$0.42&      ---        & old field  & 88.89&   78.6$\pm$30.0\\ 
1203+1810   & M0  & 12.62&  11.21& 10.331$\pm$0.015&  118.&  37.4e-12& ---             &  3.54$\pm$0.65&   8.03$\pm$0.56&      ---        & old field  & 95.07&   79.0$\pm$33.5\\ 
1206+0016   & M5  & 13.83&  11.49& 10.348$\pm$0.023&  116.&  36.8e-12& ---             &  2.34$\pm$0.37&  10.26$\pm$0.87&   $-$5.4$\pm$0.4& old field  & 93.95&   73.0$\pm$29.2\\ 
1210$-$4612 & M1  & 11.65&  10.21&  9.769$\pm$0.019&  197.&  62.8e-12& ---             &      ---      & 149.14$\pm$7.38&      ---        & old field  & 98.99&   24.5$\pm$12.0\\  
1218+1140   & M7.5& 20.18&  16.65& 14.185$\pm$0.032&  3.38&  1.08e-12& ---             &      ---      &   6.37$\pm$1.21&      ---        & old field  & 98.73&   31.7$\pm$12.2\\  
1222$-$8449 & M3  & 12.04&  10.45&  9.721$\pm$0.024&  206.&  65.6e-12& ---             &      ---      &  24.96$\pm$5.23&      ---        & old field  & 95.29&   41.4$\pm$19.6\\ 
1235+4450   & M4.5& 15.14&  13.50& 11.985$\pm$0.018&  25.6&  8.16e-12& ---             &      ---      &  13.07$\pm$2.64&      ---        & old field  & 98.20&   47.0$\pm$23.2\\  
1240+2047   & M7  & 19.45&  15.77& 14.106$\pm$0.024&  3.63&  1.16e-12& --- &      ---      &   ---   &  $-$14.1$\pm$1.2\tablenotemark{a}  & old field  & 99.10&   31.7$\pm$13.0\\  
1247$-$4344 & M5  & 16.69&  14.71& 13.279$\pm$0.025&  7.78&  2.48e-12& ---             &      ---      &   8.61$\pm$2.45&      ---        & old field  & 82.32&   53.4$\pm$20.4\\  
1454+0053   & M3  & 13.40&  11.40& 10.444$\pm$0.021&  106.&  33.7e-12& ---             &      ---      &   1.60$\pm$0.44&      ---        & old field  & 90.53&   37.0$\pm$15.2\\ 
1516$-$2832 & M5  & 14.39&  12.14& 10.516$\pm$0.021&  99.1&  31.6e-12& ---             &      ---      &  31.57$\pm$5.61&   $-$4.7$\pm$0.3& old field  & 76.00&   14.9$\pm$5.6 \\ 
1546$-$5534 & M8  & 16.45&  12.80& 10.209$\pm$0.022&  131.&  41.9e-12& ---             &      ---      &   ---          &   $-$6.1$\pm$0.3& Argus      & 50.17&    9.3$\pm$0.8 \\ 
1615+0336   & M7  & 19.72&  16.19& 13.817$\pm$0.026&  4.74&  1.51e-12& ---             &      ---      &   4.74$\pm$1.40&   $-$5.1$\pm$1.2& old field  & 96.99&   41.4$\pm$15.8\\ 
1634+4827   & M4  & 11.78&  10.01&  9.115$\pm$0.032&  360.&  115.e-12& ---             &      ---      &   3.39$\pm$0.43&      ---        & old field  & 91.42&   47.8$\pm$21.2\\  
1718$-$2246 & M4.5& 13.80&  12.18& 10.207$\pm$0.018&  132.&  42.0e-12&  3.6$\pm$1.4e-13&      ---      &  18.25$\pm$4.56&   $-$7.1$\pm$0.2& old field  & 87.69&   53.0$\pm$20.2\\ 
1722$-$6951A& M3  & 11.71&  10.22&  9.330$\pm$0.021&  295.&  94.1e-12& 13.9$\pm$3.7e-13&      ---      &   ---          &   $-$5.5$\pm$0.3& old field  & 94.37&   49.8$\pm$18.6\\ 
1905$-$5434 & M4  & 12.28&  10.57&  9.409$\pm$0.024&  275.&  87.5e-12& ---             &      ---      &  37.81$\pm$3.64&   $-$4.7$\pm$0.2& $\beta$ Pic& 66.68&   19.7$\pm$1.6 \\ 
2002$-$4433 & M8  & 19.37&  15.80& 13.528$\pm$0.023&  6.18&  1.97e-12& ---             &      ---      &   ---          &   $-$5.2$\pm$1.5& old field  & 80.30&   16.1$\pm$5.6 \\  
2007+7001   & M6  & 18.52&  15.78& 14.068$\pm$0.034&  3.76&  1.20e-12& ---             &  6.69$\pm$2.09&   ---          &      ---        & old field  & 98.26&   14.1$\pm$6.4 \\  
2200$-$4636 & M4.5& 17.78&  16.19& 14.323$\pm$0.021&  2.97&  9.46e-13& ---             &      ---      &  10.47$\pm$3.13&      ---        & old field  & 57.88&   84.2$\pm$35.9\\
2304+2111   & M0.5& 11.68&  10.46&  9.709$\pm$0.022&  208.&  66.4e-12& ---             &      ---      &  10.25$\pm$2.51&      ---        & AB Dor     & 89.70&   24.5$\pm$1.6 \\ 
\enddata
\tablenotetext{a}{This measurement is for the flaring spectrum shown in Figure~\ref{seq_M6_M7.1}.}
\end{deluxetable*}
\end{turnpage}

To ease comparison to published sources, we also provide in Table 8 $R$- and $I$-band photometry from the United States Naval Observatory B1.0 Catalog\footnote{This is described in \cite{monet2003} and is available at {\url http://irsa.ipac.caltech.edu}.} (USNO-B1), which has typical uncertainties of $\pm$0.3 mag\footnote{See {\url http://irsa.ipac.caltech.edu/data/USNO\_B1/usnob1\_ description.html\#usno\_b1}.}, and $J$-band photometry from 2MASS\footnote{This is also available at {\url http://irsa.ipac.caltech.edu}.}. These $J$-band magnitudes are also converted to $J$-band flux densities ($f_J$) and fluxes ($F_J$) and listed in the table for convenience. These were computed via the equations $f_J = ({f_J}_0)10^{-J/2.5}$ and $F_J = ({f_J}) c {\lambda}^{-2} {\Delta\lambda}$, where ${f_J}_0$ is the $J$-band flux density at $J$=0 mag, $J$ is the $J$-band magnitude, $c$ is the speed of light, $\lambda$ is the central wavelength of the $J$-band filter, and $\Delta\lambda$ is the $J$-band filter width. We use the values listed in the 2MASS Explanatory Supplement\footnote{See {\url http://www.ipac.caltech.edu/2mass/releases/allsky/doc /sec6\_4a.html}.} of ${f_J}_0$=1594$\pm$27.8 Jy, $\lambda$=1.235$\pm$0.006 $\mu$m, and $\Delta\lambda$= 0.162$\pm$0.001 $\mu$m.  

\begin{turnpage}
\begin{deluxetable*}{cccccccccccccccc}
\tabletypesize{\tiny}
\tablewidth{8.0in}
\tablenum{9}
\tablecaption{Flux Ratios for M Dwarfs with X-ray and Ultraviolet Detections\label{M_dwarf_activity2}}
\tablehead{
\colhead{WISE} & 
\colhead{SpType} &                         
\colhead{$R-J$} &  
\colhead{$I-J$} &     
\colhead{log($F_X/F_J$)} &
\colhead{modified\tablenotemark{a}} &
\colhead{$f_{\rm FUV}/f_J$} &
\colhead{$f_{\rm NUV}/f_J$} &
\colhead{NUV$-J$} &  
\colhead{$J-K_s$} &     
\colhead{NUV$-$W1} &  
\colhead{$J-$W2} \\     
\colhead{Object} & 
\colhead{} &                         
\colhead{(mag)} &  
\colhead{(mag)} &     
\colhead{(dex)} &
\colhead{(dex)} &
\colhead{} &
\colhead{} &
\colhead{(mag)} &  
\colhead{(mag)} &     
\colhead{(mag)} &  
\colhead{(mag)} \\     
\colhead{(1)} &                          
\colhead{(2)} &  
\colhead{(3)} &     
\colhead{(4)} &
\colhead{(5)} &
\colhead{(6)} &
\colhead{(7)} &
\colhead{(8)} &
\colhead{(9)} &
\colhead{(10)} &
\colhead{(11)} &
\colhead{(12)} 
}
\startdata
0157$-$0948 & M4      & 2.97&    ---&     ---&     ---&  ---    &   0.00503&  6.64$\pm$0.04&   0.64$\pm$0.05&   7.50$\pm$0.03&   1.08$\pm$0.04\\
0405+3719   & M5      & 4.10&   2.48&     ---&     ---&  ---    &   0.00021& 10.06$\pm$0.26&   0.87$\pm$0.03&  11.14$\pm$0.26&   1.29$\pm$0.03\\
0422+0337   & M4.5    & 4.72&   1.04&   -1.50&   -1.77&  ---    &   0.00023&  9.97$\pm$0.04&   0.86$\pm$0.03&  10.93$\pm$0.04&   1.11$\pm$0.03\\
0500+1916   & M1      & 2.09&   1.05&     ---&     ---&  ---    &   0.00006& 11.45$\pm$0.22&   0.82$\pm$0.03&  12.41$\pm$0.22&   0.91$\pm$0.03\\
0536$-$0006 & M4.5    & 1.99&   0.72&   -1.98&   -2.25&  ---    &   ---    &     ---       &   0.82$\pm$0.03&       ---      &   1.23$\pm$0.03\\
0912+2205   & M3      & 2.59&   0.86&     ---&     ---&  ---    &   0.00023& 10.00$\pm$0.06&   0.94$\pm$0.03&  11.07$\pm$0.06&   1.15$\pm$0.03\\
0935$-$0301 & M3      & 2.56&   0.82&     ---&     ---&  ---    &   0.00011& 10.76$\pm$0.23&   0.88$\pm$0.03&  11.74$\pm$0.23&   1.11$\pm$0.03\\
1019+3922   & M4.5    & 3.55&   1.74&     ---&     ---&  0.00003&   0.00011& 10.79$\pm$0.04&   0.82$\pm$0.02&  11.81$\pm$0.05&   1.19$\pm$0.02\\
1029+2545   & M4.5    & 3.12&   0.95&   -1.99&   -2.26&  0.00009&   0.00018& 10.25$\pm$0.15&   0.82$\pm$0.02&  11.24$\pm$0.15&   1.16$\pm$0.03\\
1059+1509   & M3.5    & 3.03&   0.91&     ---&     ---&  ---    &   0.00008& 11.14$\pm$0.34&   0.82$\pm$0.03&  12.10$\pm$0.34&   1.12$\pm$0.03\\
1114+5703   & M0      & 1.82&   0.53&     ---&     ---&  ---    &   0.00009& 10.98$\pm$0.12&   0.81$\pm$0.03&  11.92$\pm$0.12&   0.93$\pm$0.02\\
1202$-$0111 & M4      & 3.01&   1.46&     ---&     ---&  ---    &   0.00016& 10.38$\pm$0.36&   0.79$\pm$0.03&  11.30$\pm$0.36&   1.12$\pm$0.03\\
1203+1810   & M0      & 2.29&   0.88&     ---&     ---&  0.00003&   0.00007& 11.31$\pm$0.08&   0.80$\pm$0.02&  12.16$\pm$0.08&   0.89$\pm$0.02\\
1206+0016   & M5      & 3.48&   1.14&     ---&     ---&  0.00002&   0.00009& 11.02$\pm$0.09&   0.87$\pm$0.03&  12.09$\pm$0.09&   1.26$\pm$0.03\\
1210$-$4612 & M1      & 1.88&   0.44&     ---&     ---&  ---    &   0.00076&  8.70$\pm$0.05&   0.79$\pm$0.03&   9.65$\pm$0.05&   1.06$\pm$0.03\\
1218+1140   & M7.5    & 5.99&   2.46&     ---&     ---&  ---    &   0.00188&  7.70$\pm$0.21&   0.95$\pm$0.05&   8.90$\pm$0.21&   1.38$\pm$0.04\\
1222$-$8449 & M3      & 2.32&   0.73&     ---&     ---&  ---    &   0.00012& 10.69$\pm$0.23&   0.87$\pm$0.03&  11.66$\pm$0.23&   1.07$\pm$0.03\\
1235+4450   & M4.5    & 3.16&   1.52&     ---&     ---&  ---    &   0.00051&  9.13$\pm$0.22&   0.79$\pm$0.02&  10.06$\pm$0.22&   1.17$\pm$0.03\\
1247$-$4344 & M5      & 3.41&   1.43&     ---&     ---&  ---    &   0.00111&  8.28$\pm$0.31&   0.89$\pm$0.04&   9.37$\pm$0.31&   1.28$\pm$0.03\\
1454+0053   & M3      & 2.96&   0.96&     ---&     ---&  ---    &   0.00002& 12.95$\pm$0.30&   0.78$\pm$0.03&  13.87$\pm$0.30&   1.07$\pm$0.03\\
1516$-$2832 & M5      & 3.87&   1.62&     ---&     ---&  ---    &   0.00032&  9.63$\pm$0.19&   0.90$\pm$0.03&  10.73$\pm$0.19&   1.28$\pm$0.03\\
1615+0336   & M7      & 5.90&   2.37&     ---&     ---&  ---    &   0.00100&  8.39$\pm$0.32&   0.81$\pm$0.04&   9.44$\pm$0.32&   1.37$\pm$0.04\\
1634+4827   & M4      & 2.66&   0.90&     ---&     ---&  ---    &   0.00001& 13.45$\pm$0.14&   0.99$\pm$0.04&  14.61$\pm$0.14&   1.25$\pm$0.04\\
1718$-$2246 & M4.5    & 3.59&   1.97&   -2.07&   -2.34&  ---    &   0.00014& 10.54$\pm$0.27&   0.83$\pm$0.03&  11.53$\pm$0.27&   1.15$\pm$0.03\\
1722$-$6951A& M3      & 2.38&   0.89&   -1.83&   -2.10&  ---    &   ---    &    ---        &   0.86$\pm$0.03&       ---      &   1.09$\pm$0.03\\
1905$-$5434 & M4      & 2.87&   1.16&     ---&     ---&  ---    &   0.00014& 10.55$\pm$0.10&   0.86$\pm$0.03&  11.61$\pm$0.10&   1.16$\pm$0.03\\
2007+7001   & M6      & 4.45&   1.71&     ---&     ---&  0.00178&   ---    &    ---        &   0.61$\pm$0.07&       ---      &   1.16$\pm$0.04\\
2200$-$4636 & M4.5    & 3.46&   1.87&     ---&     ---&  ---    &   0.00353&  7.03$\pm$0.32&   0.82$\pm$0.04&   8.31$\pm$0.32&   1.83$\pm$0.03\\
2304+2111   & M0.5    & 1.97&   0.75&     ---&     ---&  ---    &   0.00005& 11.66$\pm$0.27&   0.88$\pm$0.03&  12.62$\pm$0.27&   0.97$\pm$0.03\\
\enddata
\tablenotetext{a}{This is the value of log($F_X/F_J$) to be used when comparing to the plots of \cite{shkolnik2009}. See text for details.}
\end{deluxetable*}
\end{turnpage}

We have computed flux and flux density ratios, as listed in Table 9,
to compare to published values in the literature. Figure 3 of \cite{shkolnik2009} shows the ratio of X-ray to $J$-band flux vs.\ $I-J$ color (or spectral type) for old field M dwarfs and M dwarfs in younger clusters and moving groups of various ages. In order to compare our values of log($F_X/F_J$) to the ones in that figure, though, it should be noted that the \cite{shkolnik2009} values are computed with an assumed $J$-band filter width of 0.3 $\mu$m, almost twice the width that we used in our calculation of $F_J$. This translates to a difference in 0.27 dex between our computed log($F_X/F_J$) values and theirs. For convenience, our Table 9
gives the modified value needed for direct comparison to their Figure 3. Also, we consider our spectral types to be a better indicator of the $x$-axis in this figure than the tabulated $I-J$ color, since fainter $I$-band measurements from USNO-B1 can sometimes be well outside the expected 0.3-mag uncertainty envelope. With these caveats in mind, we find that one of our five M dwarfs with X-ray emission, WISE 0422+0337, shows X-ray emission well above the norm for its spectral type. Its fractional X-ray flux is higher than all of the plotted members of the $\beta$ Pic moving group and most of the plotted members of the Pleiades. The other four X-ray detected M dwarfs have lower fractional X-ray fluxes. These four have values of ${\log(}L_X/L_{bol})$ that fall between $-$2.9 and $-$2.7, which are typical coronal saturation values for active field M dwarfs (see Figure 5 of \citealt{riaz2006}).

Figure 3 of \cite{shkolnik2011} provides a reference for the ultraviolet fluxes of typical M dwarfs in the field vs.\ those in younger groups such as the TW Hya Association. It should be noted that that figure actually plots the ratio of NUV to $J$-band flux {\it densities}, not {\it fluxes}, despite what the $y$-axis label implies. (For more on this, see footnote 7 of \citealt{shkolnik2014}.) Three of our X-ray detected stars -- WISE 0422+0337, 1029+2545, and 1718$-$2246 -- are also detected in the ultraviolet and fall within the locus of objects typically regarded as having ages below $\sim$300 Myr. All three of these show clear Balmer emission in our optical spectra as well. Five other M dwarfs in Table 9
have very high fractional levels of NUV flux density ($>$0.0001) -- WISE 0157$-$0948, 1218+1140, 1247$-$4344, 1615+0336, and 2200$-$4636 -- but none are detected in either the X-ray or the FUV. These presumably are objects caught by {\it GALEX} during a flaring event and do not maintain high levels of magnetic activity. Indeed, only one of these, the M7 dwarf WISE 1615+0336, was seen to have measurable H$\alpha$ emission when our optical spectrum was acquired. 

We can also identify objects having ultraviolet excesses indicative of high magnetic activity by creating colors across the {\it GALEX}, 2MASS, and {\it WISE} passbands. Four such colors are given in the final four columns of Table 9. Figure 5 of \cite{rodriguez2011} shows the location of TW Hya Association members in the NUV$-J$ vs.\ $J-K_s$ plane. Roughly half of our objects in Table 9 fall in the selection region used by \cite{rodriguez2011} to identify candidate objects with ultraviolet excesses. Similarly, Figure 1 of \cite{rodriguez2013} shows known members of young moving groups on the NUV$-$W1 vs.\ $J-$W2 plane. Again, roughly half of our objects fall within the selection wedge used by \cite{rodriguez2013} to select young candidates.

Given that some of our objects have magnetic signs of youth, do any of these share kinematics consistent with young moving groups known in the solar vicinity? To answer this question, we have used v1.4 of the Bayesian analysis tool BANYAN II\footnote{This is available at {\url http://www.astro.umontreal.ca/$\sim$gagne/ banyanII.php}.} (\citealt{malo2013,gagne2014}) that determines membership probabilities for  young associations within 100 pc of the Sun (TW Hya, $\beta$ Pic, Tucana-Horologium, Columba, Carina, Argus, and AB Dor) or for the general young or old field populations. Because we have only positions and proper motions for our objects, and no radial velocities or trigonometric parallaxes, we have input only those quantities into the tool. In the final three columns of Table 8, we list the group with the highest membership probabilty, 
the membership probability itself, and the object's predicted distance if the membership is real. The predicted distance ($d_{\rm pred}$) can be compared to our spectrophotometric estimate ($d_{\rm est}$ in Table 10;
see section~\ref{close_Ms}) as an independent judge of the results.

As the table values indicate, most of our M dwarfs are likely to be members of the old field population. This includes all of the aforementioned objects that appear to be young candidates on the \cite{shkolnik2009,shkolnik2011} and \cite{rodriguez2011,rodriguez2013} plots. Several other M dwarfs, on the other hand, have $>$50\% probabilities of belonging to young associations based on the BANYAN results:

\begin{itemize}

\item The M6 dwarf WISE 0005+0209 has a high probability of belonging to the AB Dor Moving Group, but secondary checks show that this is likely a chance association. The BANYAN-predicted distance of 18.1$\pm$1.0 pc differs by almost 5$\sigma$ from our spectrophotomeric estimate, and the object shows only Balmer emission with no {\it ROSAT} or {\it GALEX} detections. 

\item The M4.5 dwarf WISE 0546$-$0440 is given a 50\% chance of belonging to the young field population, but like the previous object, shows only Balmer emission with no {\it ROSAT} or {\it GALEX} detections, so it, too, is likely an older field star. 

\item The M8 dwarf SCR J1546$-$5534 (Figure~\ref{seq_nearby_late-Ms}), recognized as a nearby ($\sim$6.7 pc distant) star by \cite{boyd2011}, is given a 50\% probability of belonging to Argus; however, a comparison of its spectrum to an old field M8 and a young field M8 (Figure~\ref{seq_1546}) shows that SCR J1546$-$5534 exhibits no signs of youth and therefore could not be a member of an association as young as Argus (30-50 Myr; \citealt{zuckerman2011}). 

\item The M4 dwarf WISE 1905$-$5434 has a 67\% probability of association in the $\beta$ Pic Moving Group. Our distance estimate of 15 pc, however, is $\sim$3$\sigma$ closer than the predicted one. Although this object has ultraviolet detections by {\it GALEX}, there is no {\it ROSAT} detection. 

\item The M0.5 dwarf WISE 2304+2111, which is also detected by {\it GALEX}, has a 90\% probability of association in AB Dor, but our estimated distance is over 18$\sigma$ away from the predicted value by BANYAN.

\end{itemize}

None of our M dwarfs pass all four tests of youth -- high X-ray flux, high ultraviolet flux, Balmer emission detections, and high probability of group membership based on kinematics. We therefore conclude that all detections of magnetic activity are likely related to chance flares rather than persistent coronal and chromospheric activity. 

\subsubsection{Recently Discovered M Dwarfs with d$<$20 pc\label{close_Ms}}

\begin{deluxetable*}{ccccccccccccccccccccc}
\tabletypesize{\tiny}
\tablenum{10}
\tablecaption{Distance Estimates for M, L, and T Dwarfs from Table 4\label{MLT_distances}}
\tablehead{
\colhead{Name} & 
\colhead{Spec.} &                         
\colhead{d$_{\rm est}$} &  
\colhead{} & 
\colhead{} & 
\colhead{} & 
\colhead{Name} & 
\colhead{Spec.} &                         
\colhead{d$_{\rm est}$} &  
\colhead{} & 
\colhead{} & 
\colhead{} & 
\colhead{Name} & 
\colhead{Spec.} &                         
\colhead{d$_{\rm est}$} \\     
\colhead{} & 
\colhead{Type} &                         
\colhead{(pc)} &  
\colhead{} &   
\colhead{} & 
\colhead{} & 
\colhead{} & 
\colhead{Type} &                         
\colhead{(pc)} &  
\colhead{} &   
\colhead{} & 
\colhead{} & 
\colhead{} & 
\colhead{Type} &                         
\colhead{(pc)} \\     
\colhead{(1)} &                          
\colhead{(2)} &  
\colhead{(3)} &
\colhead{} &     
\colhead{} & 
\colhead{} & 
\colhead{(1)} &
\colhead{(2)} &
\colhead{(3)} &
\colhead{} &
\colhead{} & 
\colhead{} & 
\colhead{(1)} &
\colhead{(2)} &
\colhead{(3)}  
}
\startdata
  0005+0209&         M6&    23$\pm$2& & & &    1055$-$7356&       M7&     5$\pm$1& & & &  1615+0336&         M7&    41$\pm$3   \\
  0006$-$1319&      L7:&    27$\pm$2& & & &    1055$-$5750&       M4&    37$\pm$3& & & &  1634+4827&         M4&    13$\pm$1   \\
  0034+5513&       M4.5&   128$\pm$8& & & &    1056$-$5750&       M4&    35$\pm$3& & & &  1702+7158A&      M4.5&   151$\pm$10  \\
  0043+2221&         M8&    55$\pm$4& & & &    1059+1509&       M3.5&    44$\pm$3& & & &  1712+0645&     T2 pec&    22$\pm$2   \\
  0051$-$2251&       M8&    24$\pm$2& & & &    1114+5703&         M0&    74$\pm$5& & & &  1718$-$2245&     M3.5&    31$\pm$2   \\
  0111+1211&         M8&    58$\pm$4& & & &    1133$-$4140&     M0.5&   122$\pm$8& & & &  1718$-$2246&     M4.5&    22$\pm$2   \\
  0130$-$1047&       M9&    36$\pm$3& & & &    1140$-$0624&       M5&    42$\pm$3& & & &  1722$-$6951A&      M3&    28$\pm$2   \\
  0130$-$3836&       M5&   128$\pm$8& & & &    1201$-$1324&     M9.5&    40$\pm$3& & & &  1722$-$6951B&      M4&    58$\pm$4   \\
  0134+0525&       M4.5&   100$\pm$7& & & &    1202$-$0111&       M4&    85$\pm$6& & & &  1740$-$5507&     M2.5&   257$\pm$16  \\
  0157$-$0948&       M4&    78$\pm$5& & & &    1203+1810&         M0&    76$\pm$5& & & &  1741$-$4234&       M4&    70$\pm$5   \\
  0158+3231&       L4.5&    33$\pm$2& & & &    1206+0016&         M5&    20$\pm$2& & & &  1743+6313&         M9&    42$\pm$3   \\
  0231+2811&         M9&    41$\pm$3& & & &    1210$-$4612&       M1&    52$\pm$4& & & &  1746+5100&        L0:&    43$\pm$3   \\
  0309$-$1354&      M6:&    48$\pm$3& & & &    1218+1140&       M7.5&    45$\pm$3& & & &  1747+4008&         M4&    42$\pm$3   \\
  0404$-$6259&       M4&    39$\pm$3& & & &    1222$-$8449&       M3&    33$\pm$2& & & &  1758$-$5839&       M6&    16$\pm$1   \\
  0405+3719&         M5&    26$\pm$2& & & &    1222$-$2116&      L7:&    25$\pm$2& & & &  1800$-$1559&     L4.5&    10$\pm$1   \\
  0422+0337&       M4.5&    18$\pm$2& & & &    1223+5510&     M8 pec&    48$\pm$3& & & &  1835$-$7912&       M5&    28$\pm$2   \\
  0447+2534&         M4&    25$\pm$2& & & &    1235+4450&       M4.5&    49$\pm$3& & & &  1842+2104&       L1.5&    33$\pm$2   \\
  0500+1916&         M1&    39$\pm$3& & & &    1240+2047&         M7&    47$\pm$3& & & &  1843$-$6355&     M4.5&    92$\pm$6   \\
  0500+0442&   L0.5 pec&    24$\pm$2& & & &    1241$-$2457&    L2.5p&    49$\pm$3& & & &  1905$-$5434&       M4&    15$\pm$1   \\
  0507$-$0342&  M9 pec?&    33$\pm$2& & & &    1247$-$4344&       M5&    78$\pm$5& & & &  1940+6346&       M9.5&    39$\pm$3   \\
  0508+3319&         L2&    24$\pm$2& & & &    1305$-$1019&       M5&   104$\pm$7& & & &  2002$-$4433&       M8&    30$\pm$2   \\
  0536$-$0006&     M4.5&    26$\pm$2& & & &    1333+3744&         L5&    27$\pm$2& & & &  2004$-$2637&    M7.5:&    71$\pm$5   \\
  0546$-$0440&     M4.5&    23$\pm$2& & & &    1343$-$1216&       L4&    18$\pm$2& & & &  2007+7001&         M6&    59$\pm$4   \\
  0559+5844&         M9&    41$\pm$3& & & &    1348$-$4227&       L2&    34$\pm$2& & & &  2008+7030&        M7:&    79$\pm$5   \\
  0632+2643&         M7&    53$\pm$4& & & &    1403$-$5923&       M3&    43$\pm$3& & & &  2101$-$4907A&    M4.5&    13$\pm$1   \\
  0701$-$0137&       M4&    25$\pm$2& & & &    1404$-$5924&       M3&    42$\pm$3& & & &  2120+2652&         M7&    60$\pm$4   \\
  0705$-$1007&       M5&    19$\pm$2& & & &    1404$-$4726&       M6&    60$\pm$4& & & &  2121$-$6239&       T2&    16$\pm$1   \\
  0730$-$6335&       M4&    28$\pm$2& & & &    1411$-$1403&       M9&    45$\pm$3& & & &  2133+7319&       M4.5&   246$\pm$15  \\
  0826$-$1640&      L8:&    16$\pm$1& & & &    1423$-$1646B&    M0.5&    54$\pm$4& & & &  2135+7312&        L2:&    26$\pm$2   \\
  0852+5139&       M7.5&    40$\pm$3& & & &    1452+2723&         L0&    42$\pm$3& & & &  2200$-$4636&     M4.5&   143$\pm$9   \\
  0855$-$0233&       L7&    21$\pm$2& & & &    1454+0053B&        M9&    49$\pm$3& & & &  2201$-$4112&      M4:&   136$\pm$9   \\
  0912+2205&         M3&    53$\pm$3& & & &    1454+0053A&        M3&    46$\pm$3& & & &  2230$-$2720&      L0:&    40$\pm$3   \\
  0920$-$7557&     M4.5&    33$\pm$2& & & &    1507+6030&        L2:&    56$\pm$4& & & &  2249+3205&        L4:&    28$\pm$2   \\
  0935$-$0301&       M3&    49$\pm$3& & & &    1516$-$2832&       M5&    22$\pm$2& & & &  2304+2111A&      M0.5&    54$\pm$4   \\
  0949$-$0103&       M0&    81$\pm$5& & & &    1539$-$5352&       M4&    39$\pm$3& & & &  2304+2111B&      M4.5&    68$\pm$5   \\
  1019+3922&       M4.5&    33$\pm$2& & & &    1540$-$5101&     M6.5&     5$\pm$1& & & &  2324+1617&         M8&    57$\pm$4   \\
  1029+5715&     L6 pec&    32$\pm$2& & & &    1542$-$1007&     M7.5&    49$\pm$3& & & &  2344+1312&        M7:&    36$\pm$3   \\
  1029+2545&       M4.5&    23$\pm$2& & & &    1546$-$5254&       M5&    54$\pm$4\\
\enddata
\end{deluxetable*}

New all-sky motion surveys that probe more deeply, to smaller motions, or at different wavelengths than previous surveys allow us to test how complete our knowledge of the nearby stellar census is. We have specifically targeted with our optical spectroscopic observations previously overlooked candidates that could be within $\sim$20 pc of the Sun. In Table 10
we list spectrophotometric distance estimates for M, L, and T dwarfs from Table 4 and Table 5. These estimates, which use our measured spectral types (adopted types are given in Table 10)
and 2MASS $J$-band magnitudes, are computed using the table of M$_J$ vs.\ spectral type in \cite{kirkpatrick1994} for M dwarfs and the M$_J$ vs.\ spectral type relation of \cite{looper2008} for L and T dwarfs. In the section below we discuss those M dwarfs in Table 10
whose distance estimates place them within 20 pc, along with a few other nearby candidates recently published by others for which we have new, supporting observations. (Nearby L and T dwarfs are discussed in section~\ref{section_Ldwarfs} and section~\ref{section_Tdwarfs}.)

\begin{figure}
\figurenum{45}
\includegraphics[scale=0.40,angle=0]{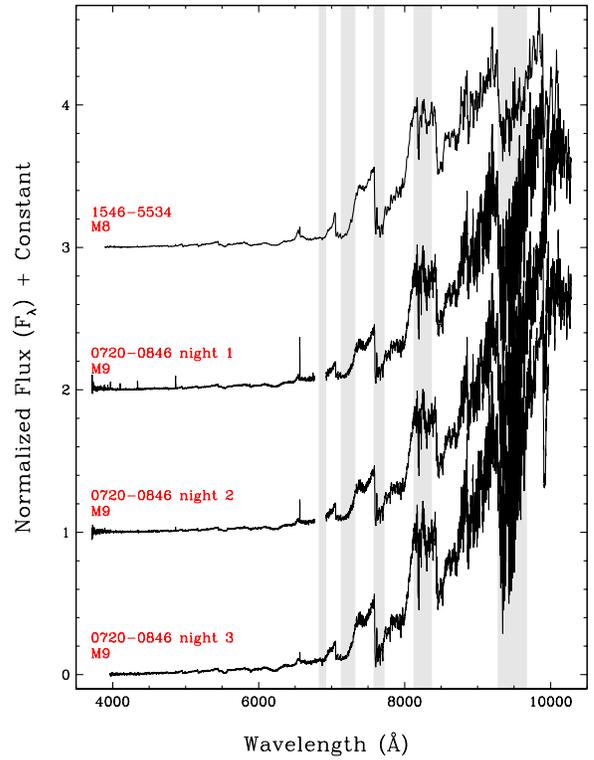}
\caption{Spectra of two known, nearby late-M dwarfs from the literature. Three spectra are shown for WISE 0720$-$0846 and represent the total integration sequence on each of our three nights of Palomar/DSpec observations. (See the caption of Figure~\ref{seq_wds1} for other details.)
\label{seq_nearby_late-Ms}}
\end{figure}

\begin{figure}
\figurenum{46}
\includegraphics[scale=0.40,angle=0]{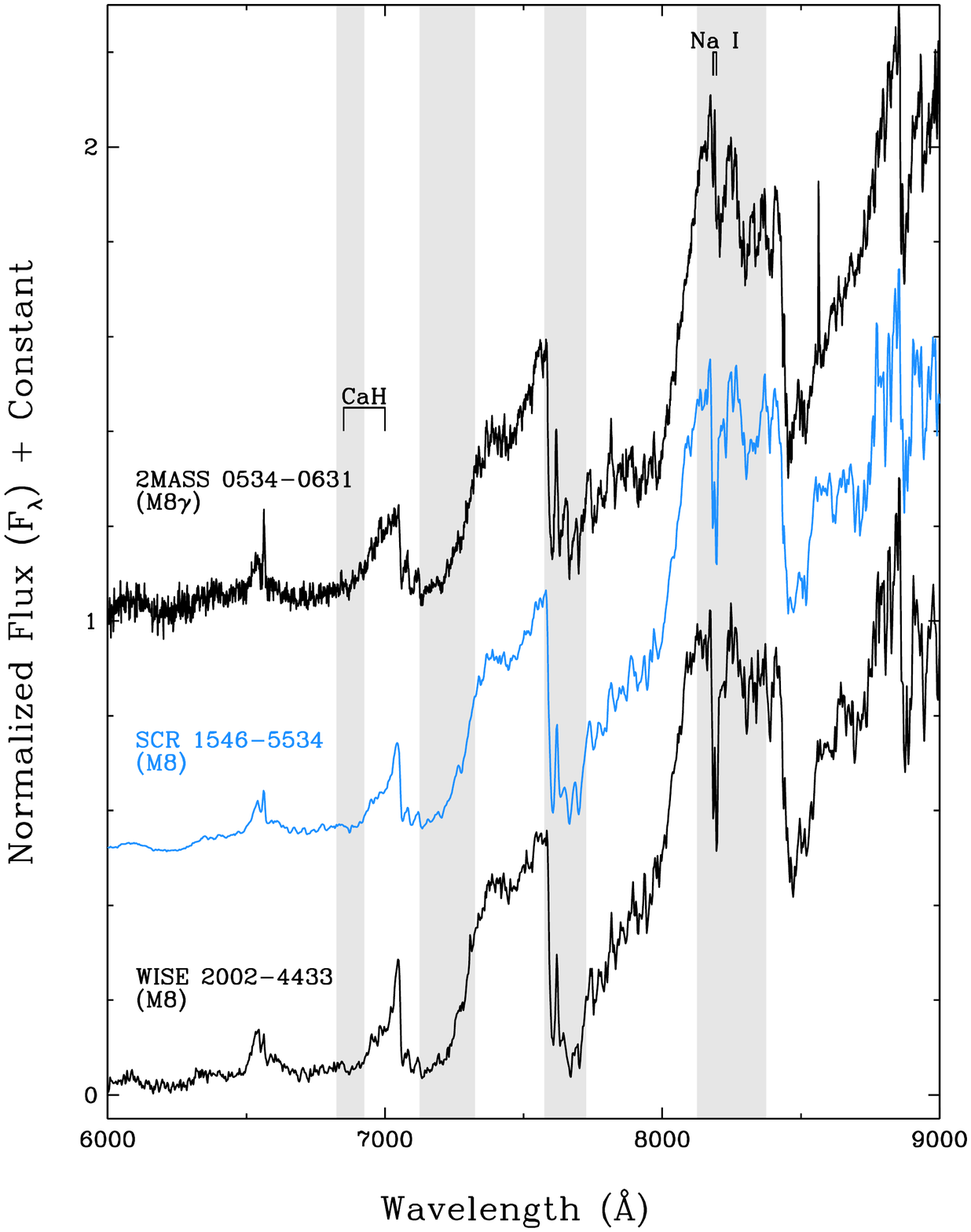}
\caption{Spectra of three M8 dwarfs: the young M8$\gamma$ dwarf 2MASS J05341594$-$0631397 (top, black; \citealt{kirkpatrick2010,gagne2014}), the M8 dwarf SCR J1546$-$5534 (middle, blue), and the old field M8 dwarf WISE 2002$-$4430 (bottom, black; this paper). The CaH band and \ion{Na}{1} doublet are very weak in the young M8$\gamma$ and are hallmarks of lower gravity; these same features are much stronger in the spectrum of SCR J1546$-$5534 and have strengths similar to the old field M8. All spectra are normalized at 8150 \AA. (See the caption of Figure~\ref{seq_wds1} for other details.)
\label{seq_1546}}
\end{figure}

\begin{figure}
\figurenum{47}
\includegraphics[scale=0.40,angle=0]{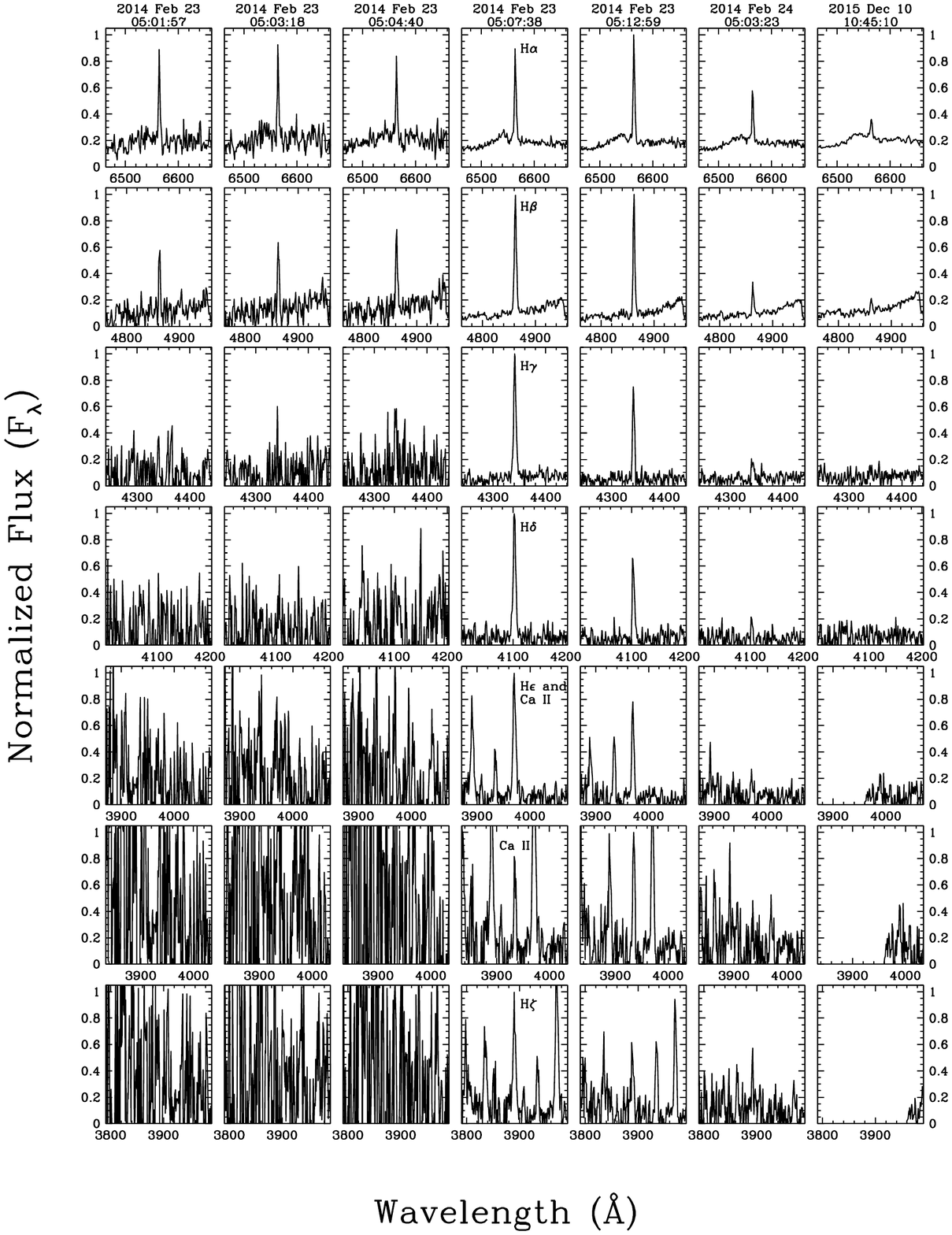}
\caption{Sequence of spectra showing the variable \ion{H}{1} and \ion{Ca}{2} emission lines in WISE 0720$-$0846. From top to bottom the rows show H$\alpha$, H$\beta$, H$\gamma$, H$\delta$, H$\epsilon$ plus \ion{Ca}{2} ``H'', \ion{Ca}{2} ``K'', and H$\zeta$. Each column represents a different spectrum, with the UT date and time of shutter opening shown at the top of the panel. From left to right, the exposure times per spectrum were 60s, 60s, 60s, 300s, 300s for the 2014 Feb 23 spectra, 660s for the 2014 Feb 24 spectrum, and 1800s for the 2015 Dec 10 spectrum. In each row, the normalization is set to one for the observation with the strongest line in the four spectra of longest integration. Note that the 2015 Dec 10 spectrum does not cover the wavelength range for \ion{Ca}{2} ``K'' or H$\zeta$.
\label{seq_0720_flare}}
\end{figure}

\begin{itemize}

\item WISE 0422+0337 was identified as an optical counterpart to a {\it ROSAT} source by \cite{zickgraf2003} and as a bright, possibly M-type motion star by \cite{frith2013}. Our spectral type appears to be the first for this object, and our resulting spectrophotometric distance estimate places this M4.5 dwarf at 18$\pm$2 pc. WISE 0442+0337 has not only an X-ray detection, but also a far ultraviolet detection from {\it GALEX} (Table 8)
and a spectrum rich with \ion{H}{1} and \ion{Ca}{2} emission (Figure~\ref{seq_M4_M45.2}). As discussed in section~\ref{young_Mdwarfs} it does not have the kinematics typical of a young star or a high probability of belonging to a known young association. The strong activity may indicate the presence of a close companion.

\item WISE 0705$-$1007 is a new source not previously cataloged. It might be associated with the faint {\it ROSAT} source 1RXS J070511.2$-$100801, which was not included in our Table~\ref{M_dwarf_activity} because the X-ray detection is $<$3$\sigma$. Our spectral type of M5 (Figure~\ref{seq_M45_M5.2}) gives it a spectrophotometric distance estimate of 19$\pm$2 pc. 

\item WISE 0720$-$0846 was first published by \cite{scholz2014}, who discovered the object using {\it WISE} and gave it a photometric classification of M9$\pm$1. Their trigonometic parallax placed it at 7.0$\pm$1.9 pc. We independently discovered the object during the AllWISE1 Motion Survey and provided the first spectral classification, a near-infrared type of M9 (Table 5 of \citealt{kirkpatrick2014}). \cite{burgasser2015} obtained an optical spectrum, typed at M9.5, which showed prominent and variable H$\alpha$ emission, and their own trigonometric parallax gave an improved distance of 6.0$\pm$1.0 pc. Through imaging with laser guide star adaptive optics, they also reported a possible companion located 0$\farcs$14 away (later confirmed as sharing common proper motion with the primary in \citealt{burgasser2015b}), which is supported by the fact that the near-infrared spectrum of WISE 0720$-$0846 is better fit by an M9 and T5 composite than by a single M (or L) dwarf. \cite{ivanov2015} also acquired optical (M9) and near-infrared (L0) spectroscopy, along with their own improved parallactic distance of 6.07$^{+1.36}_{-0.95}$ pc. \cite{mamajek2015} recognized that the radial velocity reported by \cite{burgasser2015} indicated that the system passed within $\pm$0.25 pc of the Sun roughly 70,000 years ago, making it the closest known encounter of a nearby star to our solar system. We have continued to monitor this object with optical spectroscopy (Figure~\ref{seq_nearby_late-Ms}) and find that the Balmer emission is persistent and, as reported by \cite{burgasser2015,burgasser2015b}, variable. Figure~\ref{seq_0720_flare} shows a zoom-in on the \ion{H}{1} and \ion{Ca}{2} emission lines across our individual exposures of the system. The spectra from UT 2014 Feb 23 show a flare in progress. Note that the higher order (shorter wavelength) Balmer lines decay more quickly than the lower order (longer wavelength) ones in roughly this order: H$\delta$/H$\zeta$ $\rightarrow$ H$\gamma$ $\rightarrow$ H$\beta$ $\rightarrow$ H$\alpha$. The \ion{Ca}{2} K line appears to be still peaking at the end of our spectral sequence, and may have a peak strength after the H$\alpha$ line.\footnote{Because the H$\epsilon$ and \ion{Ca}{2} ``H'' lines are blended with one another, we cannot discern the order in which these peak.} This evolution of the line strengths has been noted in M dwarf flares before (\citealt{doyle1988,hawley1991}). The timescale between the peaks of the H$\gamma$ and H$\alpha$ lines ($\sim$5 min) or between the H$\beta$ and H$\alpha$ lines ($\sim$2.5 min) indicates a relatively low-energy flare (see Figure 18 of \citealt{kowalski2013}).

\item WISE 1055$-$7356 was identified by \cite{schneider2016} as a likely M7 dwarf located within 10 pc. We provide the first spectrum for this object (Figure~\ref{seq_M6_M7.1}), which confirms the M7 classification. Our distance estimate (Table 10)
gives d=5$\pm$1 pc.

\item WISE 1206+0016 is a source not previously cataloged. It is an M5 dwarf with {\it GALEX} detections (Table 8)
and \ion{H}{1} and \ion{Ca}{2} emission (Figure~\ref{seq_M45_M5.2}). Our spectrophotomeric distance estimate places it at 20$\pm$2 pc.

\item WISE 1540$-$5101 was discovered by \cite{kirkpatrick2014} as part of the AllWISE1 Motion Survey. The optical spectrum was typed as an M6, and a crude trigonometric parallax gave a distance of $6.1^{+2.0}_{-1.2}$ pc. After our announcement, \cite{perez2014} reported that they had uncovered the object independently and typed it as M7 in both the optical and near-infrared. Their trigonometric parallax determination places it even closer, at d = $4.4^{+0.5}_{-0.4}$ pc, although a re-analysis of their astrometric data by one of us (ELW) does not support such a close distance or such small error bars but rather d = $5.2^{+1.1}_{-0.8}$ pc. Regardless, the astrometric uncertainies would be greatly improved through a dedicated monitoring program. We have obtained a new optical spectrum to improve upon the one in \cite{kirkpatrick2014} and find a type of M6.5 (Figure~\ref{seq_M6_M7.1}), intermediate between our previous classification and that of \cite{perez2014}. Our duPont/BCSpec spectrum hints at H$\alpha$ emission (pseudo equivalent width of ${\sim}-0.8$ \AA), but the low resolution of these data render such a measurement insignificant since blending by nearby TiO bands can confound interpretation. Nonetheless, our observation is consisent with the quoted \cite{perez2014} value of $-1.0{\pm}0.2$ \AA\ from their much higher resolution spectrum, and this may indicate persistent, although low level, activity in this late-M dwarf. Running the object's sky position, measured proper motion, and rough distance through v1.4 of the BANYAN II tool (\citealt{malo2013}, \citealt{gagne2014}) gives a 99.66\% probability of this object belonging to the old field population, a conclusion that is supported by the \cite{perez2014} analysis of gravity diagnostics and lack of a lithium detection in their high-resolution spectrum.

\item SCR J1546$-$5534 was discovered by \cite{boyd2011} and recognized through a photomeric distance estimate to be $\sim$6.7 pc from the Sun. \cite{rajpurohit2013} find an optical spectral type of M7.5, although the spectrum itself is not plotted. We type our optical spectrum, shown in Figure~\ref{seq_nearby_late-Ms}, as M8. As discussed in section~\ref{young_Mdwarfs}, this appears to be another member of the old field population.

\item WISE 1634+4827 was cataloged as proper motion star G 202-59 (\citealt{giclas1971}) and as LSPM J1634+4827 (\citealt{lepine2005}). \cite{frith2013} also cataloged it as a motion star. Our spectral type of M4 (Figure~\ref{seq_M4_M4.2}) appears to be the first published classification, which gives a distance estimate of 13$\pm$1 pc. 

\item WISE 1758$-$5839 is a source not previously cataloged. It is an M6 dwarf (Figure~\ref{seq_M5_M6.2}) whose spectrophotometric distance estimate places it at 16$\pm$1 pc.

\item WISE 1905$-$5434 was first uncovered as proper motion star WT 625 by \cite{wroblewski1994} and later also identified as a motion source by \cite{frith2013}. Our type of M4 is the first spectral classification, which gives it a distance estimate of 15$\pm$1 pc. Although this source is not detected by {\it ROSAT}, it is detected by {\it GALEX} (Table~\ref{M_dwarf_activity}) and has a rich emission line spectrum (Figure~\ref{seq_M4_M45.1}). As discussed in section~\ref{young_Mdwarfs}, there is a 67\% probability of membership in the $\beta$ Pic Moving Group, but only if the distance is closer to $\sim$20 pc. Interestingly, providing the BANYAN II software with our motion measurement {\it and} our best distance estimate for the object gives a 33\% probability of membership in the $\beta$ Pic Association and a 67\% probability of membership in the Argus Association. This object is worthy of additional scrutiny.

\item WISE 2101$-$4907 is the previously cataloged motion star WT 766 (\citealt{wroblewski1994}), which shares common proper motion with the star WISE 2101$-$4906, also known as WT 765. \cite{reyle2002} also identified it as a proper motion star (APMPM J2101$-$4907) and computed a crude photometric distance estimate of 16.5 pc. \cite{reyle2006} then obtained a spectrum, whose classification of M3.5 provided a spectrophotometric distance estimate of 13.7$\pm$4.6 pc. This object was also identified as a motion star by \cite{frith2013} and classified as an M4 by \cite{rajpurohit2013}. Our classification of M4.5 (Figure~\ref{seq_M45_M45.2}) yields a spectrophotometric distance estimate of 13.1$\pm$1 pc. We also classify the companion as a possible DA white dwarf (Figure~\ref{seq_wds1}).

\end{itemize}
                                     
\subsection{L Dwarfs\label{section_Ldwarfs}}

Optical spectra of L dwarfs were classified using the optical standards of \cite{kirkpatrick1999}. Our spectra and resulting types are shown in Figure~\ref{seq_L0_L2.1} through Figure~\ref{seq_L4_L8.2}. Near-infrared spectra were classified using the near-infrared standards and methodology established in \cite{kirkpatrick2010}. These spectra and resulting types are shown in Figure~\ref{seq_Ls_IR}. A few of these L dwarfs deserve special mention:

\begin{figure}
\figurenum{48}
\includegraphics[scale=0.375,angle=0]{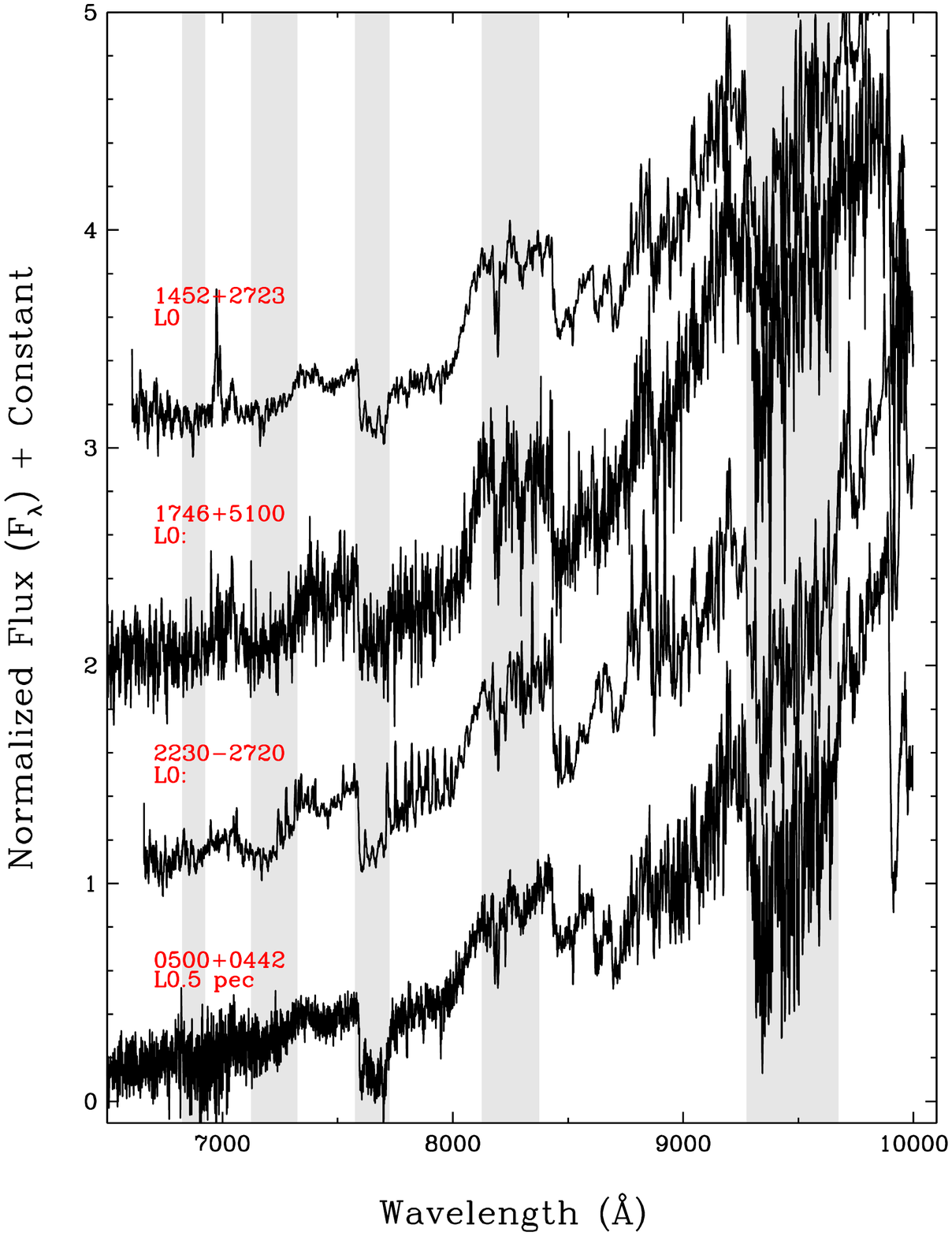}
\caption{Spectra of dwarfs with types of L0 through L0.5. Spectra have been normalized at 8250 \AA\ and offsets added to separate the spectra vertically. The light grey bands indicate wavelength zones with uncorrected telluric absorption.
\label{seq_L0_L2.1}}
\end{figure}

\begin{figure}
\figurenum{49}
\includegraphics[scale=0.375,angle=0]{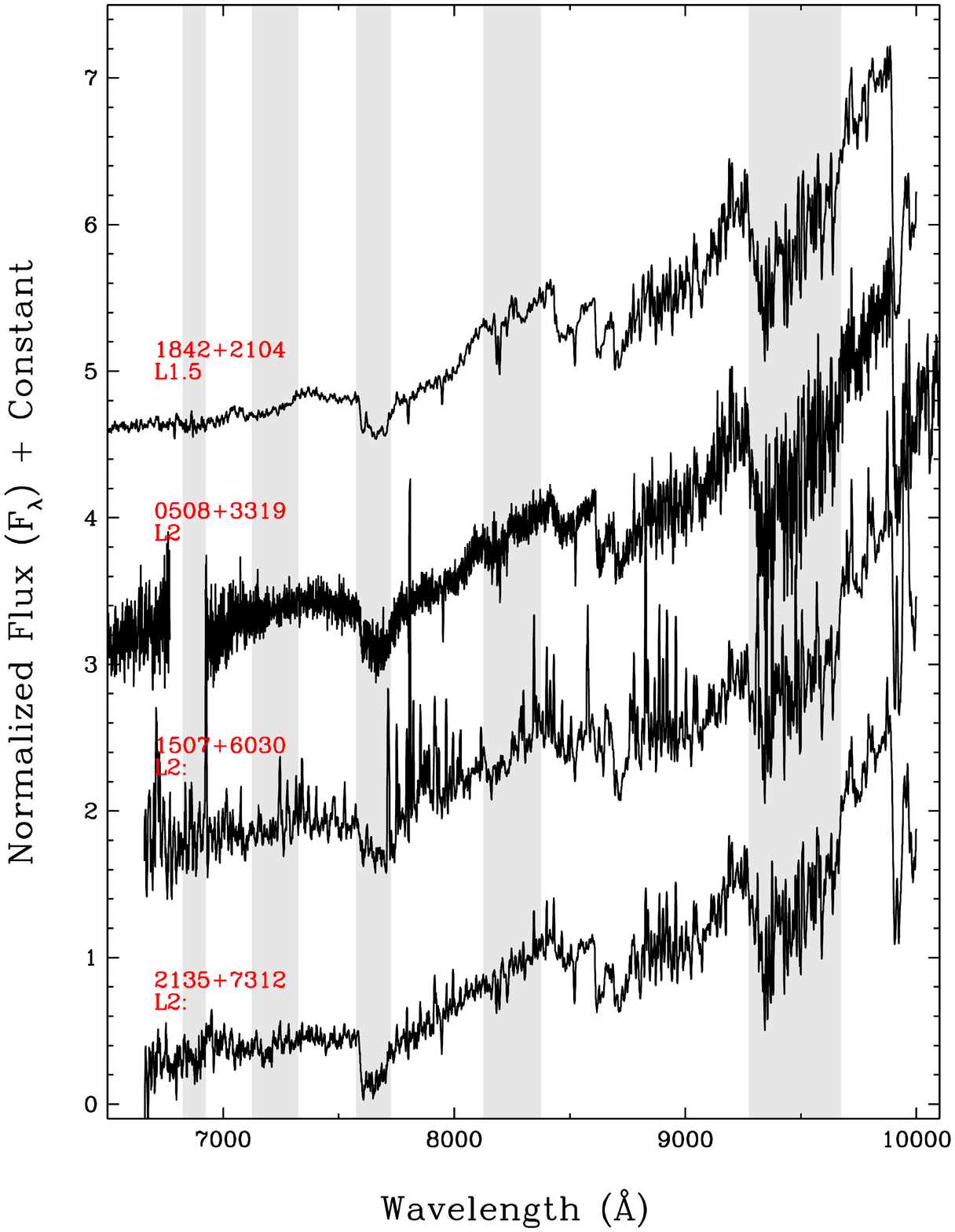}
\caption{Spectra of dwarfs with types of L1.5 through L2. (See the caption of Figure~\ref{seq_L0_L2.1} for other details.)
\label{seq_L0_L2.2}}
\end{figure}

\begin{figure}
\figurenum{50}
\includegraphics[scale=0.375,angle=0]{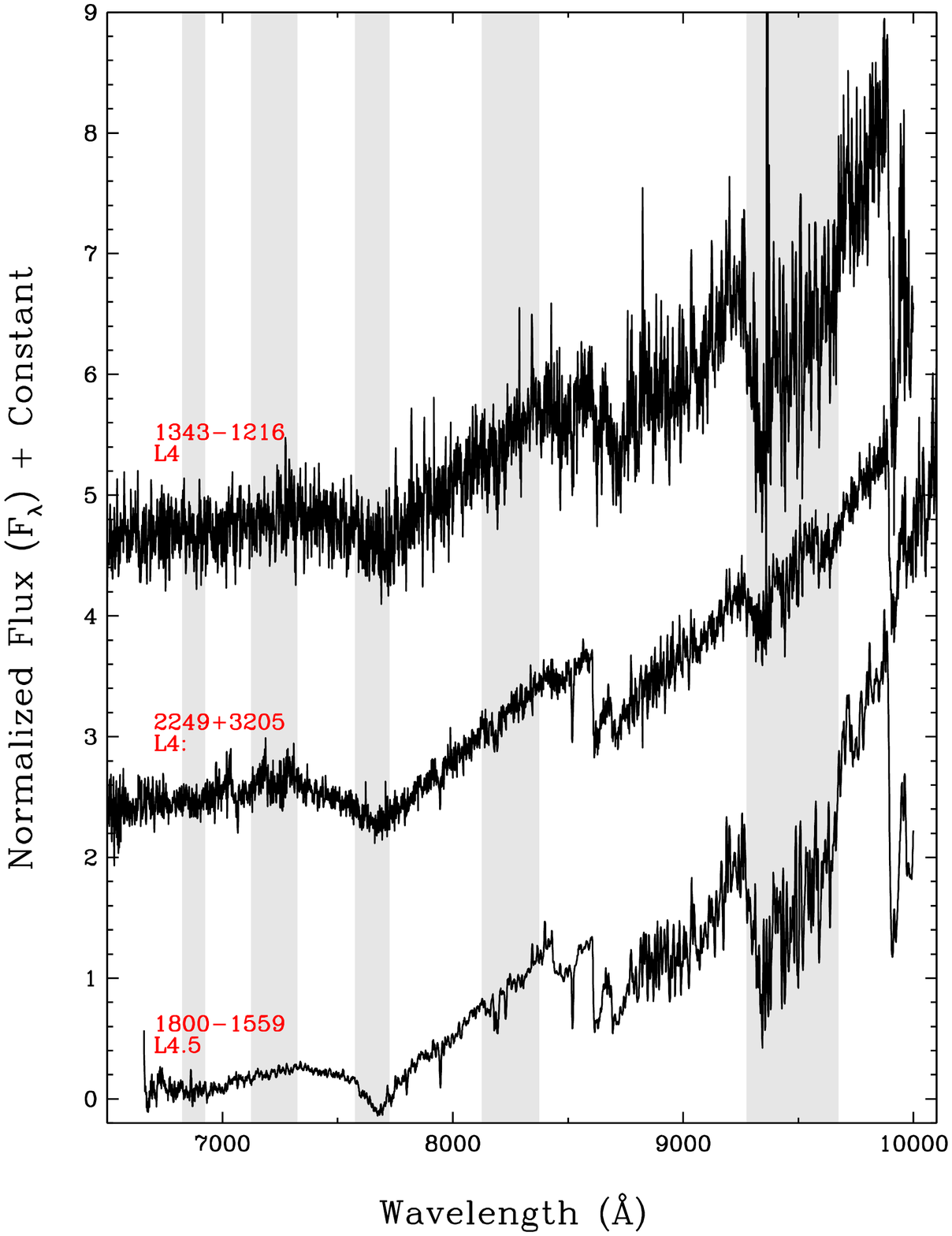}
\caption{Spectra of dwarfs with types of L4 through L4.5. (See the caption of Figure~\ref{seq_L0_L2.1} for other details.)
\label{seq_L4_L8.1}}
\end{figure}

\begin{figure}
\figurenum{51}
\includegraphics[scale=0.375,angle=0]{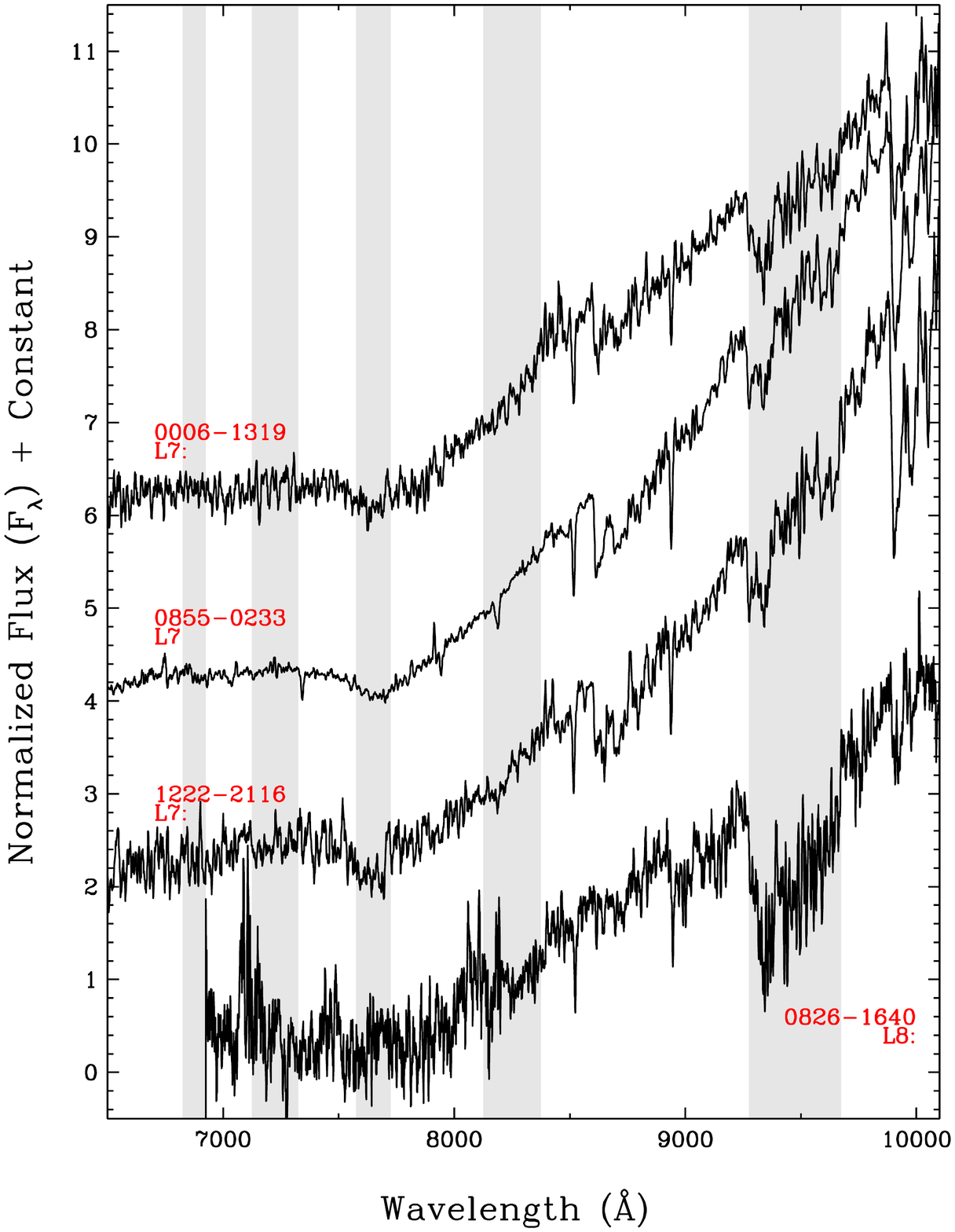}
\caption{Spectra of dwarfs with types of L7 through L8. (See the caption of Figure~\ref{seq_L0_L2.1} for other details.)
\label{seq_L4_L8.2}}
\end{figure}

\begin{itemize}

\item WISE 0500+0442 is an enigmatic object. It is a poor match to the late-M and early-L standards, all known late-M and early-L subdwarfs, and all known late-M and early-L low-gravity objects. It comes closest to matching a normal-gravity, normal metallicity L0 or L1 dwarf, which is why we classify it as an L0.5 pec (Figure~\ref{seq_L0_L2.1}). The main discrepancy with the L dwarf standards is elevated flux relative to the standards below 9000 \AA. A white light flare could add the needed flux to the blue, but there are no accompanying emission lines to indicate that a flare was in progress during our observation. An unresolved, white dwarf companion might provide a better explanation, but we have no other supporting evidence that this is the case. A third alternative is also possible. Figure~\ref{seq_Ls_0500} compares WISE 0500+0442 to another odd early-L dwarf found during the 2MASS Motion Survey of \cite{kirkpatrick2010}. This L dwarf, 2MASS J19495702+6222440, shows excess blue flux shortward of 9000 \AA\ as well. Although the comparison is far from perfect -- because WISE 0500+0442 and 2MASS 1949+6222 have their best fits to standards of slightly different subtypes -- both objects seem to compare well to the hydride (CrH and FeH) strengths in the standard while showing far less absorption by TiO. The lack of TiO absorption in 2MASS 1949+6222 is even more extreme than that of WISE 0500+0442. This may indicate that the rainout of molecular TiO (\citealt{lodders1999,burrows1999,allard2001}) into other titanium oxides (TiO$_2$, Ti$_2$O$_3$, Ti$_3$O$_5$, Ti$_4$O$_7$) and perovskite compounds (CaTiO$_3$, Ca$_3$Ti$_2$O$_7$, Ca$_4$Ti$_3$O$_{10}$) has been accelerated in these objects relative to the norm, perhaps because of higher calcium or oxygen abundances relative to titanium. Models with element-to-element abundance tweaks have yet to be produced for L dwarfs, so it is not clear how large an elemental abundance difference is needed for significant rainout to occur earlier in the objects' evolution. Higher signal-to-noise spectra of both these peculiar objects, particularly at higher spectra resolution, would help to shed more light on the anomalies seen. 

\item WISE 0826$-$1640, which has not been cataloged previously, receives an optical classification of L8: based on our Palomar/DSpec spectrum (Figure~\ref{seq_L4_L8.2}). Given that this is one of the latest type objects for which we have ever obtained a classifiable optical spectrum from the 200-inch, it is not surprising that its distance estimate of 16$\pm$1 pc (Table~\ref{MLT_distances}) places it relatively close to the Sun.

\item WISE 1029+5715 fits the L6 standard well at $J$-band, but is suppressed at $H$- and $K$-bands relative to the L6. We classify it in the near-infrared (Figure~\ref{seq_Ls_IR}) as an L6 pec (blue) and add it to the growing population of anomalously blue L dwarfs (section 6.4 of \citealt{kirkpatrick2010}) whose physical explanation is not yet fully understood (\citealt{burgasser2015c}).

\item WISE 1241$-$2457 fits the $J$-band spectra of the L2 and L3 standards equally well but shows suppressed $H$- and $K$-band spectra relative to both of those standards (Figure~\ref{seq_Ls_IR}). We classify it in the near-infrared as L2.5 pec (slightly blue). The spectrum also has a peaky $H$-band morphology, one of the hallmark features in low-gravity L dwarfs (e.g., \citealt{kirkpatrick2006}, \citealt{allers2013}). However, low-$g$ objects are redder, not bluer, than the best-fitting spectral standards, and the VO molecular bands in the $Y$- and $J$-band windows are generally stronger than in the standards, which these are not. Both of these facts suggest that low gravity is not the explanation of the $H$-band morphology. A slightly earlier type object with similar $H$-band morphology and a bluer-than-average continuum, Gl 660.1B, has been extensively analyzed by \cite{aganze2015}, who conclude that the object is somewhat metal poor.

\item WISE 1333+3744 shows a divot at the top of its $H$-band peak (Figure~\ref{seq_Ls_IR}). This is sometimes indicative of a spectral binary, the 1.6 $\mu$m divot being caused by methane absorption in a T dwarf secondary (\citealt{burgasser2007b}). We have used the spectra of our synthetic binaries (see section~\ref{section_Tdwarfs}) to find possible matches to this $H$-band morphology. The best fitting hybrid is an L5+T8 binary, although it does not fit as well at the top of the $J$-band as the spectrum of the L5 standard alone (Figure~\ref{seq_Ls_1333}). We therefore classify this object as a normal L5.

\item WISE 1343$-$1216 has an optical type of L4 based on our Palomar/DSpec spectrum (Figure~\ref{seq_L4_L8.1}). This object has not been classified previously and yet appears to be relatively near the Sun, our estimate (Table~\ref{MLT_distances}) placing it at 18$\pm$2 pc.

\item WISE 1800$-$1559 was originally identified as an L dwarf candidate by \cite{folkes2012}, who estimated a type of L5.5$\pm$1.5 and a distance of 9.1$\pm$2.1 pc. \cite{bardalez2014} classify this object as L4.3 in the near-infrared based on an IRTF/SpeX spectrum. We type this object in the optical as an L4.5 using our Palomar/DSpec spectrum (Figure~\ref{seq_L4_L8.1}), which gives it a spectrophotometric distance estimate of 10$\pm$1 pc, very close to the \cite{folkes2012} result even though our type is one subclass earlier. This object may be a member of the 10 pc census.

\end{itemize}

\begin{figure}
\figurenum{52}
\includegraphics[scale=0.40,angle=0]{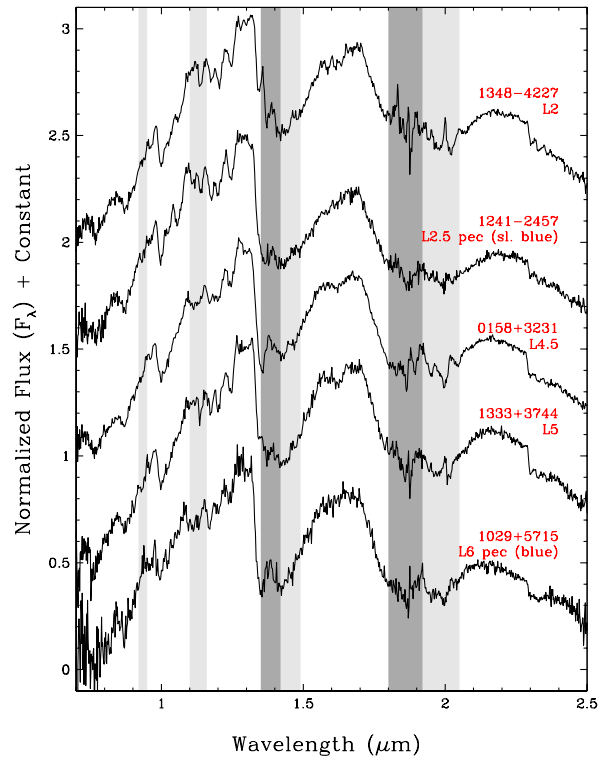}
\caption{Near-infrared spectra of L dwarfs. Spectra are normalized at 1.28 $\mu$m and offsets added to separate the spectra vertically. Telluric zones are color coded as described in Figure~\ref{seq_reddened_IR}.
\label{seq_Ls_IR}}
\end{figure} 

\begin{figure}
\figurenum{53}
\includegraphics[scale=0.325,angle=270]{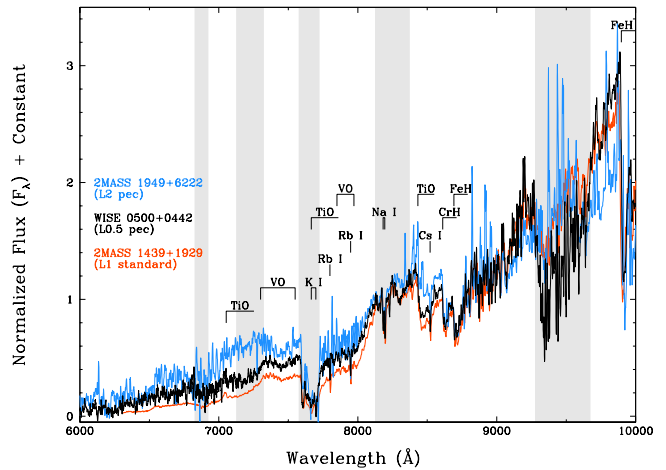}
\caption{Comparison of the optical L1 standard to the peculiar L0.5 dwarf WISE 0500+0442 and the peculiar L2 dwarf 2MASS 1949+6222, showing the non-standard oxide feature strengths in the latter two objects. All spectra are normalized at 8150 \AA\ and those of 2MASS 1949+6222 and WISE 0500+0442 have been smoothed with a 5-pixel boxcar to improve signal-to-noise per plotted point.
\label{seq_Ls_0500}}
\end{figure} 

\begin{figure}
\figurenum{54}
\includegraphics[scale=0.40,angle=0]{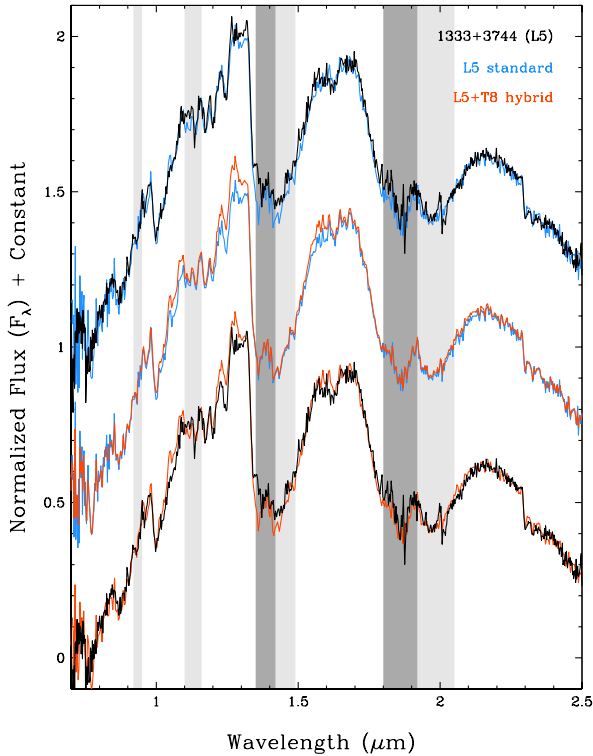}
\caption{Comparison of the best fitting standard, the L5 (top), and the best fitting synthetic binary, the L5+T8 (bottom), to the spectrum of WISE 1333+3744. (See the caption of Figure~\ref{seq_Ls_IR} for other details.)
\label{seq_Ls_1333}}
\end{figure} 

\subsection{T Dwarfs\label{section_Tdwarfs}}

\begin{figure}
\figurenum{55}
\includegraphics[scale=0.40,angle=0]{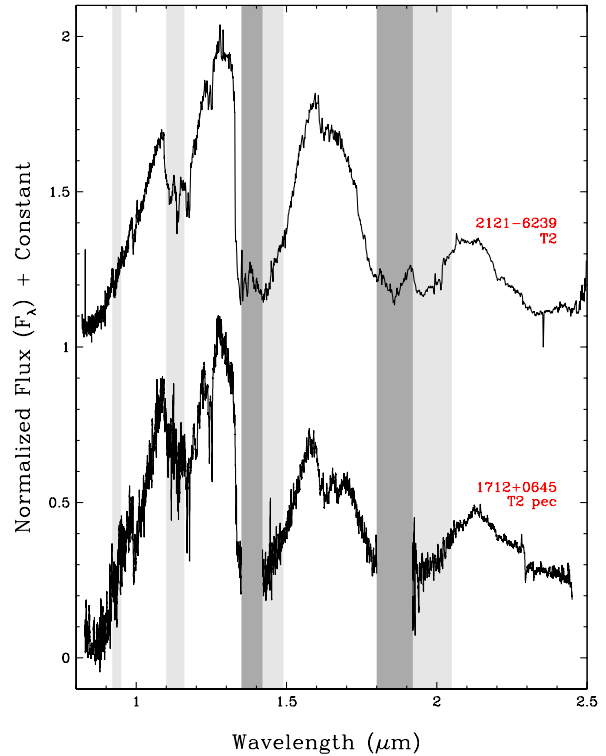}
\caption{Spectra of T dwarfs. (See the caption of Figure~\ref{seq_Ls_IR} for other details.)
\label{seq_Ts_IR}}
\end{figure}

The near-infrared spectra of T dwarfs were classified by eye against the T0-T8 spectral standards established by \cite{burgasser2006}. We have only two objects with T dwarf classifications, both of which have types of early-T (Figure~\ref{seq_Ts_IR}):

\begin{itemize}

\item WISE 2121$-$6239, one of the T dwarf candidates from \cite{kirkpatrick2014} that lacked spectroscopic verification at the time of publication, appears to be a normal T2 dwarf, as independently confirmed by \cite{beamin2015}. These authors find a spectrophotometric distance of 14$\pm$2 pc, which compares well to our value of 16$\pm$1 pc. 

\item WISE 1712+0645 does not fit any of the spectral standards well but comes closest to matching the T2 standard. Further analysis is given below.

\end{itemize}

The spectrum of WISE 1712+0645 shows strong methane absorption at $H$ and $K$ bands, although CO bands are also present at $K$. There is also strong H$_2$O and CH$_4$ absoprtion between the $Y$ and $J$ bands along with a strong Wing-Ford band of FeH at 9896 \AA. Normally, the two rival carbon-bearing molecules do not co-exist at these strengths in the spectrum of a brown dwarf, nor does a brown dwarf with strong CH$_4$ absorption normally show strong FeH. These observations strongly suggest that this is a spectral binary comprised of two unresolved brown dwarfs of type L and T. Various papers, such as those by \cite{burgasser2007}, \cite{burgasser2010}, and \cite{bardalez2014}, have created hybrid binaries in an attempt to explain spectra with unusual features, and we follow a similar prescription here. First, we consider the near-infrared spectra of the L0 through L9 spectral standards established in \cite{kirkpatrick2010} along with near-infrared spectra of the T0 through T8 spectral standards established in \cite{burgasser2006}, further extended to type T9 by \cite{cushing2011}. Second, synthetic photometry for each of those spectra was measured at $H$-band using a rectangular, flat-topped bandpass closely approximating the 2MASS $H$-band filter, with edges corresponding to the half-power points\footnote{See Figure 2 of \url{http://www.ipac.caltech.edu/2mass/releases/ allsky/doc/sec6\_4a.html}.} of 1.52 and 1.78 $\mu$m. Third, the spectra were then rescaled relative to one another to place all objects at the same distance, using the M$_H$ versus spectral type relation of \cite{looper2008} for types L0 through T4 and the relation of \cite{kirkpatrick2012} (the one shown in red in their Figure 12) for types T5 through T9. Fourth, hybrid binaries were created for all possible combinations, and the resulting spectra were compared by eye to the observed spectrum of WISE 1712+0645. 

\begin{figure}
\figurenum{56}
\includegraphics[scale=0.40,angle=0]{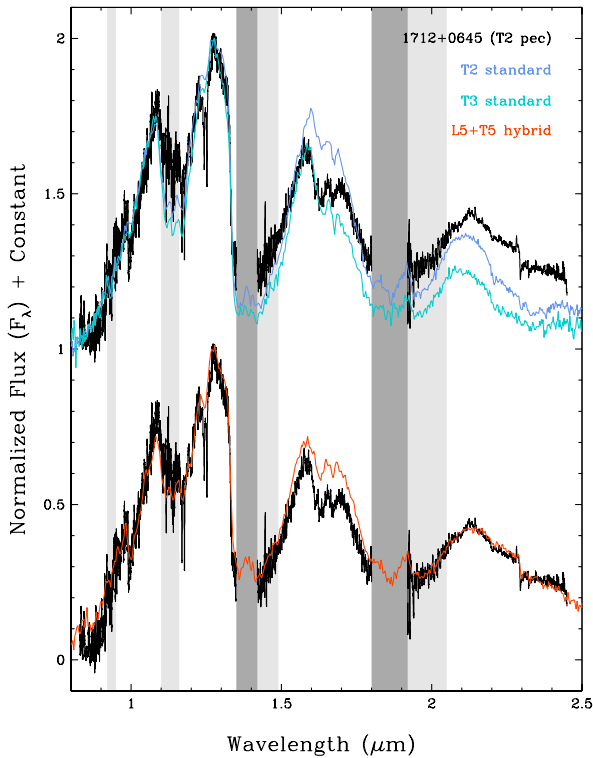}
\caption{Spectra of the unusual T dwarf WISE 1712+0645 compared to the best fitting T dwarfs standards and the best fitting hybrid binary. (See the caption of Figure~\ref{seq_Ls_IR} for other details.)
\label{seq_Ts_WISE1712}}
\end{figure}

Figure~\ref{seq_Ts_WISE1712} shows a comparison of WISE 1712+0645 to the best-fit standard as well as the hybrid binary that provides the best match. Note that the hybrid binary provides a much better overall fit to the spectrum than does the single standard and is able to account for most of the unusual features noted above. Although the overall morphology of the observed spectrum and the hybrid binary are markedly similar, there is a deficit in the overall $H$-band flux of the observed spectrum relative to the hybrid. Still, we conclude that the best explanation is that WISE 1712+0645 is an unresolved binary with types of approximately L5 and T5.  

\subsection{Subdwarfs (Low-metallicity Dwarfs)\label{subdwarf_analysis}}

The coldest subdwarfs are especially intriguing. They give us probes of the Milky Way's earliest star formation at low masses. Because they are old, the subdwarf brown dwarfs serve as time capsules of unprocessed, low-metallicity material from the Galaxy's early history. Because they are cold, isolated, and metal-poor, these objects also represent the simplest test cases against which to test exoplanet atmospheric theory. Furthermore, they also serve as direct checks of theoretical cooling rates via the ``subdwarf gap'' discussed in \cite{kirkpatrick2014}. Because the AllWISE2 survey samples brighter magnitudes than AllWISE1, we did not expect to find a sufficient number of later L subdwarfs with which to further explore the gap. Indeed, although a small number of new early- to mid-L subdwarfs were identified, no later L subdwarfs were found.

Subdwarfs spectroscopically observed for this paper were classified using the revised spectral typing scheme of \cite{lepine2007}, which expands upon the scheme introduced by \cite{gizis1997}. Whereas the \cite{gizis1997} scheme divided subdwarfs into two classes, the \cite{lepine2007} system uses three -- normal subdwarfs (sd), extreme subdwarfs (esd), and ultra subdwarfs (usd) -- which at a given late-K or M subtype are believed to correspond to increasing levels of metal deficiency. We have typed our new discoveries by using overplots of our spectra with spectra of the standards given in Table 2 of \cite{lepine2007}. 

This classification system, however, predates the discovery of subdwarfs of type M9 and later. The latest \cite{lepine2007} standards in each class are sdM8, esdM8.5, and usdM8.5. In the subsections that follow, we classify our new discoveries on the \cite{lepine2007} system where possible then examine ways to extend the system to later subclasses using both our discoveries and other late-type subdwarfs announced in the literature. 

\begin{figure}
\figurenum{57}
\includegraphics[scale=0.40,angle=0]{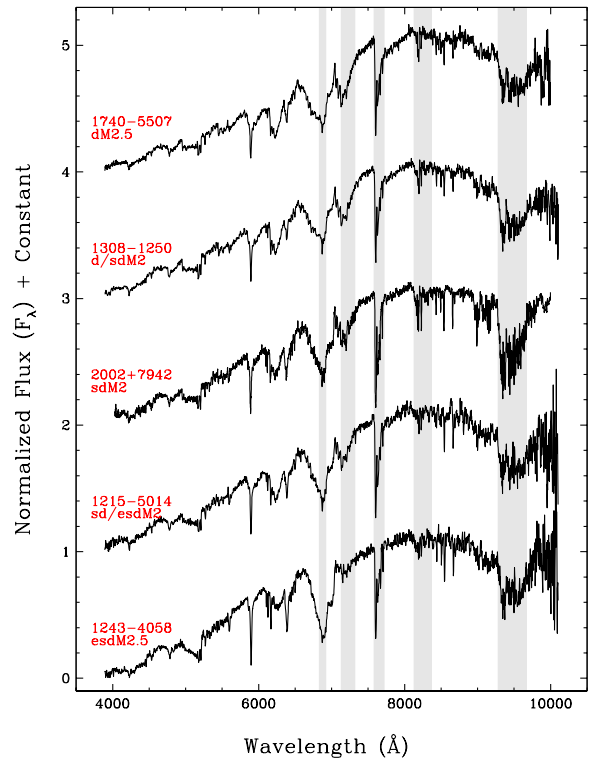}
\caption{Spectra of objects with core types of M2 or M2.5 and with prefix types ranging from d (normal dwarf) to esd (extreme subdwarf). (See the caption of Figure~\ref{seq_wds1} for other details.)
\label{seq_dM2_esdM2}}
\end{figure}

Note that in some relatively rare cases, an object will fall intermediate between metallicity classes, so a class such as d/sd (meaning halfway between the normal M dwarf classification and the sdM classification) or sd/esd is given. Figure~\ref{seq_dM2_esdM2} shows a sequence around subtype M2 where these intermediate classes are warranted.

\subsubsection{Normal Subdwarfs}

\begin{figure}
\figurenum{58}
\includegraphics[scale=0.375,angle=0]{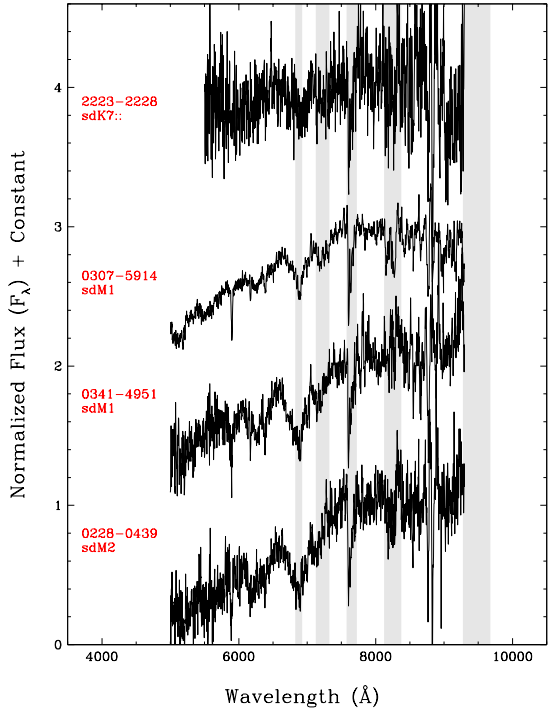}
\caption{Spectra of normal subdwarfs with types of sdK7 through sdM2. (See the caption of Figure~\ref{seq_wds1} for other details.)
\label{seq_sdK7_sdM5.1}}
\end{figure}

\begin{figure}
\figurenum{59}
\includegraphics[scale=0.375,angle=0]{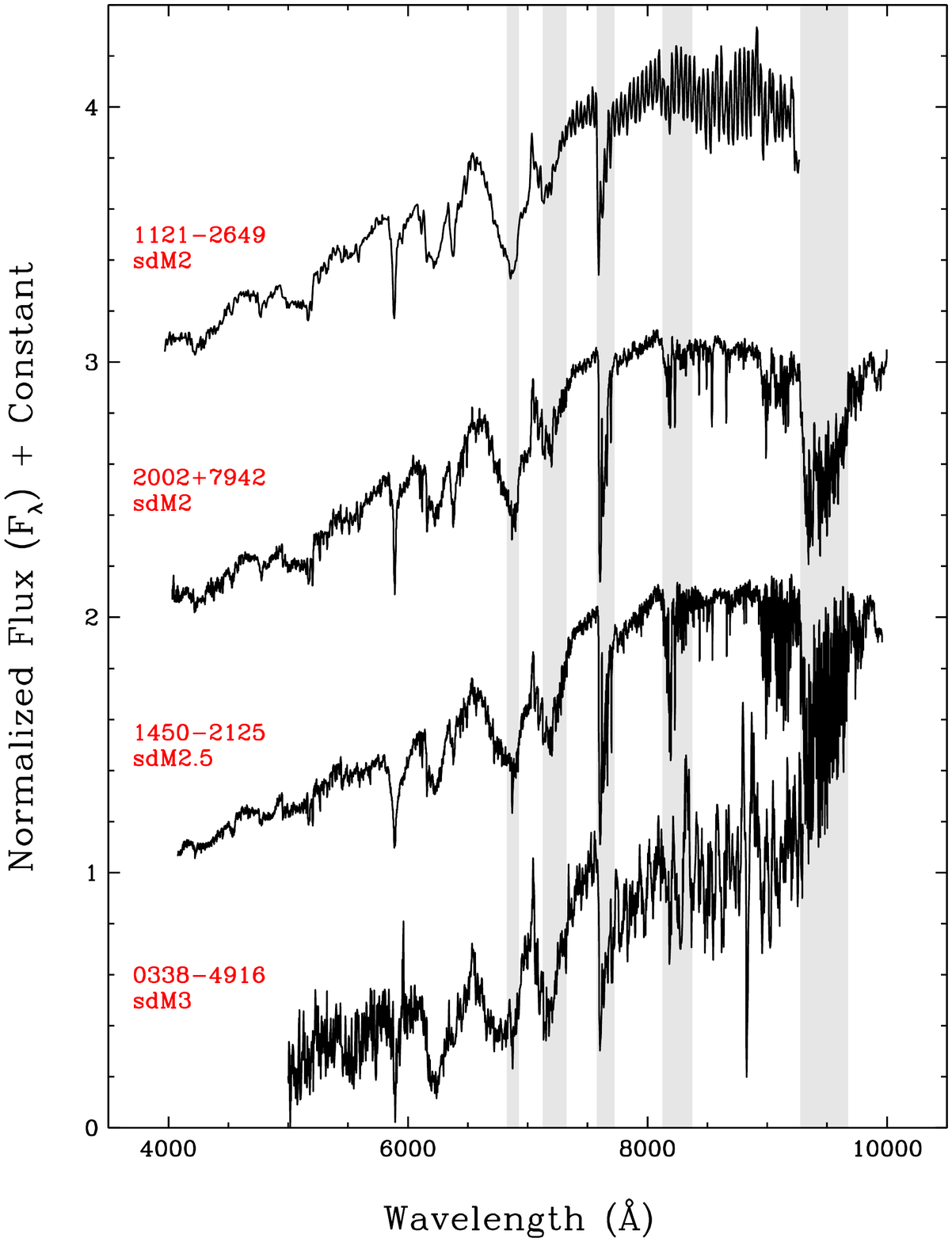}
\caption{Spectra of normal subdwarfs with types of sdM2 through sdM3. (See the caption of Figure~\ref{seq_wds1} for other details.)
\label{seq_sdK7_sdM5.2}}
\end{figure}

\begin{figure}
\figurenum{60}
\includegraphics[scale=0.375,angle=0]{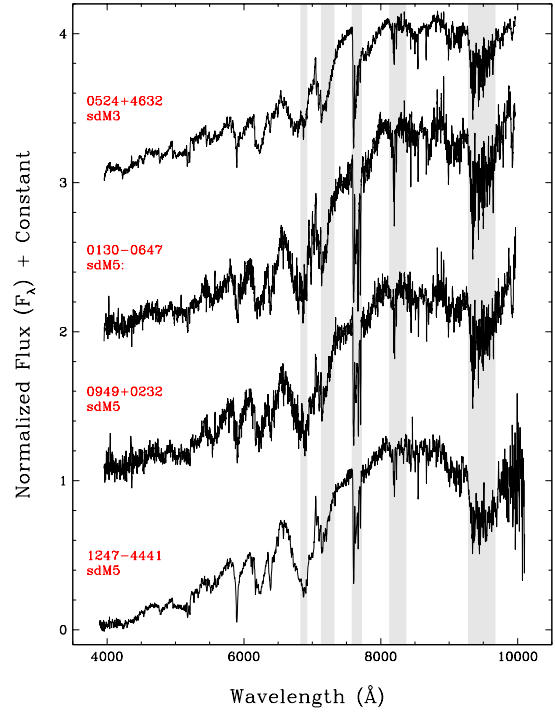}
\caption{Spectra of normal subdwarfs with types of sdM3 through sdM5. For WISE 1247$-$4441, the duPont/BCSpec spectrum is the one shown. (See the caption of Figure~\ref{seq_wds1} for other details.) 
\label{seq_sdK7_sdM5.3}}
\end{figure}

\begin{figure}
\figurenum{61}
\includegraphics[scale=0.375,angle=0]{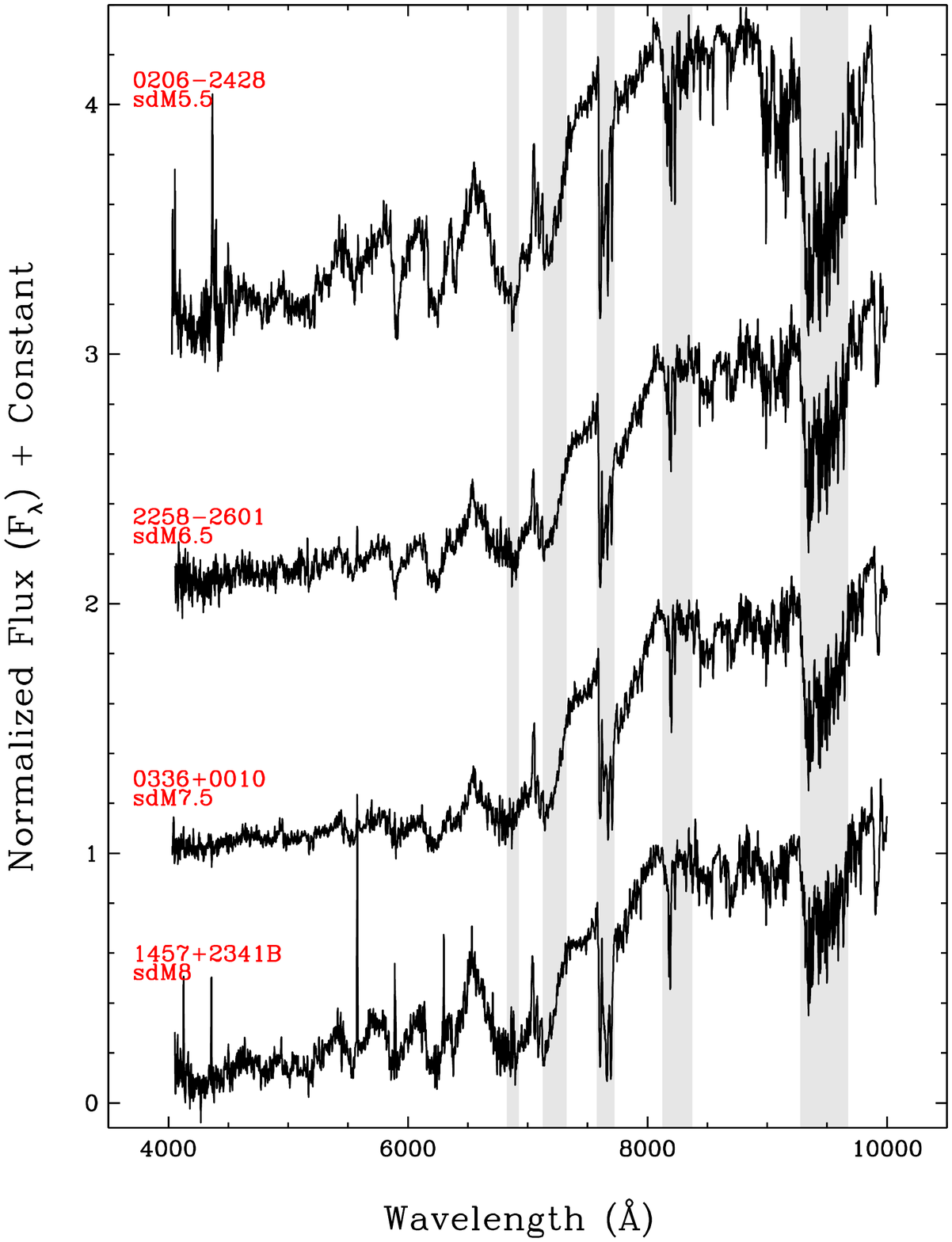}
\caption{Spectra of normal subdwarfs with types of sdM5.5 through sdM8. (See the caption of Figure~\ref{seq_wds1} for other details.)
\label{seq_sdM65_sdM9.1}}
\end{figure}

\begin{figure}
\figurenum{62}
\includegraphics[scale=0.375,angle=0]{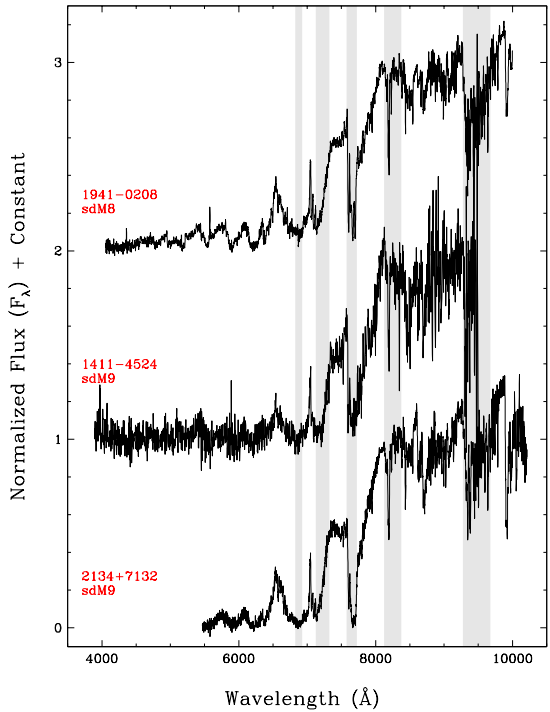}
\caption{Spectra of normal subdwarfs with types of sdM8 through sdM9. (See the caption of Figure~\ref{seq_wds1} for other details.) For WISE 1411$-$4524, the spectrum shown is the one from duPont/BCSpec, and for WISE 2134+7132 it is the one from Keck/LRIS.
\label{seq_sdM65_sdM9.2}}
\end{figure}

Objects classified as normal subdwarfs are shown in Figure~\ref{seq_sdK7_sdM5.1} through Figure~\ref{seq_sdM65_sdM9.2}. Four of these objects fall at or below the bottom of the normal subdwarf scale as defined by \cite{lepine2007}: WISE 1457+2341B, 1941$-$0208, 1411$-$4524, and 2134+7132. (As we discuss later in this section, WISE 1013$-$7246 and 1355$-$8258, which have only near-infrared spectra, are likely to be new additions to this list as well.)

\begin{figure*}
\figurenum{63}
\includegraphics[scale=0.85,angle=0]{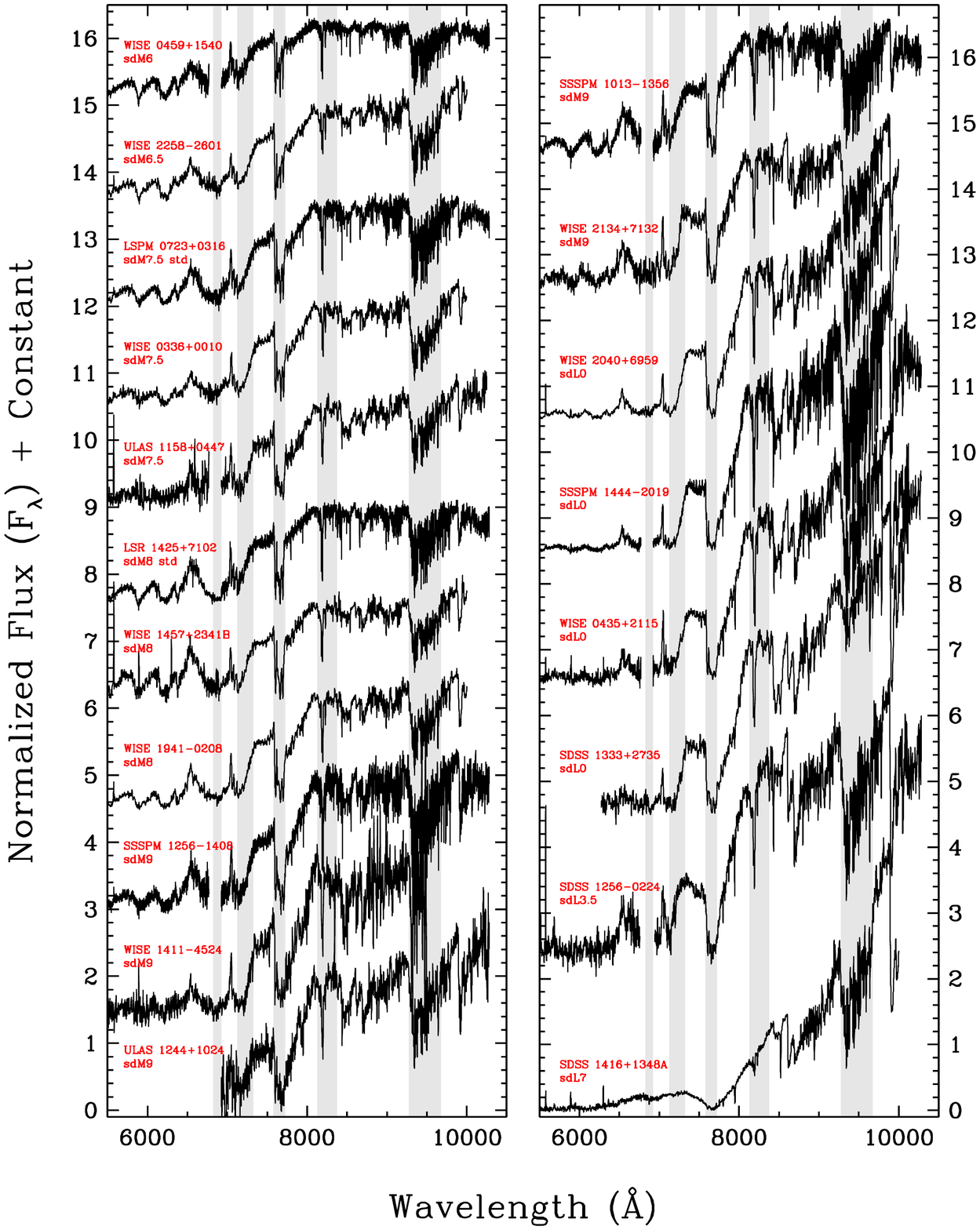}
\caption{Optical spectral sequence extending through and beyond the bottom of the normal subdwarf sequence established by \cite{lepine2007}. (See the caption of Figure~\ref{seq_wds1} for other details.)
\label{seq_sdM_extension}}
\end{figure*}

The optical spectra of these late-type subdwarfs are shown in context with other late-sdM and sdL objects from the literature in Figure~\ref{seq_sdM_extension} using spectra reported in Table 4
and Table~\ref{spectra_literature}\footnote{Our Palomar/DSpec spectra of spectral standards LSPM J0723+0316 (sdM7.5) and LSR J1425+7102 (sdM8) have a small gap in coverage between 6770 and 6920 \AA, coincident with the atmospheric B band but also in the middle of the telltale CaH absorption feature. For these two objects, we filled the gap using spectra of these standards kindly provided by S\'ebastien L\'epine.}. In this figure we provide an ordering of the spectra, starting at sdM6 and proceding to later types, that shows the smoothest progression of spectral features from mid-sdM through mid-sdL. The sdM7.5 LSR J0723+0316 and the sdM8 LSR J1425+7102 from the \cite{lepine2007} system are used as anchor points. Later objects are anchored by the relatively bright objects SSSPM J1013$-$1356 and WISE 2040+6959, which we tentatively assign as the sdM9 and sdL0 standards, respectively. Two later L subdwarfs are shown for comparison.

\begin{figure}
\figurenum{64}
\includegraphics[scale=0.40,angle=0]{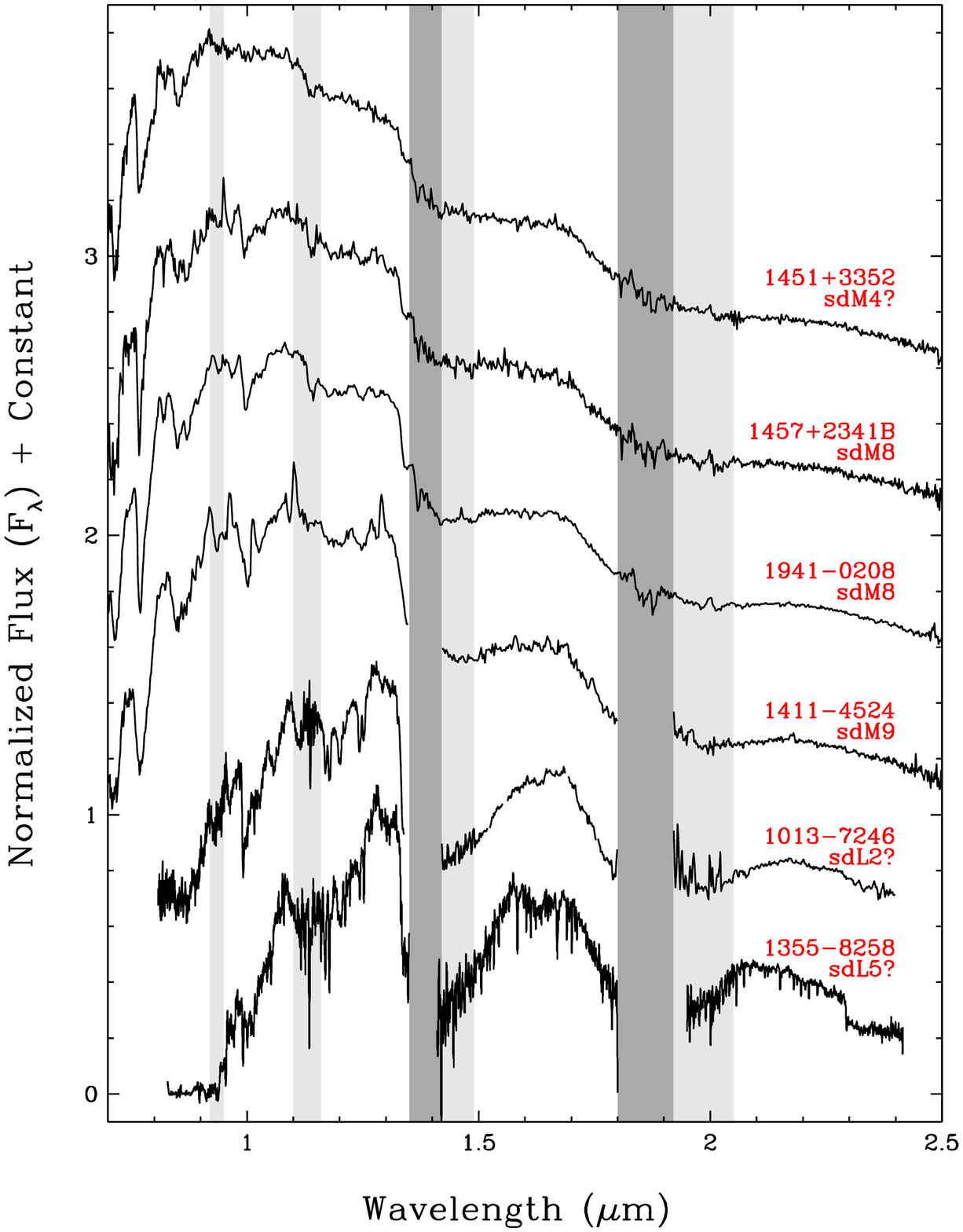}
\caption{Near-infrared spectra of normal subdwarfs. (See the caption of Figure~\ref{seq_Ls_IR} for other details.)
\label{seq_sdM_sdL_IR}}
\end{figure}

Notes on the classification of previously published objects are given below:

\begin{itemize}

\item WISE 0459+1540 was classified by \cite{kirkpatrick2014} as an sdL0 based only on a near-infrared spectrum. The optical spectrum obtained here has a classification of sdM6.

\item ULAS J115826.26+044746.8 was classified by \cite{lodieu2012} as an sdM9.5 by comparing to a very low-resolution IRTF/SpeX spectrum of an unspecified star reported to be an sdM9.5. Our classification pushes this object earlier, to a type of sdM7.5, which is more in line with the spectral type that \cite{lodieu2012} derives when extending the index definitions of \cite{gizis1997} and \cite{lepine2007} to later types.

\item SSSPM J1256$-$1408 was announced by \cite{schilbach2009}; although no spectrum had been obtained, its placement on color-magnitude diagrams suggested a type near sdM8. We classify the object as sdM9.

\item ULAS J124425.90+102441.9 was classified by \cite{lodieu2012} as an sdL0.5 by comparing to a very low-resolution IRTF/SpeX spectrum of an unspecified star reported to be an sdM9.5. Our classification pushes this object earlier, to a type of sdM9, which is more in line with the spectral type that \cite{lodieu2012} derives when extending the index definitions of \cite{gizis1997} and \cite{lepine2007} to later types.

\item SSSPM J1013$-$1356 was discovered by \cite{scholz2004a}, who classified it as an sdM9.5. Although this was an earlier discovery pre-dating the \cite{lepine2007} scheme, we classify it very similarly, at sdM9, and use it as our sdM9 standard.

\item WISE 2040+6959 was classified by \cite{kirkpatrick2014} in the optical as an sdL0, and we use it as the sdL0 spectral anchor here. \cite{luhman2014} classified it in the near-infrared as an sdM9. 

\item SSSPM J1444$-$2019 was discovered by \cite{scholz2004b}, who classified it as an sdM9 with features similar to an L dwarf. They mention that the object is redder than SSSPM J1013$-$1356 and likely an L subdwarf, although there were few similar objects to compare to when this discovery was made. Our classification indeed places it at sdL0. This object was mentioned by \cite{lepine2007} but no attempt was made to include it into the new classification scheme. 

\item WISE 0435+2115 was classified by \cite{kirkpatrick2014} in the optical as an sdL0, and we retain that classification here. \cite{luhman2014} classified it in the near-infrared as an sdM9. 

\item SDSS J133348.24+273508.8 was announced by Zhang et al.\footnote{See http://www.mpia.de/homes/joergens/ringberg2012\_proc/ zhang.pdf.}, where it was classified as an sdL3 in the optical. Our spectrum shows that its spectral morphology is nearly identical to WISE 2040+6959, so we classify it as an sdL0.

\item SDSS J125637.13$-$022452.4 was published by \cite{sivarani2009} as an sdL4. \cite{burgasser2009} classify it as an sdL3.5, and we retain that classification here.

\item SDSS J141624.08+134826.7 was discovered simultaneously by several teams including \cite{schmidt2010}, \cite{burningham2010}, and \cite{bowler2010}. It also has a T dwarf companion, ULAS J141623.94+134836.3 (\citealt{burningham2010,scholz2010,burgasser2010b}). We regard both as subdwarfs and retain the sdL7 classification from \cite{kirkpatrick2010} for the brighter component.

\end{itemize}

Despite our attempts to provide the most natural ordering of these spectra, object-to-object differences within the same subclass remain. For example, the spectrum of the new object WISE 2134+7132 is unique in that it has features not shared by the other sdM9's on the plot. Notably, the plateau between 7250 and 7550 \AA\ slopes downward for longer wavelengths whereas for the other sdM9 objects the plateau slopes upward. Also, the sharp upward spike caused by the opacity hole between the 8432 \AA\ TiO band and the 8611 \AA\ CrH band is much more pronounced in WISE 2134+7132. These features resemble those in the later subdwarf, the sdL3.5 SDSS 1256$-$0224. The presence of this feature may indicate increased metal deficiency, in which case spectra with this morphology may be separable into a distinct class of their own. However, we leave discussion of further subdivision of subdwarf types to a later time when a larger set of these objects is available for study. 

\begin{figure}
\figurenum{65}
\includegraphics[scale=0.40,angle=0]{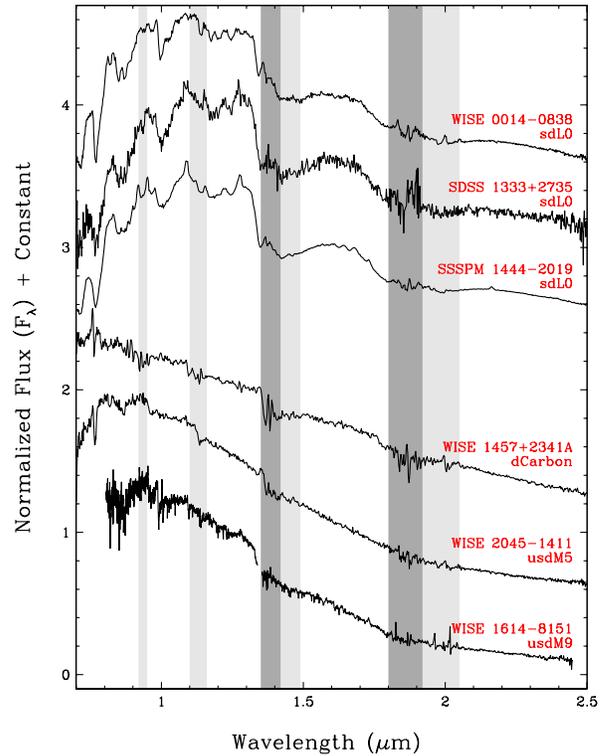}
\caption{Near-infrared spectra of L0 subdwarfs, M ultra subdwarfs, and a carbon dwarf. (See the caption of Figure~\ref{seq_Ls_IR} for other details.)
\label{seq_sdL_usdM_dCarbon_IR}}
\end{figure}

For some of these objects we obtained near-infrared spectroscopy, shown in Figure~\ref{seq_sdM_sdL_IR}, with which to classify other objects either too faint for optical spectroscopy or lacking optical spectral follow-up. For the near-infrared spectra of WISE 1457+2341B, 1941$-$0208, and 1411$-$5424, optical spectra are also available, so we adopt the optical classifications as those for the near-infrared. Near-infrared spectra of three previously identified, optical L subdwarfs are shown in Figure~\ref{seq_sdL_usdM_dCarbon_IR}; these are all sdL0's with similar, although not identical, near-infrared morphologies. 

\begin{figure}
\figurenum{66}
\includegraphics[scale=0.40,angle=0]{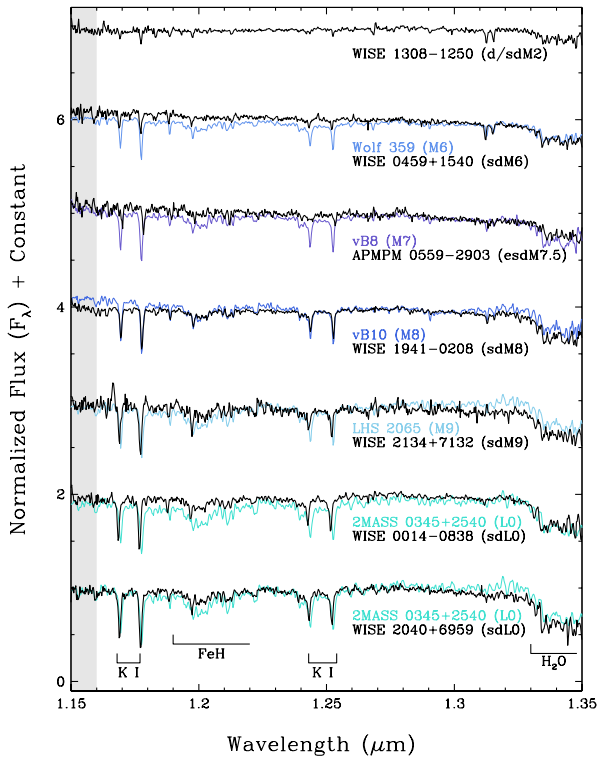}
\caption{Near-infrared spectra at $J$-band of subdwarfs and extreme subdwarfs compared to normal dwarfs of the same subclass. Although the \ion{K}{1} doublets are weaker in the sdM6 and esdM7.5 objects compared to the standards, the spectral slope and the depth of the H$_2$O band are very similar. At types of sdM8 through sdL0, however, very little difference is seen with the standards despite dramatic changes in the spectral morphology in the optical.
\label{seq_NIRSPEC}}
\end{figure}

Unfortunately, the three remaining subdwarfs that need classification -- WISE 1451+3352, 1013$-$7246, and 1355$-$8258 -- fall earlier or later than the set for which we have classifications tied to the optical. WISE 1451+3352 receives a type of ``sdM4?''; comparison to the near-infrared M dwarf standards of \cite{kirkpatrick2010} confirms that this object, though matching the M4 standard well at $J$ band, has a suppressed spectrum at $H$ and $K$ bands relative to the standard. At low to moderate resolution, the near-infrared portion of the spectrum -- even the relative feature-rich $J$ band (Figure~\ref{seq_NIRSPEC}) -- offers few other diagnostics against which to distinguish metallicity effects that are obvious in the optical.

\begin{figure}
\figurenum{67}
\includegraphics[scale=0.325,angle=270]{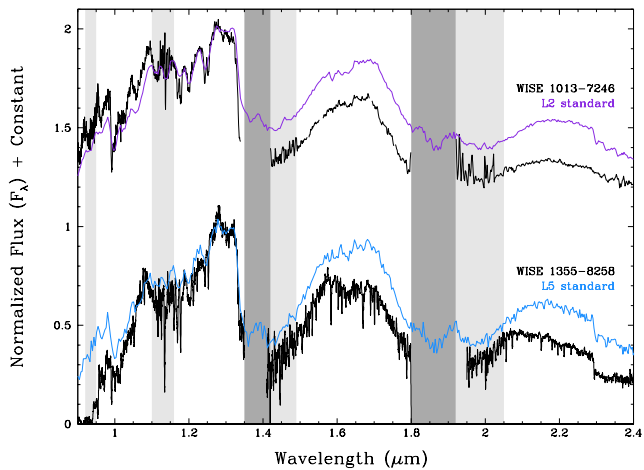}
\caption{Comparison of the near-infrared spectra of WISE 1013$-$7246 and WISE 1355$-$8252 to the best matching L dwarf standards at $J$ band. (See the caption of Figure~\ref{seq_Ls_IR} for other details.)
\label{seq_1013_1355_IR}}
\end{figure}

The two remaining near-infrared-classified objects are L subdwarfs:

\begin{itemize}

\item WISE 1013$-$7246 has a spectrum that best matches an L2 in the \cite{kirkpatrick2010} scheme at $J$ band but is considerably bluer than the standard itself (Figure~\ref{seq_1013_1355_IR}). Also apparent are the much stronger bands of FeH at 9896 \AA, with which we justify our classification of this object as a subdwarf and not just a ``blue'' L dwarf. Note, however, that the 1.19-1.24 $\mu$m FeH bands are not noticeably stronger than the standard, further highlighting the difficulty (Figure~\ref{seq_NIRSPEC}) in classifying late-M and early-L subdwarfs at $J$ band.

\item WISE 1355$-$8258 has a spectrum that best matches an L5 in the \cite{kirkpatrick2010} scheme at $J$ band but is much bluer, as shown in Figure~\ref{seq_1013_1355_IR}. Unlike the L5 standard, the $H$-band plateau slopes downward at longer wavelengths and the plateau at $K$-band is noticeably flattened. Both of these characteristics are hallmarks of the increased relative importance of collision-induced absorption by H$_2$, a trademark of L subdwarfs. Moreover, relative to the L5 standard, WISE 1355$-$8258 has much stronger bands of FeH at 1.19-1.24 $\mu$m and possibly at 1.58-1.64 $\mu$m. The interpretation of this object would be more secure with the acquisition of an optical spectrum providing additional checks on metal deficiency and temperature. 

\end{itemize}

\begin{figure}
\figurenum{68}
\includegraphics[scale=0.40,angle=0]{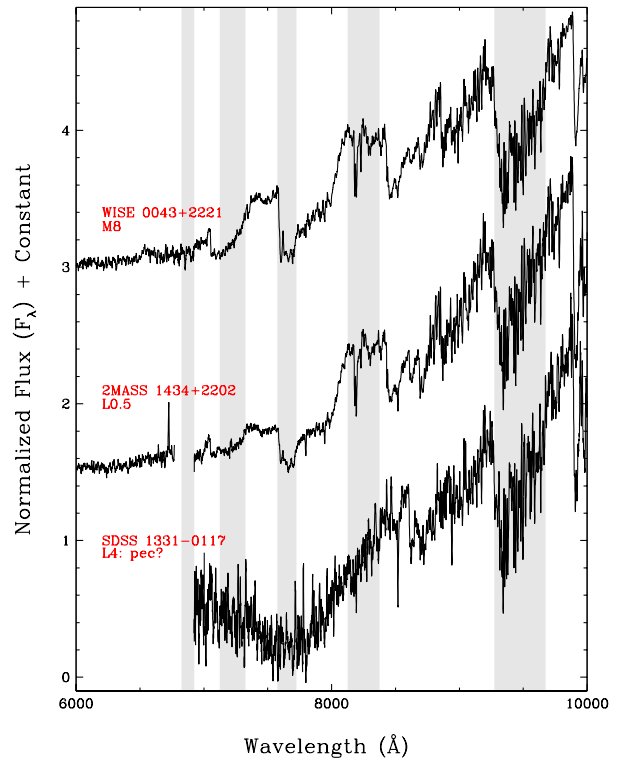}
\caption{Optical spectra of three objects published as subdwarfs but showing no low-metallicity signatures in our spectra. (See the caption of Figure~\ref{seq_wds1} for other details.)
\label{seq_former_subdwarfs}}
\end{figure}

Finally, our spectroscopic follow-up of three published subdwarfs from the literature shows no obvious subdwarf characteristics in the optical, contrary to prior classifications (Figure~\ref{seq_former_subdwarfs}). WISE 0043+2221 was classified in the near-infrared as an sdL1 (\citealt{kirkpatrick2014}), but this object has the morphology of a normal M8 dwarf in the optical. Its near-infrared spectrum is distinctly peculiar -- and most like the sdL1 (or blue L1) 2MASS J17561080+2815238 -- but the cause must not be related to metallicity. 2MASS J14343661+2202463 is classified in the near-infrared by \cite{sheppard2009} as a possible sdM9. Our optical spectrum is that of a normal L0.5. SDSS J133148.90$-$011651.4 has appeared at various times as a normal optical L6 dwarf (\citealt{hawley2002}), a near-infrared L8 with uncertain classification (\citealt{knapp2004}), a near-infrared T0 (\citealt{schneider2014}), and a peculiar blue L1 (\citealt{marocco2013}). We classify our optical spectrum as a possibly peculiar L4 with no obvious subdwarf signature, but because of the low signal-to-noise we cannot rule out the slightly metal-poor hypothesis advocated by \cite{marocco2013}.

\subsubsection{Extreme Subdwarfs}

\begin{figure}
\figurenum{69}
\includegraphics[scale=0.375,angle=0]{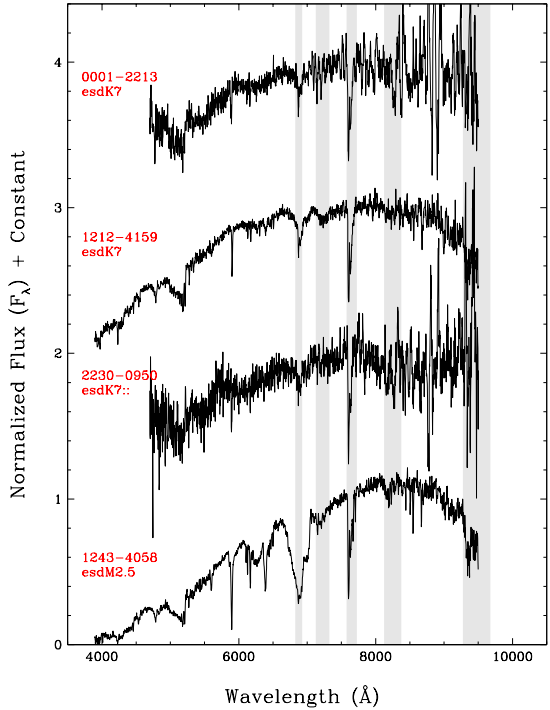}
\caption{Spectra of extreme subdwarfs with types of esdK7 through esdM2.5. (See the caption of Figure~\ref{seq_wds1} for other details.)
\label{seq_esdK7_esdM8.1}}
\end{figure}

\begin{figure}
\figurenum{70}
\includegraphics[scale=0.375,angle=0]{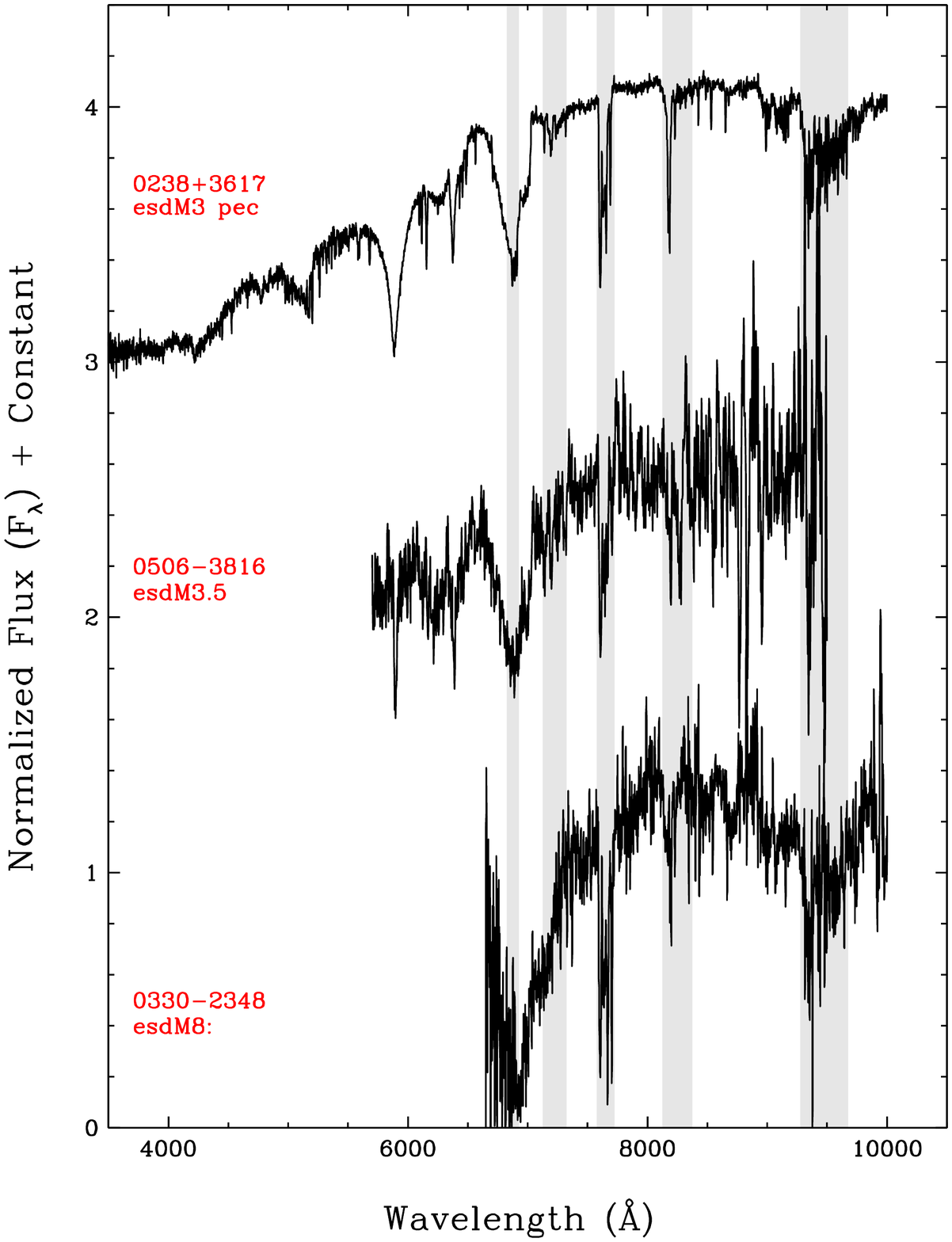}
\caption{Spectra of extreme subdwarfs with types of esdM3 through esdM8. (See the caption of Figure~\ref{seq_wds1} for other details.)
\label{seq_esdK7_esdM8.2}}
\end{figure}

\begin{figure}
\figurenum{71}
\includegraphics[scale=0.375,angle=0]{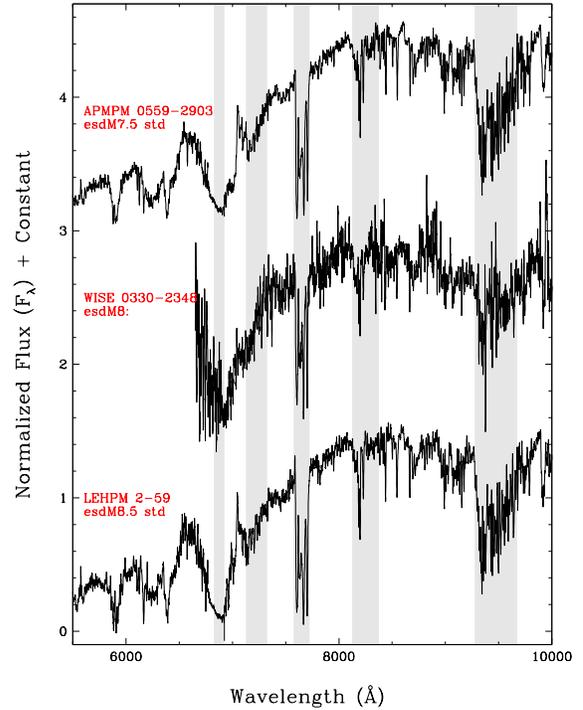}
\caption{Optical spectral sequence at the bottom of the extreme subdwarf sequence established by \cite{lepine2007}. (See the caption of Figure~\ref{seq_wds1} for other details.)
\label{seq_esdM_extension}}
\end{figure}

Objects classified as extreme subdwarfs are shown in Figure~\ref{seq_esdK7_esdM8.1} and Figure~\ref{seq_esdK7_esdM8.2}. One of these objects falls close to the bottom of the extreme subdwarf scale as defined by \cite{lepine2007}. This object, WISE 0330$-$2348, is shown in context with the two latest esdM standards in Figure~\ref{seq_esdM_extension}. We have tentatively classified it as esdM8: pending a higher resolution spectrum with which to get a more solid classification.

\subsubsection{Ultra Subdwarfs}

\begin{figure}
\figurenum{72}
\includegraphics[scale=0.375,angle=0]{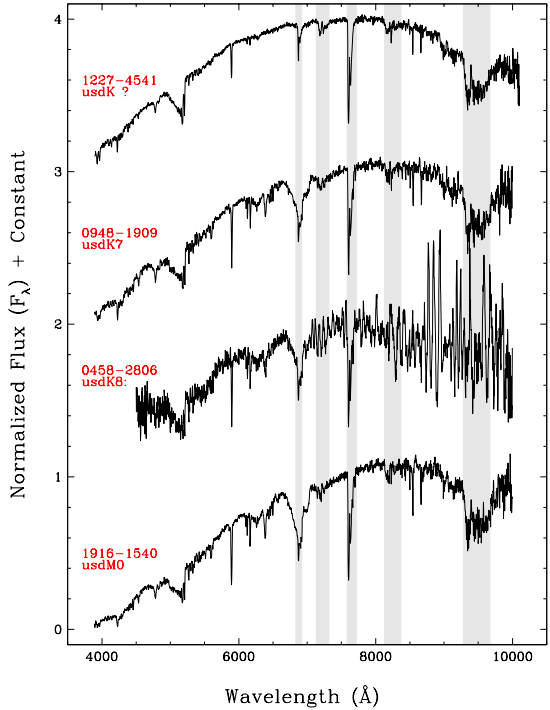}
\caption{Spectra of ultra subdwarfs with types of usdK through usdM0. (See the caption of Figure~\ref{seq_wds1} for other details.)
\label{seq_usdK7_usdM9.1}}
\end{figure}

\begin{figure}
\figurenum{73}
\includegraphics[scale=0.375,angle=0]{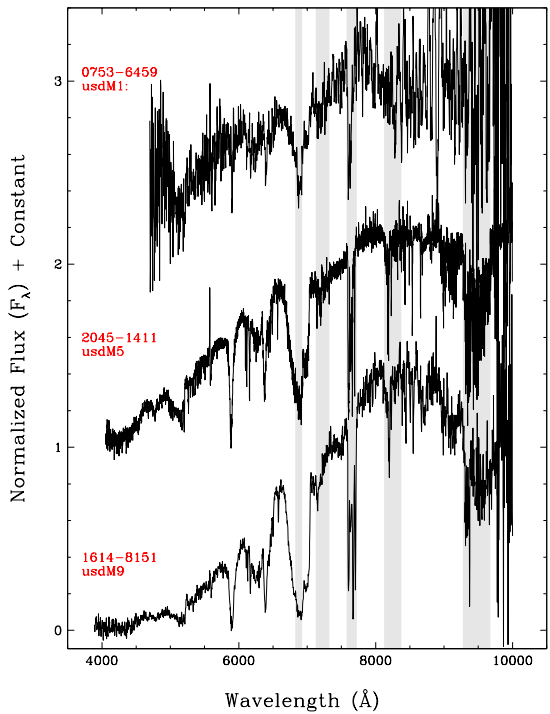}
\caption{Spectra of ultra subdwarfs with types of usdM1 through usdM9. (See the caption of Figure~\ref{seq_wds1} for other details.) For WISE 1614$-$8151, the spectrum shown is the one from duPont/BCSpec.
\label{seq_usdK7_usdM9.2}}
\end{figure}

\begin{figure}
\figurenum{74}
\includegraphics[scale=0.375,angle=0]{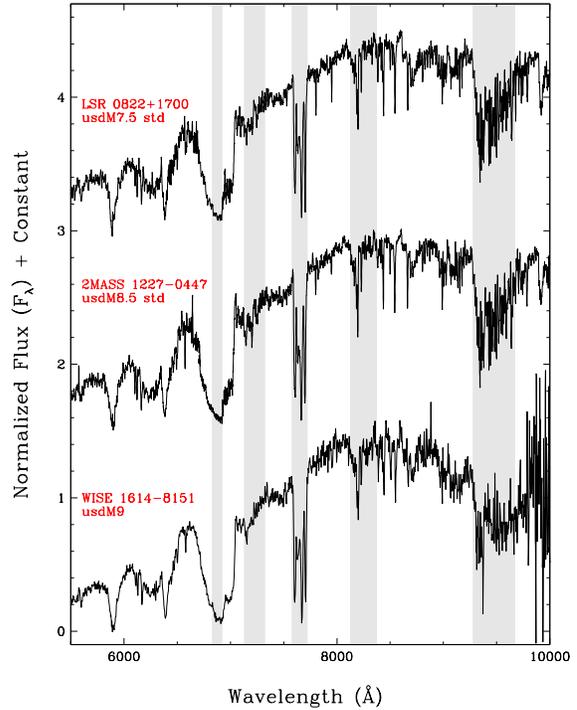}
\caption{Optical spectral sequence extending through and beyond the bottom of the ultra subdwarf sequence established by \cite{lepine2007}. (See the caption of Figure~\ref{seq_wds1} for other details.)
\label{seq_usdM_extension}}
\end{figure}

Objects classified as ultra subdwarfs are shown in Figure~\ref{seq_usdK7_usdM9.1} and Figure~\ref{seq_usdK7_usdM9.2}. One of these objects falls below the bottom of the ultra subdwarf scale as defined by \cite{lepine2007}. This object, WISE 1614$-$8151, is shown in context with the two latest usdM standards in Figure~\ref{seq_usdM_extension}. It has an extremely strong CaH band at 6750 \AA\ yet a very weak TiO band at 7053 \AA. It has a slightly redder slope than the usdM8.5 standard along with a slightly broader Na `D' doublet that continues the broadening trend seen from usdM7.5 to usdM8.5\footnote{See the following section for more discussion on the use of the Na `D' line as a metallicity diagnostic.}, so we classify it as a usdM9. This object is relatively bright ($R$=17.71 mag from the USNO-B1 and $J$=14.94 mag from 2MASS), which is even brighter than the usdM8.5 standard 2MASS J12270506$-$0447207 ($R$= 19.06 mag and $J$=15.49 mag), and so should have a robust trignonometic parallax available soon from Gaia.

We also obtained near-infrared spectra of two of our usdM discoveries, as shown in Figure~\ref{seq_sdL_usdM_dCarbon_IR}. As expected, these show blue continua that are the result of collision-induced absorption by H$_2$, which is expected to impact $K$ band more strongly than $H$, and $H$ more strongly than $J$ (\citealt{borysow1997}).

\subsubsection{Subdwarfs of Special Note}

Two subdwarfs are peculiar enough that they deserve special attention. One is a warmer ultra subdwarf and the other is a unique extreme subdwarf:

\begin{itemize}

\item WISE 1227$-$4541 (SCR J1227$-$4541) was originally identified as a high motion star by \cite{subasavage2005} and identified as a probable subdwarf. We are unable to classify our spectrum, believed to be the first taken for this object, using the usual main sequence dwarf standards or using the set of subdwarf, extreme subdwarf, and ultra subdwarf standards established by \cite{lepine2007}, which extend only as early as K7. We tentatively classify it as an earlier usdK (Figure~\ref{seq_usdK7_usdM9.1}).

Fits to the suite of PHOENIX spectroscopic models by \cite{husser2013} indicate that WISE 1227$-$4541 has [Fe/H]=$-$3.0$\pm$1.0, T$_{eff}\approx$3700K, and log($g$)$\approx$4.0 (Figure~\ref{seq_1227_model_fits}). This suggests an uncommonly low metallicity. The large uncertainty in the metallicity measurement is due to the lack of diagnostics at low resolution. Obtaining higher resolution spectra, which would not be difficult given the object's apparent brightness ($R$=14.49 mag from the USNO-B1), is strongly recommended.

\begin{figure}
\figurenum{75}
\includegraphics[scale=0.40,angle=0]{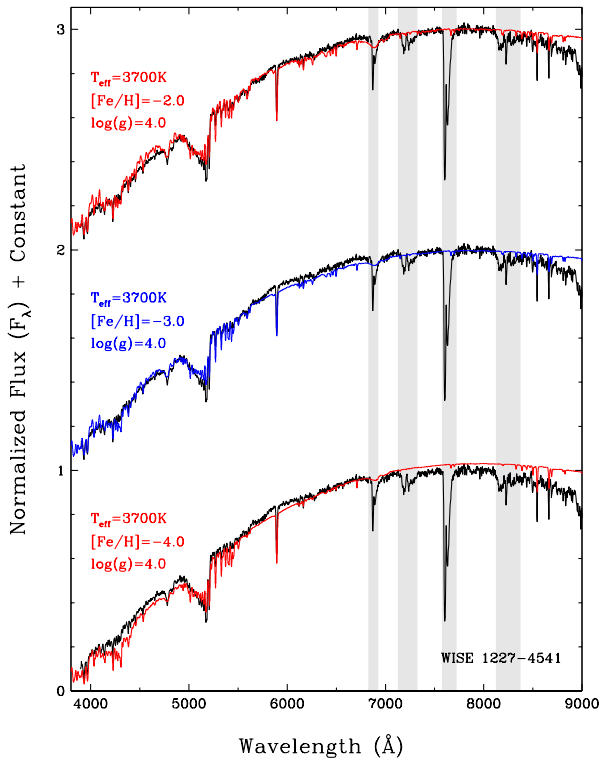}
\caption{Our three best by-eye fits of WISE 1227$-$4541 to the suite of theoretical spectra given in \cite{husser2013}. 
\label{seq_1227_model_fits}}
\end{figure}

\item WISE 0238+3617 (LP 245-62) has a unique spectrum. As shown in Figure~\ref{seq_sodium_dwarf}, its overall slope and its oxide and hydride strengths best match an early- to mid-esdM. Notably odd are its strong \ion{Na}{1} and \ion{K}{1} lines and its weaker \ion{Ca}{2} lines, as shown by the comparsion to our high signal-to-noise spectrum of the normal esdM2.5 dwarf WISE 1243$-$4058. In particular, the Na `D' doublet is deep and very broad, like that seen in early-L dwarfs (\citealt{reid2000b}). The sodium (8183 and 8195 \AA) and potassium (7665 and 7699 \AA) doublets, which strengthen at higher gravities, and the calcium infrared triplet (8498, 8542, and 8662 \AA), which weakens at higher gravities, have been used extensively as luminosity criteria to distinguish giants and dwarfs (e.g., \citealt{kirkpatrick1991}). It is therefore tempting to brand WISE 0238+3617 as an object similar in T$_{eff}$ and metallicity to WISE 1243$-$4058 but with an unsually high gravity.

\begin{figure}
\figurenum{76}
\includegraphics[scale=0.40,angle=0]{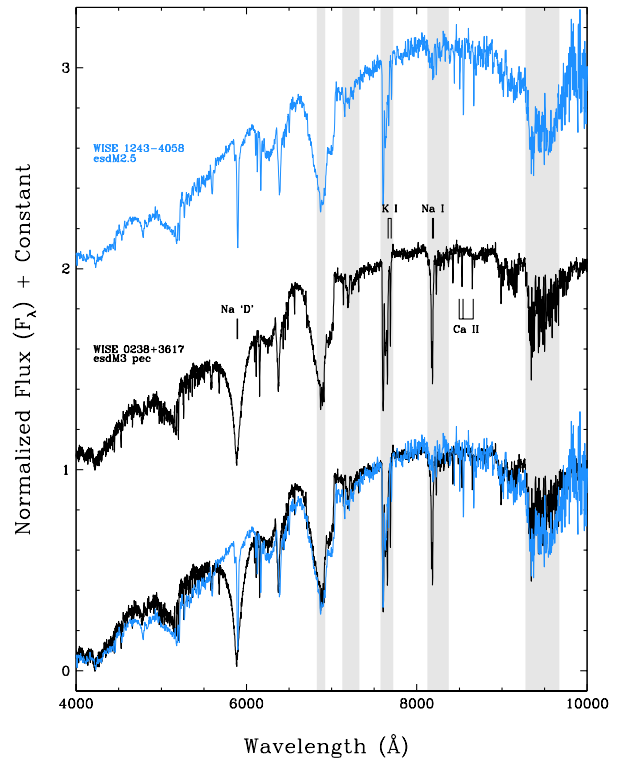}
\caption{Optical spectral sequence comparing the unusual spectrum of WISE 0238+3617 to the normal esdM dwarf WISE 1243$-$4058. Both spectra have roughly the same resolution. Features with notably different strengths between to two spectra are labeled. (See the caption of Figure~\ref{seq_wds1} for other details.)
\label{seq_sodium_dwarf}}
\end{figure}

However, this is likely too simplistic an explanation. Figure~\ref{seq_sdM_models} presents subdwarf models from the PHOENIX stellar atmosphere calculations (\citealt{husser2013}) showing a sequence of spectra that have, like WISE 0238+3617 and 1243$-$4058, similar slopes and CaH bandstrengths. For all of these models, the \ion{Na}{1}, \ion{K}{1}, and \ion{Ca}{2} line strengths are in tandem with the CaH bandstrength, and this is because CaH is {\it also} gravity sensitive.\footnote{The weakness of the CaH band has been used to distinguish giants from dwarfs (\citealt{morgan1943}) as well as old, compact M dwarfs, from young, still contracting brown dwarfs of similar type (\citealt{luhman1997}).} As these models clearly indicate, similar line and band strengths and overall continuum slope can be found across a substantial range in effective temperature, metallicity, and gravity. 

\begin{figure}
\figurenum{77}
\includegraphics[scale=0.40,angle=0]{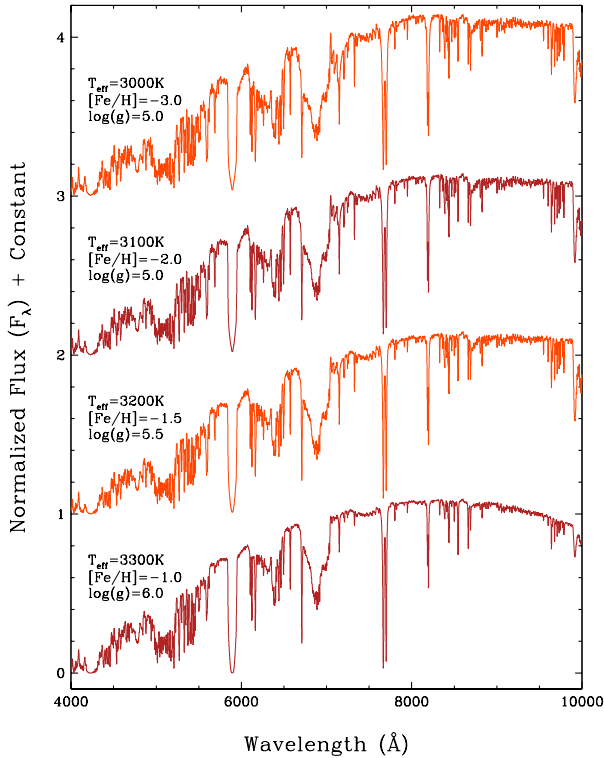}
\caption{A sequence of model spectra from \cite{husser2013} showing that spectra with similar strengths in \ion{Na}{1}, \ion{K}{1}, \ion{Ca}{2}, and CaH can result from a sizable spread in physical parameters (T$_{eff}$, [Fe/H], and log($g$)).
\label{seq_sdM_models}}
\end{figure}

The strong alkali lines may be pointing to a large column density in the visible photosphere of WISE 0238+3617, meaning that this object has an unusually low opacity. The broadened Na `D' line, like that of an early-L dwarf, may be another indication of this low opacity: as column depth increases, the high relative abundance of sodium and the increasing gas pressure at these depths gives rise to spectacular broadening of the line wings (\citealt{burrows2003}). Still, we have to explain why WISE 0238+3617 has such a low opacity.

The most viable explanation is that WISE 0238+3617 is an extremely small subdwarf of uncommonly low metallicity. Low metallicity makes for low opacity, and the resulting reduction in radiation pressure means that the star achieves hydrostatic equilibrium at a smaller radius. This would explain the high gravity and low opacity effects in the alkali lines. The spectrum, however, exhibits weak TiO bands and would therefore not even be considered as extreme in type as an ultra subdwarf. It must be cautioned, though, that mapping subdwarf spectral types into metallicity classes is not as clean as one might at first believe (e.g., \citealt{rajpurohit2014}). As the top two models of fixed gravity in Figure~\ref{seq_sdM_models} show, the more metal-poor star can have a spectrum with stronger TiO bands; in the case of the top model, substantial TiO absorption is seen even at values of [Fe/H] = $-$3.0. 

Figure~\ref{seq_0238_model_fits} shows the two best fitting models to both subdwarfs. The normal esdM2.5 WISE 1243$-$4058 has a best fit of T$_{eff}$=3150K, [Fe/H]=$-$1.5, and log($g$)=4.0; the peculiar esdM3 WISE 0238+3617 has a best fit of T$_{eff}$=3200K, [Fe/H]=$-$2.0, and log($g$)=4.5-5.0. With the exception of the \ion{Na}{1} wings, which are not adequately modeled in these examples\footnote{The \cite{husser2013} paper states only that their code accounted for non-LTE (local thermodynamic equilibrium) effects in line wings of \ion{Na}{1} (and other species). A more proper treatment of the distant wings like that of \cite{burrows2003} was clearly not incorporated, as is obvious from the non-physical Na `D' line shapes in the models in the lower panel of Figure~\ref{seq_0238_model_fits}.}, the fits to WISE 0238+3617 are at least as good as the ones for WISE 1243$-$4058, meaning that no special compensation had to be made to tweak abundances or any other parameter from the default values. The two objects are very close in effective temperature, but WISE 0238+3617 is more metal poor (by 0.5 dex) and much higher in gravity (by 0.75 dex) than WISE 1243$-$4058 despite having a similar TiO/CaH ratio. 

This ratio, which is the basis of both the \cite{gizis1997} and \cite{lepine2007} classification systems, has a fundamental shortcoming: it is sensitive not only to metallicity but also to gravity. In very low metallicity cases, the CaH will be stronger than normal {\it but} the TiO will also be stronger because of the larger column density probed in this less opaque atmosphere. The spectra presented in \cite{gizis1997} covered the wavelength region from 6000-8000 \AA\ and those in \cite{lepine2007} covered 6000-9000 \AA. Both missed the Na `D' doublet at 5896/5890 \AA. Further studies of colder M subdwarfs would benefit from including a measurement of the broadness of the Na `D' line in the classification scheme, since this appears to be a more direct measure of gravity (and hence metallicity). The \ion{Na}{1} doublet at 8183 and 8195 \AA\ could also be used, although this doublet falls in a zone of telluric absorption, making accurate measurement more problematic.

\begin{figure}
\figurenum{78}
\includegraphics[scale=0.40,angle=0]{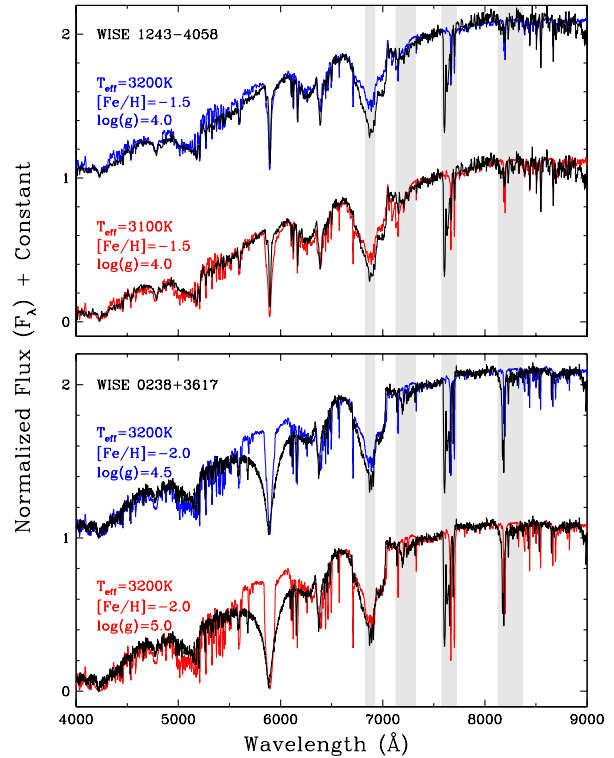}
\caption{Our two best by-eye fits of WISE 1243$-$4058 (top) and WISE 0238+3617 (bottom) to the suite of theoretical spectra given in \cite{husser2013}. Note that the broad, extended wings of the Na `D' doublet are not adequately modeled in this set of predictions.
\label{seq_0238_model_fits}}
\end{figure}

Several obvious follow-up observations can be done to further test this hypothesis. WISE 0238+3617 is quite bright (USNO-B1 $R$=16.65 mag) so soon will have a robust trigonometric parallax measurement from the Gaia mission, enabling an indirect estimate of the stellar radius. Also, higher resolution spectra for a radial velocity meaurement would allow the computation of the total space velocity and would provide possible evidence for membership in the halo population, which could further bolster the claim of extreme metal deficiency.

\end{itemize}

\subsubsection{Carbon Dwarfs\label{carbon_dwarfs}}

The existence of a class of dwarf stars with strong carbon bands was discovered by \cite{dahn1977} when they obtained a spectrum of the motion star G 77-61. Highly processed material in the photosphere of a main sequence object could only be explained via mass transfer from an object now too faint to be seen. Under this scenario, the former (no longer dominant) primary had been a much higher mass star that earlier evolved off of the main sequence. During its asymptotic giant branch phase, processed material from the primary was dredged up and transferred via a stellar wind onto the surface of the lower mass companion. As evolution proceeded, the higher mass star became a white dwarf, which eventually cooled to become the secondary in the system, leaving the now polluted carbon dwarf as the primary. This hypothesis, first proposed by \cite{dahn1977}, has survived closer scrutiny. Radial velocity variations in G 77-61 itself have proven the existence of an unseen companion (\citealt{dearborn1986}), and several other carbon dwarfs are known in which the white dwarf secondary can be identified in the composite spectrum of the system (\citealt{heber1993,liebert1994,green2013}).

Carbon dwarfs cannot be created unless the receiving star is itself low in metallicity. As \cite{dearborn1986} postulated, the spectra of carbon dwarfs appear as they do because the C/O ratio in those objects is flipped. If mass transfer occurred onto an ordinary solar-metallicity K or M dwarf, the oxygen-bearing species already present would continue to overwhelm the carbon in the transferred material unless the amount of transfer was exceedingly large. This means that carbon dwarfs are expected to be low-metallicity (subdwarf) systems. Model fits to the spectrum of G 77-61 give further credence to this idea, although this prototype object's exceptionally low metallicity, [Fe/H]${\approx}-$4  (\citealt{plez2005}; see also \citealt{gass1988}), may be unusual, even for carbon dwarfs (\citealt{green2013}).

\begin{figure}
\figurenum{79}
\includegraphics[scale=0.40,angle=0]{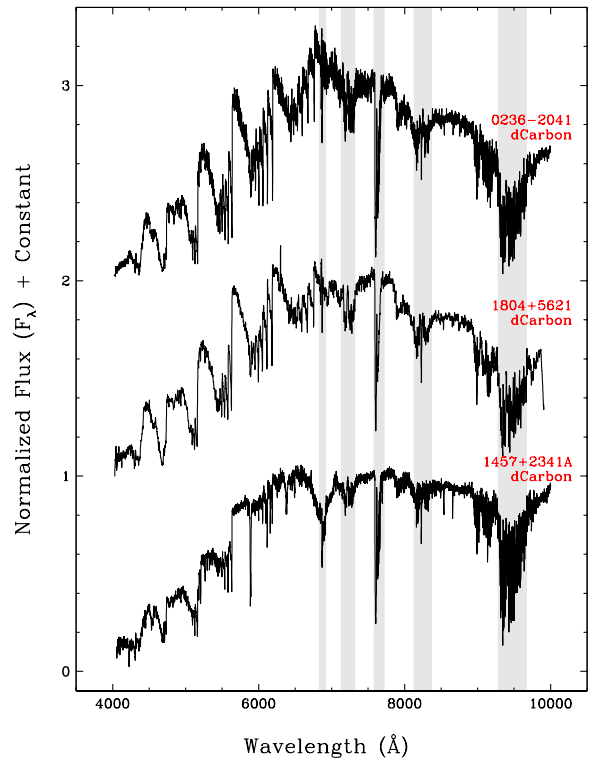}
\caption{Spectra of carbon dwarfs. (See the caption of Figure~\ref{seq_wds1} for other details.) For WISE 0236$-$2041, the spectrum shown is the one from UT 2014 Oct 24, and for WISE 1804+5621 it is the one from UT 2015 Sep 07.
\label{seq_dCarbon}}
\end{figure}

Three of our objects are identified as carbon dwarfs. Spectra of these, illustrated in Figure~\ref{seq_dCarbon}, show the hallmark C$_2$ bands in the 4500-6500 \AA\ region and CN bands at longer wavelengths. WISE 0236$-$2041 was previously cataloged as LP 830-18 but never spectroscopically characterized. The same is true of WISE 1804+5621, which was previously cataloged as NLTT 45912. The third carbon dwarf, WISE 1457+2341A, was previously identified as such through the Sloan Digital Sky Survey by \cite{green2013}. We also obtained a near-infrared spectrum of this carbon dwarf, as shown in Figure~\ref{seq_sdL_usdM_dCarbon_IR} that shows the depressed continuum at longer wavelengths, a sign of the importance of collision-induced absorption by H$_2$ as an important opacity source in these more metal-poor objects.

On building a finder chart for this latter object, shown in Figure~\ref{finder_chart}, we noticed that it had a common-proper-motion companion. Both objects had also been noted as a common-proper-motion pair by \cite{lepine2005}, who designated the system as LSPM J1457+2341NS. For the southern component, which we designate as WISE 1457+2341B (even though it was not detected separately by WISE), we obtain the spectrum shown in both Figure~\ref{seq_sdM65_sdM9.1} and Figure~\ref{seq_sdM_extension} and classify it as an sdM8. To the best of our knowledge, this is the first carbon dwarf discovered in a triple system where the third component can place constraints on the metallicity of the carbon dwarf itself. 

\begin{figure}
\figurenum{80}
\includegraphics[scale=0.325,angle=270]{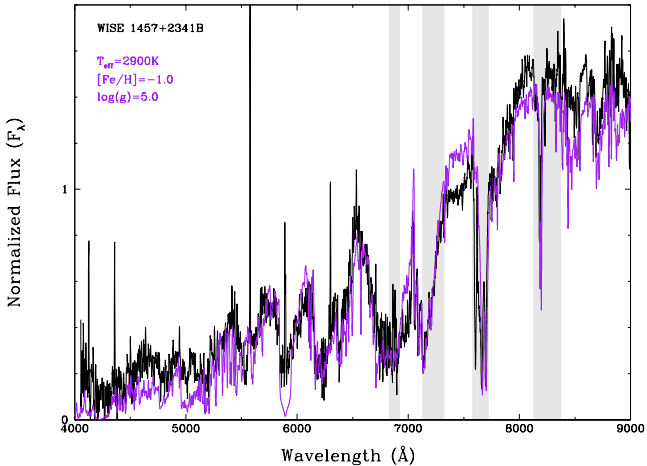}
\caption{Our best model fit to the optical spectrum of WISE 1457+2341B. This object is found to have a metallicity of [Fe/H]${\approx}-$1.0, setting stringent limits on the metallicity of the carbon dwarf primary, WISE 1457+2341A.
\label{seq_1457B_model_fits}}
\end{figure}

Figure~\ref{seq_1457B_model_fits} shows our best-fit attempt to match our spectrum of WISE 1457+2341B to the suite of models from \cite{husser2013}. We find T$_{eff}\approx$2900K, [Fe/H]${\approx}-$1.0, and log($g$)$\approx$5.0. All models fail to fit the TiO/CaH ratio while maintaining a color between 7500\AA\ and 8000\AA\ that is as red as the observed spectrum. Nonetheless, this best fit has parameters similar to the best fits (on higher resolution data) found by \cite{rajpurohit2013} for subdwarfs of similar type: T$_{eff}\approx$3100K, [Fe/H]${\approx}-$1.0, and log($g$)$\approx$5.3 for the sdM7 LHS 377 and T$_{eff}\approx$3000K, [Fe/H]${\approx}-$1.1, and log($g$)$\approx$5.5 for the sdM9.5 SSSPM J1013$-$1356. Future models of the WISE 1457+2341 system will have to account for the amount of mass needed for transfer onto WISE 1457+2341A from its unseen companion given that the system has a metallicity of only [Fe/H]${\approx}-$1.0.

\section{Common-proper-motion systems\label{cpm_systems}}

With many tens of thousands of motion objects now confirmed from the AllWISE Point Source Catalog, we can perform a systematic search for widely separated common-proper-motion systems. Such systems provide valuable checks of theory because a more easily modeled member of the common-proper-motion pair, such as a G dwarf, will have measurable parameters such as age and metallicity that presumably apply to its harder-to-model companion, such as a late-L dwarf, thus setting physical constraints on the modeling of the latter object. These so-called ``benchmark'' systems have been the subject of a large number of recent papers (e.g., \citealt{smith2015,best2015,pavlenko2015,line2015,bowler2015,crepp2015}). As previous sections have demonstrated, AllWISE identifies motion objects running the gamut of the main sequence and all of the known brown dwarf sequence to early-Y dwarfs. Thus, it is ideally suited for finding common-motion systems with disparate spectral types. 

Using the list of discoveries and re-discoveries from this paper, AllWISE1, and NEOWISER; the discovery list from \cite{luhman2014}; and the list of candidate motion objects from Table 6 of \cite{gagne2015}\footnote{The list of motion candidates from \cite{gagne2015} was selected to have 2MASS and {\it WISE} colors consistent with dwarfs of type $\ge$M5 and to be located more than 15$^\circ$ away from the Galactic Plane.}, we have constructed a master set of over 150,000 unique motion sources and motion candidates from which we can identify potentially co-moving systems. We consider a fixed angular separation for all sources since we do not have distance estimates for the vast number of these, and we set this value to be the maximum physical separation at which a G dwarf at a typical distance would be able to retain a low-mass companion for 1 Gyr or more. The simulations of \cite{weinberg1987} show that there is a $>$50\% survivability rate for such a system if the separation is 0.1 pc or less. Empirical evidence supports separations this large or much larger. The M4 dwarf Fomalhaut C lies 0.8 pc away from the A3 dwarf Fomalhaut A and 1.0 pc from the third member of the system, the K4 star Fomalhaut B (\citealt{mamajek2013}), but the age of the system is much younger and the total mass much greater than the typical systems in our survey. \cite{caballero2009}, however, identified a number of old systems having separations between 0.1 and 0.3 pc and primaries with spectral types from F to K. Many other nearby systems, with separations from 1 to 8 pc, have been tabulated by \cite{shaya2011}. So our choice of a 0.1-pc radius for our search can be considered conservative.

Choosing an apparent separation of 20 arcmin as our search radius will enable us to fully search the separation space for primaries more distant than 17.2 pc. Objects with distances closer than this will be searched to physical separations smaller than our intended 0.1-pc cutoff, however. As \cite{schneider2016} show, only 2\% (22 out of 1,006) of the discoveries from the NEOWISER Motion Survey are suspected to lie within $\sim$20 pc of the Sun. Therefore, objects with distances within 17.2 pc are few enough in number that larger areal searches can be conducted individually later as these objects become known.  

With these constraints in mind, we used the Tool for Operations on Catalogs and Tables (TopCat\footnote{See {\url http://www.star.bristol.ac.uk/$\sim$mbt/topcat/} for more information.}) to conduct a positional cross-match of sources in our master set, which resulted in over 28,000 matched groups. We then wrote code to identify group members for which both the RA and Dec components of the 2MASS-to-AllWISE motion measurements overlapped at the 2$\sigma$ level. To keep the number of spurious matches among slower moving sources to a minimum, we further required that the total motion of at least one component in the system exceeds 150 mas yr$^{-1}$. Finder charts like the one shown in Figure~\ref{finder_chart} were created for each system to verify further that there was common motion between components\footnote{By blinking between the charts for each component in the pair, one can determine whether or not the objects co-align at different epochs (i.e., share common proper motion).}. The older imaging data give a glimpse of the sky 15 to 50 years earlier than the 2MASS data and therefore provide an excellent check of the co-moving hypothesis. Objects with clearly divergent motions over this longer time baseline were therefore eliminated from the list. Also, the finder charts enabled us to test the reality of the motion candidates from \cite{gagne2015}; many objects originating from this list were found to be blended or extended sources and were also eliminated. The result is a final list of 1,039 candidate motion systems\footnote{Because the match groups are constructed before the motion hypothesis test, it is possible to identify co-moving binaries for which the separation is larger than 20 arcmin because linked members of the match group may fail the finder chart test. For example, star A and star B fall within 20 arcmin of one another, as do star B and star C. Stars A and C are more than 20 arcmin apart, but all three stars are in the same match group. If star B fails the finder chart test, then star A and C are left in the match group as a possible co-moving pair despite the fact that they fail to meet the $<$20 arcmin separation requirement. Only one such case -- system \#754 with a separation of 1346$\farcs$7 -- however, remains in our list.}, which is given in Table 11.

\begin{turnpage}
\begin{deluxetable*}{cccccccccccc}
\tabletypesize{\tiny}
\tablewidth{8.0in}
\tablenum{11}
\tablecaption{Candidate and Re-discovered Common-proper-motion Systems\label{cpm_systems}}
\tablehead{
\colhead{Sys.} &  
\colhead{Comp.} &     
\colhead{WISEA Designation} &                          
\colhead{2MASS $J$} &  
\colhead{2MASS $H$} &     
\colhead{2MASS $K_s$} &
\colhead{W1} &
\colhead{W2} &
\colhead{$\mu_\alpha$\tablenotemark{a}} &
\colhead{$\mu_\delta$\tablenotemark{a}} &
\colhead{Flag\tablenotemark{b}} &
\colhead{Sep.\tablenotemark{c}} \\
\colhead{No.} &                          
\colhead{No.} &                          
\colhead{} &                          
\colhead{(mag)} &  
\colhead{(mag)} &     
\colhead{(mag)} &
\colhead{(mag)} &
\colhead{(mag)} &
\colhead{(mas/yr)} &
\colhead{(mas/yr)} &
\colhead{} &
\colhead{(arcsec)} \\
\colhead{(1)} &                          
\colhead{(2)} &  
\colhead{(3)} &     
\colhead{(4)} &
\colhead{(5)} &
\colhead{(6)} &
\colhead{(7)} &
\colhead{(8)} &
\colhead{(9)} &
\colhead{(10)} &
\colhead{(11)} &
\colhead{(12)}
}
\startdata
  1&  1.1& J000136.86$-$010146.9& 12.363$\pm$0.021& 11.831$\pm$0.02 & 11.565$\pm$0.021& 11.374$\pm$0.022& 11.202$\pm$0.021&  -38.5$\pm$11.4& -233.1$\pm$10.5& A&    0.0 \\
---&  1.2& J000035.38$-$011248.8& 13.822$\pm$0.03 & 13.226$\pm$0.022& 12.899$\pm$0.027& 12.724$\pm$0.024& 12.506$\pm$0.024&  -24.9$\pm$9.9 & -246.1$\pm$8.2 & A& 1135.1 \\        
  2&  2.1& J000341.71$-$282347.8&   6.97$\pm$0.023&  6.621$\pm$0.024&  6.554$\pm$0.016&  6.515$\pm$0.078&  6.484$\pm$0.026&  289.7$\pm$6.7 & -158.7$\pm$6.7 & A&    0.0 \\      
---&  2.2& J000342.53$-$282242.7& 13.068$\pm$0.024& 12.376$\pm$0.028& 11.972$\pm$0.025& 11.705$\pm$0.024& 11.519$\pm$0.023&  288.5$\pm$6.9 & -150.5$\pm$6.8 & A&   66.0 \\    
  3&  3.1& J000502.68+682211.9&   12.153$\pm$0.023& 11.611$\pm$0.032& 11.287$\pm$0.023& 11.061$\pm$0.023& 10.892$\pm$0.02 &  194.6$\pm$8.1 &  -37.3$\pm$6.5 & A&    0.0 \\      
---&  3.2& J000304.52+680700.6&   14.117$\pm$0.029& 13.517$\pm$0.04 & 13.193$\pm$0.032& 13.028$\pm$0.024& 12.868$\pm$0.025&  168.7$\pm$7.6 &  -56.7$\pm$6.8 & A& 1123.4 \\      
  4&  4.1& J000537.09$-$013955.0& 11.859$\pm$0.023& 11.306$\pm$0.024& 11.072$\pm$0.025& 10.926$\pm$0.023& 10.755$\pm$0.021&  324.6$\pm$6.7 &  131.0$\pm$5.8 & A&    0.0 \\      
---&  4.2& J000536.59$-$013937.5& 12.875$\pm$0.023& 12.354$\pm$0.025& 12.064$\pm$0.027& 11.896$\pm$0.023& 11.701$\pm$0.023&  322.7$\pm$6.8 &  136.6$\pm$5.9 & N&   19.0 \\
  5&  5.1& J000556.08$-$610412.8& 10.527$\pm$0.024&  9.903$\pm$0.021&  9.679$\pm$0.021&  9.531$\pm$0.023&  9.365$\pm$0.02 &  502.9$\pm$6.9 &   49.0$\pm$6.9 & A&    0.0 \\      
---&  5.2& J000557.22$-$610354.7& 12.041$\pm$0.024& 11.432$\pm$0.021& 11.183$\pm$0.019& 10.959$\pm$0.023& 10.766$\pm$0.02 &  501.2$\pm$7.0 &   46.1$\pm$7.0 & A&   19.9 \\
  6&  6.1& J000647.41$-$085238.8& 11.967$\pm$0.023& 11.434$\pm$0.024&  11.09$\pm$0.023& 10.884$\pm$0.022& 10.66 $\pm$0.022&  -67.1$\pm$9.8 & -315.5$\pm$8.2 & A&    0.0 \\      
---&  6.2& J000649.11$-$085249.5& 14.143$\pm$0.03 & 13.551$\pm$0.028& 13.132$\pm$0.039& 12.762$\pm$0.023& 12.353$\pm$0.027&  -61.9$\pm$10.0& -324.9$\pm$8.4 & A&   27.4 \\      
  7&  7.1& J000843.53+660756.3&   11.076$\pm$0.024& 10.428$\pm$0.027& 10.223$\pm$0.021&  10.16$\pm$0.023&  10.09$\pm$0.02 &  174.8$\pm$9.0 &   24.8$\pm$7.3 & A&    0.0 \\
---&  7.2& J000838.88+660801.3&    13.88$\pm$0.03 & 13.351$\pm$0.033& 13.133$\pm$0.03 & 12.989$\pm$0.024& 12.801$\pm$0.024&  176.0$\pm$9.2 &   19.0$\pm$7.6 & A&   28.7 \\      
  8&  8.1& J000903.07+273907.3&    9.419$\pm$0.024&  8.777$\pm$0.036&  8.662$\pm$0.021&  8.575$\pm$0.023&  8.613$\pm$0.02 &  221.7$\pm$8.0 &  152.7$\pm$6.3 & A&    0.0 \\
---&  8.2& J000859.74+273959.9&   10.428$\pm$0.02 &  9.812$\pm$0.022&  9.595$\pm$0.017&  9.489$\pm$0.022&  9.403$\pm$0.02 &  222.5$\pm$6.1 &  140.6$\pm$6.0 & A&   68.8 \\      
  9&  9.1& J000945.01+235635.3&    13.19$\pm$0.02 &  12.59$\pm$0.02 &   12.3$\pm$0.02 & 12.075$\pm$0.023& 11.878$\pm$0.022&  149.5$\pm$11.2&  -13.2$\pm$18.1& G&    0.0 \\      
---&  9.2& J000946.16+235636.5&    13.24$\pm$0.02 &  12.66$\pm$0.02 &  12.37$\pm$0.02 & 12.191$\pm$0.024& 12.019$\pm$0.023&  148.8$\pm$11.2&  -18.0$\pm$18.1& G&   15.8 \\
 10& 10.1& J000948.68$-$405335.6&  7.417$\pm$0.023&   7.16$\pm$0.029&  7.059$\pm$0.02 &  6.935$\pm$0.052&  7.054$\pm$0.02 &  123.0$\pm$6.3 & -110.9$\pm$6.3 & A&    0.0 \\     
---& 10.2& J000923.50$-$410241.6&  7.703$\pm$0.025&  7.353$\pm$0.034&  7.247$\pm$0.026&   7.09$\pm$0.036&  7.254$\pm$0.019&  148.6$\pm$7.1 & -124.2$\pm$6.2 & A&  616.0 \\      
 11& 11.1& J001056.25+480637.4&    7.373$\pm$0.019&  7.143$\pm$0.018&  7.096$\pm$0.027&  6.985$\pm$0.05 &  7.078$\pm$0.02 &  168.7$\pm$6.7 &  -11.3$\pm$5.9 & A&    0.0 \\       
---& 11.2& J001058.00+480653.7&    8.848$\pm$0.021&  8.301$\pm$0.026&  8.212$\pm$0.029&   8.13$\pm$0.023&  8.143$\pm$0.019&  163.4$\pm$6.5 &  -14.9$\pm$5.8 & A&   24.0 \\
 12& 12.1& J001233.65+214245.3&    8.837$\pm$0.02 &  8.277$\pm$0.046&  8.042$\pm$0.027&  7.944$\pm$0.023&  7.905$\pm$0.02 &  190.0$\pm$12.8& -290.7$\pm$10.3& N&    0.0 \\       
---& 12.2& J001234.60+214219.7&    9.662$\pm$0.02 &  9.098$\pm$0.021&  8.863$\pm$0.021&  8.708$\pm$0.022&  8.554$\pm$0.02 &  182.8$\pm$12.8& -283.1$\pm$10.3& N&   28.7 \\
 13& 13.1& J001303.32$-$412738.4& 12.321$\pm$0.023& 11.781$\pm$0.025& 11.537$\pm$0.025& 11.453$\pm$0.022& 11.262$\pm$0.02 &  278.2$\pm$7.2 & -117.8$\pm$6.4 & A&    0.0 \\      
---& 13.2& J001257.18$-$412945.5& 14.415$\pm$0.027& 13.836$\pm$0.036& 13.755$\pm$0.051& 13.544$\pm$0.024& 13.325$\pm$0.027&  272.8$\pm$7.7 & -126.5$\pm$6.9 & N&  144.7 \\
 14& 14.1& J001339.84+803958.6&    7.756$\pm$0.034&  7.131$\pm$0.047&  6.904$\pm$0.02 &  6.651$\pm$0.068&  6.707$\pm$0.022&  260.6$\pm$8.9 &  173.6$\pm$7.1 & N&    0.0 \\       
---& 14.2& J001344.00+803951.1&   10.936$\pm$0.024&  10.37$\pm$0.032& 10.059$\pm$0.026&  9.654$\pm$0.036&  9.408$\pm$0.034&  236.8$\pm$10.4&  175.8$\pm$8.7 & N&   12.6 \\
 15& 15.1& J001505.10+425047.8&    13.03$\pm$0.02 &  12.45$\pm$0.02 &  12.07$\pm$0.02 & 11.866$\pm$0.022& 11.684$\pm$0.022&  -63.3$\pm$4.5 & -162.8$\pm$9.6 & G&    0.0 \\       
---& 15.2& J001515.40+423927.0&    14.65$\pm$0.03 &  13.77$\pm$0.03 &  13.33$\pm$0.03 & 12.905$\pm$0.024& 12.640$\pm$0.025&  -66.6$\pm$4.8 & -143.7$\pm$9.2 & G&  690.2 \\
 16& 16.1& J001506.05+295537.9&   11.094$\pm$0.022& 10.509$\pm$0.02 &  10.22$\pm$0.019& 10.049$\pm$0.021&  9.877$\pm$0.02 &  375.9$\pm$8.2 & -221.1$\pm$6.7 & A&    0.0 \\
---& 16.2& J001502.38+295929.9&   16.158$\pm$0.081& 15.226$\pm$0.084& 14.482$\pm$0.073& 13.733$\pm$0.026& 13.388$\pm$0.032&  384.2$\pm$8.8 & -218.2$\pm$8.7 & A&  236.9 \\          
\enddata
\tablecomments{Only a portion of this table is shown here to demonstrate its form and content. A machine-readable version of the full table is available online.}
\tablenotetext{a}{This is the motion measured between the 2MASS and AllWISE epochs.}
\tablenotetext{b}{This is the source for the motion measurement and association to 2MASS: A = AllWISE (\citealt{kirkpatrick2014} and this paper), N=NEOWISER (\citealt{schneider2016}), G=\cite{gagne2015}, L=\cite{luhman2014}.}
\tablenotetext{c}{This is the measured separation between this source and the brighter/brightest component at $J$-band (which has the suffix ``.1'' in the Component Number column).}
\end{deluxetable*}
\end{turnpage}

To test our methods, we have checked our results against common-proper-motion pairs identified by \cite{kirkpatrick2014}, \cite{luhmansheppard2014}, and \cite{schneider2016}. Of the 80 unique pairs tabulated by those authors, we recover 58. Of the 22 systems not recovered, 14 have only one component in our master set, 7 have motions that fail to overlap at the 2$\sigma$ level, and 1 has a total motion below our threshold of 150 mas yr$^{-1}$. That is, our technique identified 100\% of the motion pairs that it could have recovered. The fact that 14 systems had only one component in our master set is a reflection of the fact that the AllWISE-generated lists themselves are incomplete both for very bright sources with poor astrometry (or low $w1nm/w1m$ and $w2nm/w2m$ values; see section~\ref{criteria}) and for closely separated sources that are not resolved by {\it WISE}. Such motion pairs were identified by the original authors as serendipitous discoveries during finder chart checks, and not strictly through association of {\it WISE}-detected sources. (Indeed, in Table 11
we identify eight systems -- \#144, \#219, \#481, \#772, \#805, \#807, \#888, and \#1035 -- for which a third co-moving member was identifed on the finder charts but was not present in our master set.) 

We spectroscopically observed a number of these possible co-moving systems to obtain spectrophotometric distance estimates for the components:

\begin{itemize}

\item System \#464 is comprised of WISE 1055$-$5750 and 1056$-$5750, which we classify as two M4 dwarfs. Our distance estimates from Table 10
place the former at 37$\pm$3 pc and the latter at 35$\pm$3 pc, which supports physical association. The separation of 253$\farcs$8 corresponds to $\sim$9100 AU. WISE 1056$-$5750 is cataloged as UPM J1056$-$5750 by \cite{finch2010} but the other component is new.

\item System \#494 is comprised of WISE 1133$-$4139 (K0) and 1133$-$4140 (M0.5). For a K0 dwarf, \cite{pecaut2013} give $M_J = 4.29$ mag, which implies a distance estimate of $\sim$95 pc for WISE 1133$-$4139. Our estimate for WISE 1133$-$4140 (Table 10)
is 122$\pm$8 pc, which is discrepant from the primary's estimate by over 3$\sigma$. However, such estimates assume that these are single stars; a re-derivation of the distance for the primary, assuming it has an unresolved companion, places the likely range to be 95-135 pc, which encompasses the distance estimate for the secondary. The angular separation of 22$\farcs$3 implies an apparent physical separation of $\sim$2700 AU. Although the proper motion of the primary (also known as CD$-$40 6796) has been reported in \cite{hog2000}, the secondary is new.

\item System \#525 is comprised of WISE 1210$-$4619 (G0) and 1210$-$4612 (M1). For a G0 dwarf, \cite{pecaut2013} list $M_J = 3.29$ mag, which gives a distance estimate of $\sim$42 pc. Our estimate for WISE 1210$-$4612 (Table 10)
is 52$\pm$4 pc, which agrees to within 2.5$\sigma$ of the primary's distance. The apparent physical separation of the pair is 432$\farcs$8, or 20,300 AU (0.1 pc). Although the primary is a Luyten proper motion star, the secondary's proper motion is being reported for the first time.

\item System \#547, which has a large magnitude difference of ${\Delta}J = 6.71$ mag, consists of WISE 1240+2048 (K2) and 1240+2047 (M7). The absolute $J$ magnitude from \cite{pecaut2013} implies $\sim$37 pc for the K2, and our estimate for the M7 (Table 10)
is 47$\pm$3 pc. This 3$\sigma$ difference could be easily resolved if the primary is itself a close double, as a pair of equal-magnitude K2 dwarfs would push the primary's distance to $\sim$52 pc. Indeed, the primary, also known as G 59-32, is a double-lined spectroscopic binary (\citealt{pourbaix2004}) with a mass ratio of nearly unity ($q = 0.91{\pm}0.01$; \citealt{mazeh2003}). The apparent physical separation with the M7 is 112$\farcs$7, or 5300 AU. The M7 component was not noted prior to {\it WISE} observations.

\item System \#642 includes WISE 1404$-$5924 and 1403$-$5923, a pair of M3 dwarfs. Our distance estimates (Table 10)
of 42$\pm$3 and 43$\pm$3 pc, respectively, support physical association. The 78$\farcs$1 separation corresponds to 3300 AU. Although the primary is cataloged as motion star L 197-165, the secondary is an AllWISE discovery.

\item System \#705 is WISE 1454+0053 AB. The primary is Wolf 559 (M3) and the secondary is TVLM 868-20073 (M9). The distance estimates from (Table 10)
are 46$\pm$3 and 49$\pm$3 pc, which are in excellent agreement, and the 21$\farcs$6 separation corresponds to 1000 AU. \cite{smith2014} also reported this pair as a likely common-proper-motion system.

\item System \#821 is comprised of WISE 1718$-$2245 (M3.5) and 1718$-$2246 (M4.5). Our distance estimates (Table 10)
of 31$\pm$2 and 22$\pm$2 pc, respectively, are discrepant by over 4$\sigma$, and it should be noted that the brighter component in the 2MASS and {\it WISE} bands is the M4.5. If the secondary is assumed to be an equal-magnitude double, this would push its distance out to $\sim$30 pc, removing the discrepancy. The pair's separation is 54$\farcs$3, or 1600 AU. \cite{finch2012} also noted this pair as a possible common-proper-motion system and likewise found discrepant distance estimates: 25.4 pc for the primary and 13.2 pc for the secondary based on photometric relations alone. Those authors do not, however, speculate on a cause for the discrepancy.

\item System \#920 is WISE 2101$-$4907 and 2101$-$4906, which have ${\Delta}J = 5.00$ mag. The primary is an M4.5 and the secondary is a cool white dwarf. Our distance estimate to the former is 13$\pm$1 pc, placing the cool white dwarf within easy reach of trigonometric parallax monitoring. The apparent physical separation of the pair is 64$\farcs$4, or 800 AU. The only other mention of this secondary is by \cite{wroblewski1994}, who designate it as WT 765, but they do not associate it with the nearby motion star.

\end{itemize}

\subsection{Systems with L or T Dwarf Members}

We identify forty-three systems in Table 11
as having at least one component with colors indicating a possible L or T dwarf. The four systems (system \#487, \#623, \#841, and \#951) with possible T dwarf components have been previously published as benchmark systems with K or M dwarf primaries and T dwarf secondaries. Fifteen other systems have already been identified and have secondaries of either late-M or L type. These nineteen previously published systems are summarized in Table 12.

\begin{deluxetable*}{cccccc}
\tablenum{12}
\tablecaption{Previously Published Common-proper-motion Systems with Late-M, L, or T Dwarf Secondaries\label{published_cpm_systems}}
\tablehead{
\colhead{Sys.} &  
\colhead{Name} &                          
\colhead{Sep.} &  
\colhead{Prim.} &     
\colhead{Sec.} &
\colhead{Ref.} \\
\colhead{No.} &  
\colhead{} &                          
\colhead{(arcsec)} &  
\colhead{Type} &     
\colhead{Type} &
\colhead{}\\
\colhead{(1)} &                          
\colhead{(2)} &  
\colhead{(3)} &     
\colhead{(4)} &
\colhead{(5)} &
\colhead{(6)} 
}
\startdata
 16&  NLTT 730 AB         &   236.9& M4  & L7.5 pec (blue)& 1 \\
 20&  NLTT 1011 AB        &    55.5& K7  & L2             & 1 \\
 24&  K\"onigstuhl 1 AB   &    77.8& M6  & M9.5           & 2 \\
 44&  NLTT 2274 AB        &    23.0& M4  & L0             & 3 \\
125&  G 3-40 AB           &    72.3& M1.5& L3             & 3,4 \\
292&  2MASS 0525$-$74 AB  &    43.8& M3  & L2             & 5 \\
352&  2MASS 0719$-$50 AB  &    58.7& M3.5& L0             & 6 \\
397&  NLTT 20640 AB       &    15.2& M4  & L0             & 4 \\ 
487&  LHS 302 AB          &   254.9& M5  & T6             & 7 \\ 
591&  G 62-33 AB          &    66.5& K0  & L2.5           & 3 \\ 
601&  G 255-34 AB         &    43.8& K8  & L2             & 8 \\ 
623&  LHS 2803 AB         &    67.2& M4.5& T5.5           & 5,9 \\
630\tablenotemark{a}&  NLTT 35593 AB       &  1106.1& M2  & L2             & 1 \\
663&  G 200-28 AB         &   569.8& G5  & L5.5           & 3 \\ 
766&  G 225-36 AB         &   121.8& K5  & M9             & 10 \\ 
837&  G 259-20 AB         &    29.6& M2.5& L5             & 11 \\
841&  G 204-39 AB         &   196.9& M3  & T6.5           & 3 \\
951&  $\epsilon$ Ind ABaBb&   403.2& K5  & T1/T6          & 12,13,14 \\
958&  PM I22118$-$1005 AB &   204.5& M2  & L1.5           & 1 \\
\enddata
\tablecomments{References establishing physical association or possibility thereof: 
(1) \cite{deacon2014}, (2) \cite{caballero2007}, (3) \cite{faherty2010}, (4) \cite{zhang2010}, (5) \cite{muzic2012}, (6) \cite{andrei2011}, (7) \cite{kirkpatrick2011}, (8) \cite{gomes2013}, (9)  \cite{deacon2012}, (10) \cite{pinfield2006}, (11) \cite{luhman2012}, (12)  \cite{volk2003}, (13) \cite{scholz2003}, (14) \cite{mccaughrean2004}.}
\tablenotetext{a}{\cite{deacon2014} find this pair to be likely unassociated.}
\end{deluxetable*}

The remaining twenty-four candidate systems are listed in Table 13.
Notes on individual systems of interest are given below:

\begin{turnpage}
\begin{deluxetable*}{clcclccrr}
\tabletypesize{\tiny}
\tablewidth{8.0in}
\tablenum{13}
\tablecaption{Candidate Common-proper-motion Systems with Possible L Dwarf Components\label{Ltype_cpm_systems}}
\tablehead{
\colhead{Sys.} &  
\colhead{Name of Primary} &                          
\colhead{Spec.Ty.} &  
\colhead{Ref.} &     
\colhead{Name of Possible} &
\colhead{Spec.Ty.} &  
\colhead{Ref.} &     
\colhead{${\Delta}J$} &
\colhead{Sep.} \\
\colhead{No.} &  
\colhead{} &                          
\colhead{of Primary} &  
\colhead{} &     
\colhead{Companion} &
\colhead{of Comp.} &  
\colhead{} &     
\colhead{(mag)} &
\colhead{(arcsec)} \\
\colhead{(1)} &                          
\colhead{(2)} &  
\colhead{(3)} &     
\colhead{(4)} &
\colhead{(5)} &
\colhead{(6)} &
\colhead{(7)} &
\colhead{(8)} &
\colhead{(9)} 
}
\startdata
15  &     LP 192-44                  & --- &       ---&  WISEA J001515.40+423927.0  & ---            & ---&    1.62&   690.2\\
23  &     WISEA J002029.66$-$153527.6& --- &       ---&  WISEA J002050.25$-$151913.1& ---            & ---&    0.62&  1019.1\\
31  &     LP 465-20                  & --- &       ---&  2MASS J00283943+1501418    & L4.5           &   1&    3.37&   917.2\\
74  &     LEHPM 1437                 & --- &       ---&  WISEA J011959.44-423310.1  & ---            & ---&    3.83&   621.7\\
93  &     WISEA J013616.64+151319.1  & --- &       ---&  WISEA J013708.17+152549.6  & ---            & ---&    2.44&  1057.8\\
130 &     LSPM J0209+0732            & --- &       ---&  WISEA J020934.30+073219.4  & ---            & ---&    5.02&    27.0\\
146 &     WISEA J022118.51$-$191157.5& --- &       ---&  WISEA J022044.13$-$191205.1& ---            & ---&    0.42&   487.1\\
179 &     LHS 1470                   & --- &       ---&  LHS 1469                   & DQ             &   2&    1.89&    57.2\\     
220 &     G 77-56AB\tablenotemark{a} & late G&       9&  WISEA J032838.73+015517.7  & ---            & ---&    7.58&   630.8\\     
257 &     SIPS J0427$-$1547          & --- &       ---&  SIPS J0427$-$1548          & ---            & ---&    2.41&    38.8\\
345 &     TYC 4530-1008-1            & --- &       ---&  WISEA J065935.80+771457.8  & ---            & ---&    7.53&  1157.4\\
425 &     2MASS J09581512+4519524    & M4.5&         3&  SDSS J095932.74+452330.5   & L3-L4$\beta$\tablenotemark{b}&   4&    3.66&   846.7\\
455 &     LP 490-57                  & --- &       ---&  WISEA J104335.09+121312.0  & L9             &   5&    4.20&  1039.6\\          
508 &     WISEA J115043.79$-$103636.2& --- &       ---&  WISEA J114942.15$-$103426.3& ---            & ---&    1.86&   918.1\\
625 &     UPM J1349$-$4228           & --- &       ---&  WISEA J134954.75$-$422451.7& ---            & ---&    3.05&   637.2\\
... &     ...                        & ... &       ...&  WISEA J134824.42-422744.9  & L2             &   8&    5.47&   410.1\\
660 &     LSPM J1415+0626            & --- &       ---&  WISEA J141618.87+062133.9  & ---            & ---&    6.41&   986.6\\
680 &     WISEA J143228.56$-$032422.2& --- &       ---&  WISEA J143135.30$-$031311.0& ---            & ---&    2.61&  1042.4\\
685 &     G 200-51                   & --- &       ---&  WISEA J143443.19+501121.3  & ---            & ---&    6.87&   789.9\\
706 &     WISEA J145452.27$-$090106.8& --- &       ---&  WISEA J145449.83$-$090432.1& ---            & ---&    1.76&   208.4\\
814 &     LP 226-36                  & --- &       ---&  2MASS J17073334+4301304    & L0.5           &   6&    1.11&   986.5\\
918 &     WISEA J205811.51$-$371104.5& --- &       ---&  WISEA J205811.90$-$371029.5& ---            & ---&    5.48&    35.3\\
937 &     LP 818-26                  & --- &       ---&  WISEA J214335.16$-$183223.1& ---            & ---&    3.29&   635.6\\
1018&     WISEA J233448.79$-$275558.9& --- &       ---&  WISEA J233450.24$-$275559.7& ---            & ---&    1.38&    19.3\\
1032&     Gl 908.1                   & K5 V&         7&  2MASS J23512200+3010540    & L5 pec (red)   &   5&    8.76&   934.9\\
\enddata
\tablecomments{References for spectral types: (1) \cite{kirkpatrick2000}, (2) \cite{reid2005}, (3) \cite{bochanski2005}, (4) \cite{hinkley2013}, (5) \cite{kirkpatrick2010},
(6) \cite{cruz2003}, (7) \cite{stephenson1986}, (8) This paper, (9) Spectral type is inferred from the measurement of $T_{eff}$ = 5250K from \cite{latham2002}.}
\tablenotetext{a}{The primary is a double-lined spectroscopic binary and metal poor ([m/H] = $-$0.43) according to \cite{carney1994}.}
\tablenotetext{b}{The $\beta$ suffix is assumed from the comment in \cite{hinkley2013} that this object has an age of $\sim$150 Myr.}
\end{deluxetable*}
\end{turnpage}
      
\begin{itemize}

\item System \#31 is comprised of LP 465-20 and 2MASS J00283943+1501418. The 2MASS object is an optical L4.5 dwarf which was not resolved during high-resolution {\it Hubble Space Telescope} imaging by \cite{gizis2003}. Using the $M_J$ vs.\ spectral type relation of \cite{looper2008}, we estimate a distance to this L dwarf of 40$\pm$3 pc, assuming it is a single object. LP 465-20 has no spectral type in the published literature, but its colors of $J-K_s$ = 0.88$\pm$0.04, $J-$W2 = 1.27$\pm$0.04, and W1$-$W2 = 0.18$\pm$0.03 mag place it very roughly around M6-M7 according to Figure~\ref{color_color_JKs_JW2} through Figure~\ref{color_mag_J_JW2}. This places LP 465-20 at $\sim$30-38 pc, which is in rough agreement with the distance estimate of the L dwarf. If associated, these two objects have an apparent physical separation of $\sim$37,000 AU, or $\sim$0.2 pc.

\item System \#146 consists of WISEA J022118.51$-$191157.5 and WISEA J022044.13$-$191205.1. The brighter component has ($J-K_s$, $J-$W2, W1$-$W2) = (1.51$\pm$0.10, 2.27$\pm$0.07, 0.27$\pm$0.04) mag and the fainter one has ($J-K_s$, $J-$W2, W1$-$W2) = (1.13$\pm$0.22, 1.65$\pm$0.14, 0.19$\pm$0.04) mag. Note that it is the brighter component that has the redder colors. These colors suggest types of roughly L3 and M8. The distance estimates for these two objects cannot be reconciled even if we invoke the primary as an equal-magnitude double. In this case, we derive distances of 65$\pm$4 pc for the brighter component and 115$\pm$8 pc for the fainter one. We believe that these two sources are unassociated; the possible association is merely a consequence of the fact that the motion for the fainter object has very large uncertainties.

\item System \#179 is the published common-proper-motion system LHS 1470 and LHS 1469. The brighter component has no published spectral type, and the fainter component, despite superficially having the colors of an L dwarf, is a white dwarf of type DQ (\citealt{reid2005}). Specifically, the colors of LHS 1469 are ($J-K_s$, $J-$W2, W1$-$W2) = (1.27$\pm$0.04, 1.97$\pm$0.04, 0.31$\pm$0.03) mag, although the object, unlike an L dwarf, is clearly seen in the $B$ (and $R$) band images of the DSS. Because the classification spectrum from \cite{reid2005} covers only a very small wavelength range ($\sim$4000-5400 \AA\ and $\sim$6300-7500 \AA), we hypothesize that this object may be a DQ + early-L dwarf composite. Spectra across a broader swath of the optical and near-infrared are needed to confirm this.

\item System \#220 is comprised of G 77-56AB and WISEA J032838.73+015517.7. G 77-56AB is a late-G dwarf that is a single-lined spectroscopic binary (\citealt{latham2002}) and has been measured to be slightly metal poor ([m/H] = $-$0.43; \citealt{carney1994}). An L dwarf companion, if confirmed, would provide an excellent benchmark subdwarf with an established metallicity. Assuming $M_J \approx$ 4.0 mag for a late-G star (\citealt{pecaut2013}) implies a distance of $\sim$100 pc for G 77-56AB, which in turn implies an absolute $J$ magnitude of $\sim$11.6 mag for the {\it WISE} object. This matches expectations since the {\it WISE} object has the $J-K_s$ color of a late-M or early-L dwarf. We caution, however, that the motion measurement for WISEA J032838.73+015517.7 has very large uncertainties, so the pair may not represent a physical system.

\item System \#425 consists of the M4.5 dwarf 2MASS J09581512+4519524 and the young L3-L4 dwarf SDSS J095932.74+452330.5. If associated, this system could be a valuable check of theory since the L dwarf shows the hallmarks of youth. However, we note that the spectrophotometric distance estimates are discrepant for the two components: 54$\pm$3 pc for the primary and 38$\pm$6 pc for the secondary, and that the motions for the two components show disagreement near our tolerance threshold. Refined astrometry is needed to confirm/refute this pair.

\item System \#455 is made up of LP 490-57 and the L9 dwarf WISEA J104335.09+121312.0. LP 490-57 has colors of ($J-K_s$, $J-$W2, W1$-$W2) = (0.76$\pm$0.03, 1.13$\pm$0.03, 0.22$\pm$0.03) mag, which suggests a type of M3-M4 and distance spanning a rather large range of 44-87 pc. The L dwarf, however, has a distance estimate of only 17$\pm$1 pc based on the near-infrared L9 classification of \cite{kirkpatrick2010}. (The more uncertain near-infrared classification of \citealt{chiu2006} gives L7$\pm$1.) These two objects are almost certainly not related. 

\item System \#625 is a possible triple composed of UPM J1349$-$4228, WISEA J134954.75$-$422451.7, and WISEA J134824.42$-$422744.9. The latter object has a classification of L2 from Table 4 
and a distance estimate of 34$\pm$2 pc from Table 10.
\cite{finch2012} give a photometric distance estimate of 24.4 pc for UPM J1349$-$4228; we find colors of ($J-K_s$, $J-$W2, W1$-$W2) = (0.83$\pm$0.03, 1.12$\pm$0.03, 0.14$\pm$0.03) mag for this object, which implies a type between K7 and M3, which translates into a large distance range of 29-58 pc. The other component, WISEA J134954.75$-$422451.7, has colors of ($J-K_s$, $J-$W2, W1$-$W2) = (0.74$\pm$0.04, 1.12$\pm$0.03, 0.21$\pm$0.03) mag, which implies a type between K5 and M3 and a much larger distance range of 120-300 pc. We conclude that WISEA J134954.75$-$422451.7 is not part of the same system as UPM J1349$-$4228 and WISEA J134824.42$-$422744.9 but that the latter two objects need additional data to confirm/refute physical association. Using the distance estimate of the L dwarf, we find that this twosome would have an apparent physical separation of 14,000 AU if confirmed.

\item System \#814 refers to LP 226-36 and the L0.5 dwarf 2MASS J17073334+4301304. Using the \cite{looper2008} $M_J$ vs.\ spectral type relation, we find a distance estimate of 26$\pm$2 pc for the L dwarf. LP 226-36 has colors of ($J-K_s$, $J-$W2, W1$-$W2) = (0.87$\pm$0.03, 1.31$\pm$0.03, 0.21$\pm$0.03) mag, which implies a type of roughly M5-M7 and gives it a crude distance estimate of 27-64 pc. If confirmed as a true system, the apparent physical separation is 26,000 AU (0.1 pc), assuming the distance estimate for the L dwarf.

\item System \#1032 has a large magnitude difference of ${\Delta}J$ = 8.76$\pm$0.10 mag and is comprised of the K5 dwarf Gl 908.1 and the L dwarf 2MASS J23512200+3010540. The K dwarf has a Hipparchos-measured distance of 23.9$\pm$0.6 pc (\citealt{vanleeuwen2007}). The L dwarf is classified in the optical as an L5.5 and in the near-infrared as a peculiar red L5 (\citealt{kirkpatrick2010}); using the optical type, which historically yields better distance estimates, we find a distance of 24$\pm$2 pc. The two distance estimates are in excellent agreement. The on-sky separation implies an apparent physical separation of 22,300 AU (0.1 pc) for the pair. Interestingly, \cite{gagne2014} originally reported the L dwarf as a possible member of the Argus Association, but then reassigned it as a field object in \cite{gagne2015b}. Nevertheless, the association of this peculiar red L dwarf with a known K dwarf member of the Gliese nearby star sample should enable new insights into the physical cause(s) of the red L dwarf phenomenon. Moreover, at 0.1 pc, this would be among the most distant L or T dwarf companions yet identified to any star. Other very widely separated L and T dwarfs include the T6 dwarf 2MASS J16150413+1340079, a possible companion to the close G9+G9 double star 49 Ser (\citealt{raghavan2010}), and the L dwarf LP 678-45 B, a possible (though unlikely) companion to the M dwarf LP 678-45. Those two objects have separations from their primaries of $\sim$40,000 and $\sim$69,700 AU, respectively.

\end{itemize}

\subsection{Systems with White Dwarf Members}

We used the $H_J$ vs.\ $J-$W2 diagram (Figure~\ref{color_mag_HJ_JW2}) to search Table 11
for systems having a possible white dwarf component. Six such systems were found, and all are verified to have white dwarf members. Five of these systems are previously known: \#359 (Gl 283 AB; M6.5 + DAZ6), \#364 (Gl 288 AB; F9 + DC10), \#435 (LHS 2231 and WT 1759; K7 + DZ7), \#624 (Gl 1179 AB; M5 + DA10), and \#972 (WD 2226$-$754 AB; DC9 + DC12). The remaining system, \#920, is new and has already been discussed above.

\subsection{Systems with Large Magnitude Differences}

From Table 11,
 we also identify four systems having components differing in $J$-band by over 10.0 mag. These systems are discussed below:

\begin{itemize}

\item System \#128 is comprised of the A5 IV-V star 58 And and WISEA J020908.98+380849.8. The ${\Delta}J$ value is 10.27 mag. The Hipparchos parallax places 58 And at 58.7$\pm$0.9 pc (\citealt{vanleeuwen2007}). WISEA J020908.98+380849.8 has colors of ($J-K_s$, $J-$W2, W1$-$W2) = (0.87$\pm$0.07, 1.24$\pm$0.05, 0.21$\pm$0.04) mag, which imply a spectral type of M4-M6 and a distance of 86-181 pc. Unless the colors of the secondary are atypical for its class, these two objects are likely unassociated.

\item System \#364 is Gl 228 AB, with ${\Delta}J$ = 10.30 mag. This F9 dwarf + DC10 binary is mentioned in the white dwarf section above.

\item System \#663 is G 200-28 AB, a confirmed G5 + L5.5 pair listed in Table~\ref{published_cpm_systems}.

\item System \#1015 consists of the G8 III star 14 And and a new object from the NEOWISER survey, WISEA J233154.15+392429.1. The pair has ${\Delta}J$ = 11.60 mag. The giant star is located at a distance of 79.2$\pm$1.7 pc (\citealt{vanleeuwen2007}). The NEOWISER discovery has colors of ($J-K_s$, $J-$W2, W1$-$W2) = (0.81$\pm$0.07, 1.24$\pm$0.05, 0.22$\pm$0.04) mag, which suggests a spectral type of (very roughly) M3-M6 and a distance range of 75-317 pc. If verified, this system would have an apparent physical separation of 59,000 AU (0.3 pc). 

\end{itemize}

\subsection{Serendipitously Discovered Systems\label{serendipitous_systems}}

Several other candidate co-moving systems, not included in Table 11,
were serendipitously found while inspecting finder charts like those shown in Figure~\ref{finder_chart}. These are discussed below:

\begin{itemize}

\item WISE 0345$-$0348 AB is blended into a single source detection by AllWISE, which is why it does not appear as a possible co-moving pair in Table 11.
This is a double degenerate system for which both components are tentatively classified as ``DA?'' by us (Figure~\ref{seq_wds1}). The two objects are resolved in 2MASS All-Sky Point Source Catalog, where they have designations of 2MASS J03450171$-$0348444 and 2MASS J03450153$-$0348492. The northern component is the brighter of the two. The difference at $J$ band is 0.56$\pm$0.11 mag and the separation is 5$\farcs$4.

\item WISE 0632+2644 (LP 363-4) and 0632+2643 have ${\Delta}J$=5.73$\pm$0.05 mag and sep=25$\farcs$6. The brighter component does not appear in our compiled list of {\it WISE} motion sources. The \cite{lepine2005} motion measurment for the brighter component is $\mu_\alpha$=257 mas yr$^{-1}$ and $\mu_\delta$=25 mas yr$^{-1}$. This is similar to the values \cite{schneider2016} derive for the fainter component: $\mu_\alpha$=218.7$\pm$7.5 mas yr$^{-1}$ and $\mu_\delta$=35.9$\pm$7.7 mas yr$^{-1}$. We derive distances of 90$\pm$6 pc for WISE 0632+2644 and 53$\pm$4 pc for WISE 0632+2643. Invoking the fainter component as an equal-magnitude double moves its distance estimate to 75$\pm$5 pc. Given that the motions and distances are both discrepant by $>$3$\sigma$, we conclude that these two objects are unassociated despite their close on-sky separation.

\item WISE 1247$-$4343 (LTT 4892) and 1247$-$4344 (SCR J1247$-$4344B) have ${\Delta}J$ = 5.03$\pm$0.03 mag and sep = 49$\farcs$5. The brighter component does not appear in our compiled list of {\it WISE} motion sources, which is why no association was made. \cite{boyd2011} also discovered this pair serendipitously, labeled it as a possible co-moving system, and find it to be $\sim$100 pc distant. (Although their positions for the pair agree with ours, they report an erroneous separation and position angle.) Our spectral types of G0 and M5 suggest distances of 98$\pm$6 and 78$\pm$5 pc, respectively. These distances can be reconciled if the secondary is a nearly equal-magnitude double.

\item WISE 1423$-$1646 A (PPM 228725) and WISE 1423$-$1646 B (SCR J1423$-$1646B) have ${\Delta}J$ = 1.10$\pm$0.03 mag and sep = 10$\farcs$2. The brighter component does not appear in our compiled list of {\it WISE} motion sources, which is why no association was made. \cite{boyd2011} also discovered this pair serendipitously and label it as a possible co-moving system. They provide distance estimates for neither component. Our spectral types of K5 pec and M0.5 yield distance estimates of 50$\pm$3 and 54$\pm$3 pc for the primary and secondary, respectively. This appears to be a physical system with a separation of 530 AU, although the peculiarity noted for the primary spectrum (section~\ref{early_types}) remains unexplained.

\item WISE 1457+2341 AB is also known as LSPM J1457+2341 NS (\citealt{lepine2005}) and is the dwarf carbon star plus sdM8 system discussed in section~\ref{subdwarf_analysis}. The secondary (the sdM8) does not appear in our compiled list of {\it WISE} motion sources, which is why no association was made. The separation of the system is 3$\farcs$9 and ${\Delta}J$ = 1.66$\pm$0.10 mag. Both 2MASS and SDSS resolve the two components, but the time baseline between the survey data is only four years. Earlier $I$-band imaging from the DSS (epoch 1991), though showing the pair as a blend, shows the system offset from the 2MASS and DSS position and with a position angle between components that appears identical. Although harder to discern, the same appears to be true of the DSS $R$-band images from 1991 and 1950. So this appears to be a true co-moving pair, as the naming by \cite{lepine2005} indicates. Only a small number of late sdM objects have trigonometric parallax measurements in the literature. If we take the absolute $J$ mag implied by the parallax measurement of the sdM8 standard LSR J1425+7102 (\citealt{dahn2008}) to be a typical value for an sdM8, the we estimate a distance to WISE 1457+2341 B of $\sim$140 pc. This implies a separation between the AB pair of $\sim$550 AU.

\item WISE 1702+7158 AB is also known as LP 43-310 and LSPM J1702+7158N. Common-proper-motion was first noted by \cite{lepine2005}. The secondary is not detected by {\it WISE}. We classify the primary as M4.5 and the secondary as a cold white dwarf, although our spectrum of the latter has low signal-to-noise. The SDSS DR12 survey indicates an angular separation of 10$\farcs$7 and magnitude differences of ${\Delta}r$ = 1.24$\pm$0.02, ${\Delta}i$ = 2.71$\pm$0.02, and ${\Delta}z$ = 3.36$\pm$0.06 mag. The M4.5 classification of the primary suggests a distance of 151 pc, which means that the apparent physical separation is 1600 AU.

\item WISE 1722$-$6951 AB is a new candidate pair. A check of the finder chart shows that the objects appear to be co-moving. Our distance estimates for this pair (Table 10)
are wildly discrepant, however: 28$\pm$2 pc for the M3 primary and 58$\pm$4 pc for the M4 secondary. Assuming the primary is an equal-magnitude, unresolved binary results in a new distance estimate of 39$\pm$3 pc, which still falls short of resolving the discrepancy. Surprisingly, both objects fall in our list of {\it WISE} motion sources, but this pair failed to meet our tolerance thresholds for the creation of Table 11.
This is with good reason: although the motion values in Declination are identical to within 1$\sigma$, the values in Right Ascension are discrepant by 9$\sigma$. Hence, despite the near, on-sky alignment (37$\farcs$2), these sources are not associated, as both the distance estimates and motion measurements attest.

\item WISE 2304+2111 AB is another new candidate pair. Both components appear in our list of {\it WISE} motion sources, but the duo failed to meet our tolerance thresholds for the creation of Table 11
because the motions in RA and Dec are discrepant by 5$\sigma$ and 8$\sigma$, respectively. In this case, the AllWISE astrometry for the secondary may be to blame, as the pair is separated by only 15$\farcs$2. (More specifically, the astrometric uncertainties have been underestimated in AllWISE for the blended, secondary component). A check of finder charts going back to 1954 appears to show co-movement. Our Table 10
distance estimate for the M0.5 primary is 54$\pm$4 pc and that of the M4.5 secondary is 68$\pm$5 pc, which agree to within 3$\sigma$ without having to invoke binarity in the primary. 

\end{itemize}

\section{Conclusions}

The AllWISE1 and AllWISE2 Motion Surveys, despite having a time baseline of only six months over most of the sky, have uncovered 48,000 motion objects. Among the findings from our spectroscopic follow-up are new, cold white dwarfs (including the candidate in Fajardo-Acosta et al., in prep.); the ability to probe line-of-sight clumpiness in nearby molecular clouds by monitoring motion objects behind them; previously missed M and L dwarfs in the 20-pc census; the realization that the Na `D' line might be used as a spectroscopic hallmark of low metallicity in late-M subdwarfs; a carbon dwarf system whose common-proper-motion third member has a measured metallicity of [Fe/H]${\approx}-$1.0, thus placing a strong physical constraint on models of the carbon dwarf itself; and, when combined with all extant WISE-based motion surveys, the identification of over a thousand possible common-proper-motion systems.

Discoveries are important, but so is the ability to characterize known motion objects using archival data that previous motion surveys lacked. By combining the WISE photometric and astrometric data on these objects with data from 2MASS, SDSS, DSS1, and DSS2, we have been able to characterize each object and prioritize those needing further scrutiny. This characterization has led to the identification of interesting objects that have lain unnoticed and unobserved in prior catalogs. Examples are the bright carbon dwarfs LP 830-18 and NLTT 45912, the unusual and possibly very metal-poor M subdwarf LP 245-62, and the nearby (d$\approx$13 pc) M dwarf + white dwarf pair WT 765 and WT 766.

The AllWISE2 survey made these gains despite having motion sensitivity only to relatively bright sources. Its success gives a small taste for the advances that the Gaia mission will soon be making for the immediate solar vicinity and far beyond. As promising as Gaia is, it will not detect brown dwarfs over much of the known temperature range of these objects because the survey is taking place shortward of 1 $\mu$m. \cite{smart2014} predicts that Gaia will directly image 500 field L and T dwarfs to the $G$=20 mag limit, although warmer, cluster brown dwarfs and other brown dwarfs identified indirectly can be detected in greater numbers (Table 1 of \citealt{kirkpatrick2014b}). Nonetheless, Gaia's direct measurements will be sensitive to only the closest and brightest T dwarfs, none of the Y dwarfs, and almost none of the L dwarf exotica, including later-type L subdwarfs.

This portion of parameter space is, however, easily accessible to {\it WISE} now that data with longer time baselines are available. The accrued time baseline between the classic {\it WISE} mission and the continuing {\it NEOWISE} mission is currently envisioned as spanning at least six years. As the NEOWISER motion survey of \cite{schneider2016} has already shown, the expansion of the time baseline enables smaller motion objects at fainter magnitudes to be uncovered. However, the \cite{schneider2016} effort was limited to detections on single frames, since the {\it NEOWISE} mission is not producing any coadds of the individual exposures. Maximizing the potential of {\it WISE} data for proper motion detection involves building coadds, digging deep for detections, and producing searchable source lists of possible motion candidates. This data processing campaign is one that we and much of the astronomical community (see \citealt{faherty2015}) would like to see happen soon since the most interesting discoveries would be prime targets for the {\it James Webb Space Telescope}.

\vskip 12pt

Facilities: \facility{WISE}, \facility{Hale(Double Spectrograph)}, \facility{Keck:I(LRIS)}, \facility{Keck:II(DEIMOS, NIRSPEC)}, 
\facility{IRTF(SpeX)}, \facility{Magellan:Baade(FIRE)}.

\vskip 12pt

This publication makes use of data products from WISE, which is a joint project of the University of California, Los Angeles, and the Jet Propulsion Laboratory (JPL)/California 
Institute of Technology (Caltech), funded by the National Aeronautics and Space Administration (NASA). This research has made use of the NASA/IPAC Infrared Science Archive,
which is operated by JPL/Caltech, under contract with NASA. We are indebted to the SIMBAD database and the VizieR catalog access tool, provided by CDS, Strasbourg, France. This paper makes use of data from the Catalina Sky Survey, which is funded by the National Aeronautics and Space Administration under Grant No.\ NNG05GF22G issued through the Science Mission Directorate Near-Earth Objects Observations Program. JDK acknowledges fruitful discussions with Richard Gray, Lee Rottler, and Patrick Lowrance. This research has benefitted from the M, L, T, and Y dwarf compendium housed at DwarfArchives.org. We thank S\'ebastien L\'epine for providing published spectra of his subdwarf standards.

\clearpage

\end{document}